\documentclass[11pt]{article}

\usepackage{aas_macros, amsmath, amssymb, comment, cite, esint, graphicx, mathtools}
\usepackage[margin=.8in,letterpaper]{geometry}
\usepackage[colorlinks=true]{hyperref}
\usepackage[affil-it]{authblk}
\usepackage{subcaption}
\usepackage[utf8]{inputenc}
\usepackage{mathrsfs}
\usepackage{appendix}
\usepackage{float}
\usepackage{color}
\usepackage{hyperref}
\hypersetup{pageanchor=false}
\usepackage{indentfirst}
\usepackage{url}
\usepackage{caption}
\usepackage[numbers, square, comma, sort&compress, merge]{natbib}
\usepackage{overpic}
\usepackage{epsf, bbold, amsfonts, stmaryrd}
\usepackage{textcomp}
\usepackage{ulem}
\usepackage{tikz}

\numberwithin{equation}{section}
\let\originalleft\left
\let\originalright\right
\renewcommand{\left}{\mathopen{}\mathclose\bgroup\originalleft}
\renewcommand{\right}{\aftergroup\egroup\originalright}
\mathcode`\*="8000
{\catcode`\*=\active\gdef*{\mathclose{}\,\mathopen{}}}

\newcommand{\bea}{\setlength\arraycolsep{2pt} \begin{eqnarray}}
\newcommand{\eea}{\end{eqnarray}}

\def\bag{\begin{aligned}}
\def\eag{\end{aligned}}
\def\bea{\begin{eqnarray}}
\def\eea{\end{eqnarray}}
\def\ba{\begin{array}}
\def\ea{\end{array}}

\def\bc{\begin{center}}
\def\ec{\end{center}}

\begin{document}
\title{Novel extended inner shadow in images of Johannsen-Psaltis black holes with thin accretion disks}
	
\author{Xinyu Wang$^{1, 2}$, Xiaobao Wang$^{2, 3}$, Xin-Yang Wang$^{2, 4\ast}$, Minyong Guo$^{1, 2}$}
\date{}
	
\maketitle
\vspace{-10mm}

\begin{center}
{\it
$^1$ School of Physics and Astronomy, Beijing Normal University, Beijing 100875, China\\\vspace{4mm}

$^2$ Key Laboratory of Multiscale Spin Physics (Beijing Normal University), Ministry of Education, Beijing 100875, China\\\vspace{4mm}

$^3$ School of Applied Science, Beijing Information Science and Technology University, Beijing 100192, China\\\vspace{4mm}

$^4$ Faculty of Arts and Sciences, Beijing Normal University, Zhuhai 519087, China\\\vspace{4mm}
}
\end{center}

\vspace{8mm}

\begin{abstract}
    The Johannsen-Psaltis (JP) metric is constructed by introducing deviation parameters into the Kerr metric. The presence of these deviation parameters provides additional degrees of freedom for testing the extent to which astrophysical black holes are consistent with the Kerr paradigm and for examining the validity of the no-hair theorem. Since black hole imaging provides one of the most direct means of probing the properties of astrophysical black holes, a detailed investigation of the imaging characteristics of JP black holes constitutes a natural first step toward employing the JP metric to explore potential deviations from the Kerr geometry in astrophysical black hole systems. Accordingly, numerical backward ray-tracing simulations are employed to conduct a comprehensive investigation of the image structures, intensity distributions, and redshift-blueshift signatures of JP black holes with both closed and non-closed event horizons. The dependence of these imaging characteristics on the deviation parameter is also investigated systematically. Furthermore, the emergence of an extended inner shadow, distinct from the original inner shadow of the black hole, in images of JP black holes with non-closed event horizons motivated a detailed investigation of its physical formation mechanism using both numerical backward ray-tracing simulations and an approximate analytical framework.
\end{abstract}

\vfill{\footnotesize \flushleft $\ast$ Corresponding author: xinyangwang@bnu.edu.cn
}

\maketitle

\newpage
\baselineskip 18pt
\section{Introduction}\label{sec1}

Black holes represent a distinctive class of spacetime structures that arise as solutions to the gravitational field equations of general relativity (GR) proposed by Einstein. A characteristic null hypersurface, referred to as the event horizon, serves as the effective boundary of the black hole in spacetime. At the center of black holes, a spacetime singularity inevitably forms, where curvature invariants and physical quantities diverge. The divergent nature of these quantities undermines the global well-posedness of the spacetime manifold and disrupts its inherent causal structure. Consequently, based on the aforementioned considerations, the singularities at the center of black holes cannot manifest as exposed entities within the spacetime manifold. When the singularity is enclosed by the event horizon of black holes, the null nature of the horizon causally isolates the singular region from external observers, thereby preserving the coherence of the spacetime structure and maintaining the integrity of its causal order. It leads to the reasonable hypothesis that the singularity of black holes should remain permanently concealed within the event horizon and never be exposed to the external universe. This hypothesis is formalized in the Weak Cosmic Censorship Conjecture (WCCC) proposed by Penrose \cite{Penrose:1969pc}, which serves as a cornerstone in the theoretical study of black hole physics. According to the WCCC, external observers can access only information associated with the event horizon and the spacetime region outside it, while all information contained within the event horizon remains fundamentally inaccessible. Furthermore, the uniqueness theorems of classical GR establish that the four-dimensional and asymptotically flat solutions of the Einstein field equations in pure vacuum or electromagnetic vacuum spacetime are uniquely described by the Kerr-Newman (KN) spacetime family of solutions \cite{Newman:1965my}. Black holes corresponding to the KN spacetime are commonly referred to as KN black holes, which are completely characterized on the event horizon by three parameters: the mass $M$, electric charge $Q$, and angular momentum $J$. In other words, all physical information about KN black holes is encoded solely in these three parameters, while no additional independent properties are accessible to external observers. This fundamental property of KN black holes in classical GR is encapsulated in the so-called ``no-hair theorem'' \cite{Israel:1967wq, Carter:1971zc, Robinson:1975bv}.

The gravitational properties in the vicinity of compact objects such as stars and black holes in vacuum spacetimes have been extensively investigated within the framework of classical GR. In realistic astrophysical environments, however, these objects are typically embedded in surrounding matter distributions that interact with the gravitational field generated by the compact object itself. With continuous advances in observational techniques and increasing precision in measurements of the spacetime geometry and matter distributions surrounding compact astrophysical bodies, GR has been remarkably successful in describing gravitational phenomena across a wide range of scales. Nevertheless, it is widely expected that the theory may require extensions in regimes involving extremely strong gravitational fields or quantum effects in order to accurately characterize the spacetime structure near compact objects and to fundamentally describe the interactions between gravitational and matter fields in these extreme environments. Therefore, to deepen our understanding of the fundamental nature of gravity and its interaction with matter in the strong-field regime and in the presence of quantum effects, it is necessary to extend classical GR toward a broader class of gravitational theoretical frameworks. In the absence of a complete and consistent theory of quantum gravity, investigations of gravitational self-interactions and interactions between gravity and matter fields at the quantum scale must rely on appropriate approximation schemes. At present, the low-energy effective field theory (EFT) approach to quantum gravity provides the most reliable approximate framework for exploring quantum gravitational effects. In particular, the low-energy EFT description of gravity provides a systematic framework for incorporating higher-order curvature terms and additional interaction terms describing couplings between gravity and matter fields into the classical gravitational action \cite{Capozziello:2011et}. These additional contributions induced by the low-energy EFT of quantum gravity are commonly referred to as quantum correction terms. The inclusion of these quantum corrections leads to a modified gravitational Lagrangian and gives rise to a broad class of extended or modified theories of gravity. To date, the EFT framework incorporating quantum corrections has been widely employed to formulate a variety of modified gravitational theories \cite{Stelle:1976gc, Kanti:1995vq, Sotiriou:2008rp, DeFelice:2010pv}. Within these theoretical frameworks, numerous black hole solutions have been obtained that exhibit properties distinct from those predicted by GR. In these modified theories of gravity, the resulting black hole spacetimes may contain additional parameters beyond the three fundamental parameters that characterize the KN family of solutions. Black holes characterized by these additional parameters are commonly referred to as hairy black holes. Since these solutions generally lie outside the KN family, hairy black holes typically violate one or more assumptions underlying the classical no-hair theorem. Consequently, the classical no-hair theorem does not directly apply to hairy black holes.

Black holes were initially identified as special spacetime structures derived from theoretical solutions of the gravitational field equations within classical GR or various modified gravitational theories. Nevertheless, whether black holes belonging to the KN family in GR or hairy black holes arising in modified gravity theories genuinely exist in the universe remains a subject of considerable debate and an unresolved issue in contemporary research. Since the advent of GR, substantial effort has been devoted to seeking evidence across a wide range of astrophysical environments and phenomena to directly or indirectly confirm the existence of black holes in the universe. At present, however, the direct detection of information from the interior of black hole remains unattainable. This limitation arises from the null nature of the event horizon and the one-way membrane property of black holes interior, which together prevent any internal information from escaping to external observers. Nonetheless, advances in observational techniques and improvements in experimental precision have progressively enabled astrophysical methods capable of probing both the existence of black holes and the structure of spacetime in their vicinity.

The gravitational waves detected by LIGO and Virgo from binary black hole mergers provide highly persuasive evidence supporting GR, while simultaneously imposing stringent constraints on modified gravity theories \cite{LIGOScientific:2016aoc, LIGOScientific:2016emj, LIGOScientific:2017vwq}. Accordingly, gravitational wave observations enable, to some extent, the systematic identification of modified gravity models that remain consistent with experimental observations. These observationally constrained theories may therefore serve as more reliable frameworks for studying astrophysical phenomena in the real universe. Furthermore, the spacetime geometry in the vicinity of black hole event horizons can be probed with high precision through very long baseline interferometry (VLBI) observations of black hole shadows. In particular, the Event Horizon Telescope (EHT), which employs a global network of radio telescopes operating in concert via VLBI techniques, has successfully obtained the first images of black holes \cite{EventHorizonTelescope:2019pgp, EventHorizonTelescope:2019dse, EventHorizonTelescope:2019uob, EventHorizonTelescope:2019jan, EventHorizonTelescope:2019ths, EventHorizonTelescope:2019ggy, EventHorizonTelescope:2021bee}. Consequently, the detection of gravitational waves from binary mergers, together with the imaging of black holes through electromagnetic radiation emitted by surrounding hot plasma, provides indirect but compelling evidence for the existence of black holes in the universe. To distinguish observational evidence for black hole existence obtained from gravitational wave detections by LIGO and Virgo and from black hole imaging by the EHT from theoretical black hole models, the compact objects observed in astrophysical environments are commonly referred to as astrophysical black holes. The release of black hole images by the EHT has significantly renewed interest in the imaging properties of black hole solutions in a variety of gravitational theories. Consequently, extensive studies have been conducted on the shadows and accretion disk images of black holes in modified gravity theories as well as in frameworks incorporating physics beyond GR \cite{Hu:2020usx, Guo:2020zmf, Zhong:2021mty, Hou:2021okc, Hou:2022eev, Zhang:2022osx, Wang:2023nwd, Zhang:2023cuw, Zhang:2024lsf, Wang:2024uda, Yang:2024nin, He:2024amh, He:2025rjq}.

Although observations across both the gravitational and electromagnetic spectra have provided compelling evidence for the existence of astrophysical black holes, it remains uncertain whether these objects are fully characterized by the KN family of solutions in classical GR or are more accurately described by black hole models arising from modified gravitational theories. In other words, it has not yet been definitively established whether astrophysical black holes can be comprehensively explained within the framework of classical GR, or whether an accurate description requires the extended formulations provided by modified gravity theories. Moreover, although four-dimensional KN black holes strictly comply with the classical no-hair theorem as previously outlined, hairy black holes arising within modified gravity frameworks generally violate the fundamental assumptions on which the theorem is based \cite{Johannsen:2010xs, Johannsen:2010ru, Johannsen:2010bi}. This implies that the classical no-hair theorem ceases to apply to these hairy black holes. Therefore, evaluating the correspondence between astrophysical black holes and theoretical black hole models, whether the KN family of solutions or hairy black hole solutions, provides a crucial avenue for assessing the validity of the no-hair theorem in astrophysical settings. With advances in black hole imaging techniques, particularly the successful deployment of the EHT, the achievable resolution of images depicting the spacetime region surrounding black holes has greatly improved. These observations provide valuable information for inferring astrophysical black hole parameters and impose constraints on both modified gravity theories and the parameter space of the corresponding hairy black hole solutions. Measurements of black hole images indicate that astrophysical black holes are generally massive and rotating, and can therefore be characterized primarily by two parameters, i.e., the mass $M$ and angular momentum $J$, consistent with the Kerr solution of Einstein field equations in classical GR \cite{Bambi:2011mj, Cardoso:2019rvt}. Consequently, astrophysical black holes are frequently modeled as Kerr black holes. 

The Kerr solution is derived under the assumption that the spacetime exterior to a black hole is governed by the vacuum Einstein field equations. In realistic astrophysical environments, however, compact objects are typically embedded in surrounding matter configurations such as accretion flows, stellar companions, or dark matter halos. Although these environmental effects may introduce perturbative corrections to the spacetime geometry in the vicinity of the compact object, the Kerr solution nevertheless provides an excellent approximation for an isolated black hole within the framework of GR \cite{Bambi:2011mj, Cardoso:2019rvt}. To assess whether astrophysical black holes are fundamentally described by the Kerr solution or instead require more general geometries that permit intrinsic deviations from Kerr solution, potentially arising from modifications of the underlying gravitational theory, one commonly introduces a set of dimensionless phenomenological parameters into the Kerr metric \cite{Johannsen:2010xs, Bambi:2011mj}. These parameters, referred to as deviation parameters, systematically quantify potential departures of astrophysical black holes from the Kerr solution. In the limiting case where all deviation parameters vanish, the corresponding spacetime metric smoothly reduces to the Kerr metric. Metrics constructed by introducing deviation parameters into the Kerr geometry do not, in general, represent exact solutions of the vacuum Einstein field equations. Instead, these modified Kerr metrics are interpreted as effective metric frameworks designed to probe possible intrinsic deviations from the Kerr spacetime. Despite not being exact vacuum solutions, these effective metrics retain many of the fundamental structural and geometric properties of the Kerr solution and are therefore commonly referred to as Kerr-like metrics. Several Kerr-like metrics incorporating deviation parameters have been proposed to explore potential observational signatures of non-Kerr black holes \cite{Collins:2004ex, Vigeland:2009pr, Glampedakis:2005cf, Mankot1992GeneralizationsOT, Vigeland:2011ji}. If significant deviations from the Kerr geometry are detected observationally, this could indicate that the observed compact object is not described by the classical Kerr solution and might instead correspond to a horizonless compact configuration or another class of exotic compact object. Conversely, if the object is confirmed to possess an event horizon while still exhibiting deviations from the Kerr geometry, this would imply a violation of the fundamental assumptions underlying the classical no-hair theorem, thereby limiting its applicability to astrophysical black holes. Such a scenario would further suggest that either the assumptions of the no-hair theorem break down in realistic settings or that GR itself may require modification in the strong-field regime.

Since black holes described by Kerr-like metrics with deviation parameters can violate the fundamental assumption of the classical no-hair theorem, spacetime pathologies including singularities and closed timelike curves typically arise outside event horizons of these black holes \cite{Johannsen:2013szh, Johannsen:2013rqa}. In currently proposed Kerr-like spacetimes, these pathologies generally occur in regions extremely close to the event horizon of the corresponding Kerr solution. As a result, the presence of these spacetime defects imposes significant constraints on studies of the electromagnetic spectra emitted in the vicinity of Kerr-like black holes, including radiation from accretion disks and inner accretion flows. Moreover, the existence of these spacetime pathologies outside the event horizon of black holes poses substantial challenges for testing the no-hair theorem of astrophysical black holes using electromagnetic spectral observations. The extent to which these spacetime pathologies hinder tests of the no-hair theorem for Kerr-like black holes based on electromagnetic observations depends primarily on the magnitude of the spin parameter of black holes. For moderately rotating black holes, singularities and closed timelike curves typically arise in the region between the innermost stable circular orbit (ISCO) or the unstable circular photon orbit and the event horizon of black holes, with the spacetime pathological region lying closer to the event horizon than to the ISCO or circular photon orbit. In this situation, the influence of these spacetime pathologies on electromagnetic spectral analyses can be mitigated by introducing an artificial truncation with inflow boundary conditions between the pathological region and the ISCO or circular photon orbit. By contrast, for rapidly rotating black holes, the artificial truncation cannot be introduced between the ISCO and the event horizon, as the ISCO lies extremely close proximity to the event horizon in this regime. Consequently, examining the no-hair theorem using electromagnetic spectra becomes infeasible for rapidly rotating configurations. Therefore, to rigorously test the no-hair theorem for Kerr-like black holes, whose spacetime metrics include deviation parameters characterizing departures from the original Kerr geometry, using electromagnetic spectral signatures associated with the photon ring and the ISCO, it is essential to construct an asymptotically physical metric that incorporates one or more deviation parameters while simultaneously ensuring that no singularities or closed timelike curves arise outside the event horizon for any rotation configuration of black holes.

To address the limitations imposed by spacetime pathologies, such as singularities and closed timelike curves, arising outside the event horizon of Kerr-like black holes and thereby obstructing tests of the no-hair theorem for astrophysical black holes, Johannsen and Psaltis \cite{Johannsen:2011dh} introduced a Kerr-like metric, commonly referred to as the Johannsen-Psaltis (JP) metric. This metric is specifically constructed to eliminate the spacetime defects that occur outside the event horizon in previously proposed Kerr-like geometries. Moreover, the JP metric is formulated without invoking additional matter fields and incorporates a set of dimensionless deviation parameters that characterize departures from the Kerr geometry in the strong gravitational field regime. When all deviation parameters vanish, the JP metric naturally reduces to the original Kerr solution. It is important to emphasize, however, that the JP metric is still not a direct solution of the Einstein field equations but instead represents a perturbative modification of the Kerr spacetime. Accordingly, the JP metric provides an effective theoretical framework for modeling realistic astrophysical environments in which matter fields surrounding a black hole may induce deviations in the spacetime structure near the black hole from the standard Kerr geometry. Consequently, the JP metric can be regarded as a practical description of astrophysical spacetimes, and JP black holes constructed by the JP metric may be considered as asymptotical black holes in the realistic universe. Furthermore, the deviated behavior of JP metric from the Kerr solution encoded in these deviation parameters provides a valuable means for testing the validity of the no-hair theorem in the context of astrophysical black holes. Notably, in the spacetime constructed by the JP metric, no singularities or closed timelike curves arise outside the event horizon for any spin configuration of JP black hole. Therefore, the no-hair theorem for astrophysical black holes may be probed by investigating how electromagnetic phenomena associated with circular photon orbits or the ISCO around a JP black hole respond to variations in the deviation parameters of the JP metric. This necessitates first obtaining images of the circle photon orbit or the distribution of accreting matter in the vicinity of the ISCO surrounding the JP black hole, namely, the JP black hole shadow or the image of a JP black hole encompassed by an accretion disk. Since the JP metric retains the same nonvanishing components as the expression of Kerr metric, analyses of JP black hole shadows and accretion flow images can be conducted directly using established geodesic approaches, with the corresponding images obtained through analogous ray-tracing techniques. 

However, in contrast to the Kerr case, the image of a JP black hole cannot be obtained through a fully analytic approach. The JP metric belongs to Petrov type-I in the Petrov classification, rather than type-D as in the Kerr geometry. This classification implies that a Carter constant cannot be defined in the spacetime described by JP metric, meaning that the Hamilton-Jacobi equation governing photon motion in spacetime cannot be solved via separation of variables. As a result, an analytic deviation of photon trajectories is not feasible, and the numerical computation method based on of backward ray-tracing techniques are essential for accurately computing either the JP black hole shadow or the image of a JP black hole surrounded by an accretion disk. The shadow of the JP black hole has been extensively investigated in previous studies \cite{Wang:2025ihg}. Because the spacetime structure of a JP black hole depends on the values of deviation parameter $\epsilon_3$, the spacetime configuration of the JP black hole may correspond either to a Kerr-like black hole with closed event horizon or to a black hole characterized by a dumbbell-shaped and non-closed event horizon accompanied by a naked singularity. Analyses of the shadows of JP black hole associated with these two possible configurations of the event horizon show that, in the case of a Kerr-like black hole with closed event horizon, the shadow can be obtained using a semi-analytic approach. By contrast, when the configuration of event horizon is non-closed and involves a naked singularity, the JP black hole shadow can be determined solely through numerical computations employing backward ray-tracing techniques. However, the imaging properties of accretion disks surrounding JP black holes have not yet been systematically or comprehensively examined. To more thoroughly assess the applicability of the no-hair theorem to astrophysical black holes, it is therefore essential to examine the imaging characteristics of JP black holes encompassed by accretion flows. In the present work, we focus our analysis specifically on the imaging features of JP black holes surrounded by thin accretion disks.

The structure of this paper is as follows. In Sec. \ref{jpmetricjpbh}, we briefly review the construction of the JP metric, together with the fundamental properties of the associated spacetime and JP black hole solutions. In Sec. \ref{investigationjpshadow}, we systematically investigate the imaging properties of JP black holes surrounded by thin accretion disks and possessing both closed and non-closed event horizons, using a numerical backward ray-tracing method. Finally, the main conclusions of this study are presented in Sec. \ref{conlcusions}.

In this study, the fundamental constants $c$ and $G$ are set to unity for simplicity, and the spacetime metric signature is adopted as $(-,+,+,+)$.

\section{The JP metric and JP black holes}\label{jpmetricjpbh}

The formulation of the JP metric provides a theoretical framework for examining the compatibility between astrophysical black holes and the no-hair theorem. As a Kerr-like extension of the Kerr geometry, the JP metric introduces a set of deviation parameters that quantify departures from the Kerr solution. These deviation parameters provide an effective phenomenological framework for characterizing potential observational deviations between realistic astrophysical black holes and the Kerr solution of classical GR, as well as for describing spacetime perturbations induced by surrounding matter fields in astrophysical environments. Because the JP metric is constructed as a deformation of the Kerr spacetime through the introduction of these deviation parameters, it preserves many of the essential structural properties of the Kerr geometry. In particular, the spacetime described by the JP metric remains stationary, axisymmetric, and asymptotically flat. Moreover, for appropriate ranges of the deviation parameters, the JP metric admits a well-defined event horizon, indicating that black hole solutions, commonly referred to as JP black holes, can be consistently realized within this spacetime. In contrast to black holes arising from other Kerr-like metrics, JP black holes can be constructed to avoid spacetime pathologies such as singularities and closed timelike curves outside the event horizon. The absence of these exterior pathologies removes potential ambiguities that typically complicate the analysis of circular photon orbits and the electromagnetic spectra emitted by accreting matter near the ISCO. Furthermore, because the JP metric is not derived as an exact solution of an underlying gravitational theory, its explicit form is not required to satisfy the vacuum Einstein field equations, and the geodesic motion of particles in JP spacetime is not guaranteed to be fully integrable. These comparatively relaxed requirements have important implications for tests of the no-hair theorem based on gravitational wave observations. By contrast, their impact on electromagnetic tests of the no-hair theorem is considerably more limited, since such analyses depend only on the geodesic motion of timelike or null particles in the background spacetime and do not require knowledge of the underlying gravitational field equations. Accordingly, when employing electromagnetic observations to test the no-hair theorem, it is sufficient to assume the validity of the Einstein equivalence principle, which governs particle motion in curved spacetime. Therefore, analyzing the properties of circular photon orbits and the dependence of electromagnetic radiation emitted by accreting matter on the deviation parameters of the JP metric not only provides observational signatures for probing potential deviations of astrophysical black holes from the Kerr solution, but also offers a means to assess whether astrophysical black holes that deviate from the Kerr geometry continue to satisfy the no-hair theorem. To investigate the spacetime properties in the vicinity of astrophysical black holes and to test the no-hair theorem using photon orbits or electromagnetic emission from accreting matter, it is therefore necessary to compute either the shadow of a JP black hole or the image of a JP black hole illuminated by its accretion flow. Before constructing these images, we first provide a brief review of the formulation of the JP metric and the essential properties of JP black holes.

In Newtonian gravity, the gravitational potential generated by a compact object at large distances can be approximated by that of a spherically symmetric mass distribution with the same total mass. An analogous perspective applies within the framework of GR. For stationary and asymptotically flat spacetimes, the asymptotic structure at large radial distances can be effectively described by a Schwarzschild-like metric. In an appropriate gauge, this Schwarzschild-like metric, expressed in Schwarzschild coordinates $\left(t\,, r\,, \theta\,, \varphi \right)$, can be written as
\begin{equation}\label{schwarzschildlikemetric}
    \begin{split}
        \text{d} s^2 = - f (r) \left[1 + h (r) \right] \text{d} t^2 + f^{-1} (r) \left[1 + h (r) \right] \text{d} r^2 + r^2 \left(\text{d} \theta^2 + \sin^2 \theta \text{d} \varphi^2 \right)\,,
    \end{split}
\end{equation}
where the parameter $M$ denotes the mass of the compact object, and the metric function $f(r)$ is given by $f (r) = 1 - 2 M / r$. The function form of $h (r)$ is chosen as
\begin{equation}\label{expressionhr}
    \begin{split}
        h (r) \equiv \sum_{k = 0}^{\infty} \epsilon_k \left(\frac{M}{r} \right)^k\,.
    \end{split}
\end{equation}
This metric smoothly reduces to the Schwarzschild solution when the function $h (r)$ in Eq. (\ref{schwarzschildlikemetric}) vanishes. Since the Schwarzschild metric can be extended to the axisymmetric Kerr geometry via the Newman-Janis (NJ) algorithm, the Schwarzschild-like metric can be analogously generalized to the axisymmetric case by applying the same procedure, thereby yielding a Kerr-like metric. By employing the NJ algorithm in conjunction with the Newman-Penrose (NP) formalism, the quasi-Schwarzschild metric given in Eq. (\ref{schwarzschildlikemetric}) can be formally extended to construct the JP metric. The explicit form of the JP metric in Boyer-Lindquist (BL) coordinates is given by
\begin{equation}\label{expressionjpmetric}
    \begin{split}
        \text{d} s^2 = & - H \left(r\,, \theta \right) \left(1 - \frac{2 M r}{\Sigma}\right) \text{d} t^2 - \frac{4 a M r \sin^2 \theta}{\Sigma} H \left(r\,, \theta \right) \text{d} t \text{d} \phi + \frac{\Sigma H \left(r\,, \theta \right)}{\Delta + h \left(r\,, \theta \right) a^2 \sin^2 \theta} \text{d} r^2 + \Sigma \text{d} \theta^2\\
        & + \sin^2 \theta \left[\Sigma + H \left(r\,, \theta \right) \frac{a^2 \left(\Sigma + 2 M r \right) \sin^2 \theta}{\Sigma} \right] \text{d} \varphi^2\,,
    \end{split}
\end{equation}
where
\begin{equation}\label{expressiondeltasigma}
    \begin{split}
        \Delta = r^2 - 2 M r + a^2\,, \qquad \Sigma = r^2 + a^2 \cos^2 \theta\,.
    \end{split}
\end{equation}
The parameter $M$ denotes the mass of the spacetime, while $a$, defined as $a = J / M$, with $J$ denoting the angular momentum of the spacetime, specifies the angular momentum per unit mass. The function $H \left(r\,, \theta\right) = 1 + h \left(r\,, \theta\right)$, where the explicit form of the function $h \left(r\,, \theta\right)$ appearing in $H \left(r\,, \theta\right)$ is given by 
\begin{equation}\label{expressionhrtheta}
    \begin{split}
        h \left(r\,, \theta\right) \equiv \sum_{k = 0}^{\infty} \left(\epsilon_{2 k} + \epsilon_{2 k + 1} \frac{M r}{\Sigma} \right) \left(\frac{M^2}{\Sigma} \right)\,.
    \end{split}
\end{equation}
According to the explicit form of the function $h \left(r\,, \theta \right)$ given in Eq. (\ref{expressionhrtheta}), all deviation parameters $\epsilon_k$ are encoded in the function $h \left(r\,, \theta \right)$, and these parameters quantify the extent to which the JP metric deviates from the Kerr spacetime. In this sense, the function $h \left(r\,, \theta \right)$ fully characterizes the deviation between the two geometries. Owing to the nontrivial dependence of the JP metric on $h \left(r\,, \theta \right)$, the deviation parameters entering this function directly influence the structure of circular photon orbits and the ISCO in the vicinity of JP black holes, and these parameters also affect the electromagnetic signals emitted from these orbits as observed by distant observers. Accordingly, the role of the function $h \left(r\,, \theta \right)$ in the JP metric provides an essential mechanism for probing the consistency of astrophysical black holes with the no-hair theorem through electromagnetic observations. Furthermore, observational constraints may impose restrictions on the functional form of $h \left(r\,, \theta\right)$, thereby allowing it to be further simplified.

As previously noted, within the framework of GR, a stationary and axisymmetric spacetime can be approximated in the asymptotic region by a spherically symmetric Schwarzschild-like metric in the large-radius limit for the purpose of analyzing its asymptotic properties. For more general spacetimes, even when the metric does not correspond to an exact solution of the Einstein field equations, the spacetime geometry described by such a metric is still expected to exhibit the same asymptotic behavior in this large-radius regime. Consequently, in the asymptotic limit $r \gg M$ and $r \gg a$, the JP metric can be approximated by the following asymptotic form
\begin{equation}\label{asymptoticaljpmetric}
    \begin{split}
        \text{d} s^2 \approx & - \left[1 - \frac{2 M}{r} + h (r) \right] \text{d} t^2 + \left[1 - \frac{2 M}{r} + h (r) \right] \text{d} r^2 - \frac{4 a \left[1 + h (r) \right]}{r} \sin^2 \theta \text{d} t \text{d} \varphi\\
        & + r^2 \left(\text{d} \theta^2 + \sin^2 \theta \text{d} \varphi^2 \right)\,,
    \end{split}
\end{equation}
where the function $h (r)$ retains the form provided in Eq. (\ref{expressionhr}). In the asymptotic expression of the JP metric shown in Eq. (\ref{asymptoticaljpmetric}), the coefficients of the $t-$ and $r-$ components contain contributions of order $1 / r^{n}$ with $n = 1\,, 2$, in addition to the deviation encoded in the function $h (r)$. Consequently, based on the structure of the function $h (r)$ as defined in Eq. (\ref{expressionhr}), the first two terms in the expression of the function $h (r)$ do not contribute to the asymptotic form of the JP metric. Therefore, in the asymptotic limit described by Eq. (\ref{asymptoticaljpmetric}), the function $h (r)$ contains only the deviation parameters $\epsilon_k$ with $k \geq 2$, while the first two deviation parameters $\epsilon_0$ and $\epsilon_1$ in Eq. (\ref{expressionhr}) should satisfy $\epsilon_0 = \epsilon_1 = 0$. Within the framework of the parameterized post-Newtonian (PPN) approximation, observational constraints on deviations from GR in the weak-field regime impose bounds on the parameter $\epsilon_{2}$ appearing in the function $h (r)$. In this context, $\epsilon_{2}$ can be expressed as
\begin{equation}\label{epsilon2beta}
    \epsilon_2 = 2 \left(\beta - 1 \right)\,,
\end{equation}
where $\beta$ is one of the parameterized post-Newtonian (PPN) parameters. Lunar Laser Ranging (LLR) experiments impose stringent constraints on its value, leading to the following allowed range for $\beta$ as
\begin{equation}\label{betamines1}
    \begin{split}
        \left|\beta - 1 \right| \leq 2.3 \times 10^{-4}\,.
    \end{split}
\end{equation}
After combining Eqs. (\ref{epsilon2beta}) and (\ref{betamines1}), the deviation parameter $\epsilon_{2}$ is constrained to lie within the interval as
\begin{equation}\label{internalepsilon2}
    \begin{split}
        \left|\epsilon_2 \right| \leq 4.6 \times 10^{-4}\,.
    \end{split}
\end{equation}
Based on the constraints on the deviation parameter $\epsilon_2$ inferred from the LLR experiments, the allowed range of $\epsilon_2$ is given in Eq. (\ref{internalepsilon2}). This result indicates that $\epsilon_2$ is restricted to an extremely small interval, implying that its contribution to the metric can be safely neglected. Therefore, it is reasonable to set $\epsilon_2 = 0$ in the subsequent analysis. Furthermore, since $\epsilon_3$ is not constrained by current experimental tests of GR, only $\epsilon_3$ is retained in the expression for $h \left(r\,, \theta \right)$, while higher-order deviation parameters $\epsilon_k$ with $k > 3$ are set to zero. Consequently, the final form of the function $h \left(r\,, \theta \right)$ in Eq. (\ref{expressionhrtheta}) can be written as
\begin{equation}\label{finalexpressionhrtheta}
    \begin{split}
        h \left(r\,, \theta \right) = \epsilon_3 \frac{M^3 r}{\Sigma^2}\,.
    \end{split}
\end{equation}

From the JP metric given in Eq. (\ref{expressionjpmetric}), together with the function $h \left(r\,, \theta \right)$ defined in Eq. (\ref{finalexpressionhrtheta}), the location of the event horizon can be determined by solving the following equation as 
\begin{equation}
    \begin{split}
        g_{t \varphi}^2 - g_{t t} g_{\varphi \varphi} = 0\,,
    \end{split}
\end{equation}
and the corresponding radial location of the event horizon is denoted by $r_h$. The presence of an event horizon indicates that the spacetime described by the JP metric represents a black hole, and such solutions are conventionally referred to as JP black holes. The JP metric is constructed as a deformation of the Kerr spacetime, with the dominant deviation encoded in the parameter $\epsilon_3$. The magnitude of $\epsilon_3$ determines the extent to which the JP geometry departs from the Kerr metric, and the JP metric smoothly reduces to the Kerr solution in the limit $\epsilon_3 = 0$. The nontrivial dependence of the JP metric on $\epsilon_3$ leads to qualitative differences in the properties of its event horizon compared with those of a Kerr black hole. In Kerr spacetime, the event horizon of a black hole is determined solely by the mass $M$ and the specific angular momentum $a$. The WCCC requires that the condition $a \leq M$ be satisfied in order to ensure the existence of an event horizon, thereby preserving causal consistency in spacetime. If this condition is violated, i.e., if $a > M$, the event horizon ceases to exist, exposing the central singularity and leading to a breakdown of the causal structure. For JP black holes, however, the dependence of the event horizon on the deviation parameter $\epsilon_3$ is considerably more intricate, as this parameter simultaneously determines both its radial location and geometric configuration. Variations in the position and geometry of the event horizon induced by the deviation parameter $\epsilon_3$ can significantly influence the structure of circular photon orbits as well as the properties of the surrounding accretion flow. Consequently, to assess the validity of the no-hair theorem for astrophysical black holes using electromagnetic observables associated with photon rings or accretion flows, it is essential to examine in detail the location and geometric of the event horizon of JP black hole, as well as the dependence of circular photon orbits and accretion flow properties on the deviation parameter $\epsilon_3$. For simplicity, and without loss of generality, we set the mass of the JP black hole to $M = 1$ in the following analysis.

\begin{figure}[htbp]
    \centering
    \begin{subfigure}[c]{0.52\textwidth}
        \includegraphics[width=\textwidth]{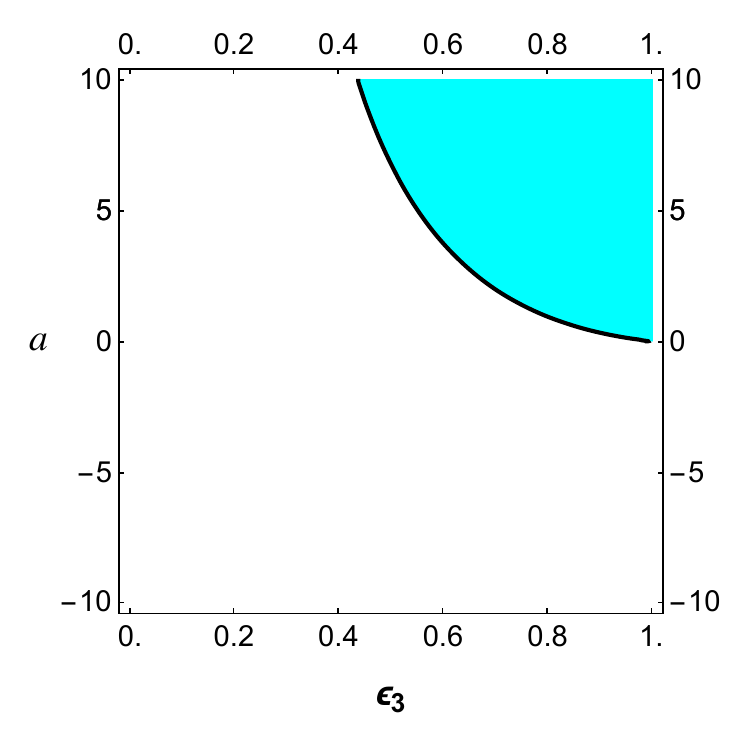}
        \caption{}
        \label{}
    \end{subfigure}
    \begin{subfigure}[c]{0.45\textwidth}
        \includegraphics[width=\textwidth]{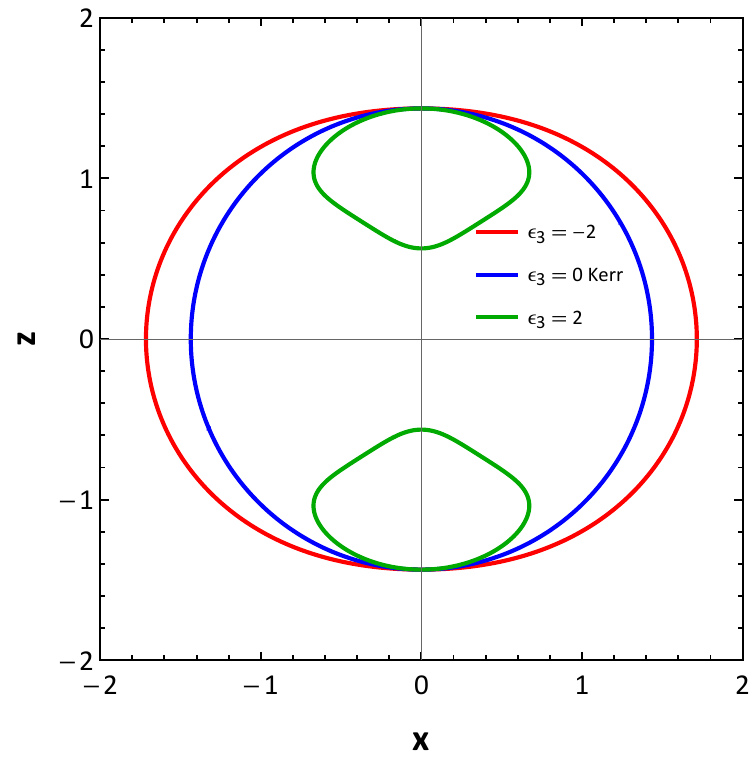}
        \caption{}
        \label{}
    \end{subfigure}
    \caption{(a) Parameter space of the JP metric in the $(a\,, \epsilon_3)$ plane. The white region corresponds to JP black holes with closed event horizons, while the light blue region corresponds to configurations with non-closed event horizons. The black solid curve denotes the critical boundary separating the two regimes. (b) Projections of the event horizon geometry onto the $xz$-plane for $\epsilon_3 = -2\,, 0$, and $2$.}
    \label{fig1}
\end{figure}

The dependence of the event horizon geometry of JP black holes on the dimensionless spin parameter $a$ and the deviation parameter $\epsilon_3$ is shown in Fig. \ref{fig1} (a). The corresponding projection of the event horizon onto the $x z$-plane is presented in Fig. \ref{fig1} (b).
The parameter space spanned by the spin parameter $a$ and the deviation parameter $\epsilon_3$, as shown in Fig. \ref{fig1} (a), illustrates the evolution of the event horizon geometry of JP black holes as functions of these parameters. As shown in Fig. \ref{fig1} (a), the event horizon exhibits two distinct topological configurations as the spin parameter $a$ and the deviation parameter $\epsilon_3$ increase from smaller to larger values. The first configuration corresponds to a closed event horizon analogous to that of a Kerr black hole and is represented by the white region in Fig. \ref{fig1} (a). The second configuration corresponds to a dumbbell-shaped, non-closed event horizon, in which the central singularity becomes exposed to the external spacetime, and this configuration is represented by the cyan region in Fig. \ref{fig1} (a). The boundary separating these two regimes is indicated by the black curve, which delineates the critical transition between the closed and non-closed event horizon configurations of JP black holes within the parameter space spanned by the spin parameter $a$ and the deviation parameter $\epsilon_3$. This result demonstrates that the topology of the JP black hole event horizon undergoes a transition between two qualitatively distinct configurations as both the spin parameter $a$ and the deviation parameter $\epsilon_3$ vary. A more detailed examination of the parameter space in Fig. \ref{fig1} (a) shows that, when the deviation parameter satisfies $\epsilon_3 < 0$, the JP black hole always possesses a closed event horizon for all admissible values of the spin parameter $a \in [0\,, 1]$. In this regime, the horizon geometry remains qualitatively similar to that of a Kerr black hole, although a nonzero deviation parameter $\epsilon_3$ introduces a small but discernible deformation. By contrast, when $\epsilon_3 > 0$, the behavior of the event horizon of the JP black hole becomes considerably more intricate. For $\epsilon_3 > 0$ and $a \in [0\,, 0.4]$, the event horizon of JP black hole remains closed. However, for $\epsilon_3 > 0$ and the value of the spin parameter $a$ in the range $a \in (0.4\,, 1]$, the event horizon undergoes a transition from a closed configuration to a non-closed configuration. This transition highlights the rich horizon topology induced by the deviation parameter $\epsilon_3$, which has no counterpart in the Kerr spacetime. This result indicates that the deviation parameter $\epsilon_3$ fundamentally alters the global structure of the spacetime, leading to novel horizon topologies absent in the Kerr geometry.

To provide a clearer and more intuitive characterization of the JP black hole event horizon for different values of the deviation parameter $\epsilon_3$, the configurations of the event horizon are projected onto the $x z$-plane, and their geometric configurations are presented in Fig. \ref{fig1} (b). As shown by the red curve in Fig. \ref{fig1} (b), when the deviation parameter satisfies $\epsilon_3 < 0$, the event horizon of the JP black hole exhibits an oblate, elliptical morphology. In this regime, the geometric configuration of the event horizon progressively deviates from that of the Kerr black hole as the value of $\epsilon_3$ decreases. Specifically, as indicated by the red curve in Fig. \ref{fig1} (b), the contour of the event horizon becomes increasingly elongated along the $x$-axis while remaining nearly unchanged along the $z$-axis, resulting in a more pronounced elliptical deformation relative to the Kerr case. As the deviation parameter increases from negative value toward $\epsilon_3 = 0$, the event horizon continuously approaches that of the Kerr black hole and coincides with it exactly at $\epsilon_3 = 0$, as shown by the blue curve in Fig. \ref{fig1} (b). For positive deviation parameters $\epsilon_3 > 0$, the event horizon of the JP black hole undergoes the previously described transition from a closed to a non-closed configuration as $\epsilon_3$ increases. In particular, as illustrated by the green curve in Fig. \ref{fig1} (b), when the deviation parameter is set to $\epsilon_3 = 2$, the event horizon becomes a non-closed, dumbbell-shaped structure elongated along the $z$-axis. The central constricted region of this configuration corresponds to the location of the singularity of the JP black hole. The singularity is no longer enclosed by the event horizon but instead becomes exposed to the external spacetime.

An examination of Fig. \ref{fig1} reveals that both decreasing and increasing the deviation parameter $\epsilon_3$ lead to significant departures of the JP black hole event horizon configuration from that of the Kerr black hole, giving rise to two distinct topological structures, namely closed and non-closed event horizons. When the deviation parameter $\epsilon_3$ takes negative values, the event horizon of the JP black hole is displaced relative to that of a Kerr black hole, whereas for $\epsilon_3 > 0$, it undergoes a transition from a closed to a non-closed configuration. Consequently, variations in the deviation parameter $\epsilon_3$ induce substantial modifications in the geometric structure of the event horizon. These geometric changes significantly influence the properties of circular photon orbits and the behavior of the surrounding accretion flow, thereby leading to corresponding variations in the electromagnetic signals emitted by photon rings and accreting matter. By systematically analyzing how the geometric configuration of the JP black hole event horizon evolves with different values of the deviation parameter $\epsilon_3$, and how circular photon orbits and accretion disks respond to the resulting changes in the local spacetime structure, one can investigate the associated modifications in observable electromagnetic signal. Examining how these electromagnetic signals respond to variations in the spacetime geometry near the event horizon provides an effective means of evaluating the extent to which the no-hair theorem remains valid for astrophysical black holes. It is noteworthy that when the event horizon transitions from a closed to a non-closed configuration, the central singularity becomes exposed within the spacetime described by the JP metric. The appearance of such a naked singularity formally violates the WCCC and, in principle, threatens the causal structure of the spacetime. However, previous studies of JP black hole shadows based on backward ray-tracing numerical methods have shown that photons originating from the vicinity of the naked singularity require an infinite coordinate time to reach an observer at spatial infinity. This implies that information emitted from the naked singularity cannot be observed through electromagnetic signals. Consequently, the presence of a naked singularity in the spacetime governed by the JP metric does not lead to observable violations of the WCCC. In other words, although the singularity becomes exposed in the case of a non-closed event horizon, this does not hinder the study of electromagnetic radiation emitted from circular photon orbits and accretion flows in the vicinity of JP black holes. This result further emphasizes that observational signatures of JP black holes are primarily governed by the spacetime structure outside the horizon, rather than the detailed properties of the singularity.

\section{Image of JP black hole with thin accretion disks}\label{investigationjpshadow}

Based on the preceding analysis, a fundamental prerequisite for theoretically examining the compatibility between astrophysical black holes and the no-hair theorem is the establishment of a theoretical framework capable of accurately modeling the physical properties of astrophysical black holes. Observational results from gravitational-wave detections by the LIGO and Virgo collaborations, together with the imaging of supermassive black holes by the EHT, strongly suggest that astrophysical black holes can be effectively described as Kerr spacetimes surrounded by accreting matter flows. Accordingly, an appropriate Kerr-like metric incorporating a set of deviation parameters that quantify departures from the Kerr solution can be employed to model the spacetime geometry in the vicinity of astrophysical black holes. These deviation parameters can be constrained through observational data, thereby enabling tests of whether astrophysical black holes are accurately described by the Kerr metric and whether the no-hair theorem within the framework of classical GR remains valid in astrophysical environments. In this context, the JP metric is specifically formulated to describe the spacetime geometry near astrophysical black holes while ensuring that spacetime pathologies, such as singularities and closed timelike curves, do not arise in the region exterior to the event horizon. Since the exterior spacetime of the JP black hole is free of pathologies for a wide range of spin parameters, the dependence of electromagnetic signals emitted from photon rings and accretion disks on the deviation parameters can be investigated without geometric restrictions. On this basis, the JP metric provides a reliable framework for analyzing how deviations from the Kerr geometry modify circular photon orbits and the electromagnetic radiation from accretion flows. These results are crucial for assessing the degree of consistency between astrophysical black holes and the Kerr solution, as well as for evaluating the validity of the no-hair theorem in realistic astrophysical environments. Furthermore, since the influence of deviation parameters on electromagnetic emission is directly encoded in black hole images, these observational signatures offer a powerful probe of potential deviations from the Kerr geometry and of the applicability of the no-hair theorem. Accordingly, constructing images of JP black holes surrounded by accretion disks is of central importance. In this study, we therefore investigate images of JP black holes illuminated by thin accretion disks and analyze the influence of deviation parameters in the JP metric on the resulting imaging properties.

\subsection{Calculation method of the image of JP black hole with thin accretion disk}\label{theoreticalframework}

In the exterior spacetime of the JP black hole, the surrounding accretion disk is typically assumed to lie in the equatorial plane, with its vertical thickness treated as geometrically negligible. Consequently, the accretion disk is modeled as a geometrically thin accretion disk confined to the equatorial plane. According to the form of the JP metric in BL coordinates given in Eq. (\ref{expressionjpmetric}), the thin accretion disk is located at $\theta = \pi / 2$. The material constituting the thin accretion disk is modeled as electrically neutral plasma moving along timelike trajectories in the equatorial plane of the spacetime described by the JP metric. Although the JP metric is constructed by introducing the deviation parameter $\epsilon_3$ into the Kerr metric, thereby producing deviations from Kerr spacetime, it nevertheless preserves the fundamental geometric properties and symmetry characteristics of the Kerr solution. In particular, as in Kerr spacetime, JP spacetime remains stationary and axisymmetric. Consequently, the spacetime described by the JP metric admits two Killing vector fields associated with these symmetries, namely $\partial_t$ and $\partial_\varphi$, corresponding to invariance under time translations and rotations about the symmetry axis, respectively. These Killing vector fields give rise to two conserved quantities, the energy $\mathcal{E}$ and the angular momentum $\mathcal{L}$, which play a fundamental role in determining the orbital motion of particles in the accretion disk. For massive electrically neutral particles undergoing timelike motion in the accretion disk, the four-velocity can be denoted by $u^{a}$. Accordingly, the conserved energy $\mathcal{E}$ and angular momentum $\mathcal{L}$ of these particles can be expressed in terms of the four-velocity as
\begin{equation}
    \begin{split}
        \mathcal{E} = - \left(\partial_t\right)^a u_a = - u_t\,, \qquad \mathcal{L} = \left(\partial_\varphi\right)^a u_a = u_\varphi\,.
    \end{split}
\end{equation}
Furthermore, by imposing the normalization condition on the four-velocity of massive particles, $u^{a} u_{a} = -1$, the radial component of the equation of motion can be given as
\begin{equation}
    \begin{split}
        u^r = - \sqrt{- \frac{V \left(r\,, \mathcal{E}\,, \mathcal{L} \right)}{g_{rr}}}\,,
    \end{split}
\end{equation}
where $V \left(r\,, \mathcal{E}\,, \mathcal{L} \right)$ denotes the effective potential, defined by
\begin{equation}\label{effectivepotential}
    \begin{split}
        V \left(r\,, \mathcal{E}\,, \mathcal{L} \right) = \left. \left(1 + g^{t t} \mathcal{E}^2 + g^{\varphi \varphi} \mathcal{L}^2 - 2 g^{t \varphi} \mathcal{E} \mathcal{L} \right)\right|_{\theta = \frac{\pi}{2}}\,.
    \end{split}
\end{equation}

On the other hand, although the thin accretion disk is geometrically thin in the vertical direction, it possesses a finite radial extent in the exterior spacetime of a JP black hole. The spacetime near the event horizon admits a characteristic radius known as the ISCO, which represents the smallest radius at which a massive particle can maintain a stable circular orbit around the black hole. In the region exterior to the ISCO, massive particles can sustain stable circular motion. Once a particle crosses the ISCO and moves inward, however, stable circular motion is no longer possible, and the particle inevitably transitions into a critical plunging orbits toward the event horizon. Accordingly, the ISCO serves as a natural boundary separating the accretion disk into two distinct radial regions. In the region exterior to the ISCO, massive neutral plasma particles do not strictly follow geodesic motion. This deviation arises because adjacent circular orbits possess different angular velocities, leading to differential rotation that generates magnetic and viscous dissipative forces between neighboring fluid elements. As a result, the energy $\mathcal{E}$ and angular momentum $\mathcal{L}$ of the plasma are not strictly identical to the conserved quantities associated with the Killing symmetries of the spacetime. Nevertheless, at each individual circular orbit, the motion of plasma particles can be approximated as geodesic. Consequently, the energy $\mathcal{E}$ and angular momentum $\mathcal{L}$ of plasma particles depend only on the radial coordinate $r$ of the circular orbit and can be written as $\mathcal{E} = \mathcal{E}_{\text{circ}} (r)$ and $\mathcal{L} = \mathcal{L}_{\text{circ}} (r)$. For particles on stable circular orbits outside the ISCO, the conditions, $V = \partial_r V = 0$, must be satisfied. The functions $\mathcal{E}_{\text{circ}} (r)$ and $\mathcal{L}_{\text{circ}} (r)$ can therefore be determined by solving these equations using the expression of the effective potential $V \left(r\,, \mathcal{E}\,, \mathcal{L} \right)$ defined in Eq. (\ref{effectivepotential}). The stability of circular orbits is ensured by the condition $\partial_r^2 V > 0$, while the ISCO is characterized by the marginal stability condition $\partial_r^2 V = 0$. In the region interior to the ISCO, the inner edge of the thin accretion disk is typically assumed to extend down to the event horizon. However, stable circular orbits are no longer supported in this region, and the plasma instead plunges inward toward the event horizon along critical plunging orbits. During this plunging phase, particle trajectories can be approximated as geodesic, and the energy and angular momentum remain conserved throughout the motion. Since the particles originate at the ISCO and subsequently follow geodesic trajectories toward the event horizon, the energy $\mathcal{E}$ and angular momentum $\mathcal{L}$ of these particles during the plunging motion can be taken to be equal to those at the ISCO, namely $\mathcal{E} = \mathcal{E}_{\text{ISCO}}$ and $\mathcal{L} = \mathcal{L}_{\mathrm{ISCO}}$. Accordingly, by substituting these conserved quantities into the radial equation of motion, the radial component of the four-velocity of plasma particles plunging from the ISCO toward the event horizon can be expressed as
\begin{equation}\label{fourvelocityurc}
    \begin{split}
        u_c^r = - \sqrt{- \frac{V \left(r\,, \mathcal{E}_{\text{ISCO}}\,, \mathcal{L}_{\text{ISCO}} \right)}{g_{rr}}}\,,
    \end{split}
\end{equation}

To investigate how the electromagnetic signals emitted by a thin accretion disk evolve as the JP black hole departs from the Kerr geometry, it is necessary to introduce a distant observer located at spatial infinity in the JP spacetime. This setup enables the construction of images of the JP black hole illuminated by the surrounding thin accretion disk in the asymptotic region. The radiation detected by this observer consists of photons emitted from the accretion disk in the immediate vicinity of the black hole and subsequently propagating to spatial infinity. To characterize the spacetime properties in the neighborhood of the observer, an appropriate tetrad system is introduced at the location of the observer, which provides the local inertial frame of the observer. By exploiting the spacetime symmetries associated with the Killing vector fields $\partial_t$ and $\partial_\varphi$ in the JP spacetime, the observer is conveniently specified by the coordinates $(t_0\,, r_0\,, \theta_0\,, \varphi_0) = (0\,, r_0\,, \theta_0\,, 0)$, where $r_0$ denotes the radial position of the observer and $\theta_0$ represents the inclination angle between the observer's line of sight and the rotation axis of the JP black hole. To describe the local spacetime geometry in the vicinity of the observer, a coordinate system centered on the observer is constructed, and the associated orthonormal tetrad basis vectors are defined as
\begin{equation}\label{zamocoordinatesbasis}
    \begin{split}
        \hat{e}_{(0)} = \zeta \partial_t + \gamma \partial_\varphi\,, \quad \hat{e}_{(1)} = - \frac{\partial_r}{\sqrt{g_{r r}}}\,, \quad \hat{e}_{(2)} = \frac{\partial_\theta}{\sqrt{g_{\theta \theta}}}\,, \quad \hat{e}_{(3)} = - \frac{\partial_\varphi}{\sqrt{g_{\varphi \varphi}}}\,,
    \end{split}
\end{equation}
where
\begin{equation}
    \begin{split}
        \zeta = \frac{g_{\varphi \varphi}}{\sqrt{g_{\varphi \varphi} \left(g_{t \varphi}^2 - g_{t t} g_{\varphi \varphi}\right)}}\,, \qquad \gamma = \frac{- g_{t \varphi}}{\sqrt{g_{\varphi \varphi} \left(g_{t \varphi}^2 - g_{t t} g_{\varphi \varphi}\right)}}\,.
    \end{split}
\end{equation}
In this coordinate system, the basis vector $\hat{e}_{(0)}$ is timelike and is chosen to coincide with the four-velocity of the observer at spatial infinity. The remaining basis vectors span the spacelike directions. All basis vectors form an orthonormal set, and the corresponding orthonormality conditions are given by
\begin{equation}\label{orthonormalconditions}
    \begin{split}
        \hat{e}_{(0)} \cdot \hat{e}_{(0)} = -1\,, \quad \hat{e}_{(0)} \cdot \hat{e}_{(i)} = 0\,, \quad \hat{e}_{(i)} \cdot \hat{e}_{(j)} = \delta_{i j}\,, \quad i\,, j = 1\,, 2\,, 3\,.
    \end{split}
\end{equation}
The asymptotic observer is assumed to be locally stationary. In the constructed coordinate system, the timelike basis vector $\hat{e}_{(0)}$ is chosen to be orthogonal to the Killing vector $\partial_\varphi$, ensuring that the observer has vanishing angular momentum. Accordingly, this frame is referred to as the zero-angular-momentum observer (ZAMO) frame.

The trajectories of photons emitted from the accretion disk in the vicinity of a JP black hole and propagating to spatial infinity are determined by the four-momentum of photons, which is tangent to the corresponding null geodesics. For photons reaching a distant observer, the propagation direction is characterized by the spatial projection of the photon four-momentum onto the ZAMO frame, thereby defining the photon three-momentum. A two-dimensional image plane can be introduced on the spatial hypersurface of the ZAMO frame at the infinity observer, serving as the screen on which the image of the JP black hole and its accretion disk is formed. By projecting the three-momentum of the photon onto this image plane, the corresponding pixel coordinates of each photon on the image plane in the ZAMO coordinates can be determined. Repeating this procedure for all photons emitted from the accretion disk that reach the observer allows the construction of the image of the black hole with a thin accretion disk as seen at infinity.

The four-momentum of a photon in the JP spacetime, expressed in BL coordinates, can be written as $p^\mu = \left(\dot{t}\,, \dot{r}\,, \dot{\theta}\,, \dot{\varphi} \right)$, where the overdot denotes differentiation with respect to the affine parameter $\lambda$ along the null geodesic. For any quantity $X$, the notation $\dot{X}$ represents $\text{d} X / \text{d} \lambda$. From this expression, the conserved quantities along null geodesics, namely the energy $E$ and angular momentum $L$ associated with the Killing vectors $\partial_t$ and $\partial_\varphi$ in JP spacetime, are defined as $E = - p_t$ and $L = p_\varphi$. The four-momentum of photon in the ZAMO frame at the location of the observer can be expressed as $p^a = p^{(\mu)} e_{(\mu)}^a$, where indices enclosed in parentheses denote components of vectors measured in the ZAMO frame. To obtain an explicit description of photon propagation from the accretion disk to a distant observer, it is necessary to establish a correspondence between the photon four-momentum in BL coordinates and that in the ZAMO frame. A convenient approach is to introduce celestial coordinates $(\Theta\,, \Psi)$ as an intermediate representation connecting these two descriptions. According to Ref. \cite{Hou:2022eev}, the celestial coordinates $(\Theta\,, \Psi)$ can be expressed in terms of the photon four-momentum in the ZAMO frame as
\begin{equation}
    \begin{split}
        \cos \Theta = \frac{p^{(1)}}{p^{(0)}}\,, \qquad \tan \Psi = \frac{p^{(3)}}{p^{(2)}}\,.
    \end{split}
\end{equation}
Subsequently, using the stereographic projection formalism in the ZAMO frame introduced in Ref. \cite{Hu:2020usx}, the coordinates of the pixel point $p$ on the image plane, corresponding to photons propagating from the accretion disk to a distant observer, can be expressed as
\begin{equation}\label{imageplanecooridinateswithceleestialcoordinates}
    \begin{split}
        x_{p} = - 2 \left| \overrightarrow{OP} \right| \tan \left(\frac{\Theta}{2} \right) \sin \Psi\,, \qquad y_{p} = - 2 \left| \overrightarrow{OP} \right| \tan \left(\frac{\Theta}{2} \right) \cos \Psi\,,
    \end{split}
\end{equation}
where $\left| \overrightarrow{OP} \right|$ denotes the magnitude of the photon three-momentum in the ZAMO frame, which can be written as
\begin{equation}
    \begin{split}
        \left| \overrightarrow{OP} \right| = \gamma L - \zeta E\,.
    \end{split}
\end{equation}

In the Petrov classification, the Kerr metric belongs to type D, a property associated with the existence of the Carter constant. The presence of this constant permits the complete separation of variables in the Hamilton-Jacobi equation governing photon motion, thereby enabling a fully analytic description of null geodesics in the Kerr spacetime. Consequently, the shadow boundary of a Kerr black hole can be determined analytically, and the corresponding geodesic structure can be investigated in a fully self-consistent manner. By contrast, the JP metric introduces a deviation parameter $\epsilon_3$ that generates departures from the Kerr solution while reducing smoothly to the Kerr metric in the limit $\epsilon_3 \to 0$. The introduction of this deviation parameter generally changes the algebraic structure of the spacetime from Petrov type D to the more general Petrov type I, thereby eliminating the hidden symmetry associated with the Kerr spacetime and, in general, preventing the existence of an exact Carter constant. As a consequence, the Hamilton-Jacobi equation governing photon geodesics in the JP spacetime is no longer exactly separable, which precludes a fully analytic treatment of photon motion. However, in our previous study of JP black hole shadows, a function $\tilde{I}(r\,, \theta)$ was identified in the Hamilton-Jacobi equation governing photon geodesics in the JP spacetime. Because this function depends simultaneously on the radial coordinate $r$ and the polar coordinate $\theta$ in BL coordinates, it constitutes the fundamental source of the non-separability of the Hamilton-Jacobi equation. An analysis of the coordinate dependence of $\tilde{I}(r\,, \theta)$ shows that, at radial distances sufficiently far from the event horizon, the function is predominantly governed by the radial coordinate $r$, whereas its dependence on $\theta$ becomes comparatively weak. Under this approximation, $\tilde{I}(r\,, \theta)$ may be treated as a function of $r$ alone, which allows the introduction of an approximate Carter constant and renders the Hamilton-Jacobi equation approximately separable. This approximation enables the application of semi-analytic techniques to construct the shadows of JP black holes with closed event horizons. However, when the event horizon assumes a non-closed configuration, the approximation that treats $\tilde{I}(r\,, \theta)$ as predominantly radial becomes inadequate, and the resulting semi-analytic treatment no longer reproduces the exact shadow morphology accurately. By contrast, backward ray-tracing simulations provide robust and reliable results irrespective of the topology of the event horizon. Accordingly, all subsequent analyses presented in this work are based exclusively on numerical backward ray-tracing simulations for constructing and investigating the images of JP black holes surrounded by thin accretion disks.

The theoretical framework for modeling photon propagation in the spacetime described by the JP metric using backward ray-tracing numerical methods has been presented in detail in Ref. \cite{Wang:2025ihg}, where it was employed to construct images of JP black hole shadows. When extending this method to construct images of JP black holes surrounded by thin accretion disks, the initial stage of photon propagation, namely the backward ray-tracing of photons from the image plane of a distant observer through the JP spacetime toward the black hole, follows the same procedure as that used in calculations of JP black hole shadows. However, when applying the backward ray-tracing method to construct images of JP black holes surrounded by thin accretion disks, once photons propagate into the strong-field region near the black hole, gravitational lensing effects significantly deflect their trajectories, causing photons to undergo orbital motion around the black hole. As a result, photons may intersect the thin accretion disk located in the equatorial plane and interact with the accreting matter within the accretion disk. From a theoretical perspective, these interactions not only modify the geometric structure of the resulting image but also play a crucial role in determining the observed intensity distribution. Therefore, when applying backward ray-tracing methods to construct images of JP black holes surrounded by thin accretion disks, it is essential to incorporate interactions between photons and the accretion disk into the numerical framework. In particular, radiative transfer along photon trajectories, including emission, absorption, and frequency shift effects, should be taken into account. Accordingly, the remainder of this subsection is devoted to outlining the theoretical formalism required to incorporate these interactions into the backward ray-tracing calculations.

\begin{figure}[htbp]
    \centering
    \includegraphics[width=0.6\textwidth]{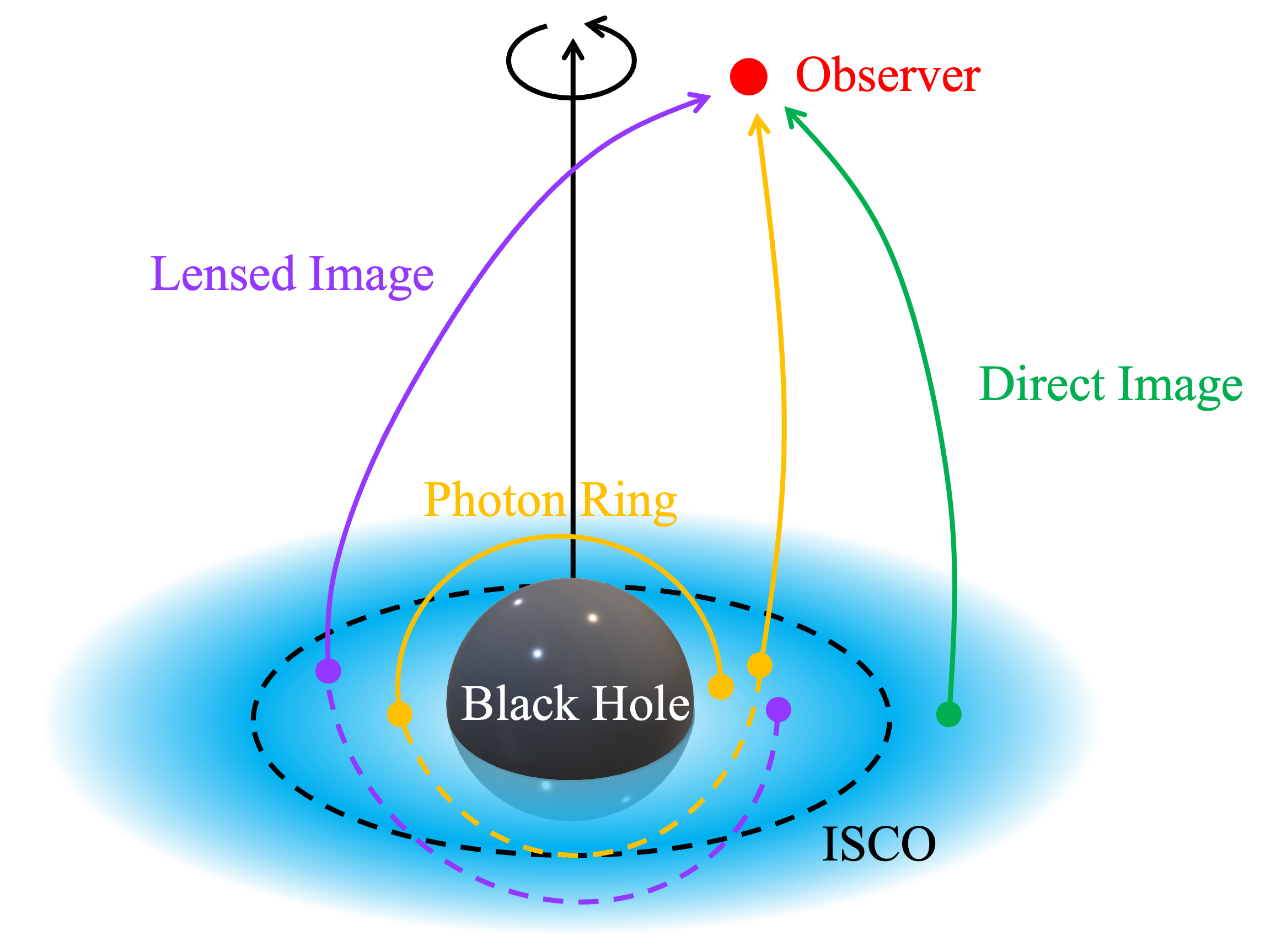}
    \caption{Three classes of photon trajectories emitted from the accretion disk in the vicinity of the black hole.}
    \label{fig2}
\end{figure}

When a photon trajectory is traced backward along a null geodesic in the spacetime described by the JP metric to its emission point on the accretion disk in the equatorial plane, strong gravitational lensing near the black hole significantly deflects the photon trajectory. As a consequence, the photon may intersect the accretion disk at least once, and possibly multiple times, along this trajectory of photon. The radii at which the photon trajectory intersects the accretion disk are denoted by $r_n (x\,, y)$, where the index $n$ labels the interaction order and takes discrete values $n = 1\,, 2\,, \cdots N_{\text{max}} \left(x\,, y \right)$. Here, $N_{\text{max}} \left(x\,, y \right)$ represents the maximum number of accretion disk intersections for photons arriving at the pixel $(x\,, y)$ on the image plane. Because $n$ is discrete, the function $r_{n} (x\,, y)$ partitions the image plane into discrete regions corresponding to different intersection orders. This function therefore provides an effective characterization of the imaging features on the image plane in the ZAMO frame, associated with photons that have crossed the accretion disk $n$ times during their propagation through the spacetime. Accordingly, $r_{n} (x\,, y)$ is referred to as the transfer function. This relation further indicates that the observable image of a black hole illuminated by an accretion disk depends sensitively on the number $n$ of intersections between the photon and the accretion disk.
\begin{itemize}
    \item For $n = 1$, photons emitted from the accretion disk reach the observer directly without further interaction with the accretion disk. The resulting image is referred to as the direct image.

    \item For $n = 2$, photons emitted from the accretion disk undergo gravitational lensing, orbit the black hole, and intersect the disk once before reaching the observer. The resulting image is referred to as the lensed image. 

    \item For $n \geq 3$, photons undergo increasingly strong gravitational lensing, orbiting the black hole multiple times and intersecting the disk three or more times before reaching the observer. The resulting images constitute what is commonly referred to as the photon ring. 
\end{itemize}
The three types of photon trajectories in the vicinity of the black hole, classified by the number $n$ of intersections between the photon trajectory and the accretion disk, are shown in Fig. \ref{fig2}.
In this figure, the black sphere represents the black hole, while the blue disk denotes the surrounding thin accretion disk. The dashed black circle indicates the location of the ISCO. The solid green curve corresponds to photon trajectories that propagate directly from the accretion disk to the observer, producing the direct image. The purple curve represents photon trajectories that orbit the black hole due to gravitational lensing and intersect the accretion disk twice before reaching the observer, forming the lensed image. The yellow curve illustrates photon trajectories undergoing stronger gravitational lensing, crossing the accretion disk three times before reaching the observer and forming the photon ring on the image plane of observer.

On the other hand, photons emitted from the accretion disk interact with the neutral plasma whenever trajectories of photons intersect the accretion disk in the vicinity of the black hole. These interactions modify the photon intensity as the radiation propagates through curved spacetime and ultimately reaches a distant observer, thereby influencing the intensity distribution of the resulting black hole images illuminated by the accretion disk on the image plane located on the distant observer. Accordingly, in the following analysis, we provide a concise overview of the relationship between the interaction of photons with the neutral plasma in the accretion disk and the resulting intensity distribution of the reconstructed image of black holes in the ZAMO frame. The number of intersections between a photon trajectory emitted from the accretion disk and the accretion disk is closely related to the observed intensity distribution on the image plane, as photons propagate along null geodesics through the spacetime and reach the distant observer. This relationship highlights the importance of accounting for multiple accretion disk intersections when modeling black hole images. It is therefore necessary to briefly summarize the theoretical framework describing how these interactions modify the photon intensity and how such modifications are encoded in the observed intensity in the black hole image. Within the numerical backward ray-tracing approach, the complete photon trajectory connecting the accretion disk to the image plane in the ZAMO frame is taken into account.

As noted above, when a photon trajectory is traced backward from the image plane to the accretion disk, gravitational lensing in the strong-field region near the black hole can cause the photon to intersect the disk multiple times, with the photon intensity being modified at each intersection. For simplicity, refraction effects in the disk medium are assumed to be negligible. Under this assumption, the variation of the photon intensity due to interactions with the accretion disk is governed by the expression as
\begin{equation}\label{radiativetransferequation}
    \begin{split}
        \frac{\text{d}}{\text{d} \lambda} \left(\frac{I_\nu}{\nu^3} \right) = \frac{J_\nu - \kappa_\nu I_\nu}{\nu^2}\,,
    \end{split}
\end{equation}
where $I_\nu$, $J_\nu$, and $\kappa_\nu$ denote the specific intensity, emissivity, and absorption coefficient at photon frequency $\nu$, respectively. When a photon propagates through spacetime without interacting with the accretion disk or other matter fields, both the emissivity $J_\nu$ and the absorption coefficient $\kappa_\nu$ vanish. Under these conditions, the quantity $I_\nu / \nu^3$ remains invariant and is conserved along the photon geodesic. Based on the preceding analysis, Eq. (\ref{radiativetransferequation}) provides a quantitative description of the variation of photon intensity due to interactions with the accreting material as the photon traverses the accretion disk surrounding the black hole. This equation is commonly referred to as the radiative transfer equation.

Since the accretion disk surrounding the black hole is assumed to be geometrically thin, the emissivity $J_\nu$ and the absorption coefficient $\kappa_\nu$ can be treated as constant during photon propagation through the accretion disk. To obtain the specific intensity at each pixel on the image plane in the ZAMO frame using the numerical backward ray-tracing method, the radiative transfer equation is integrated along the photon trajectories in the reverse direction. In performing this integration, and in determining the specific intensity associated with each pixel on the image plane, the intensity, emissivity, and absorption coefficient are expressed in terms of invariant quantities as
\begin{equation}\label{ijklorentzinvariant}
    \begin{split}
        \mathcal{I} = \frac{I_\nu}{\nu^3}\,, \qquad \mathcal{J} = \frac{J_\nu}{\nu^2}\,, \qquad \mathcal{K} = \nu \kappa_\nu\,.
    \end{split}
\end{equation}
The quantities $\mathcal{I}$, $\mathcal{J}$, and $\mathcal{K}$ are Lorentz invariants. In the backward ray-tracing numerical technique used to construct black hole images, photon propagation is treated opposite to the physical direction. Consequently, the affine parameter $\lambda$ parametrizing the photon geodesic is transformed as $\lambda \rightarrow -\lambda$ in the numerical integration. Under this reversal of the affine parameter, the emissivity $J_\nu$ and absorption coefficient $\kappa_\nu$ in the radiative transfer equation transform as $J_\nu \rightarrow - J_\nu$ and $\kappa_\nu \rightarrow \kappa_\nu$, respectively, ensuring consistency of the backward ray-tracing integration scheme. To describe the interactions between photons and the accretion disk, a locally comoving rest frame $\{F_n\}$ is introduced at the radial position $r_n$ on the accretion disk. The subscript $n = 1, \cdots, N_{\text{max}}$ labels the number of interactions between the photon and the accretion disk and enumerates successive interactions. In the subsequent analysis, quantities carrying the subscript $n$ are understood to be evaluated in the corresponding local rest frame $\{F_n\}$. When a backward-traced photon intersects the accretion disk at $r_{n}$, it undergoes a single emission and absorption process. The resulting variation of the photon intensity is therefore governed by the radiative transfer equation expressed in terms of the Lorentz invariant quantities $\mathcal{I}$, $\mathcal{J}$, and $\kappa$ as
\begin{equation}
    \begin{split}
        \frac{\text{d} \mathcal{I}}{\text{d} \lambda} = - \mathcal{J}_n + \mathcal{K}_n \mathcal{I}\,.
    \end{split}
\end{equation}
After the $n$-th interaction between the photon and the accretion disk, the corresponding Lorentz invariant intensity is denoted by $\mathcal{I}_{n}$. This quantity is determined by the intensity $\mathcal{I}_{n-1}$ associated with the $(n-1)$-th interaction between the photon and the accretion disk. The initial value of the Lorentz invariant intensity, referred to as $\mathcal{I}_0$, is defined in terms of the observed intensity $I_{0}$ on the image plane at the observer, where the explicit form of $I_{0}$ is given by
\begin{equation}
    \begin{split}
        \mathcal{I}_0 = \frac{I_0}{\nu_0^3}\,.
    \end{split}
\end{equation}
The accretion disk surrounding the JP black hole is assumed to be thin, and this thin disk approximation applies not only to its geometric structure in spatial dimensions but also to its optical properties. Under the optically thin accretion disk approximation, the emissivity and absorption coefficient in the radiative transfer equation can be treated as constants. Consequently, each interaction between the photon and the accretion disk produces the same modification of the photon intensity due to emission and absorption processes. Accordingly, the variation of the photon intensity before and after traversing the accretion disk, as prescribed by the radiative transfer equation, can be expressed as
\begin{equation}\label{intensityvariationequation}
    \begin{split}
        \mathcal{I}_n = e_n \mathcal{I}_{n - 1} + \delta_n\,,
    \end{split}
\end{equation}
and the coefficients $e_n$ and $\delta_n$ are defined as
\begin{equation}\label{expressionsendeltan}
    \begin{split}
        e_n = e^{\mathcal{K}_n \Delta \lambda_n}\,, \qquad \delta_n = \left(1- e_n \right) \frac{\mathcal{J}_n}{\mathcal{K}_n}\,.
    \end{split}
\end{equation}
By repeatedly iterating Eq. (\ref{intensityvariationequation}), the final expression for the photon intensity after multiple interactions with the accretion disk during the backward ray-tracing procedure can be written as
\begin{equation}\label{inrelationi0}
    \begin{split}
        \mathcal{I}_{n} = e_n \mathcal{I}_{n - 1} + \delta_n = e_n \cdots e_1 \mathcal{I}_0 + e_n \cdots e_2 \delta_1 + \cdots + e_n \delta_{n - 1} + \delta_n\,.
    \end{split}
\end{equation}
When the subscript $n$ in Eq. (\ref{inrelationi0}) is set to $n = N_{\text{max}}$, the equation describes the photon intensity after all interactions between the photon and the accretion disk along the entire photon trajectory. In backward ray-tracing simulations of black hole images, the reverse integration of photon trajectories terminates at the point corresponding to the physical emission source. When the emission point is located either near the event horizon of the black hole or at spatial infinity, the photon intensity associated with these regions is typically assumed to vanish. This implies that the photon intensity at the endpoint of the backward integration should satisfy the terminal condition $\mathcal{I}_{N_{\text{max}}} = 0$. Imposing the terminal condition and substituting the expressions of $e_n$ and $\delta_n$ from Eq. (\ref{expressionsendeltan}), Eq. (\ref{inrelationi0}) can be rearranged to yield the Lorentz invariant intensity $\mathcal{I}_0$ associated with the observed intensity on the image plane as
\begin{equation}\label{i0withlorentzinvariant}
    \begin{split}
        \mathcal{I}_0 = - \sum_{n = 1}^{N_{\text{max}}} \frac{\left(1 - e^{\mathcal{K}_n \Delta \lambda_n} \right) \frac{\mathcal{J}_n}{\mathcal{K}_n}}{e^{\mathcal{K}_n \Delta \lambda_n} \cdots e^{\mathcal{K}_1 \Delta \lambda_1}}
    \end{split}
\end{equation}
Furthermore, for notational convenience, a ``fudge factor'' $f_{n}$ can be introduced to simplify the expression for the intensity on the image plane at the location of the observer obtained from the integration of the radiative transfer equation. This factor is defined as
\begin{equation}\label{expressionfn}
    \begin{split}
        f_n = \nu_n \Delta \lambda_n\,.
    \end{split}
\end{equation}
The explicit form of the fudge factor depends on the adopted accretion disk model and can be specified only after the physical properties of the accretion disk surrounding the black hole have been determined. In Eq. (\ref{expressionfn}), $\nu_n$ denotes the photon frequency measured in the comoving frame $F_n$ at the $n$-th intersection with the accretion disk. Based on the four-momentum of photon and the four-velocity of the accretion flow in the accretion disk, the photon frequency measured in the comoving frame can be written as $\nu_n = - p_a u^a |_{r = r_n}$. The quantity $\Delta \lambda_n$ represents the corresponding change in the affine parameter as the photon crosses the accretion disk at $r_n$. According to the definition of the fudge factor $f_n$, the optical depth $\tau_n$ associated with the $n$-th intersection between the photon and the accretion disk characterizes the cumulative absorption experienced by the photon along its trajectory up to the point of initial emission. The optical depth is defined as
\begin{equation}
    \begin{split}
        \tau_m =
        \begin{cases}
            \exp\left[ \sum_{n=1}^m \kappa_n f_n \right], & m \ge 1, \\
            1, & m = 0.
        \end{cases}
    \end{split}
\end{equation}
Substituting the expressions for the fudge factor $f_n$ and the optical depth $\tau_n$ into Eq. (\ref{i0withlorentzinvariant}), and using the expressions of the Lorentz invariant quantities $\mathcal{I}$, $\mathcal{J}$, and $\mathcal{K}$ defined in Eq. (\ref{ijklorentzinvariant}), the intensity at each pixel on the image plane can be expressed as
\begin{equation}
    \begin{split}
        I_0 = \sum_{n = 1}^{N_{\text{max}}} \left(\frac{\nu_0}{\nu_n} \right)^3 \frac{J_n}{\tau_{n - 1}} \left(\frac{1 - e^{- \kappa_n f_n}}{\kappa_n} \right)
    \end{split}
\end{equation}
where $\nu_0 = - p_{(0)} = - p_a e_{(0)}^a |_{r= r_0}$ denotes the photon frequency measured in the ZAMO frame of the observer at spatial infinity. As assumed above, the accretion disk is regarded as optically thin, so that absorption effects arising from interactions between photons and the accretion flow within the accretion disk can be neglected during photon propagation. Consequently, in applying the radiative transfer equation to evaluate the variation of photon intensity as photons traverse the accretion disk, the absorption coefficient can be approximated as $\kappa_n \approx 0$, which simplifies the radiative transfer equation governing the variation of photon intensity. Furthermore, it is necessary to introduce the redshift factor $g_{n}$, defined as the ratio between the photon frequency measured in the comoving frame of the accretion disk $\{ F_n\}$ and that measured in the ZAMO frame of the distant observer. The redshift factor is defined as
\begin{equation}\label{definitionredshift}
    \begin{split}
        g_n = \frac{\nu_0}{\nu_n}
    \end{split}
\end{equation}
Using the definition of the redshift factor $g_n$ together with the approximation of the absorption coefficient $\kappa_{n} \approx 0$, the integrated radiative transfer equation determining the specific intensity at each pixel on the image plane in the ZAMO frame at the distant observer can be simplified as
\begin{equation}\label{finalsimplei0}
    \begin{split}
        I_{0} = \sum_{n = 1}^{N_{\text{max}}} f_n g_n^3 J_n\,.
    \end{split}
\end{equation}

Under the assumption that the accretion disk is considered as the optically thin disk, the absorption coefficient can be taken as $\kappa_n \approx 0$, which leads to a simplified form of the radiative transfer equation governing the variation of photon intensity resulting from interactions between photons and the accretion disk surrounding the black hole. This simplified form of the radiative transfer equation has been widely adopted in previous studies of black hole images illuminated by thin accretion disks. To evaluate the variation in photon intensity as photons traverses the accretion disk using this simplified radiative transfer equation, it is necessary to specify the fudge factor $f_n$, the redshift factor $g_n$, and the emissivity coefficient $J_n$ in Eq. (\ref{finalsimplei0}). Motivated by the fact that the images of the supermassive black holes at the centers of M87* and Sgr A* were obtained at a wavelength of $1.3\,\text{mm}$ (corresponding to a frequency of $230\,\text{GHz}$), the emissivity coefficient $J_n$ is modeled as an exponential function with a quadratic dependence in logarithmic space. This phenomenological form provides an effective description of synchrotron emission, capturing the wavelength-dependent radiative properties of the accretion flow relevant for millimeter-wave observations. The emissivity coefficient $J_n$ in the continuous limit is given by
\begin{equation}
    \begin{split}
        J = \exp \left(- \frac{1}{2} z^2 - 2 z \right)\,, \qquad z = \log \frac{r}{r_H} 
    \end{split}
\end{equation}
This emissivity model has been employed in Ref. \cite{Chael:2021rjo} to fit black hole images observed at $230\,\text{GHz}$, demonstrating its effectiveness in reproducing the millimeter-wave brightness profiles of accretion flows around supermassive black holes. It is assumed that the radiative properties of the accretion disk are axisymmetric and isotropic, with the emission intensity decreasing rapidly with increasing radial distance from the black hole event horizon $r_h$. Furthermore, variations in the value of the fudge factor $f_n$ primarily affect the brightness of narrow photon rings while having a negligible impact on the overall image luminosity. Therefore, for computational convenience, the factor can be normalized to unity $f_n = 1$ when constructing images of JP black holes with thin accretion disks. Under these assumptions about the radiative properties and dynamical behavior of the accretion disk, all quantities appearing in the simplified radiative transfer equation in Eq. (\ref{finalsimplei0}), including the emissivity $J_n$, the redshift factor $g_n$, and the fudge factor $f_n$, are fully specified. Consequently, the simplified radiative transfer equation can be directly employed in backward ray-tracing calculations to compute the photon intensity distribution of JP black hole images.

In addition to modifying the photon intensity, interactions between photons and the accretion disk also induce frequency shifts during photon propagation. Therefore, in constructing black hole images illuminated by thin accretion disks and analyzing their observational signatures, it is essential to account for the evolution of photon frequency along the photon trajectory. The photon frequency measured in the comoving frame of the accretion disk $\{F_n \}$, modified by interactions with the accreting matter in the accretion disk, is intrinsically related to that measured by the distant observer in the ZAMO frame through the redshift factor $g_n$ defined in Eq. (\ref{definitionredshift}). A detailed characterization of the redshift factor $g_n$ is therefore required to accurately describe these frequency shifts. To this end, an explicit expression for the redshift factor in terms of the photon four-momentum and the relevant conserved quantities governing photon motion should be further derived. In the thin accretion disk model considered in the present study, the accreting material is treated as electrically neutral plasma moving along timelike geodesics around the black hole, with conserved energy $E$ and angular momentum $L$. Since the dynamical behavior of the plasma differs significantly between the regions inside and outside the ISCO, the corresponding expression for the redshift factor should be determined separately in these two regimes. For the radius location $r_n$, representing the point that the position of the interaction between the photon and the accretion disk, satisfies $r_n \geq r_{\text{ISCO}}$, the plasma follows circular orbits around the black hole, and its angular velocity can be defined as
\begin{equation}\label{omegandefinition}
    \begin{split}
        \Omega_n (r) = \left. \frac{u^\varphi}{u^t} \right|_{r = r_n}\,.
    \end{split}
\end{equation}

Since the accreting neutral plasma is confined to the equatorial plane and undergoes circular motion outside the ISCO, the radial and polar components of the four-velocity in this regime satisfy $u^{r} = 0$ and $u^{\theta} = 0$. This constraint implies that the plasma undergoes purely azimuthal motion within the equatorial plane, consistent with the assumption of a geometrically thin and dynamically stable accretion disk. Using the definitions of the photon frequency $\nu_0$ measured in the ZAMO frame at the location of the observer and $\nu_n$ measured in the comoving frame $\{F_n \}$, together with the angular velocity $\Omega_n$ defined in Eq. (\ref{omegandefinition}), the redshift factor defined in Eq. (\ref{definitionredshift}) can be expressed as
\begin{equation}\label{finalredshiftexpression}
    \begin{split}
        g_n = \frac{e}{u^t \left(1 - b \Omega_n \right)}\,,
    \end{split}
\end{equation}
where the quantity $b$ denotes the impact parameter, defined as $b = L / E$, where $L$ and $E$ are the conserved angular momentum and energy of the photon, respectively. Using the relations between these conserved quantities and the photon four-momentum components, the impact parameter $b$ can be written as 
\begin{equation}\label{definitionb}
    \begin{split}
        b = \frac{L}{E} = - \frac{p_\varphi}{p_t}\,.
    \end{split}
\end{equation}
Moreover, the parameter $e$ in Eq. (\ref{finalredshiftexpression}) is defined as the ratio of the photon energy measured in the ZAMO frame to the conserved energy of photon along the null geodesic. It can be written as
\begin{equation}
    \begin{split}
        e = \frac{\nu_0}{E} = \zeta \left(1 + b \gamma \right)\,.
    \end{split}
\end{equation}
In asymptotically flat spacetimes, the parameter $e$ can be set to unity, $e = 1$. This choice also applies to the spacetime described by the JP metric, which shares the same asymptotic behavior as the Kerr metric. The component $u^t$ in Eq. (\ref{finalredshiftexpression}) is determined from the normalization condition of the four-velocity, $u_a u^a = - 1$. Using this normalization condition, together with the definition of the angular velocity $\Omega_n$ and the circular motion constraints $u^r = 0$ and $u^\theta = 0$, the explicit expression for $u^t$ is obtained as
\begin{equation}
    \begin{split}
        u^t = \left. \sqrt{\frac{-1}{g_{t t} + 2 g_{t \varphi} \Omega_n + g_{\varphi \varphi} \Omega^2_n}} \right|_{r = r_n}\,.
    \end{split}
\end{equation}

In contrast, within the region between the ISCO and the event horizon, the accreting plasma in the accretion disk follows critical plunging timelike geodesics and falls toward the event horizon of black hole. In this regime, the radial component of the four-velocity for the accreting plasma is nonvanishing $u^{r} \neq 0$, and the specific expression of the $u^{r}$ is given by the infall velocity $u^{r}_{c}$ defined in Eq. (\ref{fourvelocityurc}). The polar component of the plasma four-velocity remains vanishing $u^{\theta} = 0$. Although the plasma motion inside the ISCO differ from the circular orbits outside the ISCO, a comoving frame $\{F_n \}$ can still be introduced in the interior region of ISCO to describe the interactions between the photon and the accreting plasma at the radius location $r_n$ within the numerical backward ray-tracing scheme. Using the expressions for $\nu_0$ and $\nu_n$, together with $u^\theta = 0$ and the relations between $u^t$, $u^{\varphi}$, and the conserved quantities $E$ and $L$, the redshift factor in the region $r_n < r_{\text{ISCO}}$ can be expressed as
\begin{equation}\label{redshiftexpressioninisco}
    \begin{split}
        g_n = - \frac{e}{u_c^r p_r / E + E_{\text{ISCO}} \left(g^{t t} - g^{t \varphi} b \right) + L_{\text{ISCO}} \left(g^{\varphi \varphi} b - g^{t \varphi} \right)}\,.
    \end{split}
\end{equation}
Therefore, Eqs. (\ref{finalredshiftexpression}) and (\ref{redshiftexpressioninisco}) provide a unified description of the redshift effects for photons originating from both the accretion disk outside the ISCO and the plunging region between the ISCO and the event horizon, forming the basis for backward ray-tracing calculations.

\subsection{Imaging of JP black holes with closed event horizons and thin accretion disks}\label{jpblackholewithclosedeh}

As discussed above, assessing whether astrophysical black holes can be described by the Kerr metric within classical GR can be achieved by introducing parametric deviations from the Kerr solution. These deviation parameters quantify deviations of the spacetime geometry from the Kerr solution and provide a systematic framework for testing the Kerr hypothesis of astrophysical black holes. In realistic astrophysical environments, black holes are not isolated vacuum objects but are typically embedded in surrounding matter fields, which can perturb the spacetime geometry and induce deviations from the Kerr solution. Consequently, the spacetime around astrophysical black holes may not be exactly described by the Kerr metric, and deviation parameters provide an effective description of these departures. A physically viable Kerr-like metric must reduce smoothly to the Kerr solution in the limit of vanishing deviation parameters. This framework enables observational tests of whether the spacetime geometry around astrophysical black holes is consistent with the Kerr metric. Within this context, observational constraints on the deviation parameters determine whether the Kerr metric provides an adequate description or whether nonzero deviations are required. Moreover, the introduction of deviation parameters generally relaxes the assumptions underlying the black hole no-hair theorem in classical GR. While Kerr-like spacetimes with nonzero deviations may violate the theorem, the standard no-hair properties are recovered in the Kerr limit. Therefore, constraining these deviation parameters through observations provides a direct means to test both the Kerr hypothesis and the validity of the no-hair theorem. Accordingly, constructing appropriate Kerr-like metrics with deviation parameters is essential for probing the nature of astrophysical black holes and testing fundamental predictions of GR.

In many previously proposed Kerr-like metrics, the relaxation of the fundamental assumptions underlying the black hole no-hair theorem, without imposing sufficient regularity and causality conditions on the spacetime exterior to the black hole event horizon, can lead to pathological features outside the event horizon, such as curvature singularities or closed timelike curves. These spacetime pathologies are not expected in physically realistic spacetimes and significantly limit the feasibility of testing the no-hair theorem through electromagnetic observations. It is therefore essential that any viable Kerr-like metric remain free of these pathologies in the region exterior to the event horizon. A physically consistent Kerr-like metric should both accurately describe astrophysical black holes and their surrounding spacetime, and enable observational tests of the no-hair theorem, while avoiding unphysical behavior. In this respect, the JP metric satisfies these requirements and provides a suitable framework for exploring the spacetime structure of astrophysical black holes and testing the no-hair theorem in realistic environments.

In the preceding sections, the properties of the JP spacetime and the corresponding black holes have been comprehensively analyzed. In contrast to the Kerr metric, the presence of the deviation parameter $\epsilon_3$ in the JP metric introduces measurable departures from the Kerr geometry and modifies the Petrov classification of the spacetime. The Kerr spacetime belongs to Petrov type D and admits a Carter constant associated with a rank-$2$ Killing tensor. As a result, photon motion is governed by a separable Hamilton-Jacobi equation, allowing for analytic determination of photon trajectories and black hole images. By contrast, the introduction of the deviation parameter $\epsilon_3$ transforms the Kerr metric into the JP metric and changes the Petrov classification of spacetime from type D to type I, eliminating the Carter constant and rendering the Hamilton-Jacobi equation non-separable. Consequently, photon trajectories and black hole images in the JP spacetime cannot be obtained analytically. Accurate determination of photon trajectories and the construction of black hole images therefore require numerical backward ray-tracing techniques rather than analytic methods. This approach enables precise computation of photon trajectories and the resulting images of JP black holes surrounded by thin accretion disks. Since the previous subsection developed a backward ray-tracing framework for modeling images of black holes with thin accretion disks, this method is adopted in the following analysis to investigate the imaging properties of JP black holes.

Information encoded in black hole images obtained from electromagnetic observations at large distances can be used to assess the consistency of astrophysical black holes with the Kerr solution and to test the no-hair theorem within classical GR. Within the JP metric framework, a necessary first step in investigating the spacetime structure and testing the no-hair theorem for astrophysical black holes is the construction of black hole images as observed by a distant observer, enabling a detailed analysis of their imaging characteristics. The deviation from the Kerr spacetime in the JP metric is governed by the parameter $\epsilon_3$, whose variation modifies the spacetime geometry and, consequently, the observable black hole images. Since the deviation parameter $\epsilon_3$ in the JP metric can be constrained observationally, the resulting bounds on $\epsilon_3$ provide a criterion for determining whether astrophysical black holes are consistent with the Kerr metric and satisfy the no-hair theorem. Therefore, in the subsequent analysis, we systematically investigate the influence of $\epsilon_3$ on the imaging properties of JP black holes surrounded by thin accretion disks, based on images constructed using numerical backward ray-tracing approaches. Before performing the numerical calculations, it is necessary to specify the observer location and the radial extent of the accretion disk in the equatorial plane. To ensure that the observer is located sufficiently far from the JP black hole, the observer is placed at a radial coordinate $r_0 = 100$, while the accretion disk extends from the event horizon, $r_{\text{in}} = r_h$, to an outer radius $r_{\text{out}} = 20$.

\begin{figure}[htbp]
    \centering
    \begin{subfigure}[b]{0.18\textwidth}
        \includegraphics[width=\textwidth]{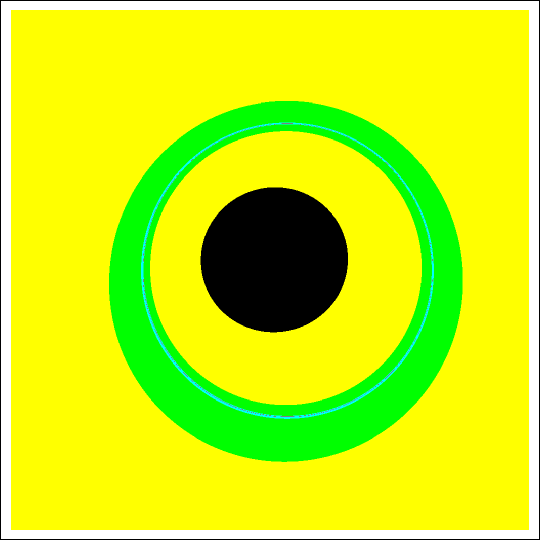}
        \caption{}
    \end{subfigure}
    \begin{subfigure}[b]{0.18\textwidth}
        \includegraphics[width=\textwidth]{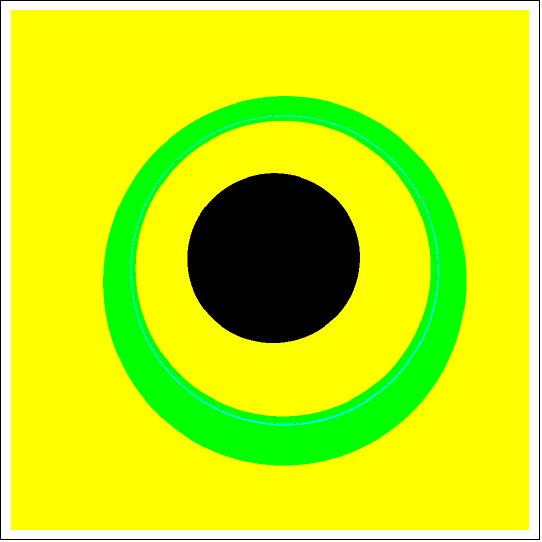}
        \caption{}
    \end{subfigure}
    \begin{subfigure}[b]{0.18\textwidth}
        \includegraphics[width=\textwidth]{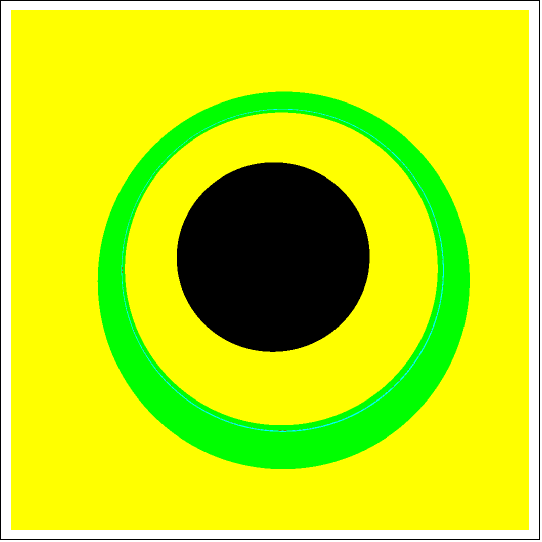}
        \caption{}
    \end{subfigure}
    \begin{subfigure}[b]{0.18\textwidth}
        \includegraphics[width=\textwidth]{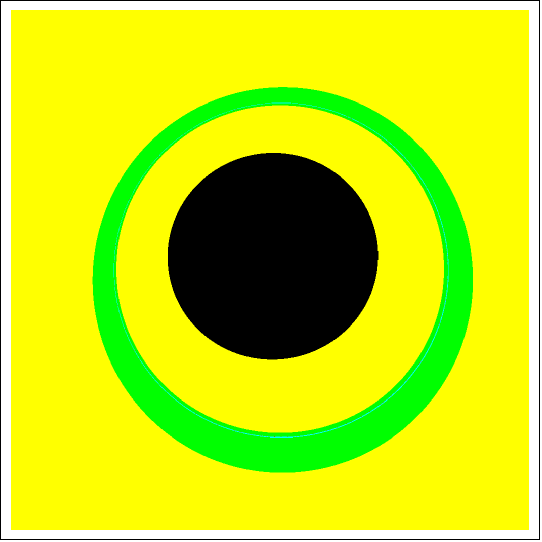}
        \caption{}
    \end{subfigure}
    \begin{subfigure}[b]{0.18\textwidth}
        \includegraphics[width=\textwidth]{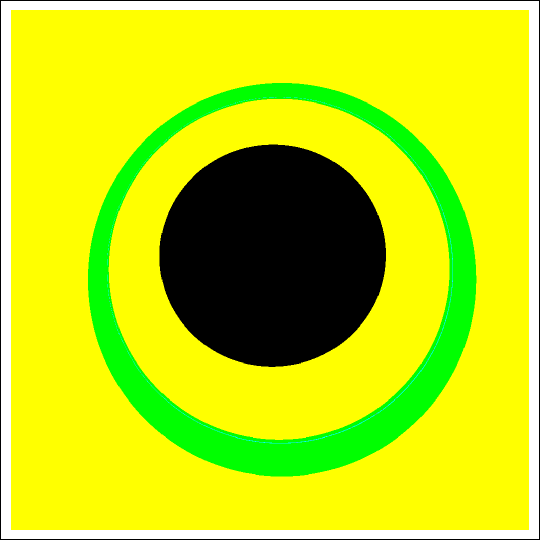}
        \caption{}
    \end{subfigure}
    \begin{subfigure}[b]{0.18\textwidth}
        \includegraphics[width=\textwidth]{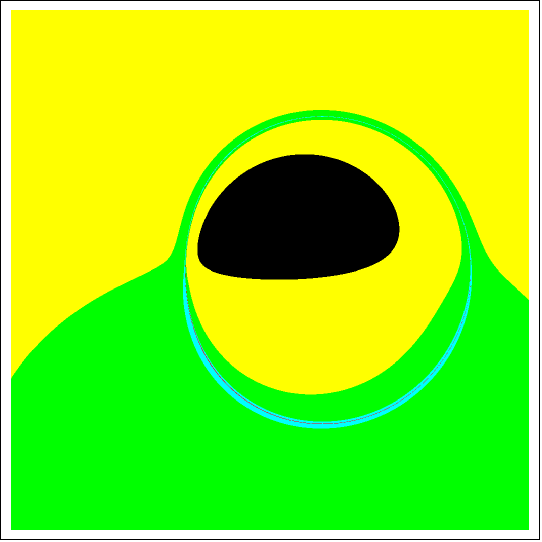}
        \caption{}
    \end{subfigure}
    \begin{subfigure}[b]{0.18\textwidth}
        \includegraphics[width=\textwidth]{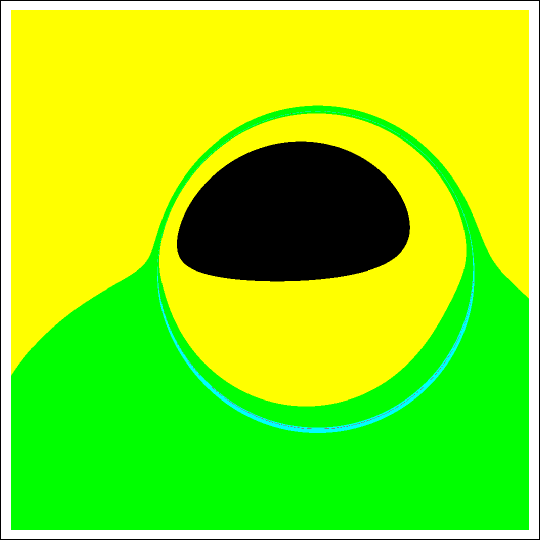}
        \caption{}
    \end{subfigure}
    \begin{subfigure}[b]{0.18\textwidth}
        \includegraphics[width=\textwidth]{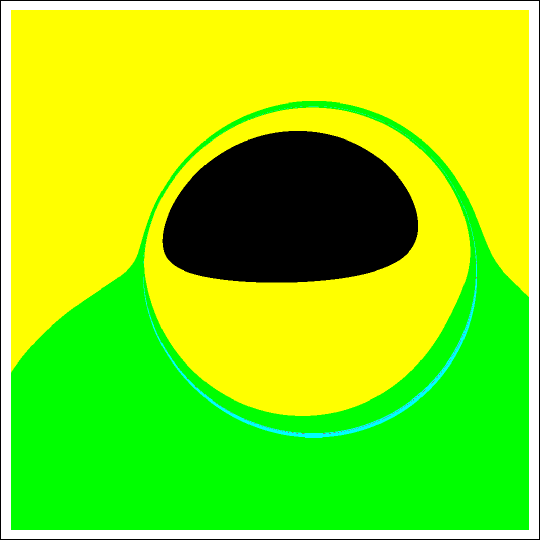}
        \caption{}
    \end{subfigure}
    \begin{subfigure}[b]{0.18\textwidth}
        \includegraphics[width=\textwidth]{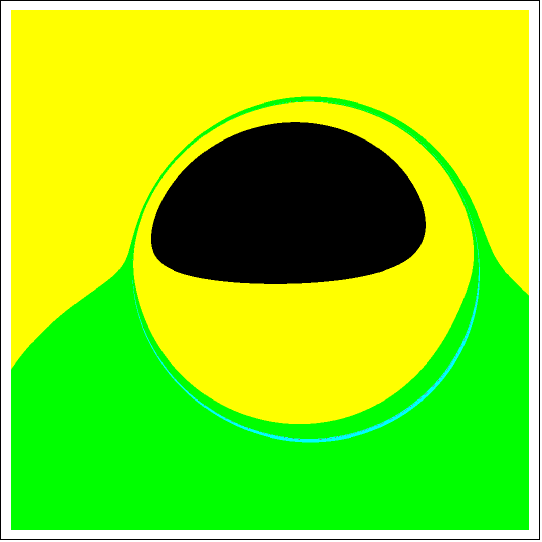}
        \caption{}
    \end{subfigure}
    \begin{subfigure}[b]{0.18\textwidth}
        \includegraphics[width=\textwidth]{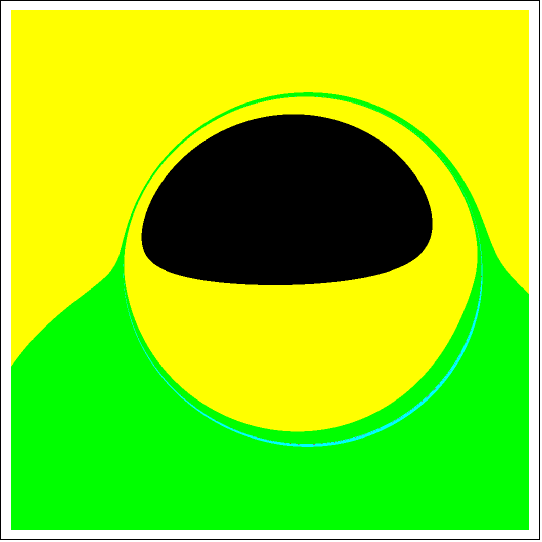}
        \caption{}
    \end{subfigure}
    \caption{Images of JP black holes with closed event horizons. Different colors denote photon trajectories classified according to the number of intersections with the accretion disk. The black, yellow, green, blue, and red regions correspond to the inner shadow, direct image, lensed image, and photon ring structures, respectively. Panels (a)-(e) show images with $a = 0.9$ and $\theta_0 = 17^\circ$ for $\epsilon_3 = 0$, $-2$, $-4$, $-6$, and $-8$, respectively. Panels (f)-(j) show the corresponding images for $\theta_0 = 80^\circ$.}
    \label{fig3}
\end{figure}

By specifying the radial coordinate of observer in the JP spacetime and the radial extent of the accretion disk surrounding the JP black hole, backward ray-tracing techniques based on the numerical framework introduced in the previous section are employed to construct images of JP black holes illuminated by thin accretion disks. This approach ensures the reliability of the resulting images and enables a precise characterization of the spacetime geometry near the JP black hole, as well as the deviations in the spacetime structure induced by variations of the parameter $\epsilon_3$ in the JP metric. Fig. \ref{fig3} shows images of JP black holes with closed event horizons, illuminated by thin accretion disks and computed using numerical backward ray-tracing, as seen by a distant observer.
To enable a systematic examination of the imaging characteristics of JP black holes illuminated by thin accretion disks, images of JP black holes constructed using numerical backward ray-tracing methods are segmented into distinct regions distinguished by different colors. In each panel of Fig. \ref{fig3}, the color-coded representation indicates the regions associated with photons that have undergone different numbers of interactions with the accretion disk during the backward ray-tracing procedure. As discussed in the previous section, the number of interactions between photons and the thin accretion disk not only determines the classification of image components but also significantly influences the corresponding intensity distribution. This number is denoted by $n$, where $n = 1$ corresponds to the direct image, $n = 2$ to the lensed image, and $n \geq 3$ to photon ring structures. A schematic illustration of photon trajectories with different values of $n$ is shown in Fig. \ref{fig2}. This color-coded representation provides an intuitive means of identifying how photons with different interaction histories contribute to the observed images. In Fig. \ref{fig3}, the central black region corresponds to the inner shadow of the JP black hole. The adjacent yellow region represents photons emitted from the accretion disk that do not intersect the accretion disk again before reaching the distant observer. The yellow region corresponds to photons that interact with the accretion disk once, denoted by $n = 1$, and represents the direct image formed by photons propagating directly from the disk to the observer. The surrounding green region corresponds to photons that intersect the accretion disk twice, denoted by $n = 2$, producing the lensed image. Within this region, additional substructures arise from photons that cross the disk three or more times before reaching the observer. These substructures are represented by the cyan and red regions, corresponding to $n = 3$ and $n \geq 4$, respectively. It is evident from Fig. \ref{fig3} that, as the number of disk intersections increases, the corresponding image features become progressively more circular. As noted above, images formed by photons with $n \geq 3$ are collectively referred to as photon rings. The yellow and green regions located in the outer parts of the image likewise correspond to the direct and lensed image components, respectively. Fig. \ref{fig3} shows images of JP black holes surrounded by prograde thin accretion disk for a spin parameter $a = 0.9$. The first row corresponds to an inclination angle $\theta_0 = 17^\circ$, while the second row shows the case $\theta_0 = 80^\circ$. From left to right, the five columns correspond to deviation parameters $\epsilon_3 = 0\,, -2\,, -4\,, -6\,,$ and $-8$, respectively. Furthermore, the accretion flow can exhibit two distinct dynamical configurations, namely prograde and retrograde motion relative to the black hole. These configurations affect only the intensity distribution of the observed images and do not alter the geometric structure of the color-coded regions associated with different values of $n$. Accordingly, the color-coded representation of the JP black hole images shown in Fig. \ref{fig3} do not include intensity variations arising from the accretion flow dynamics. The impact of these dynamical effects on the intensity distribution will be examined in detail in the subsequent analysis.

As shown in Fig. \ref{fig3}, with increasing magnitude of the deviation parameter $|\epsilon_3|$, the inner shadow of the JP black hole exhibits a gradual enlargement. This behavior is observed for both inclination angles, $\theta_0 = 17^\circ$ and $\theta_0 = 80^\circ$, indicating that the expansion of the inner shadow is largely independent of the inclination angle of the observer at infinity. For the fixed inclination angle of $\theta_0 = 17^\circ$, as the inner shadow expands with increasing $|\epsilon_3|$, both the region of direct image adjacent to the inner shadow and the photon ring shift outward. Consequently, the circumference of the photon ring increases as it moves to larger radii. However, the expansion rate of the inner shadow exceeds that of the photon ring. As a result, with increasing $|\epsilon_3|$, the region of the direct image between the inner shadow and the photon ring decreases monotonically as these two features progressively approach each other. For the inclination angle fixed at $\theta_0 = 80^\circ$, the inner shadow occupies only the upper portion of the region enclosed by the photon ring. As the deviation parameter $|\epsilon_3|$ increases, the inner shadow undergoes a similar outward expansion to that observed in the case of $\theta_0 = 17^\circ$, accompanied by a corresponding increase in the size of the photon ring. Because the inner shadow expands more rapidly than the photon ring, it progressively fills the upper portion of the enclosed region. Consequently, the region of the direct image between the inner shadow and the photon ring in the upper half diminishes, while the lower portion of the region enclosed by the photon ring continues to expand as the ring moves outward.

To quantitatively investigate the dependence of the inner shadow and photon ring in JP black hole images on the deviation parameter $\epsilon_3$, the average radii of these features are computed to analyze their variation with the parameter $\epsilon_3$. In the numerical construction of JP black hole images using backward ray-tracing techniques, the image plane is defined in the ZAMO frame of the distant observer. The observed image of the JP black hole can be constructed by projecting onto the image plane the four-momenta of photons emitted from the accretion disk and propagating toward the observer. In the numerical implementation, a two-dimensional Cartesian coordinate system is introduced on the image plane to label the position of each pixel on the black hole image. This coordinate system is oriented along directions opposite to the ZAMO basis vectors $\hat{e}_{(2)}$ and $\hat{e}_{(3)}$. Within this framework, the average radii of the inner shadow and the photon ring are defined using this Cartesian coordinate system, enabling a quantitative analysis of these two feature structures in images of JP black holes dependence on the deviation parameter $\epsilon_3$. The geometric center of these structures on the image plane are first determined by the coordinates $(x_c, y_c)$, defined as
\begin{equation}
    \begin{split}
        x_c = \frac{1}{A} \iint_D x \, \text{d} A\,, \qquad y_c = \frac{1}{A} \iint_D y \, \text{d} A\,,
    \end{split}
\end{equation}
where 
\begin{equation}
    \begin{split}
        A = \iint_D \text{d} A
    \end{split}
\end{equation}
is the area of the region $D$ corresponding to the inner shadow or the photon ring on the image plane. Furthermore, polar coordinates $(\rho, \alpha)$ are introduced on the image plane to define the average radius $\bar{r}$, with the origin of the coordinate system located at the geometric center $(x_c, y_c)$. In this coordinate system, $\rho$ measures the radial distance from the origin, while $\alpha$ specifies the angular direction on the image plane. Within this framework, the average radius $\bar{r}$ of the inner shadow and the photon ring is given by
\begin{equation}\label{averageradiusdefinition}
    \begin{split}
        \bar{r} = \frac{1}{2 \pi} \int_0^{2 \pi} \rho (\alpha) d \alpha\,.
    \end{split}
\end{equation}

As the JP metric is constructed by introducing the deviation parameter $\epsilon_3$ into the Kerr metric, the resulting spacetime geometry differs from that of the Kerr solution. Consequently, the images of JP black holes illuminated by accretion disks, as observed at infinity, exhibit measurable deviations from those of Kerr black holes. Following the definition given in Eq. (\ref{averageradiusdefinition}), an average radius $\bar{r}_{\text{Kerr}}$ is defined for Kerr black hole images to characterize the sizes of the inner shadow and photon ring. To quantify the departures from the Kerr case induced by $\epsilon_3$, the average radii of these features in JP black hole images are compared with the corresponding Kerr values defined using the same prescription. Accordingly, the dimensionless quantity $\delta = \bar{r} / \bar{r}_{\text{Kerr}}$ is introduced to quantify the relative deviation of the average radius of the inner shadow or photon ring in a JP black hole image from the corresponding Kerr value as the deviation parameter $\epsilon_3$ varies. In this definition, $\bar{r}$ denotes the average radius of the inner shadow or photon ring in the JP black hole image, whereas $\bar{r}_{\text{Kerr}}$ denotes the corresponding average radius in the Kerr black hole image. This parameter therefore provides a quantitative measure of the deviations of the inner shadow and photon ring in JP black hole images from the corresponding Kerr structures as the deviation parameter $\epsilon_3$ varies. For consistency, the reference radius $\bar{r}_{\text{Kerr}}$ is computed from Kerr black hole images with spin parameter $a = 0.9$ and adopted as a common normalization throughout the analysis. This normalization enables a systematic examination of the dependence of the average radii of the inner shadow and photon ring in JP black hole images on the spin parameter $a$ and the observer inclination angle $\theta_0$, while simultaneously quantifying the deviations induced by the parameter $\epsilon_3$. Furthermore, because prograde and retrograde accretion flows influence only the intensity distribution of the image while leaving its geometric structure unchanged, consideration of either accretion-flow configuration is sufficient to characterize the dependence of the average radii of the inner shadow and photon ring in JP black hole images on the spin parameter $a$, the observer inclination angle $\theta_0$, and the deviation parameter $\epsilon_3$.

\begin{figure}[htbp]
    \centering
    \includegraphics[width=0.8\textwidth]{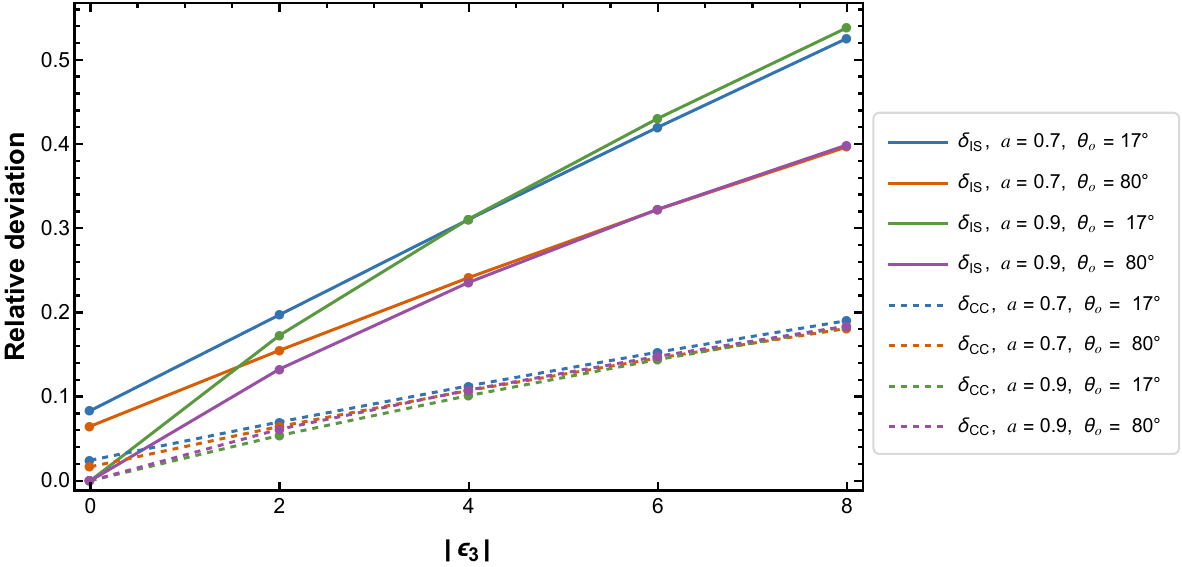}
    \caption{The evolution of the relative deviations of the average radii of the inner shadow and critical curve of JP black holes as functions of the deviation parameter $|\epsilon_3|$, measured with respect to the corresponding average radii of the inner shadow and critical curve of a Kerr black hole with spin parameter $a=0.9$. $\delta_{\text{IS}}$ denotes the relative deviation of the average radius of the JP black hole inner shadow from that of the corresponding Kerr black hole inner shadow, while $\delta_{\text{CC}}$ denotes the relative deviation of the average radius of the JP black hole critical curve from that of the corresponding Kerr black hole critical curve.}
    \label{fig4}
\end{figure}
Fig. \ref{fig4} illustrates the variation of the relative deviation parameter $\delta$ as a function of the deviation parameter $|\epsilon_3|$ for spin parameters $a = 0.7$ and $a = 0.9$, with observer inclination angles fixed at $\theta_0 = 17^\circ$ and $\theta_0 = 80^\circ$. In this figure, the average radius of the photon ring in both JP and Kerr black hole images is identified with that of the corresponding critical curve. The critical curve, defined as the locus on the image plane separating photon trajectories that fall into the black hole from those that escape to infinity, serves as a robust proxy for the photon ring. This choice facilitates a more accurate analysis of the dependence of the photon ring on the deviation parameter $\epsilon_3$. Accordingly, the evolution of the critical curve with respect to $\epsilon_3$ is used to characterize the corresponding variation of the photon ring. In Fig. \ref{fig4}, the relative deviation of the average radius of the inner shadow in JP black hole images, measured with respect to that of a Kerr black hole with $a = 0.9$, is denoted by $\delta_{\text{IS}}$, while the corresponding relative deviation of the critical curve is denoted by $\delta_{\text{CC}}$. The solid curves represent the variation of $\delta_{\text{IS}}$ with the deviation parameter $|\epsilon_3|$, with colors assigned according to the spin parameter $a$ and inclination angle $\theta_0$. Specifically, blue and green correspond to $a = 0.7$ and $a = 0.9$ at $\theta_0 = 17^\circ$, whereas red and purple curves denote the corresponding cases at $\theta_0 = 80^\circ$. The dashed curves represent $\delta_{\text{CC}}$ and follow the same color scheme.

As shown in Fig. \ref{fig4}, for a fixed observer inclination angle of $\theta_0 = 17^\circ$, the relative deviation of the inner shadow, $\delta_{\text{IS}}$, increases approximately linearly with the deviation parameter $|\epsilon_3|$ for JP black holes with the spin parameter $a = 0.7$. By contrast, for the spin parameter $a =0.9$, the evolution of the relative deviation $\delta_{\text{IS}}$ with the deviation parameter $|\epsilon_3|$ exhibits two distinct approximately linear growth regimes. In the interval $|\epsilon_3| \in [0\,, 4]$, the $\delta_{\text{IS}}$ curve corresponding to $a = 0.9$ has a noticeably steeper slope than that for $a = 0.7$, indicating a more rapid increase in the average radius of the inner shadow. However, in the interval $|\epsilon_3| \in [4\,, 8]$, the slope of the $\delta_{\text{IS}}$ curve for $a = 0.9$ decreases and becomes nearly indistinguishable from that for $a = 0.7$. This behavior indicates that, within the interval $|\epsilon_3| \in [0\,, 4]$, the average radius of the inner shadow in JP black hole images with $a = 0.9$ grows more rapidly than that for $a = 0.7$ as $|\epsilon_3|$ increases. For sufficiently large values of $|\epsilon_3|$, however, the average radii of the inner shadows for $a = 0.7$ and $a = 0.9$ become increasingly similar and subsequently increase at nearly the same rate as $|\epsilon_3|$ continues to grow. A similar trend is observed for an observer inclination angle of $\theta_0 = 80^\circ$. For $a = 0.7$, $\delta_{\text{IS}}$ increases approximately linearly with increasing $|\epsilon_3|$. For $a = 0.9$, $\delta_{\text{IS}}$ also increases nearly linearly over the interval $|\epsilon_3| \in [0\,, 4]$, with a steeper slope, indicating a more rapid growth of the average radius of the inner shadow for the higher spin case. In the interval $|\epsilon_3| \in [4\,, 8]$, the slopes of the $\delta_{\text{IS}}$ curves for the two spin parameters converge, and the two curves become nearly coincident. This behavior indicates that, within $|\epsilon_3| \in [4\,, 8]$, the average radii of the inner shadows for $a = 0.7$ and $a = 0.9$ increase at nearly identical rates, while the corresponding average radii become effectively indistinguishable. Furthermore, Fig. \ref{fig4} shows that, for both spin parameters, the growth rate of $\delta_{\text{IS}}$ with increasing $|\epsilon_3|$ is more pronounced at $\theta_0 = 17^\circ$ than at $\theta_0 = 80^\circ$. In addition, the overlap between the $\delta_{\text{IS}}$ curves corresponding to different spin parameters is more significant at $\theta_0 = 80^\circ$, particularly in the interval $|\epsilon_3| \in [4\,,8]$. This result indicates that, at higher inclination angles, the average radii of the inner shadows for $a = 0.7$ and $a = 0.9$ become more similar and exhibit a more uniform dependence on the deviation parameter. Therefore, as $|\epsilon_3|$ increases, the influence of the spin parameter on the average radius of the inner shadow becomes weaker at $\theta_0 = 80^\circ$ than at $\theta_0 = 17^\circ$. 

Meanwhile, the dashed curves in Fig. \ref{fig4} represent the relative deviation of the critical curve, $\delta_{\text{CC}}$. For different spin parameters and observer inclination angles, the dependence of $\delta_{\text{CC}}$ on the deviation parameter $|\epsilon_3|$ exhibits evolutionary behaviors distinct from those of the relative deviation of the inner shadow, $\delta_{\text{IS}}$. For an observer inclination angle of $\theta_0 = 17^\circ$, the $\delta_{\text{CC}}$ curves corresponding to $a = 0.7$ and $a = 0.9$ both increase approximately linearly with increasing $|\epsilon_3|$. Within the interval $|\epsilon_3| \in [0\,, 4]$, the curve for $a = 0.9$ exhibits a slightly steeper slope than that for $a = 0.7$, indicating a marginally faster increase in the average radius of the critical curve for the larger spin parameter. However, in the interval $|\epsilon_3| \in [4\,, 8]$, the slopes of the two $\delta_{\text{CC}}$ curves converge and become nearly indistinguishable. Consequently, the two curves $\delta_{\text{CC}}$ approach one another, implying that the average radii of the critical curves for the two spin parameters become increasingly similar and subsequently grow at nearly the same rate as $|\epsilon_3|$ increases. A similar trend is observed for an observer inclination angle of $\theta_0 = 80^\circ$. In the interval $|\epsilon_3| \in [0\,, 4]$, the $\delta_{\text{CC}}$ curves for both spin parameters again exhibit approximately linear growth, with the slope for $a = 0.9$ remaining slightly larger than that for $a = 0.7$, indicating a modestly faster growth of the critical curve radius at higher spin. For $|\epsilon_3| \in [4\,, 8]$, the two curves progressively merge and remain nearly coincident, demonstrating that the average radii of the critical curves associated with the two spin parameters become effectively indistinguishable and evolve at nearly identical rates with respect to $|\epsilon_3|$. A comparison between the results obtained for $\theta_0 = 17^\circ$ and $\theta_0 = 80^\circ$ reveals that, within the interval $|\epsilon_3| \in [4\,, 8]$, the overlap between the $\delta_{\text{CC}}$ curves corresponding to $a = 0.7$ and $a = 0.9$ is more pronounced at $\theta_0 = 80^\circ$. In particular, for $\theta_0 = 80^\circ$, the two curves become almost indistinguishable throughout this interval. This behavior indicates that, as $|\epsilon_3|$ becomes sufficiently large, the dependence of the average radius of the critical curve on the spin parameter becomes progressively weaker, especially at larger observer inclination angles. Furthermore, unlike the relative deviation of the inner shadow, $\delta_{\text{IS}}$, the four $\delta_{\text{CC}}$ curves corresponding to $a = 0.7$ and $a = 0.9$ at observer inclination angles of $\theta_0 = 17^\circ$ and $\theta_0 = 80^\circ$ almost completely overlap within the interval $|\epsilon_3| \in [4\,, 8]$. This result indicates that, for sufficiently large values of $|\epsilon_3|$, the evolution of the average radius of the critical curve becomes nearly independent of both the spin parameter and the observer inclination angle. In this regime, the evolution of the average radius of the critical curve is governed primarily by the deviation parameter $|\epsilon_3|$, while the effects of spin parameter and inclination angle become comparatively subdominant.

Finally, Fig. \ref{fig4} indicates that, for both $\theta_0 = 17^\circ$ and $\theta_0 = 80^\circ$, the relative deviations of the average radii of the inner shadow, $\delta_{\text{IS}}$, and the critical curve, $\delta_{\text{CC}}$, increase approximately monotonically with the deviation parameter $|\epsilon_3|$. The slope of the $\delta_{\text{IS}}$ curve consistently exceeds that of the $\delta_{\text{CC}}$, indicating that the average radius of the inner shadow grows more rapidly than that of the critical curve as $|\epsilon_3|$ increases. Consequently, the boundary of the inner shadow progressively approaches the critical curve associated with the photon ring with increasing the deviation parameter $|\epsilon_3|$, in agreement with the morphological evolution shown in Fig. \ref{fig3}. Furthermore, the growth rate of the average radius of the inner shadow with respect to $|\epsilon_3|$ is larger for $\theta_0 = 17^\circ$ than for $\theta_0 = 80^\circ$, whereas the growth rate of the average radius of the critical curve remains nearly identical for the two observer inclination angles. As a result, the separation between the inner shadow and the critical curve decreases more rapidly with increasing $|\epsilon_3|$ for $\theta_0 = 17^\circ$ than for $\theta_0 = 80^\circ$.

In the preceding analysis of the imaging properties of JP black holes surrounded by thin accretion disks as observed at infinity, the focus has been on the structural features of the resulting images. In particular, the spatial distributions of the direct and lensed image components have been examined, and the concept of an average radius has been introduced to characterize the characteristic scales of the inner shadow and the photon ring, as well as their dependence on the deviation parameter $\epsilon_3$. A quantitative assessment has been performed by evaluating the average radii of the inner shadow and the photon ring, thereby determining how these characteristic scales vary with the observer inclination angles $\theta_0 = 17^\circ$ and $\theta_0 = 80^\circ$, the black hole spin parameters $a =0.7$ and $a = 0.9$, and the deviation parameter $\epsilon_3$ in the JP metric. However, although the preceding analysis has systematically explored the dependence of image morphology of JP black holes on the parameters $a$, $\theta_0$, and $\epsilon_3$, the resulting JP black hole images encode not only the geometric distribution of structures on the image plane but also the associated photon intensity distribution. Consequently, a comprehensive characterization of the imaging properties requires a further analysis of the intensity profiles. Since deviations from the Kerr solution in the JP metric are governed by the deviation parameter $\epsilon_3$, which affects both the imaging morphology and the photon intensity distribution, it is necessary to investigate how the intensity distribution depends on $\epsilon_3$. Accordingly, the following analysis focuses on the intensity patterns in JP black hole images and the dependence of the intensity distribution of the JP black hole images on the deviation parameter $\epsilon_3$. Furthermore, since the accretion flow in a thin disk around a JP black hole with a closed event horizon can be either prograde or retrograde, these distinct dynamical configurations can significantly influence the observed intensity distribution. Therefore, the effects of prograde and retrograde accretion flows are examined separately in the analysis of the intensity profiles.

\begin{figure}[htbp]
    \centering
    \begin{subfigure}[b]{0.4\textwidth}
        \includegraphics[width=\textwidth]{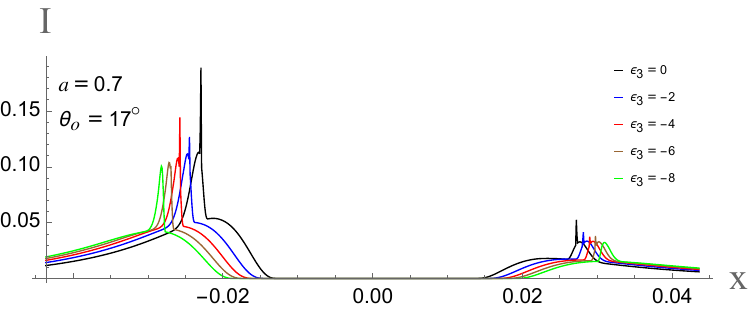}
        \caption{}
    \end{subfigure}
    \begin{subfigure}[b]{0.4\textwidth}
        \includegraphics[width=\textwidth]{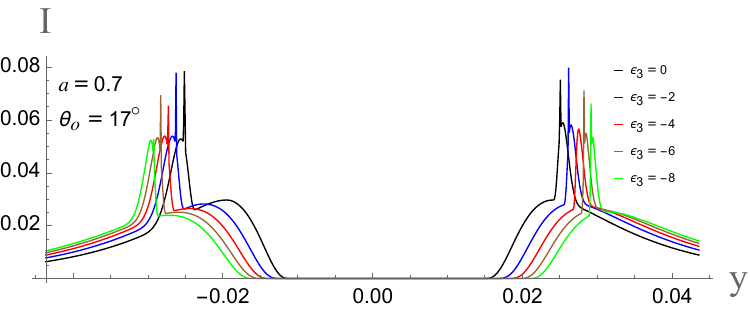}
        \caption{}
    \end{subfigure}
    \begin{subfigure}[b]{0.4\textwidth}
        \includegraphics[width=\textwidth]{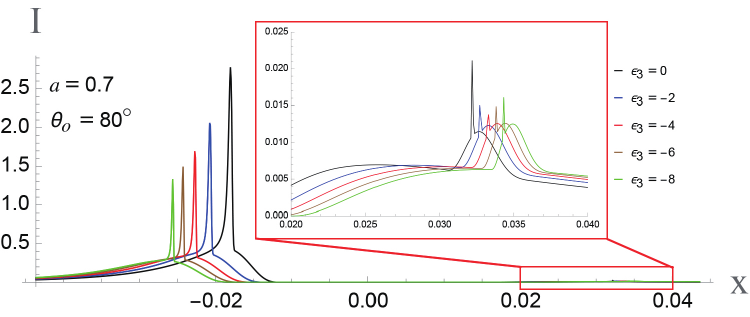}
        \caption{}
    \end{subfigure}
    \begin{subfigure}[b]{0.4\textwidth}
        \includegraphics[width=\textwidth]{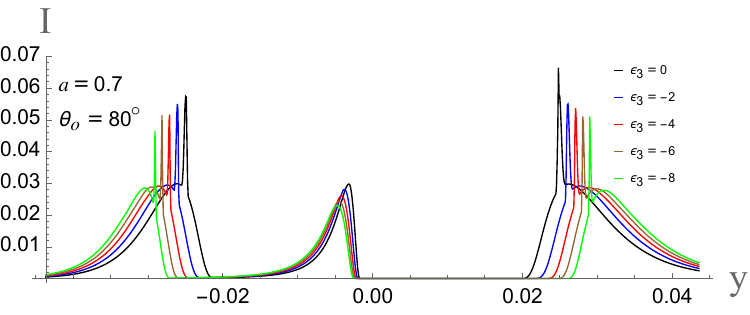}
        \caption{}
    \end{subfigure}
    \begin{subfigure}[b]{0.4\textwidth}
        \includegraphics[width=\textwidth]{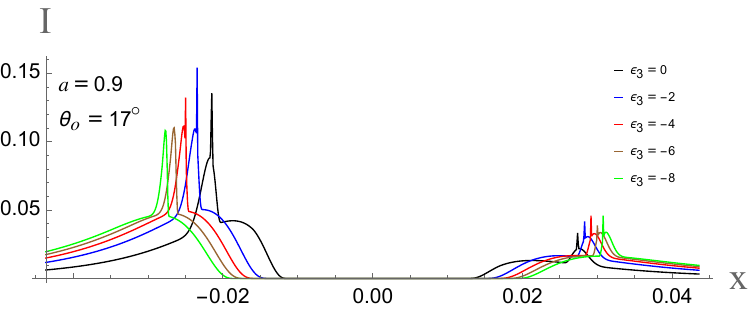}
        \caption{}
    \end{subfigure}
    \begin{subfigure}[b]{0.4\textwidth}
        \includegraphics[width=\textwidth]{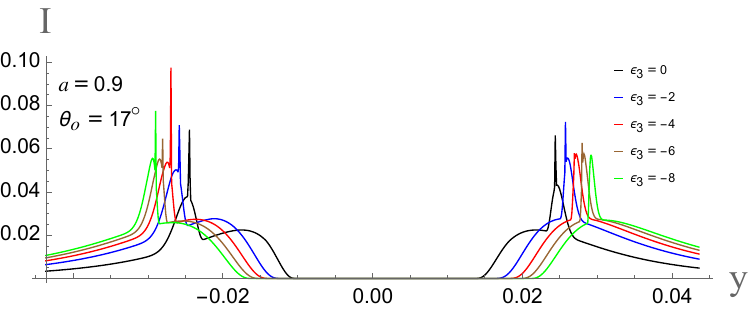}
        \caption{}
    \end{subfigure}
    \begin{subfigure}[b]{0.4\textwidth}
        \includegraphics[width=\textwidth]{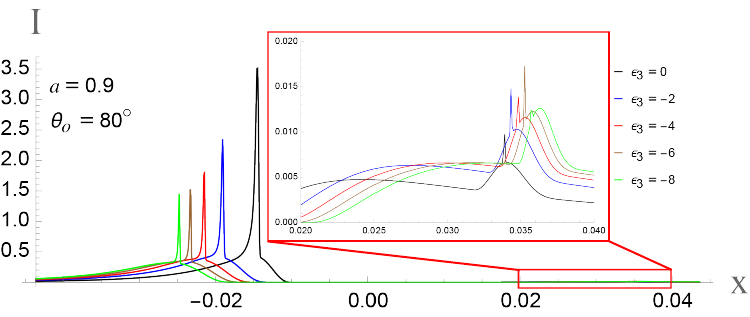}
        \caption{}
    \end{subfigure}
    \begin{subfigure}[b]{0.4\textwidth}
        \includegraphics[width=\textwidth]{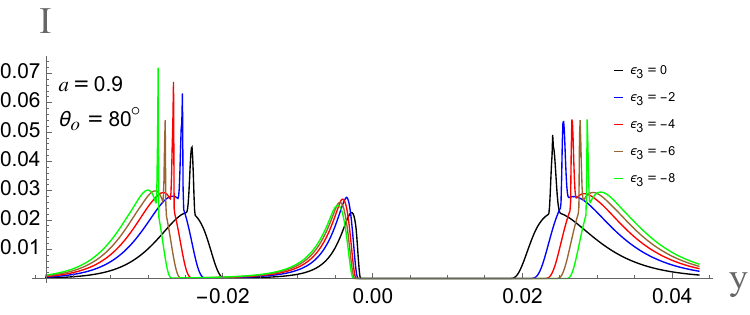}
        \caption{}
    \end{subfigure}
    \caption{The evolution of the intensity distributions along the $x$- and $y$-directions on the image plane for JP black holes with spin parameters $a = 0.7$ and $a = 0.9$, illuminated by a prograde accretion disk and observed at inclination angles of $\theta_0 = 17^\circ$ and $\theta_0 = 80^\circ$, for increasing values of the deviation parameter $|\epsilon_3|$.}
    \label{fig5}
\end{figure}

Fig. \ref{fig5} depicts the intensity distributions of JP black hole images illuminated by a prograde thin accretion disk for various values of the spin parameter $a$ and the observer inclination angle $\theta_0$. Each panel illustrates the dependence of the intensity distribution on the deviation parameter $\epsilon_3$ for the specified values of $a$ and $\theta_0$.
For fixed observer inclination angles $\theta_0 = 17^\circ$ and $\theta_0 = 80^\circ$, the intensity profiles for JP black hole images with spin parameters $a = 0.7$ and $a = 0.9$ exhibit qualitatively similar behavior as the deviation parameter $|\epsilon_3|$ increases. In particular, the regions of peak intensity progressively shift outward along both the $x$- and $y$-axes with increasing $|\epsilon_3|$. This behavior is consistent with the preceding analysis of the average radii of the inner shadow and the photon ring, both of which expand outward as $|\epsilon_3|$ increases. The associated geometric expansion leads to a corresponding outward extension of the intensity distribution along both axes.

Furthermore, as shown in the intensity distributions in Fig. \ref{fig5}, when the observer inclination angle is fixed at $\theta_0 = 17^\circ$, a comparison of the intensity distributions along the $x$- and $y$-axes for the JP black hole images with spin parameters $a = 0.7$ and $a = 0.9$ reveals that, as the deviation parameter $|\epsilon_3|$ increases, the locations of the intensity peaks along both axes in the image of the JP black hole with $a = 0.7$ progressively converge toward those in the image of the JP black hole with $a = 0.9$. This indicates that, with increasing $|\epsilon_3|$, the positions of the intensity maxima along the Cartesian axes on the image plane defined in the ZAMO frame for images of JP black holes with spin parameters $a = 0.7$ and $a = 0.9$ become increasingly aligned. For example, by comparing Figs. \ref{fig5} (a) and \ref{fig5} (e), one observes that when $|\epsilon_3| = 0$, the positions of the intensity peaks along the $x$-axis, represented by the black curves, differ between the JP black hole images with spin parameters $a = 0.7$ and $a = 0.9$ in these two figures. However, as the deviation parameter $|\epsilon_3|$ increases and reaches $|\epsilon_3| = 8$, the intensity peak positions along the $x$-axis for both JP black hole images with spin parameters $a = 0.7$ and $a = 0.9$ converge to nearly the same location, as indicated by the green curves in these two figures. A similar trend is observed for the intensity distributions along the $y$-axis in the images of JP black hole. By comparing Figs. \ref{fig5} (b) and \ref{fig5} (f), one finds that for $|\epsilon_3| = 0$, the positions of the intensity peaks along the $y$-axis differ between the JP black hole images with spin parameters $a = 0.7$ and $a = 0.9$. As the deviation parameter $|\epsilon_3|$ increases to $|\epsilon_3| = 8$, the locations of the intensity peaks along the $y$-axis in JP black hole images with $a = 0.7$ and $a = 0.9$ gradually converge and eventually coincide. Moreover, for an observer inclination angle of $\theta_0 = 80^\circ$, a detailed comparison of the intensity distributions and the corresponding peak positions in Figs. \ref{fig5} (c) and \ref{fig5} (g), as well as Figs. \ref{fig5} (d) and \ref{fig5} (h), shows that for JP black holes with spin parameters $a = 0.7$ and $a = 0.9$, the coordinates of the intensity maxima along both the $x$- and $y$-axes progressively converge as the deviation parameter $|\epsilon_3|$ increases. In particular, when $|\epsilon_3| = 8$, the intensity peak positions along both axes in the JP black hole images become nearly indistinguishable for the two spin values. As discussed previously in the analysis of the variations of the inner shadow and the photon ring in JP black hole images with changes in the deviation parameter $|\epsilon_3|$, this behavior arises because increasing $|\epsilon_3|$ causes both the average radii and the growth rates of the inner shadow and the critical curve for $a = 0.7$ and $a = 0.9$ to converge. Consequently, the intensity peak positions in JP black hole images with different spin parameters become nearly identical along both the $x-$ and $y-$ axes as the deviation parameter $|\epsilon_3|$ increases.

Building upon these results, the intensity distribution characteristics of JP black hole images illuminated by a prograde thin accretion disk can be further explored by examining how the intensity distributions shown in Fig. \ref{fig5} vary with the black hole spin parameter $a$, the observer inclination angle $\theta_0$, and the deviation parameter $\epsilon_3$. Before investigating the properties of the intensity distributions in JP black hole images and their dependence on various parameters, it is necessary to clarify the correspondence between the peaks appearing in the intensity profiles and the image structures of the JP black hole. Taking the black curve in Fig. \ref{fig5} (a) as an example, this curve represents the intensity profile along the $x-$ axis of the image plane for a JP black hole with spin parameter $a = 0.7$, observer inclination angle $\theta_0 = 17^\circ$, and deviation parameter $\epsilon_3 = 0$. In the region $x < 0$, the intensity profile is characterized by two distinct peaks, namely a sharp and narrow peak and an adjacent arc-shaped peak. A comparison between the intensity profile and the corresponding JP black hole image shows that the position and amplitude of the sharp peak correspond to those of the photon ring, whereas the arc-shaped peak corresponds to the intensity enhancement produced by the superposition of the direct and lensed images immediately adjacent to the photon ring.Further numerical investigations based on backward ray-tracing calculations indicate that the amplitude of the photon ring intensity peak is highly sensitive to the numerical resolution adopted in the image construction. Consequently, the height of the photon ring peak may vary significantly with changes in numerical accuracy. For this reason, the photon ring peak cannot be employed to reliably characterize the dependence of the intensity distribution on the black hole spin parameter $a$, the observer inclination angle $\theta_0$, and the deviation parameter $\epsilon_3$. By contrast, an analysis of the arc-shaped peak produced by the superposition of the direct and lensed images demonstrates that its intensity is nearly independent of the numerical resolution. Therefore, in the following discussion of the intensity distribution properties of JP black hole images with closed event horizons and their evolution with respect to $a$, $\theta_0$, and $\epsilon_3$, attention is restricted to the evolution of the position and intensity of the arc-shaped peak generated by the superposition of the direct and lensed images on the image plane.

From the intensity distributions of the JP black hole images with spin parameter $a = 0.7$ shown in panels (a)-(d) of Fig. \ref{fig5}, it can be seen that, for both observer inclination angles $\theta_0 = 17^\circ$ and $\theta_0 = 80^\circ$, the intensity peaks on both sides of the $x-$ and $y-$ axes generally decrease monotonically with increasing $|\epsilon_3|$. The only exception occurs in Fig. \ref{fig5} (c), which corresponds to the intensity distribution along the $x-$ axis for $\theta_0 = 80^\circ$. As revealed by the enlarged view of the intensity profile, the peak in the region $x > 0$ exhibits a slight increase as the deviation parameter $|\epsilon_3|$ increases. In addition, for JP black hole images with spin parameter $a = 0.7$ and an observer inclination angle of $\theta_0 = 80^\circ$, an additional intensity peak emerges near the center of the intensity distribution along the $y-$ axis, as shown in Fig. \ref{fig5} (d). The intensity of this central peak likewise exhibits an overall decreasing trend with increasing $|\epsilon_3|$. For the intensity distributions of the JP black hole images with spin parameter $a = 0.9$ shown in panels (e)-(h) of Fig. \ref{fig5}, the left-hand intensity peak along the $x-$ axis in Fig. \ref{fig5}(e), corresponding to $\theta_0 = 17^\circ$, exhibits a non-monotonic dependence on $|\epsilon_3|$, increasing initially and subsequently decreasing as $|\epsilon_3|$ increases. However, the corresponding peak in Fig. \ref{fig5} (g), obtained for an observer inclination angle of $\theta_0 = 80^\circ$, decreases monotonically with increasing $|\epsilon_3|$. The right-hand intensity peak along the $x-$ axis, on the other hand, exhibits a slight but systematic increase with increasing $|\epsilon_3|$ for both observer inclination angles $\theta_0 = 17^\circ$ and $\theta_0 = 80^\circ$. For a spin parameter of $a = 0.9$, the evolution of the intensity peaks on both sides of the $y$-axis with increasing $|\epsilon_3|$ differs qualitatively from that observed in the corresponding $a = 0.7$ cases for both observer inclination angles $\theta_0 = 17^\circ$ and $\theta_0 = 80^\circ$. As $|\epsilon_3|$ increases, the peak intensities on both sides of the $y-$ axis generally exhibit an increasing trend. In particular, for the observer inclination angle fixed at $\theta_0 = 80^\circ$, the evolution of the central intensity peak along the $y-$ axis in the JP black hole images with $a = 0.9$, shown in Fig. \ref{fig5} (h), differs markedly from that observed in the corresponding $a = 0.7$ case shown in Fig. \ref{fig5} (d). For the intensity distribution of the JP black hole image with $a = 0.7$, the central peak exhibits the same dependence on the deviation parameter $|\epsilon_3|$ as the off-center peaks along the $y-$ axis, decreasing monotonically as $|\epsilon_3|$ increases. By contrast, for JP black hole images with spin parameter $a = 0.9$, the central peak intensity along the $y$-axis first increases and then decreases with increasing $|\epsilon_3|$. This behavior differs from that observed for $a = 0.7$ and suggests the existence of a distinct non-monotonic feature in the central intensity distribution of rapidly rotating JP black holes viewed at high inclination angles.

Moreover, the intensity distribution in the image of a JP black hole illuminated by a thin accretion disk is influenced by the dynamical state of the accreting material within the accretion disk. Therefore, following the analysis of the intensity distribution in the prograde accretion disk case, it is necessary to further examine in detail the corresponding intensity distribution in the case where the accreting material undergoes retrograde motion.
\begin{figure}[htbp]
    \centering
    \begin{subfigure}[b]{0.4\textwidth}
        \includegraphics[width=\textwidth]{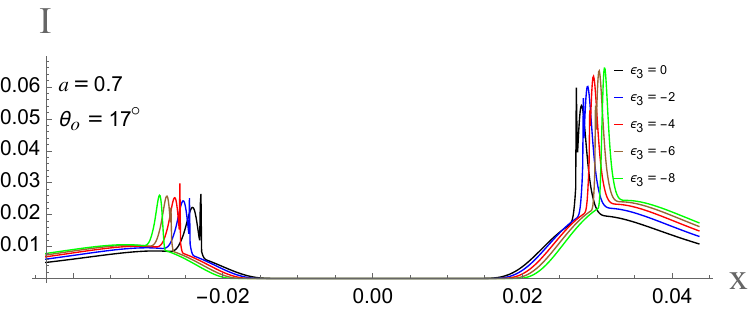}
        \caption{}
    \end{subfigure}
    \begin{subfigure}[b]{0.4\textwidth}
        \includegraphics[width=\textwidth]{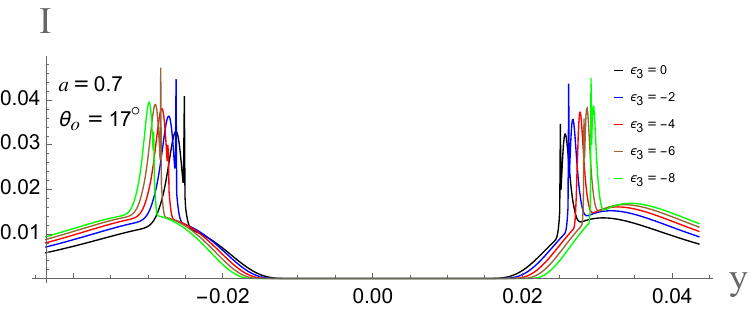}
        \caption{}
    \end{subfigure}
    \begin{subfigure}[b]{0.4\textwidth}
        \includegraphics[width=\textwidth]{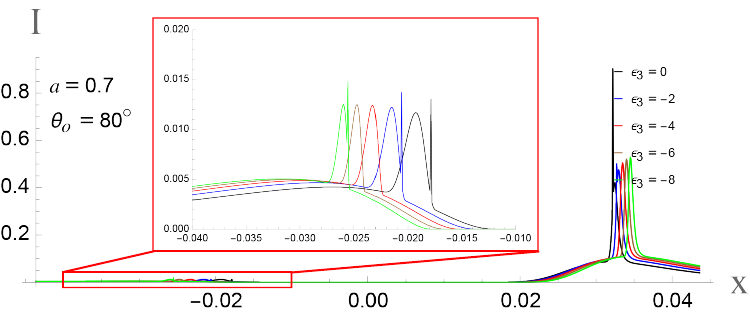}
        \caption{}
    \end{subfigure}
    \begin{subfigure}[b]{0.4\textwidth}
        \includegraphics[width=\textwidth]{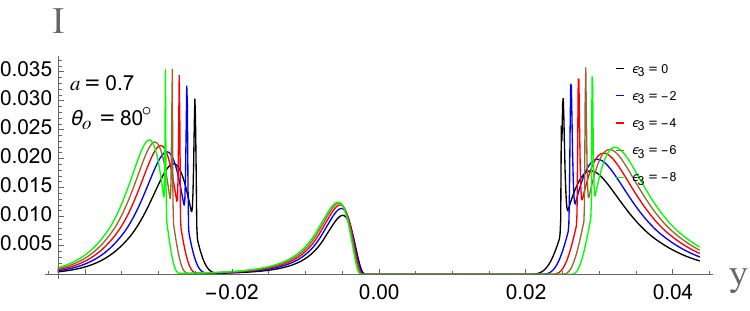}
        \caption{}
    \end{subfigure}
    \begin{subfigure}[b]{0.4\textwidth}
        \includegraphics[width=\textwidth]{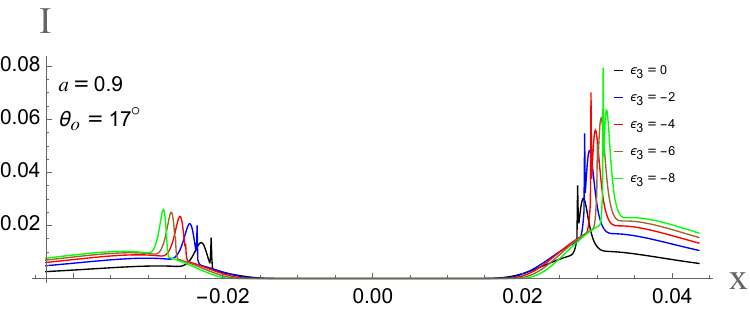}
        \caption{}
    \end{subfigure}
    \begin{subfigure}[b]{0.4\textwidth}
        \includegraphics[width=\textwidth]{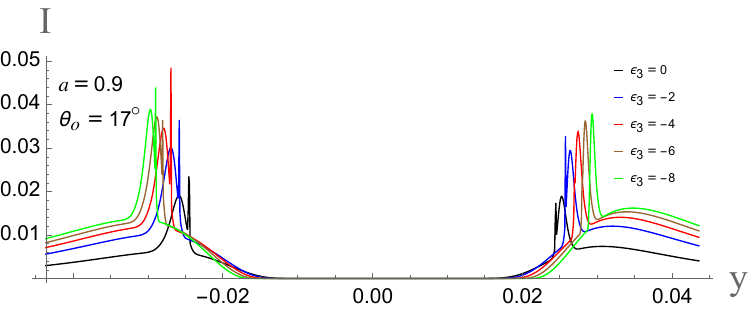}
        \caption{}
    \end{subfigure}
    \begin{subfigure}[b]{0.4\textwidth}
        \includegraphics[width=\textwidth]{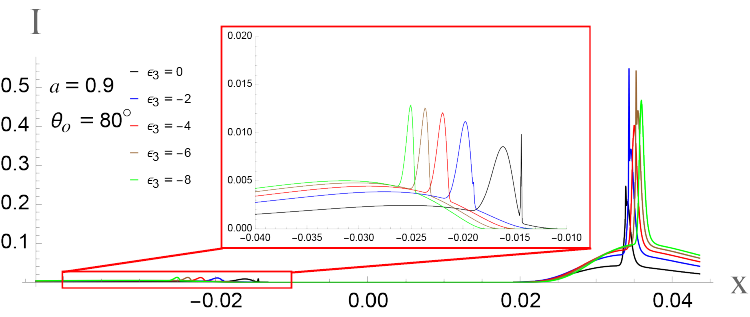}
        \caption{}
    \end{subfigure}
    \begin{subfigure}[b]{0.4\textwidth}
        \includegraphics[width=\textwidth]{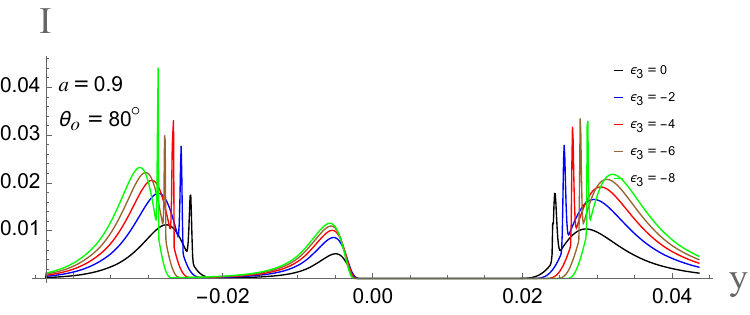}
        \caption{}
    \end{subfigure}
    \caption{The evolution of the intensity distributions along the $x$- and $y$-directions on the image plane for JP black holes with spin parameters $a = 0.7$ and $a = 0.9$, illuminated by a retrograde accretion disk and viewed at inclination angles $\theta_0 = 17^\circ$ and $\theta_0 = 80^\circ$, as the magnitude of the deviation parameter $|\epsilon_3|$ increases.}
    \label{fig6}
\end{figure}
Fig. \ref{fig6} presents the intensity distributions in images of JP black holes illuminated by retrograde accreting matter in the accretion disk, as observed by a distant observer. By comparing Figs. \ref{fig6} (a) and \ref{fig6} (e), as well as Figs. \ref{fig6} (b) and \ref{fig6} (f), it is observed that, when the observer inclination angle is fixed at $\theta_0 = 17^\circ$, the positions of intensity peaks along both the $x-$ and $y-$ axes on the image plane for JP black hole images with spin parameters $a = 0.7$ and $a = 0.9$ progressively converge as the deviation parameter $|\epsilon_3|$ increases. When $|\epsilon_3|$ reaches a value of $8$, the coordinates of the intensity peaks along the $x-$ and $y-$ axes in the Cartesian coordinate system on the image plane become nearly identical, indicating that the intensity distributions along these two axes are effectively indistinguishable. Similarly, for the observer inclination angle fixed at $\theta_0 = 80^\circ$, a comparison of Figs. \ref{fig6} (c) and \ref{fig6} (g), as well as Figs. \ref{fig6} (d) and \ref{fig6} (h), shows that as the deviation parameter $|\epsilon_3|$ increases, the positions of the intensity peaks along the $x-$ and $y-$ axes in JP black hole images with spin parameter $a = 0.7$ gradually converge toward those in the corresponding JP black hole images with spin parameter $a = 0.9$. This behavior indicates that, for an observer inclination angle of $\theta_0 = 80^\circ$, the dependence of the intensity distribution on the deviation parameter $|\epsilon_3|$ exhibits the same qualitative trend as that observed at $\theta_0 = 17^\circ$. The evolution of the intensity peak positions with increasing $|\epsilon_3|$ in JP black hole images illuminated by retrograde accretion flows follows the same pattern as in the prograde case. The observed convergence of the intensity distributions in JP black hole images corresponding to different spin parameters can be attributed to the fact that, for a fixed observer inclination, the average radii of the inner shadow and the photon ring exhibit nearly identical growth behaviors as the deviation parameter $|\epsilon_3|$ increases, largely independent of the black hole spin. Since the average radii of the inner shadow and the photon ring vary with the deviation parameter $|\epsilon_3|$ independently of the direction of the accretion flow, the evolution of the intensity peak positions with increasing $|\epsilon_3|$ in JP black hole images illuminated by retrograde accretion flows closely mirrors that in images illuminated by prograde accretion flows. Accordingly, because the average radii of the inner shadow and the photon ring are insensitive to the direction of the accretion flow, the intensity peak distributions in JP black hole images illuminated by prograde accretion flows are expected to be nearly identical to those in images illuminated by retrograde accretion flows. This expectation is confirmed by a direct comparison of the intensity peak locations along the $x-$ and $y-$ axes in the corresponding panels of Figs. \ref{fig5} and \ref{fig6}.

Furthermore, Fig. \ref{fig6} enables a more detailed examination of the dependence of the intensity distribution in JP black hole images illuminated by retrograde accretion flows on the magnitude of the deviation parameter $\epsilon_3$. As shown in Fig. \ref{fig6}, for both observer inclination angles $\theta_0 = 17^\circ$ and $\theta_0 = 80^\circ$, the peak intensities in the intensity distributions of JP black hole images with spin parameters $a = 0.7$ and $a = 0.9$ increases monotonically with increasing $|\epsilon_3|$. In particular, as illustrated in Figs. \ref{fig6} (f) and \ref{fig6} (h), corresponding to observer inclination angles $\theta_0 = 17^\circ$ and $\theta_0 = 80^\circ$, respectively, the evolution of the intensity distribution along the $y-$ axis in JP black hole images with spin parameter $a = 0.9$ differs significantly from that observed in the prograde accretion case. Specifically, for both inclination angles, the intensity distribution along the $y-$ axis in the JP black hole images with $a = 0.9$ does not exhibit the anomalous behavior observed in the prograde scenario as $|\epsilon_3|$ increases. Instead, the intensity peaks follow a monotonic increasing trend with $|\epsilon_3|$, consistent with the behavior observed along both the $x-$ and $y-$ axes for the intensity distributions corresponding to other inclination angles and spin parameters. Moreover, as shown in Figs. \ref{fig6} (d) and \ref{fig6} (h), for an observer inclination angle of $\theta_0 = 80^\circ$, an additional intensity peak emerges near the center of the intensity profiles along the $y-$ axis in JP black hole images with spin parameters $a = 0.7$ and $a = 0.9$. For JP black hole images with both spin values illuminated by retrograde accretion flows, the central intensity peak increases monotonically as $|\epsilon_3|$ increases. The dependence of this central peak on $|\epsilon_3|$ is consistent with the general trend observed for other inclination angles and spin parameters, namely that the intensity along both the $x-$ and $y-$ axes increases as the deviation parameter $|\epsilon_3|$ increases. In particular, a comparison between the intensity distributions shown in Fig. \ref{fig6} (h) and Fig. \ref{fig5} (h) indicates that, when the observer inclination angle is fixed at $\theta_0 = 80^\circ$ and the spin parameter of the JP black hole is $a = 0.9$, the evolution of the central intensity peak with increasing $|\epsilon_3|$ differs qualitatively between the retrograde and prograde accretion scenarios. In contrast to the prograde case, the anomalous behavior characterized by an initial increase followed by a subsequent decrease in the central intensity is absent in the retrograde accretion scenario. Instead, the central intensity exhibits a monotonic increase with increasing deviation parameter $|\epsilon_3|$.

In the preceding analysis of the inner shadow and photon ring configurations in JP black hole images with spin parameters $a = 0.7$ and $a = 0.9$, observed at inclination angles of $\theta_0 = 17^\circ$ and $\theta_0 = 80^\circ$, the average radius was introduced to characterize the relative deviations of the characteristic scales of the inner shadow and photon ring with respect to those of a Kerr black hole with spin parameter $a = 0.9$ as functions of the deviation parameter $|\epsilon_3|$ in Fig. \ref{fig5}. However, the evolution of the average radii with increasing $|\epsilon_3|$ reflects only the global variation of the inner shadow and photon ring configurations and does not distinguish whether these changes are primarily governed by variations along the specific directions on the image plane. Since the intensity distributions of the resulting JP black hole images are evaluated along the $x-$ and $y-$ directions of the Cartesian coordinate system defined on the image plane, and since the characteristic scales of the inner shadow and photon ring along these two directions can be clearly identified from the corresponding intensity profiles, the intensity distribution curves in JP black hole images illuminated by prograde and retrograde accretion flows illustrated along the $x-$ and $y-$ axes in Figs. \ref{fig5} and \ref{fig6} can be further utilized to determine whether the evolution of the inner shadow and photon ring with increasing deviation parameter $\epsilon_3$ is predominantly governed by variations along the $x-$ direction or the $y-$ direction. Therefore, a more detailed analysis of the characteristic scales of the inner shadow and photon ring along the Cartesian $x-$ and $y-$ directions, measured relative to the corresponding Kerr configurations, is required in order to determine whether the evolution of the inner shadow and photon ring with increasing deviation parameter $\epsilon_3$ is predominantly governed by variations along the $x-$ axis or the $y-$ axis. This analysis method therefore provides further insight into the physical mechanisms underlying the evolution of the inner shadow and photon ring in JP black hole images as functions of the deviation parameter $\epsilon_3$. 

\begin{figure}[htbp]
    \centering
    \begin{subfigure}[b]{0.4\textwidth}
        \includegraphics[width=\textwidth]{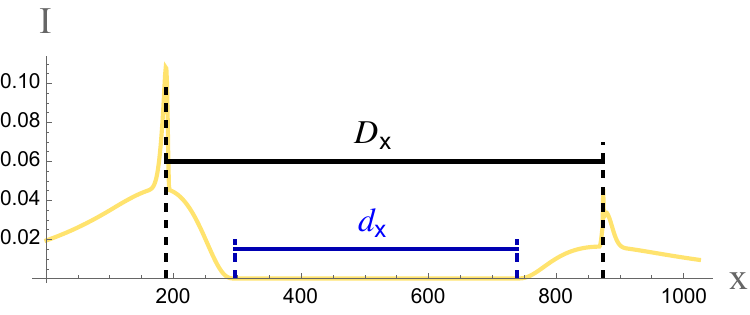}
        \caption{}
    \end{subfigure}
    \begin{subfigure}[b]{0.4\textwidth}
        \includegraphics[width=\textwidth]{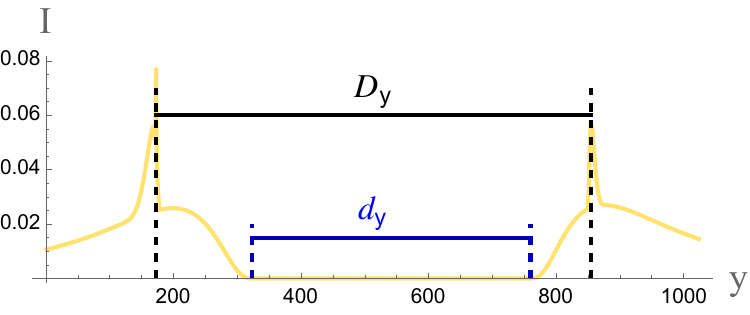}
        \caption{}
    \end{subfigure}
    \caption{Definition of the characteristic scales of the inner shadow and the photon ring on the image plane inferred from the intensity distributions of JP black hole images. $d_x$ and $d_y$ denote the characteristic scales of the inner shadow along the $x$- and $y$-directions, respectively, whereas $D_x$ and $D_y$ denote the corresponding characteristic scales of the photon ring.}
    \label{fig7}
\end{figure}

The intensity distribution shown in Fig. \ref{fig7} corresponds to the intensity distribution in an image of a JP black hole with a definite spin parameter, observer inclination angle, and deviation parameter, measured along the Cartesian $x-$ and $y-$ axes on the image plane.
As illustrated in Fig. \ref{fig7}, the characteristic scales of the inner shadow and the photon ring along the directions of the Cartesian $x-$ and $y-$ axes in JP black hole images can be quantified through the intensity distributions. Fig. \ref{fig7} (a) presents the intensity distribution along the $x-$ axis, in which the two intensity maxima correspond to the boundaries of the photon ring in the direction along the $x-$ axis. The separation between these two positions of intensity maxima is therefore characterized by the quantity $D_x$, which represents the characteristic scale of the photon ring along the $x-$ axis. Meanwhile, the zero intensity region enclosed between the two intensity peaks corresponds to the inner shadow, and the boundaries of this region correspond to the boundaries of the inner shadow along the $x-$ axis. Consequently, the quantity $d_x$ is introduced to characterize the scale of the inner shadow along the $x-$ axis. Analogously, Fig. \ref{fig7} (b) shows the intensity distribution of the JP black hole image along the $y-$ axis. In this case, the intensity maxima are defined as the boundaries of the photon ring along the direction of the $y-$ axis, whereas the boundaries of the region within which the intensity vanishes and which is enclosed by the two intensity peaks correspond to those of the inner shadow along the same axis. Consequently, the quantity $D_y$ is introduced to characterize the scale of the photon ring along the $y-$ axis, while $d_y$ characterizes the corresponding scale of the inner shadow.

By employing the characteristic scales $d_x$, $d_y$, $D_x$, and $D_y$, defined for the inner shadow and photon ring along the Cartesian $x-$ and $y-$ axes on the image plane, the evolutionary properties of the configurations of the inner shadow and photon ring with respect to the deviation parameter $\epsilon_3$ can be further investigated. To quantify the extent of the variations of these characteristic scales, an approach analogous to that employed in the analysis of the average radii of the inner shadow and photon ring is adopted. Specifically, the inner shadow and photon ring of a Kerr black hole with spin parameter $a = 0.9$ are chosen as reference configurations.The relative variations of the characteristic scales of the inner shadow and photon ring in the JP black hole images along the $x-$ and $y-$ directions, measured with respect to the corresponding Kerr configurations, are then examined as functions of the deviation parameter $\epsilon_3$. This framework enables a quantitative characterization of the magnitude and evolution of the scale and morphological changes of the inner shadow and photon ring in JP black hole images.

\begin{figure}[htbp]
    \centering
    \begin{subfigure}[b]{0.4\textwidth}
        \includegraphics[width=\textwidth]{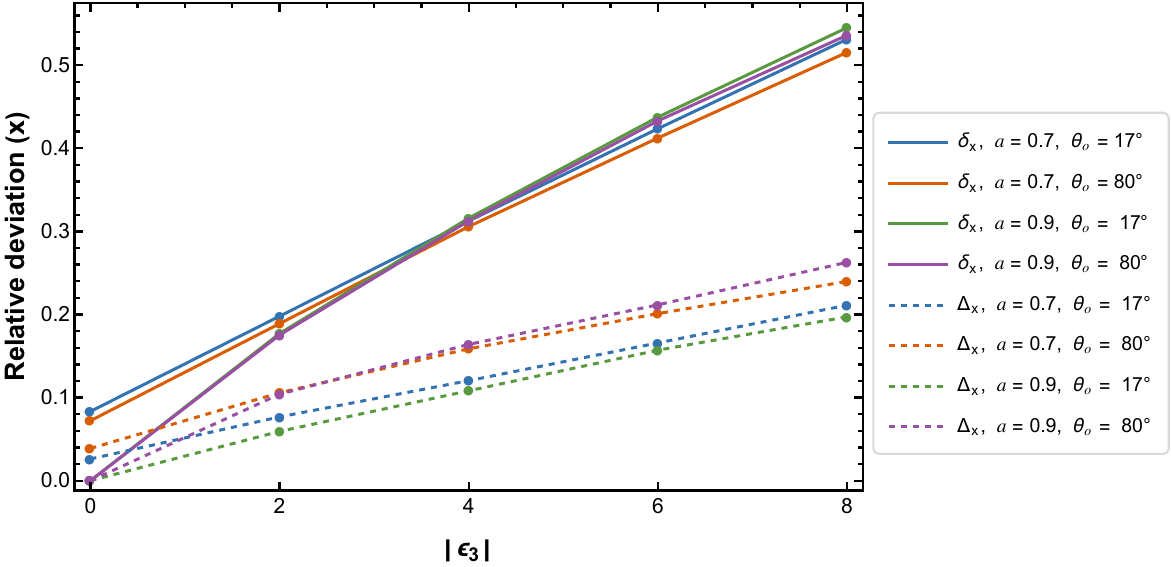}
        \caption{}
    \end{subfigure}
    \begin{subfigure}[b]{0.4\textwidth}
        \includegraphics[width=\textwidth]{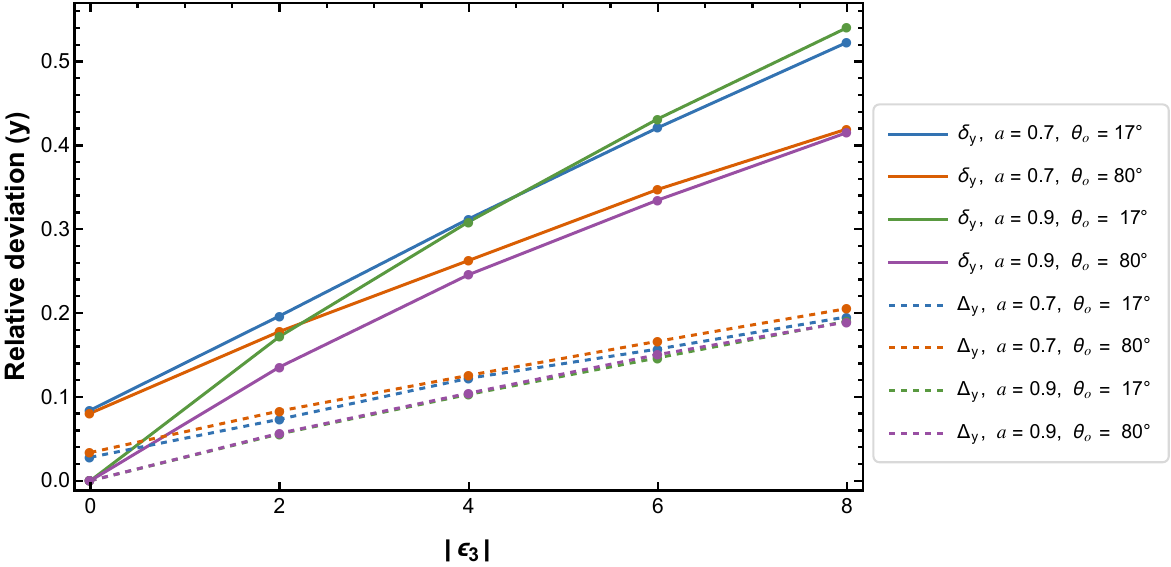}
        \caption{}
    \end{subfigure}
    \caption{The evolution of the relative deviations of the characteristic scales of the inner shadow and photon ring, defined from the intensity distributions of JP black hole images illuminated by a prograde accretion disk, as functions of the deviation parameter $|\epsilon_3|$. The relative deviations are measured with respect to the corresponding characteristic scales of the inner shadow and photon ring in the image of a Kerr black hole with spin parameter $a = 0.9$. $\delta=d/d_{\text{Kerr}}$ denotes the relative deviation of the characteristic scale of the inner shadow, while $\Delta=D/D_{\text{Kerr}}$ denotes the corresponding relative deviation of the characteristic scale of the photon ring. Panel (a) shows the evolution of the relative deviations measured along the $x$-direction of the image plane, whereas panel (b) presents the corresponding evolution measured along the $y$-direction.}
    \label{fig8}
\end{figure}
Figs. \ref{fig8} (a) and \ref{fig8} (b) illustrate the dependence of the relative deviations of the inner shadow boundary scale $d$ and the photon ring boundary scale $D$ on the deviation parameter $|\epsilon_3|$, where these relative deviations are defined along the Cartesian $x-$ and $y-$ axes of the image plane in JP black hole images illuminated by prograde accretion flows and are measured relative to the corresponding boundary scales $d_{\text{Kerr}}$ and $D_{\text{Kerr}}$ in the image of a Kerr black hole with spin parameter $a = 0.9$. In Fig. \ref{fig8}, the solid curves depict the evolution, with increasing deviation parameter $|\epsilon_3|$, of the relative deviations of the characteristic scale $d$ of the inner shadow in JP black hole images along the $x-$ and $y-$ axes on the image plane, measured relative to the corresponding characteristic scales $d_{\text{Kerr}}$ of the inner shadow in Kerr black hole images with spin parameter $a = 0.9$. These relative deviations are quantified by the parameter $\delta = d / d_{\text{Kerr}}$, and the subscripts $x$ and $y$ associated with the parameter $\delta$ denote the relative deviations of the characteristic scales of the inner shadow measured along the $x-$ and $y-$ directions on the image plane, respectively. The dashed curves represent the relative deviations of the characteristic scale $D$ of the photon ring in JP black hole images along the $x-$ and $y-$ axes on the image plane, measured relative to the corresponding characteristic scales $D_{\text{Kerr}}$ of the photon ring in Kerr black hole images with spin parameter $a = 0.9$. These relative deviations are quantified by the parameter $\Delta = D / D_{\text{Kerr}}$, and the subscripts $x$ and $y$ attached to $\Delta$ denote the relative deviations of the characteristic scales of the photon ring measured along the $x-$ and $y-$ directions on the image plane, respectively. Furthermore, the blue and orange solid and dashed curves correspond to the relative deviations $\delta$ and $\Delta$ obtained for JP black holes with spin parameter $a = 0.7$ at observer inclination angles $\theta_0 = 17^\circ$ and $\theta_0 = 80^\circ$, respectively, whereas the green and purple solid and dashed curves represent the corresponding quantities for $a = 0.9$ at the same inclination angles.

A comparison between Figs. \ref{fig8} (a) and \ref{fig8} (b) and Fig. \ref{fig5} reveals that the evolution of the relative deviations of the inner shadow and photon ring measured along the $y-$ direction, shown in Fig. \ref{fig8} (b) for JP black holes with spin parameters $a = 0.7$ and $a = 0.9$ observed at inclination angles of $\theta_0 = 17^\circ$ and $\theta_0 = 80^\circ$, closely follows the evolution of the relative deviations of the corresponding average radii presented in Fig. \ref{fig5}. By contrast, the relative deviations measured along the $x-$ direction, shown in Fig. \ref{fig8} (a), exhibit evolutionary behaviors that differ substantially from those of the corresponding average radii. This result indicates that the morphological evolution of the inner shadow and photon ring in JP black hole images with increasing $|\epsilon_3|$ is governed primarily by variations along the $y-$ direction of the image plane, whereas the variations along the $x-$ direction play a comparatively secondary role and do not dominate the overall scaling behavior of these image structures. The origin of this behavior can be understood from two complementary considerations.

The first consideration is that the deviation of the JP metric from the Kerr metric originates from the introduction of the deviation parameter $\epsilon_3$ through the function $h(r\,, \theta)$, whose explicit form is given in Eq. (\ref{finalexpressionhrtheta}). This implies that the departure of the JP metric from the Kerr geometry is entirely encoded in the function $h(r\,, \theta)$, and the JP metric smoothly reduces to the Kerr metric when $h(r\,, \theta)$ vanishes. The function $h(r\,, \theta)$ depends explicitly on the function $\Sigma$ appearing in Eq. (\ref{expressiondeltasigma}), with $\Sigma$ appearing in the denominator of $h(r\,, \theta)$. Consequently, for a fixed radial coordinate $r$, $h(r\,, \theta)$ attains its minimum value at $\theta = 0$ and its maximum value at $\theta = \pi/2$. This behavior indicates that the influence of the deviation parameter on the spacetime geometry is weakest near the polar axis and strongest near the equatorial plane. Therefore, the impact of the deviation parameter on photon motion is inherently anisotropic. As photons propagate through different angular regions of the JP spacetime, the angular dependence of $h(r\,, \theta)$ induces anisotropic modifications to the local spacetime geometry and, consequently, to the corresponding photon trajectories. Near the polar axis of the JP black hole, where $\theta \approx 0$ or $\theta \approx \pi$, the contribution of $h(r,\theta)$ is strongly suppressed. Consequently, the photon geodesics in this region exhibit only minor deviations from the corresponding Kerr geodesics, reflecting the relatively weak influence of the deviation parameter $\epsilon_3$. By contrast, near the equatorial plane, where $\theta \approx \pi/2$, the influence of $h(r\,, \theta)$ becomes maximal, causing photon trajectories to be substantially more sensitive to the deviation parameter. This behavior provides a natural explanation for the different evolutionary trends observed along the two principal directions of the image plane. The image structures measured along the $y-$ direction predominantly encode information from photon trajectories propagating through regions closer to the polar axis, where the influence of $h(r\,, \theta)$ remains comparatively weak. Consequently, increasing $|\epsilon_3|$ mainly produces an approximately uniform rescaling of the inner shadow and photon ring along the $y-$ direction while preserving, to a large extent, the characteristic morphology of the Kerr inner shadow and photon ring. In contrast, the image structures measured along the $x-$ direction are more strongly influenced by photon trajectories that probe the vicinity of the equatorial plane, where the influence of $h(r\,, \theta)$ is strongest. As a result, the geometric deformation induced by the deviation parameter modifies the evolution of the inner shadow and photon ring along the $x-$ direction more significantly than along the $y-$ direction. A second consideration is that photons propagating near the equatorial plane are additionally affected by the frame-dragging effects associated with the rotation of the JP black hole. Since both the deformation encoded in $h(r\,, \theta)$ and the frame-dragging effect attain their greatest influence in the vicinity of the equatorial plane, the interplay between these two effects becomes most pronounced in this region. Consequently, photon trajectories near the equatorial plane are governed not only by the geometric deformation introduced by the deviation parameter $\epsilon_3$ but also by the rotational dragging of inertial frames. As a consequence, photon motion near the equatorial plane is modified more strongly, leading to additional distortions in the image structures measured along the $x-$ direction and causing their evolution with increasing $|\epsilon_3|$ to depart from the nearly uniform global rescaling exhibited by the average radii of the inner shadow and photon ring. As a result, the evolutionary behavior of the relative deviations measured along the $x-$ direction differs substantially from that of the corresponding average radii as the deviation parameter $|\epsilon_3|$ increases.

However, since accretion flows may exist in either prograde or retrograde dynamical states, and these distinct configurations modify the intensity distributions of the resulting black hole images, the different dynamical states of the accretion flows may also influence the determination of the characteristic boundary scales of the inner shadow and photon ring. Therefore, building upon the preceding analysis of the relative deviations of the boundary scales of the inner shadow and photon ring in JP black hole images illuminated by prograde accretion flows, it is necessary to further investigate the influences of the deviation parameter $\epsilon_3$ on the relative deviations of the boundary scales of the inner shadow and photon ring in JP black hole images illuminated by retrograde accretion flows. Accordingly, following the methodology adopted in the analysis of the evolutionary behavior of the boundary scales of the inner shadow and photon ring along the Cartesian $x-$ and $y-$ axes on the image plane in JP black hole images illuminated by prograde accretion flows, the subsequent discussion focuses primarily on the relative evolution, with increasing deviation parameter $|\epsilon_3|$, of the boundary scales of the inner shadow and photon ring along the Cartesian $x-$ and $y-$ directions in JP black hole images illuminated by retrograde accretion flows, measured relative to the corresponding boundary scales in Kerr black hole images with spin parameter $a = 0.9$.
\begin{figure}[htbp]
    \centering
    \begin{subfigure}[b]{0.4\textwidth}
        \includegraphics[width=\textwidth]{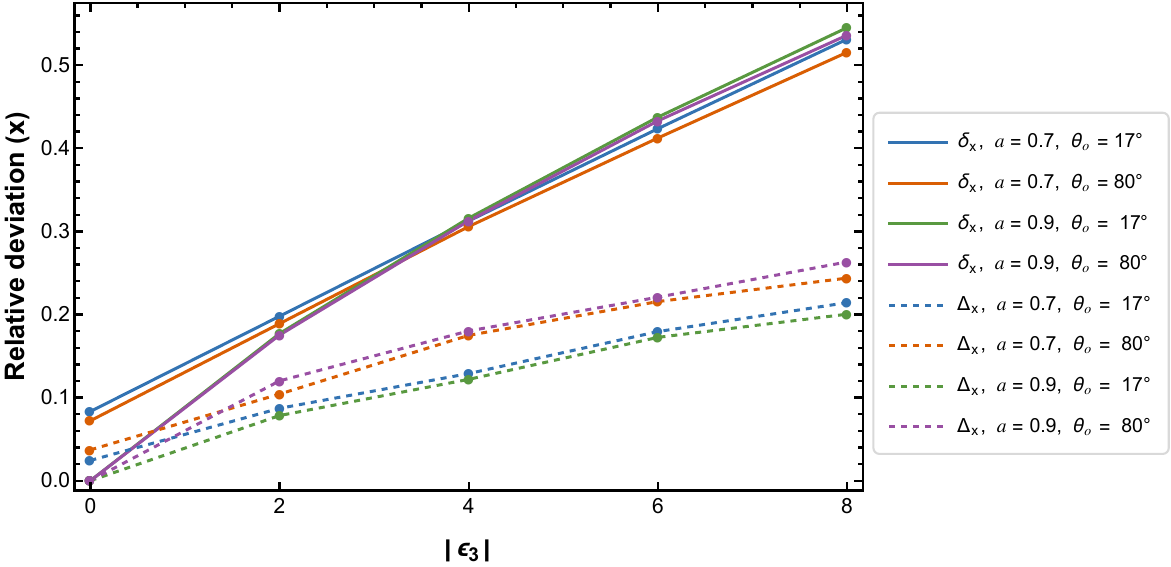}
        \caption{}
    \end{subfigure}
    \begin{subfigure}[b]{0.4\textwidth}
        \includegraphics[width=\textwidth]{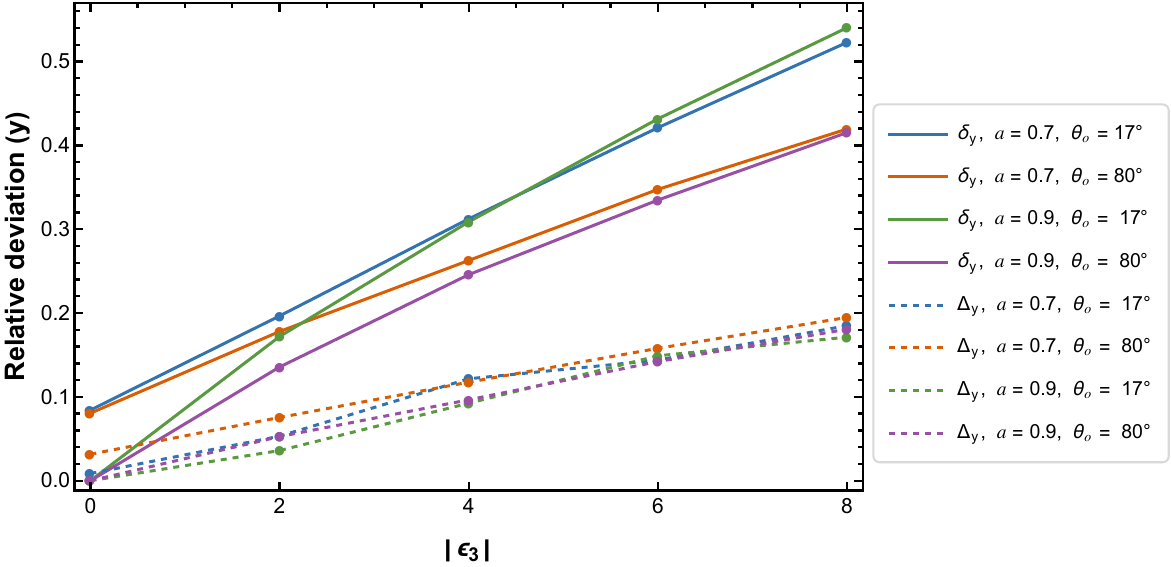}
        \caption{}
    \end{subfigure}
    \caption{The evolution of the relative deviations of the characteristic scales of the inner shadow and photon ring, defined from the intensity distributions of JP black hole images illuminated by a retrograde accretion disk, as functions of the deviation parameter $|\epsilon_3|$. The relative deviations are measured with respect to the corresponding characteristic scales of the inner shadow and photon ring in the image of a Kerr black hole with spin parameter $a = 0.9$. Panel (a) illustrates the evolution of the relative deviations measured along the $x$-direction of the image plane, whereas panel (b) presents the corresponding evolution measured along the $y$-direction.}
    \label{fig9}
\end{figure}
Figs. \ref{fig9} (a) and \ref{fig9} (b) illustrate the evolution of the relative deviations of the inner shadow and photon ring measured along the $x-$ and $y-$ directions, respectively, for JP black holes illuminated by retrograde accretion flows. The meanings of the solid and dashed curves, together with the associated color scheme, are identical to those adopted in Fig. \ref{fig8}. A comparison between Figs. \ref{fig9} (a) and \ref{fig9} (b) and Fig. \ref{fig5} shows that, similarly to the prograde accretion case, the relative deviations of the inner shadow and photon ring measured along the $y-$ direction evolve with increasing $|\epsilon_3|$ in a manner that closely follows the evolution of the corresponding average radii. By contrast, the relative deviations measured along the $x-$ direction exhibit evolutionary trends that differ substantially from those of the average radii. The physical origin of this behavior is essentially the same as that discussed for the prograde accretion case. Along the $y-$ direction on the image plane, the corresponding photon trajectories predominantly probe regions where the influence of the function $h(r\,, \theta)$ remains comparatively weak. Consequently, increasing $|\epsilon_3|$ mainly produces an approximately uniform rescaling of the inner shadow and photon ring while preserving, to a large extent, the characteristic morphology of the corresponding structures in Kerr black hole images. In contrast, the image structures measured along the $x-$ direction are more strongly influenced by photon trajectories that probe the vicinity of the equatorial plane, where the influence of $h(r,\theta)$ is strongest. In this region, the geometric deformation introduced by the deviation parameter acts simultaneously with the frame-dragging effects associated with the rotation of the JP black hole. The interplay between these two effects leads to more pronounced modifications of photon trajectories and introduces additional distortions into the resulting image structures. Consequently, the evolution of the inner shadow and photon ring along the $x-$ direction departs from the simple global scaling behavior implied by their average radii. As a result, the evolutionary trends of the relative deviations measured along the $x-$ direction differ significantly from those exhibited by the corresponding average radii as the deviation parameter $|\epsilon_3|$ increases.

By employing the quantities $d$ and $D$, defined from the intensity profiles of JP black hole images to characterize the boundary scales of the inner shadow and photon ring along the Cartesian $x-$ and $y-$ directions on the image plane, the evolution of the corresponding relative deviations with respect to the inner shadow and photon ring in the image of a Kerr black hole with spin parameter $a = 0.9$ can be systematically analyzed as functions of the deviation parameter $\epsilon_3$. This approach further reveals the physical origin of the distinct evolutionary behaviors of the inner shadow and photon ring along the $x-$ and $y-$ directions as $|\epsilon_3|$ increases. The results indicate that the evolution of the inner shadow and photon ring along the $y-$ direction can be approximately interpreted as a global rescaling process that largely preserves the characteristic morphology of the corresponding structures in Kerr black hole images. This behavior suggests that the image structures measured along the $y-$ direction primarily reflect properties of the underlying spacetime geometry that remain relatively close to those of the Kerr spacetime. By contrast, along the $x-$ direction on the image plane, the influence of the geometric deformation induced by the deviation parameter $\epsilon_3$ becomes significantly more pronounced, while the surrounding spacetime is simultaneously affected by the frame-dragging effects associated with the rotation of the JP black hole. The interplay between the geometric deformation and the frame-dragging effect produces substantially stronger modifications of photon trajectories near the equatorial plane. Consequently, the inner shadow and photon ring measured along the $x-$ direction no longer preserve the characteristic morphology of their Kerr counterparts and therefore do not exhibit the simple scaling behavior observed along the $y-$ direction as $|\epsilon_3|$ increases. These results demonstrate that an analysis based solely on the evolution of the average radii of the inner shadow and photon ring with the deviation parameter $\epsilon_3$ is insufficient to identify the physical origin underlying the morphological evolution of these characteristic image structures. Instead, it is necessary to investigate separately the evolutionary behaviors of the inner shadow and photon ring along the Cartesian $x-$ and $y-$ directions in JP black hole images as functions of the deviation parameter $\epsilon_3$. Such a directional analysis provides a clearer physical interpretation of how the deviation parameter modifies the morphology of the inner shadow and photon ring in JP black hole images.

Furthermore, the boundary scales of the inner shadow and photon ring, $d$ and $D$, defined from the intensity profiles of JP black hole images, provide a framework for quantitatively characterizing the global morphological evolution of the inner shadow and photon ring with increasing $|\epsilon_3|$. Since the boundary scales of the inner shadow and photon ring along the $x-$ and $y-$ axes are characterized by $d$ and $D$, respectively, the ellipticities of the inner shadow and photon ring can be further defined as $d_y/d_x$ and $D_y/D_x$. By analyzing the evolutionary trends of the defined ellipticities of the inner shadow and photon ring as functions of the deviation parameter $\epsilon_3$, the morphological evolution of the inner shadow and photon ring in JP black hole images with increasing $|\epsilon_3|$ can be quantitatively characterized. According to the analysis of the intensity distributions in JP black hole images, the observed intensity profiles depend sensitively on the dynamical state of the accretion flow within the accretion disk. Since the accretion flow may occur in either prograde or retrograde motion, a quantitative investigation of the evolution of the inner shadow and photon ring morphologies with the deviation parameter $\epsilon_3$, based on the ellipticities defined from the intensity distributions, likewise requires separate analyses of JP black hole images illuminated by prograde and retrograde accretion flows, respectively.

\begin{figure}[htbp]
    \centering
    \begin{subfigure}[b]{0.4\textwidth}
        \includegraphics[width=\textwidth]{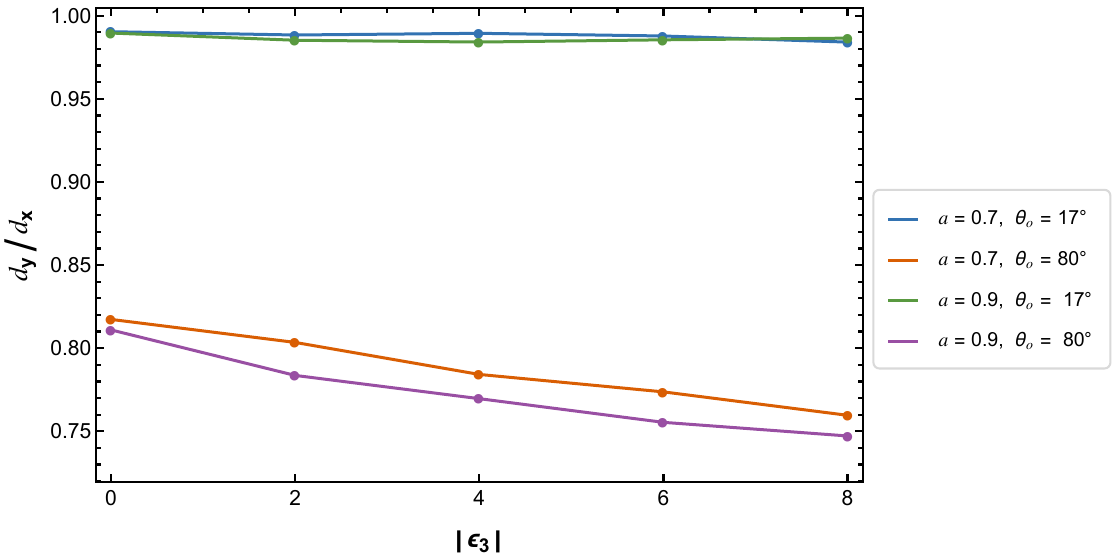}
        \caption{}
    \end{subfigure}
    \begin{subfigure}[b]{0.4\textwidth}
        \includegraphics[width=\textwidth]{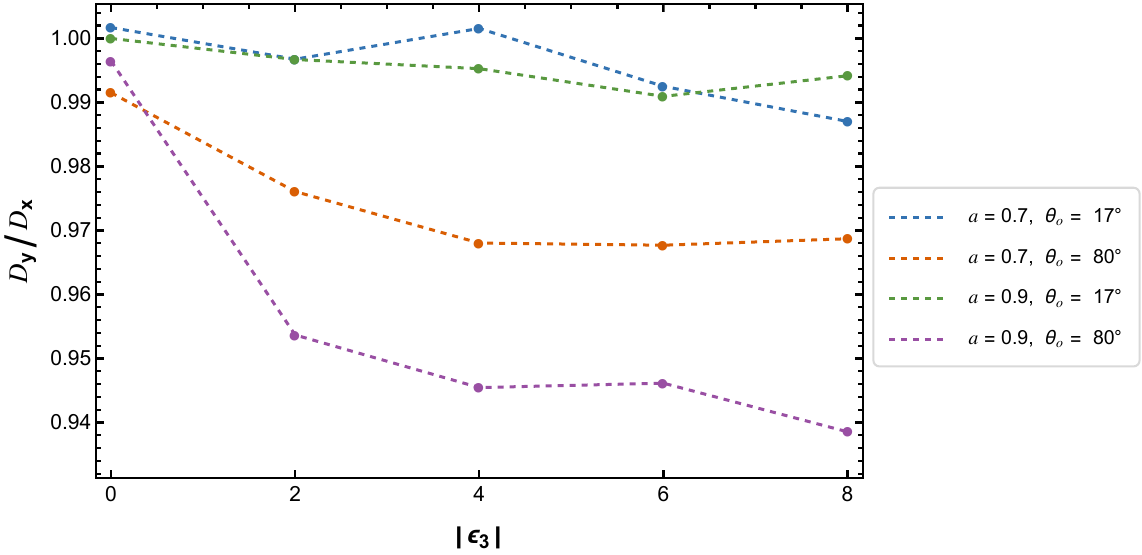}
        \caption{}
    \end{subfigure}
    \caption{The evolution of the ellipticities of the inner shadow and photon ring in JP black hole images illuminated by a prograde accretion flow, as functions of the deviation parameter $|\epsilon_3|$. The ellipticities are defined from the characteristic scales measured along the $x$- and $y$-directions of the image plane, namely $d_y/d_x$ for the inner shadow and $D_y/D_x$ for the photon ring. Panel (a) illustrates the evolution of the inner shadow ellipticity with increasing $|\epsilon_3|$, whereas panel (b) shows the corresponding evolution of the photon ring ellipticity.}
    \label{fig10}
\end{figure}
Figs. \ref{fig10} (a) and \ref{fig10} (b) present the evolution of the ellipticities of the inner shadow and photon ring, respectively, as functions of the deviation parameter $|\epsilon_3|$. The solid and dashed curves represent the ellipticities of the inner shadow and photon ring, respectively. For JP black holes with spin parameter $a = 0.7$, the ellipticities measured at observer inclination angles $\theta_0 = 17^\circ$ and $\theta_0 = 80^\circ$ are represented by the blue and orange curves, respectively, whereas for JP black holes with spin parameter $a = 0.9$, the corresponding quantities are represented by the green and purple curves. Figure \ref{fig10} (a) shows that, for an observer inclination angle of $\theta_0 = 17^\circ$, the ellipticity of the inner shadow for JP black holes with spin parameters $a = 0.7$ and $a = 0.9$ remains nearly constant as the deviation parameter $|\epsilon_3|$ increases. Moreover, for JP black holes with both spin values, the ellipticity remains close to unity, implying that the inner shadow retains an approximately circular morphology and exhibits little dependence on $|\epsilon_3|$. In contrast, for an inclination angle of $\theta_0 = 80^\circ$, the ellipticity of the inner shadow for JP black holes with both spin parameters decreases approximately linearly with increasing $|\epsilon_3|$. In this case, the observed inner shadow assumes an elliptical morphology with its major axis aligned along the $x-$ direction. As $|\epsilon_3|$ increases, the major axis along the $x-$ direction becomes progressively larger, while the minor axis along the $y-$ direction becomes progressively smaller. Equivalently, the elongation of the inner shadow along the $x-$ direction increases with increasing $|\epsilon_3|$. 

The evolution of the photon ring morphology with increasing $|\epsilon_3|$ is presented in Fig. \ref{fig10} (b). As shown in Fig. \ref{fig10} (b), for an observer inclination angle of $\theta_0 = 17^\circ$, the photon ring in the JP black hole image with spin parameter $a = 0.7$ does not exhibit a perfectly circular morphology at $|\epsilon_3| = 0$, but instead forms an elliptical structure with its major axis aligned along the $y-$ direction. As the deviation parameter $|\epsilon_3|$ increases within the interval $|\epsilon_3| \in [0\,, 2]$, the morphology of the photon ring gradually evolves from an ellipse with its major axis aligned along the $y-$ direction into an ellipse elongated along the $x-$ direction. Subsequently, within the interval $|\epsilon_3| \in [2\,, 4]$, the morphology of the photon ring gradually transitions, with increasing deviation parameter $|\epsilon_3|$, from an ellipse elongated along the $x-$ direction back into an ellipse whose major axis is aligned along the $y-$ direction. When the deviation parameter exceeds $4$, the photon ring morphology once again evolves into an ellipse elongated along the $x-$ direction, and as the deviation parameter $|\epsilon_3|$ increases further, the major axis of the elliptical photon ring along the $x-$ axis continues to expand. Nevertheless, throughout this evolutionary process, the ellipticity remains close to unity, indicating that the photon ring configurations can still be regarded as approximately circular despite the change in the orientation of the major axis. For the photon ring corresponding to $a = 0.9$, the configuration is initially perfectly circular at $|\epsilon_3| = 0$. With increasing $|\epsilon_3|$, the photon ring gradually deforms into an ellipse whose major axis is aligned along the $x-$ direction. Within the interval $|\epsilon_3| \in [0\,, 6]$, the ellipticity decreases approximately linearly, indicating a progressive enhancement of the elongation along the $x-$ axis. However, for $|\epsilon_3| \in [6\,, 8]$, the ellipticity gradually increases, implying that the photon ring tends to recover a more circular morphology. For an observer inclination angle of $\theta_0 = 80^\circ$, the photon ring corresponding to the JP black hole with $a = 0.7$ exhibits a rapid decrease in ellipticity as $|\epsilon_3|$ increases within the interval $|\epsilon_3| \in [0\,, 4]$, indicating a rapid enhancement of the elongation along the $x-$ direction. By contrast, within the range $|\epsilon_3| \in [4\,, 8]$, the ellipticity remains nearly constant, implying that the morphology of the photon ring undergoes little further evolution in this regime. For the case of the JP black hole with $a = 0.9$, the photon ring at $|\epsilon_3| = 0$ already possesses an elliptical morphology with its major axis aligned along the $y-$ direction. Within the interval $|\epsilon_3| \in [0\,, 4]$, the ellipticity decreases rapidly, indicating that the scale of the major axis along the $x-$ direction progressively increases and that the ellipticity of the photon ring becomes increasingly pronounced. In the range $|\epsilon_3| \in [4\,, 6]$, the ellipticity remains nearly unchanged, suggesting that the morphology of the photon ring is approximately stable. Finally, for $|\epsilon_3| \in [6\,, 8]$, the ellipticity of the photon ring decreases further, causing the scale of the major axis along the $x-$ direction to increase correspondingly. This behavior indicates that the elliptical deformation of the photon ring further enhances. 

\begin{figure}[htbp]
    \centering
    \begin{subfigure}[b]{0.4\textwidth}
        \includegraphics[width=\textwidth]{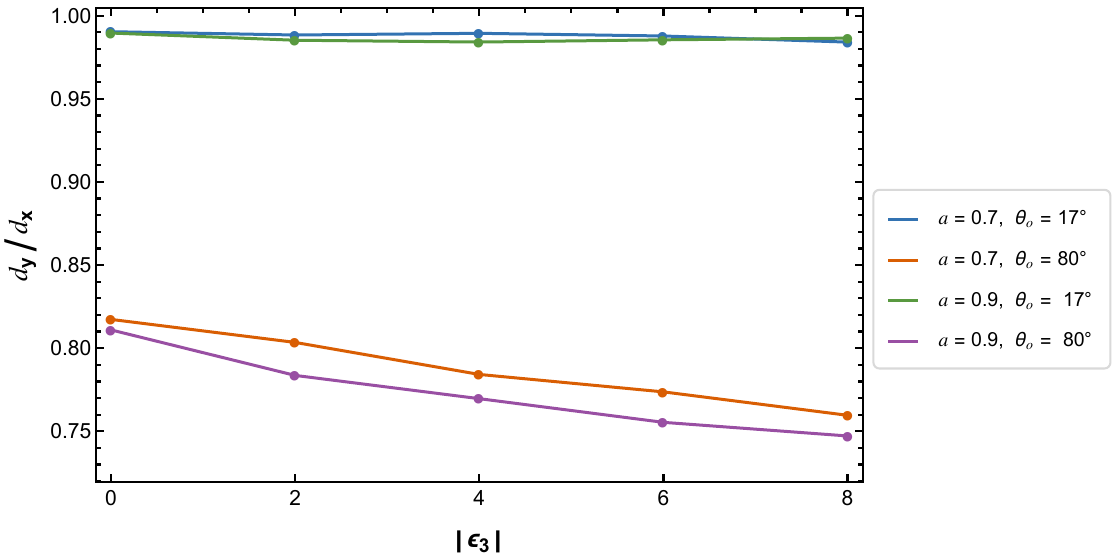}
        \caption{}
    \end{subfigure}
    \begin{subfigure}[b]{0.4\textwidth}
        \includegraphics[width=\textwidth]{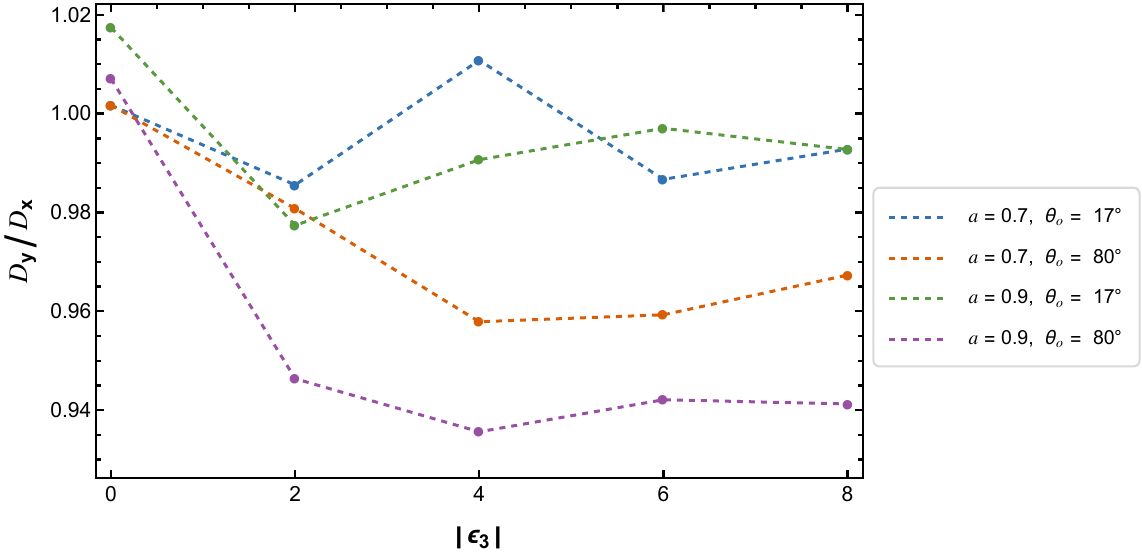}
        \caption{}
    \end{subfigure}
    \caption{The evolution of the ellipticities of the inner shadow and photon ring in JP black hole images illuminated by a retrograde accretion flow, as functions of the deviation parameter $|\epsilon_3|$. Panel (a) illustrates the evolution of the inner shadow ellipticity with increasing $|\epsilon_3|$, whereas panel (b) shows the corresponding evolution of the photon ring ellipticity.}
    \label{fig11}
\end{figure}
The evolution of the ellipticities of the inner shadow and photon ring in JP black hole images illuminated by retrograde accretion flows as functions of the deviation parameter $|\epsilon_3|$ is presented in Figs. \ref{fig11} (a) and \ref{fig11} (b), respectively. The meanings of the solid and dashed curves, together with the color-coding scheme corresponding to the spin parameters and observer inclination angles, are identical to those adopted in Fig. \ref{fig10}. A comparison between Figs. \ref{fig11} (a) and \ref{fig10} (a) demonstrates that the evolution of the inner shadow ellipticity in images illuminated by retrograde accretion flows is essentially identical to that obtained in the prograde accretion case. This result further indicates that the morphology of the inner shadow is largely independent of the dynamical state of the accretion flow in the illuminating accretion disk. In contrast, a comparison between Figs. \ref{fig11} (b) and \ref{fig10} (b) reveals that the evolution of the photon ring ellipticity with increasing $|\epsilon_3|$ depends sensitively on the dynamical state of the accretion flow. As shown in Fig. \ref{fig11} (b), for an observer inclination angle of $\theta_0 = 17^\circ$, the photon ring corresponding to the JP black hole with $a = 0.7$ possesses an ellipticity close to unity at $|\epsilon_3| = 0$, indicating an approximately circular morphology. Within the interval $|\epsilon_3| \in [0\,, 2]$, the ellipticity of the photon ring decreases gradually, implying that the photon ring evolves from a nearly circular configuration into an ellipse elongated along the $x-$ direction. As $|\epsilon_3|$ increases from $2$ to $4$, the photon ring subsequently evolves from an ellipse elongated along the $x-$ direction into one elongated along the $y-$ direction. When $|\epsilon_3|$ increases further beyond $4$, the photon ring again transforms into an ellipse elongated along the $x-$ direction. Finally, as $|\epsilon_3|$ approaches $8$, the ellipticity returns to a value close to unity, indicating that the photon ring morphology becomes nearly circular once again. For the case of the JP black hole with $a = 0.9$, the photon ring ellipticity at $|\epsilon_3| = 0$ exceeds unity, implying that the initial photon ring morphology corresponds to an ellipse elongated along the $y-$ direction rather than a perfect circle. Within the interval $|\epsilon_3| \in [0\,, 2]$, the ellipticity decreases rapidly, indicating a transition from a $y-$ elongated ellipse into an ellipse elongated along the $x-$ direction. In the range $|\epsilon_3| \in [2\,, 6]$, the ellipticity increases gradually, causing the photon ring morphology to evolve from an $x-$ elongated ellipse toward a more circular configuration. However, for $|\epsilon_3| \in [6\,, 8]$, the ellipticity decreases slowly once again and ultimately approaches the same value as that obtained for the $a = 0.7$ case. This indicates that, at $\epsilon_3 = 8$, the photon rings corresponding to the two spin parameters possess nearly identical morphologies, both taking the form of ellipses elongated along the $x-$ direction. For an observer inclination angle of $\theta_0 = 80^\circ$, the photon ring corresponding to $a = 0.7$ again exhibits an ellipticity close to unity at $|\epsilon_3| = 0$, similar to the corresponding case at $\theta_0 = 17^\circ$, indicating an approximately circular morphology. As $|\epsilon_3|$ increases within the interval $|\epsilon_3| \in [0\,, 4]$, the ellipticity decreases rapidly, implying that the photon ring quickly evolves from a nearly circular configuration into an ellipse elongated along the $x-$ direction. Within the range $|\epsilon_3| \in [4\,, 6]$, the ellipticity of the photon ring remains approximately constant, indicating that the morphology of the photon ring exhibits little variation as the deviation parameter $|\epsilon_3|$ increases. In the interval $|\epsilon_3| \in [6\,, 8]$, the ellipticity of the photon ring increases with increasing deviation parameter $|\epsilon_3|$. This behavior indicates that the major axis of the ellipse elongated along the $x-$ direction gradually decreases, causing the morphology of the photon ring to evolve from an $x-$ elongated elliptical configuration toward a more circular structure. For the case of the JP black hole with $a = 0.9$, the photon ring ellipticity at $|\epsilon_3| = 0$ is again greater than unity, implying that the initial morphology corresponds to an ellipse elongated along the $y-$ direction. Within the interval $|\epsilon_3| \in [0\,, 2]$, the ellipticity decreases rapidly, indicating a rapid transition from a $y-$ elongated ellipse into an ellipse elongated along the $x-$ direction. Subsequently, for $|\epsilon_3| \in [2\,, 8]$, the ellipticity exhibits small fluctuations while remaining approximately constant overall, indicating that the photon ring morphology remains nearly unchanged throughout this interval as $|\epsilon_3|$ increases. 

In the foregoing analysis, the image morphology and the intensity distributions of JP black holes with closed event horizons, as observed by a distant observer and illuminated by surrounding thin accretion disks, have been systematically investigated. Building upon this foundation, the dependence of the characteristic image structures of JP black holes, specifically the inner shadow and the photon ring, on the deviation parameter $|\epsilon_3|$ in the JP metric has been examined in detail. In addition, the variations in the intensity distributions of these images as functions of $|\epsilon_3|$ have been explored. Furthermore, the intensity distributions in JP black hole images were utilized to investigate the evolutionary behaviors and properties of the inner shadow and photon ring along the $x-$ and $y-$ directions of the Cartesian coordinate system on the image plane, thereby elucidating the physical mechanisms underlying the morphological evolution of the inner shadow and photon ring with the deviation parameter $\epsilon_3$. However, since JP black hole images illuminated by thin accretion disks encode not only geometric information but also dynamical information associated with the motion of the accreting matter, and since the dynamical effects of the accretion flow are manifested through the redshift and blueshift features observed in JP black hole images, it is necessary to further investigate the redshift and blueshift properties exhibited in the resulting images after photons interact with the accreting matter surrounding the black hole. Consequently, a more comprehensive characterization of the imaging properties of JP black holes with closed event horizons requires extending the analysis beyond image morphology and intensity distributions to include a detailed examination of redshift and blueshift features. Furthermore, the relationship between the redshift and blueshift characteristics and the deviation parameter $|\epsilon_3|$ should be systematically investigated. Similar to the methodology adopted previously in the analysis of the image structures and intensity distributions of JP black hole images with closed event horizons illuminated by thin accretion disks, the redshift and blueshift effects observed in the resulting JP black hole images are closely associated with the dynamical state of the accretion flow within the surrounding accretion disk. Consequently, investigations of the redshift and blueshift properties in JP black hole images likewise require separate consideration of the prograde and retrograde motion states of the accreting flows in the thin accretion disk. Since explicit expressions for the redshift factor $g_n$ have been derived in Eqs. (\ref{finalredshiftexpression}) and (\ref{redshiftexpressioninisco}) under appropriate theoretical approximations, these expressions can be employed to numerically compute the redshift and blueshift distributions associated with accreting matter in JP black hole images.

\begin{figure}[htbp]
    \centering
    \begin{subfigure}[b]{0.3\textwidth}
        \includegraphics[width=\textwidth]{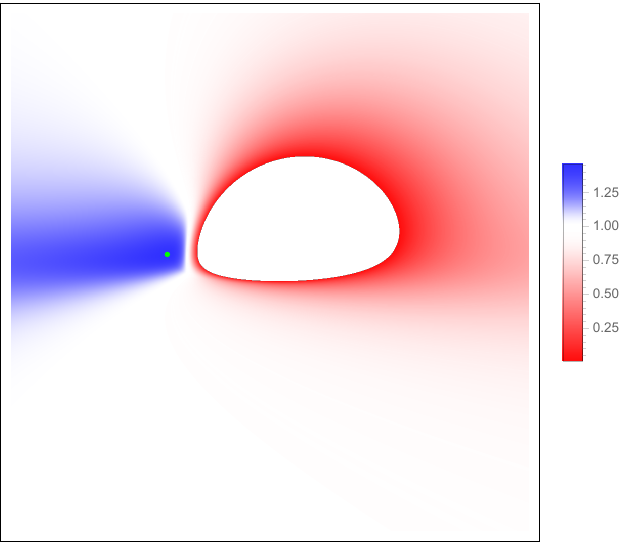}
        \caption{}
    \end{subfigure}
    \begin{subfigure}[b]{0.3\textwidth}
        \includegraphics[width=\textwidth]{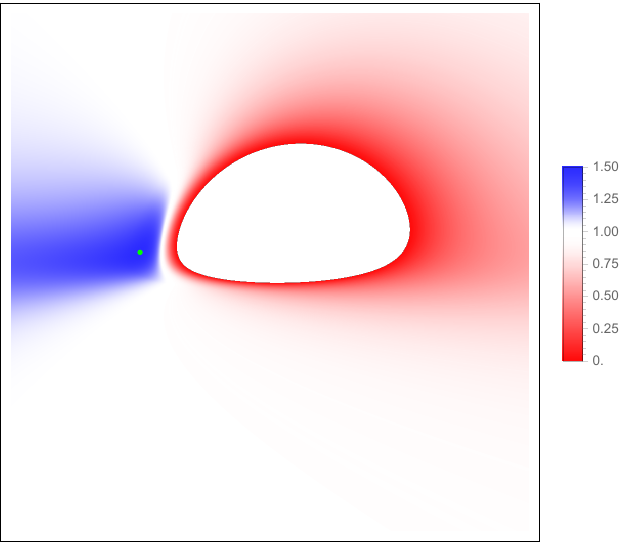}
        \caption{}
    \end{subfigure}
    \begin{subfigure}[b]{0.3\textwidth}
        \includegraphics[width=\textwidth]{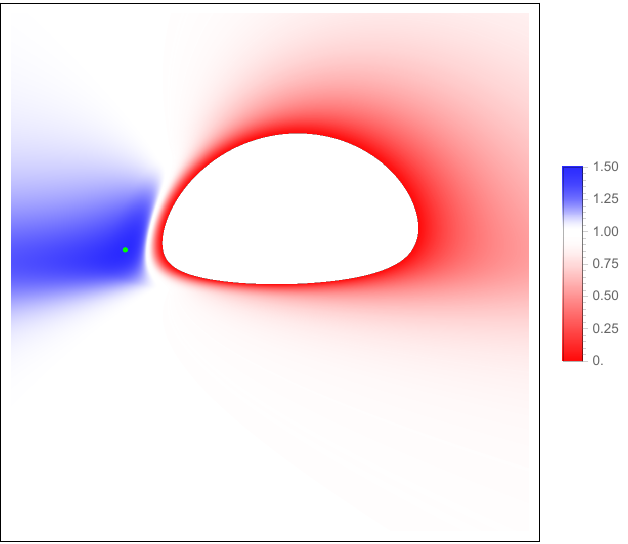}
        \caption{}
    \end{subfigure}
    \begin{subfigure}[b]{0.3\textwidth}
        \includegraphics[width=\textwidth]{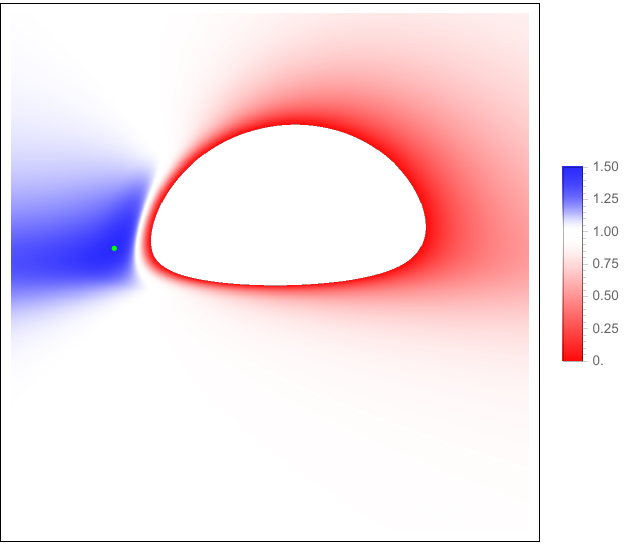}
        \caption{}
    \end{subfigure}
    \begin{subfigure}[b]{0.3\textwidth}
        \includegraphics[width=\textwidth]{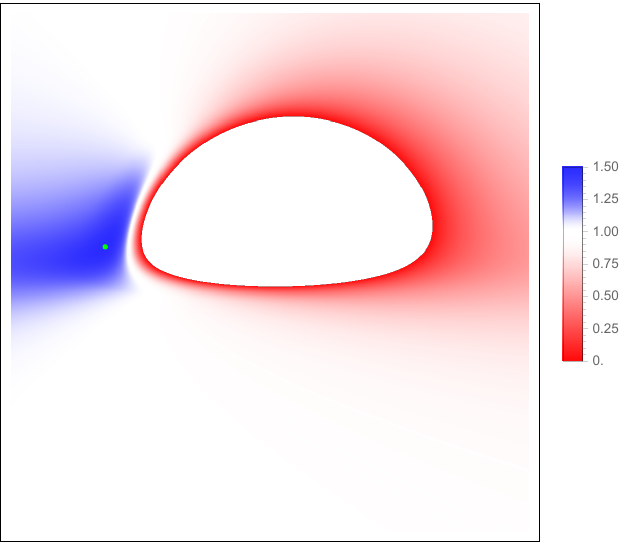}
        \caption{}
    \end{subfigure}
    \caption{The distributions of redshift and blueshift corresponding to the direct image of JP black hole images illuminated by a prograde accretion flow, for a spin parameter $a = 0.9$ and an observer inclination angle $\theta_0 = 80^\circ$. The green markers indicate the locations of the maximum blueshift. Panels (a)-(e) show the distributions for $\epsilon_3 = 0\,, -2\,, -4\,, -6$, and $-8$, respectively.}
    \label{fig12}
\end{figure}

Fig. \ref{fig12} illustrates the redshift and blueshift distributions in the direct images of a JP black hole with spin parameter $a = 0.9$, produced by the prograde motion of accreting flows in the surrounding thin accretion disk. Panels (a)-(e) in Fig. \ref{fig12} correspond to deviation parameter values $\epsilon_3 = 0\,, -2\,, -4\,, -6$, and $-8$, respectively. These panels depict the evolution of the spatial regions associated with redshift and blueshift in the direct images of the JP black hole with spin parameter $a = 0.9$ as the deviation parameter $\epsilon_3$ varies.
In Fig. \ref{fig12}, the red regions indicate locations where the emission is strongly redshifted, whereas the blue regions correspond to areas exhibiting pronounced blueshift. Moreover, the color intensity of the red and blue regions reflects the relative strengths of the redshift and blueshift, respectively, in the direct image of the JP black hole. The white region surrounded by the redshifted area at the center of each panel in Fig. \ref{fig12} corresponds to the inner shadow of the JP black hole. Since the JP metric is constructed by introducing the deviation parameter $\epsilon_3$ into the Kerr metric, the spacetime geometry described by the JP metric deviates from that of the Kerr solution by an amount that depends explicitly on $\epsilon_3$. The magnitude of the deviation parameter $\epsilon_3$ therefore provides a quantitative measure of the departure of the JP spacetime from the Kerr spacetime, with larger values of $|\epsilon_3|$ corresponding to stronger deviations from the Kerr geometry. In the limiting case $\epsilon_3 = 0$, the deviation parameter vanishes from the JP metric, which then reduces smoothly to the Kerr metric, implying that the JP spacetime continuously coincides with the Kerr spacetime in this limit. Accordingly, the spacetime structure of a JP black hole can be regarded as a deformation of the Kerr black hole geometry induced by the deviation parameter $\epsilon_3$. In the Kerr spacetime, the presence of angular momentum gives rise to an ergosphere that separates the infinite-redshift surface from the event horizon, while the infinite-redshift surface remains in close proximity to the event horizon. Since the JP black hole is constructed as a deformation of the Kerr black hole through the introduction of the deviation parameter $\epsilon_3$, it is reasonable to expect that, within the allowed range of $\epsilon_3$, the event horizon and the infinite-redshift surface remain separated by an ergosphere, and that the infinite-redshift surface continues to lie near the event horizon. Consequently, the strongest gravitational redshift is expected to occur in the immediate vicinity of the event horizon. Furthermore, the thin accretion disk surrounding the JP black hole is assumed to extend down to the event horizon. As a result, in the redshift-blueshift images associated with the direct images of JP black holes, the regions of maximum redshift are concentrated around the inner shadow. Accordingly, the areas adjacent to the inner shadow boundary appear as the darkest red regions in the redshift-blueshift images. In addition, the location of the maximum blueshift of the accreting matter within the blueshifted region in the redshift-blueshift image is indicated by the green dot in Fig. \ref{fig12}. Furthermore, as shown in Fig. \ref{fig12}, the redshifted region surrounding the inner shadow of the JP black hole expands outward as the deviation parameter $|\epsilon_3|$ increases. This behavior arises because the average radius of the inner shadow increases with increasing $|\epsilon_3|$, as discussed previously. Correspondingly, the blueshifted region and the location of maximum blueshift in the redshift-blueshift images also shift outward as $|\epsilon_3|$ increases, owing to the outward expansion of the inner shadow. For deviation parameter values in the interval $|\epsilon_3| \in [0\,, 4]$, the spatial extent of the blueshift dominated region increases slightly along the $y$-axis of the image plane. When $|\epsilon_3| > 4$, the area covered by the blueshifted region remains nearly constant, and the primary observable change is a leftward displacement of the blueshifted region induced by the continued expansion of the inner shadow.

On the basis of the foregoing analysis of the image structure of JP black holes illuminated by thin prograde accretion disks, it is evident that, in addition to the inner shadow and the photon ring, the observed images also contain both direct and lensed image components. Although the redshift and blueshift distributions arising in the direct images of JP black hole images through the interaction between photons and the prograde accretion flow in the accretion disk have already been investigated in detail, a comprehensive characterization of the redshift and blueshift features encoded by the accretion flow in JP black hole images further requires an investigation of the corresponding redshift and blueshift properties of the lensed images in JP black hole images.
\begin{figure}[htbp]
    \centering
    \begin{subfigure}[b]{0.3\textwidth}
        \includegraphics[width=\textwidth]{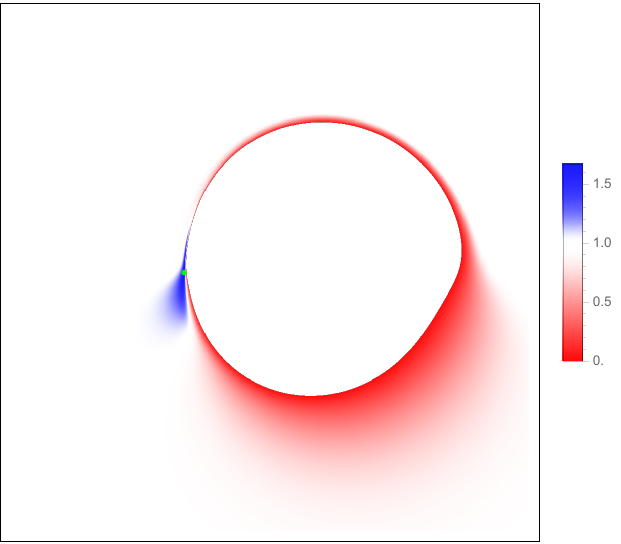}
        \caption{}
    \end{subfigure}
    \begin{subfigure}[b]{0.3\textwidth}
        \includegraphics[width=\textwidth]{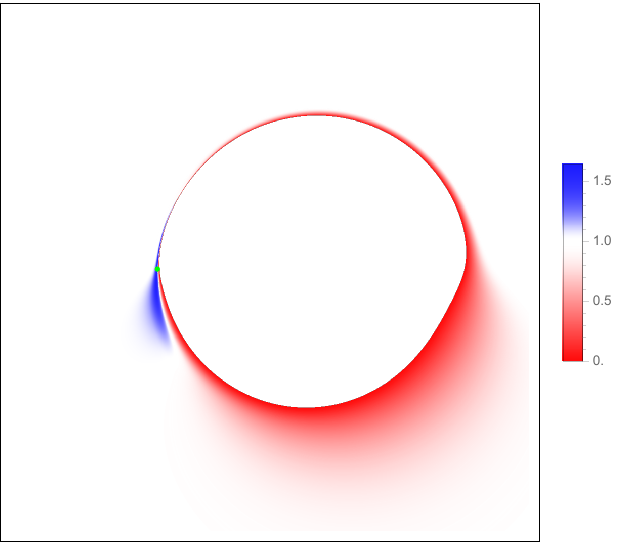}
        \caption{}
    \end{subfigure}
    \begin{subfigure}[b]{0.3\textwidth}
        \includegraphics[width=\textwidth]{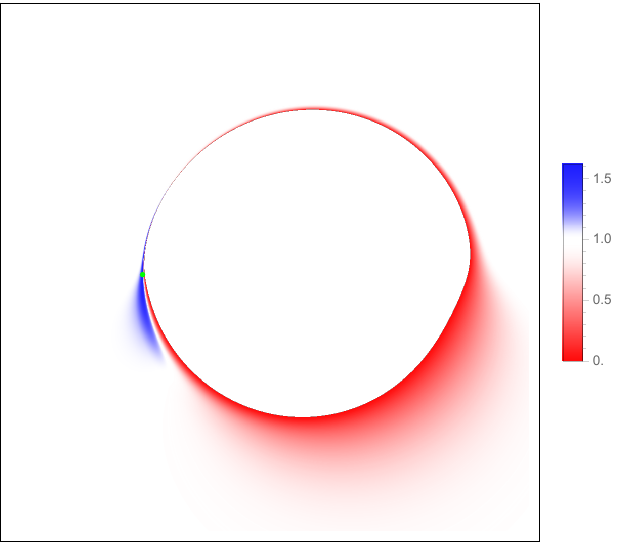}
        \caption{}
    \end{subfigure}
    \begin{subfigure}[b]{0.3\textwidth}
        \includegraphics[width=\textwidth]{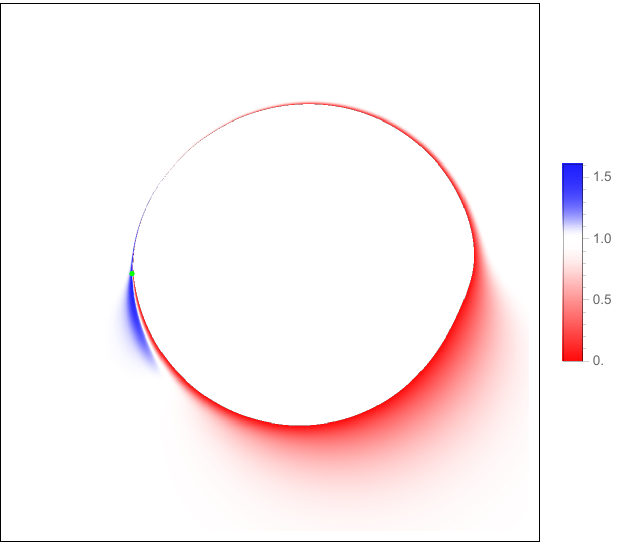}
        \caption{}
    \end{subfigure}
    \begin{subfigure}[b]{0.3\textwidth}
        \includegraphics[width=\textwidth]{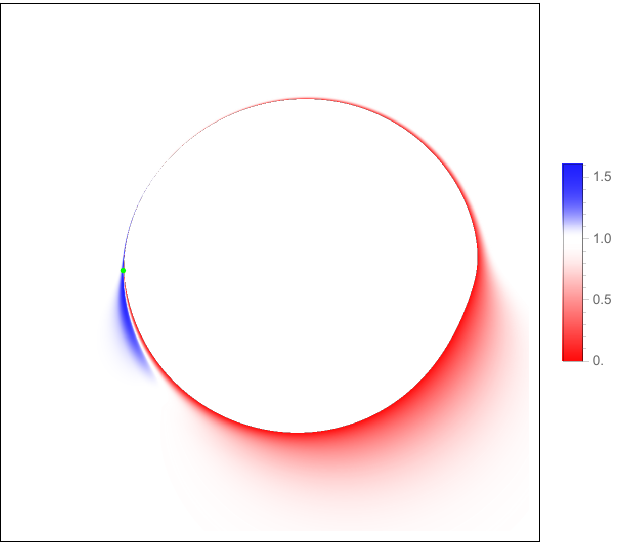}
        \caption{}
    \end{subfigure}
    \caption{The distributions of redshift and blueshift corresponding to the lensed image of JP black hole images illuminated by a prograde accretion flow, for a spin parameter $a = 0.9$ and an observer inclination angle $\theta_0 = 80^\circ$. The green markers indicate the locations of the maximum blueshift. Panels (a)-(e) show the distributions for $\epsilon_3 = 0\,, -2\,, -4\,, -6$, and $-8$, respectively.}
    \label{fig13}
\end{figure}
Fig. \ref{fig13} presents the redshift and blueshift distributions within the lensed image regions of JP black hole images generated through the interaction between photons and the prograde accretion flow in the thin accretion disk surrounding the JP black hole. Panels (a)-(e) in Fig. \ref{fig13} present the redshift and blueshift distributions in the lensed images of JP black holes corresponding to deviation parameter values $\epsilon_3 = 0\,, -2\,, -4\,, -6$, and $-8$, respectively. These panels collectively illustrate the evolution of the redshift and blueshift patterns of the lensed image as the deviation parameter $\epsilon_3$ varies. A comparative examination of these panels clearly demonstrates the systematic dependence of the redshift and blueshift distributions in the lensed images on the deviation parameter $\epsilon_3$. As shown in Fig. \ref{fig13}, the regions of maximum redshift in the lensed images produced by prograde accretion flows remain localized in the vicinity of the inner shadow of the JP black hole. This behavior is primarily determined by the location of the infinite-redshift surface, which lies in close proximity to the event horizon. In addition, the redshifted region located on the lower-right side of the inner shadow is more extended than those located in other surrounding regions. This asymmetry arises mainly from Doppler effects associated with the prograde rotational motion of the accreting plasma within the accretion disk. Furthermore, as the inner shadow of the JP black hole expands radially on the image plane with increasing magnitude of the deviation parameter $|\epsilon_3|$, the redshifted regions associated with the lensed image also undergo a corresponding outward expansion. Consequently, the evolution of the redshifted regions in the lensed image with increasing $|\epsilon_3|$ closely parallels the behavior observed in the redshifted regions of the direct image. In contrast, the blueshifted regions in the lensed image exhibit a distinct evolutionary behavior. The locations of maximum blueshift can likewise be identified within the blueshifted regions of the lensed images in JP black hole images, and these positions are indicated by green dots in Fig. \ref{fig13}. For $\epsilon_3 = 0$, the blueshifted emission is predominantly confined to the left side of the inner shadow. This feature closely corresponds to the blueshifted regions observed in the lensed images of Kerr black holes illuminated by prograde thin accretion disks. As the magnitude of the deviation parameter $|\epsilon_3|$ increases, the blueshifted region immediately adjacent to the left side of the inner shadow gradually becomes narrower and extends counterclockwise along the boundary of the inner shadow, thereby forming an increasingly elongated blueshifted structure within the region occupied by the lensed image. Simultaneously, the location of the maximum blueshift gradually shifts toward the left side of the image plane as the inner shadow expands, while the relative position of the maximum blueshift within the blueshifted region remains nearly unchanged.

\begin{figure}[htbp]
    \centering
    \includegraphics[width=0.5\textwidth]{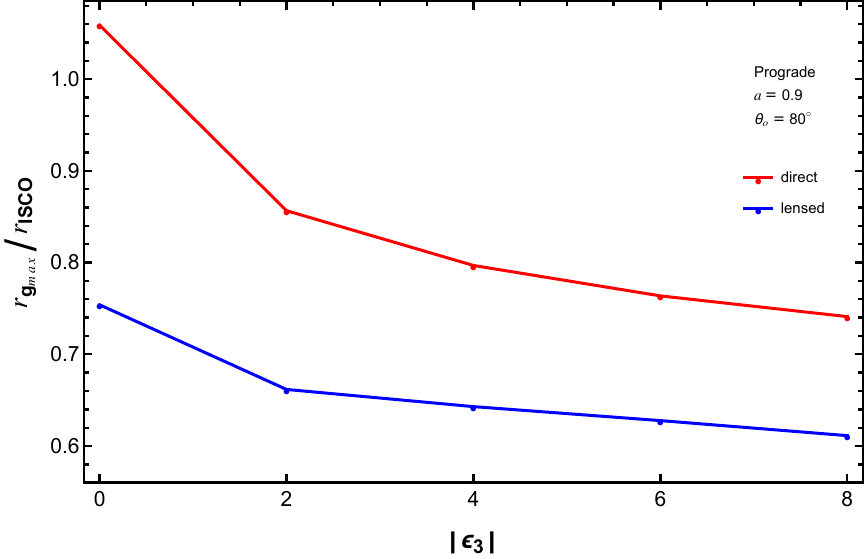}
    \caption{The evolution of the relative radial position of the maximum blueshift, $r_{g_{\text{max}}}/r_{\text{ISCO}}$, as a function of the deviation parameter $|\epsilon_3|$, for the direct and lensed images of JP black holes illuminated by a prograde accretion disk.}
    \label{fig14}
\end{figure}

In JP black hole images illuminated by a prograde thin accretion disk, the location of maximum redshift consistently lies in the vicinity of the inner shadow boundary, irrespective of whether it originates from the direct or lensed image. This behavior is governed by the radial location of the infinite-redshift surface in the JP spacetime, which resides close to the event horizon of the JP black hole, as well as by the physical properties of the infinite-redshift surface. In contrast to the redshifted regions in both the direct and lensed images produced by prograde accretion flows in the accretion disk, the blueshifted regions associated with the accreting matter in the direct and lensed images are typically located near the inner shadow of the JP black hole but do not lie immediately adjacent to the boundary of the inner shadow. Moreover, within the blueshifted regions present in both the direct and lensed images, a well-defined location of maximum blueshift can be identified. Consequently, to systematically investigate how the position of maximum blueshift associated with prograde accreting matter depends on the deviation parameter $\epsilon_3$, it is necessary to analyze in detail the variation of this location as a function of $\epsilon_3$. A characteristic circular orbit exists in the vicinity of the black hole event horizon, known as the ISCO, which represents the smallest radial distance from the black hole at which accreting matter in the surrounding accretion disk can maintain stable circular motion around the black hole. The radial position of the ISCO can therefore be employed as a natural reference scale against which the radial location of the maximum blueshift is compared, thereby enabling a quantitative analysis of the dependence of the maximum blueshift position on the deviation parameter $\epsilon_3$. The resulting dependence of the maximum blueshift position on $|\epsilon_3|$ for prograde accretion flows around JP black holes is shown in Fig. \ref{fig14}.
In Fig. \ref{fig14}, $r_{g_{\max}}$ denotes the radial location of the maximum blueshift within the blueshifted regions associated with both the direct and lensed images produced by prograde accretion flows in JP black hole images, while $r_{\text{ISCO}}$ represents the radial position of the ISCO. The vertical axis of Fig. \ref{fig14} shows the ratio $r_{g_{\max}} / r_{\text{ISCO}}$, which characterizes the radial position of the maximum blueshift relative to the ISCO. In Fig. \ref{fig14}, the red curve illustrates the dependence of the normalized radial position of the maximum blueshift in the direct image produced by prograde accretion flows on the deviation parameter $|\epsilon_3|$, while the blue curve shows the corresponding evolution of the normalized radial position of the maximum blueshift in the lensed image as $|\epsilon_3|$ increases. For the variation of the radial position of the maximum blueshift relative to the ISCO with the deviation parameter $\epsilon_3$ in the direct image generated by prograde accretion flows in JP black hole images, as represented by the red curve in Fig. \ref{fig14}, the location of the maximum blueshift lies slightly outside the ISCO and in close proximity to it when $\epsilon_3 = 0$. This behavior is consistent with the corresponding behavior observed in Kerr black hole images illuminated by prograde thin accretion disks, where the maximum blueshift position is likewise located just outside the ISCO. As the magnitude of the deviation parameter $|\epsilon_3|$ increases, the position of the maximum blueshift in the direct image, which is initially located outside the ISCO, gradually crosses the ISCO and subsequently migrates rapidly inward. For $|\epsilon_3| > 4$, the maximum blueshift position continues to move further inward as $|\epsilon_3|$ increases. However, within this regime, the radial separation between the maximum blueshift location and the ISCO grows approximately linearly with $|\epsilon_3|$. In other words, for $|\epsilon_3| > 4$, the inward migration of the maximum blueshift position relative to the ISCO proceeds at an approximately constant rate as the deviation parameter $|\epsilon_3|$ increases. In contrast, for the variation of the radial position of the maximum blueshift in the lensed image produced by prograde accretion flows around a JP black hole as a function of the deviation parameter $|\epsilon_3|$, represented by the blue curve in Fig. \ref{fig14}, the location of the maximum blueshift coincides with that in the lensed image of a Kerr black hole illuminated by a prograde thin accretion disk when $\epsilon_3 = 0$. As indicated in Fig. \ref{fig14}, unlike that in the direct image of a Kerr black hole, the maximum blueshift position in the lensed image is initially located inside the ISCO. As the deviation parameter increases within the interval $|\epsilon_3| \in [0,\, 4]$, the maximum blueshift position in the lensed image continues to migrate inward relative to the ISCO. For $|\epsilon_3| > 4$, however, the radial position of the maximum blueshift relative to the ISCO remains nearly constant as $|\epsilon_3|$ increases further. This behavior suggests that, in the regime $|\epsilon_3| > 4$, the location of the maximum blueshift in the lensed image becomes largely insensitive to variations in the deviation parameter.

The redshifted and blueshifted regions in both the direct and lensed images produced by prograde accretion flows were systematically examined, including the spatial extents of these regions, the locations of the maximum redshift and blueshift, and the evolution of these regions and of the corresponding extrema with respect to the deviation parameter $\epsilon_3$. As discussed previously, the redshift and blueshift effects produced on the image plane through the interaction between photons and the accretion flow in the accretion disk surrounding the JP black hole are strongly governed by the dynamical state of the accretion flow. Consequently, following the investigation of the redshift and blueshift properties generated through the interaction between photons and the prograde accretion flow surrounding the JP black hole, it is necessary to further investigate the corresponding redshift and blueshift properties arising from the interaction between photons and the retrograde accretion flow in the accretion disk surrounding the JP black hole. The redshift and blueshift effects manifested in JP black hole images arising from accretion flows in the surrounding accretion disk are strongly governed by the kinematic states of the accreting flows. Since retrograde accretion flows in accretion disks exhibit qualitatively different kinematic configurations, the redshift and blueshift signatures in JP black hole images produced by retrograde accretion flows are expected to display behaviors distinct from those in the prograde case. Based on the preceding discussion of the redshift and blueshift effects generated by prograde accretion flows in the direct and lensed images of JP black holes, the differences between the redshift and blueshift effects produced by retrograde and prograde accretion flows are expected to be particularly pronounced within the blueshift-dominated regions. This behavior primarily originates from the strong sensitivity of the locations of the blueshifted regions in black hole images to the direction of the orbital motion of the accretion flow within the accretion disk. This sensitivity suggests that retrograde accretion flows may imprint qualitatively different blueshift signatures on JP black hole images. Therefore, a more comprehensive characterization of the redshift and blueshift signatures in JP black hole images generated by accretion flows in thin accretion disks, as well as the dependence of the redshift and blueshift signatures on the deviation parameter $\epsilon_3$, requires a further investigation of the corresponding redshift and blueshift effects produced by retrograde accretion flows in both the direct and lensed images of JP black holes.

\begin{figure}[htbp]
    \centering
    \begin{subfigure}[b]{0.3\textwidth}
        \includegraphics[width=\textwidth]{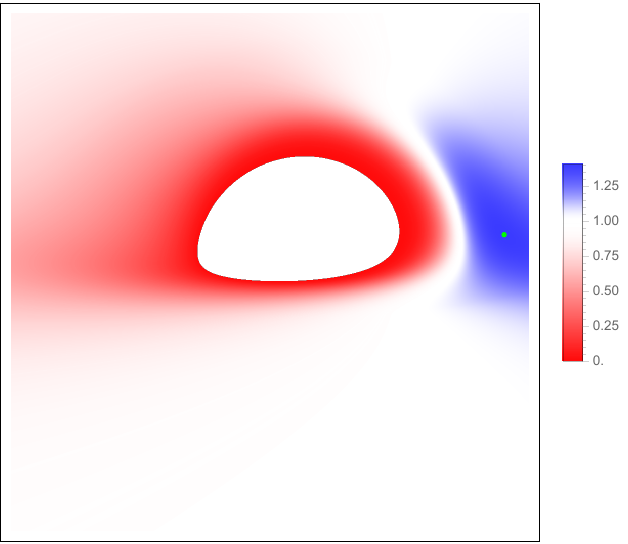}
        \caption{}
    \end{subfigure}
    \begin{subfigure}[b]{0.3\textwidth}
        \includegraphics[width=\textwidth]{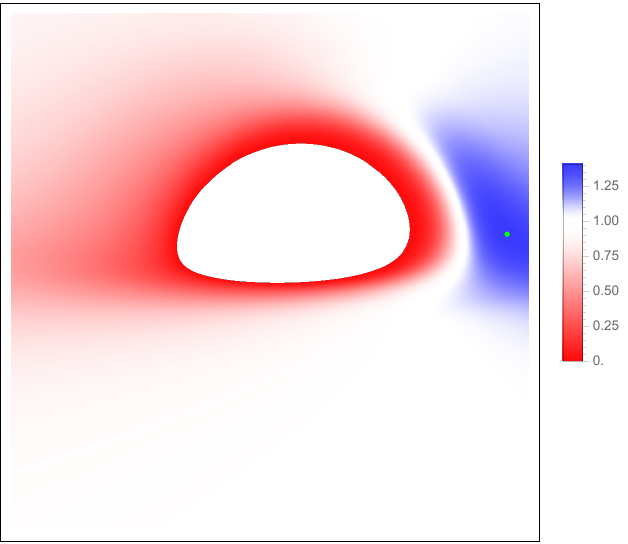}
        \caption{}
    \end{subfigure}
    \begin{subfigure}[b]{0.3\textwidth}
        \includegraphics[width=\textwidth]{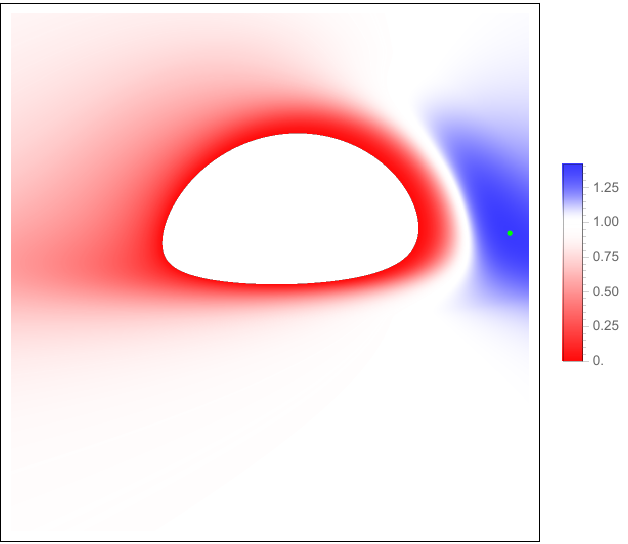}
        \caption{}
    \end{subfigure}
    \begin{subfigure}[b]{0.3\textwidth}
        \includegraphics[width=\textwidth]{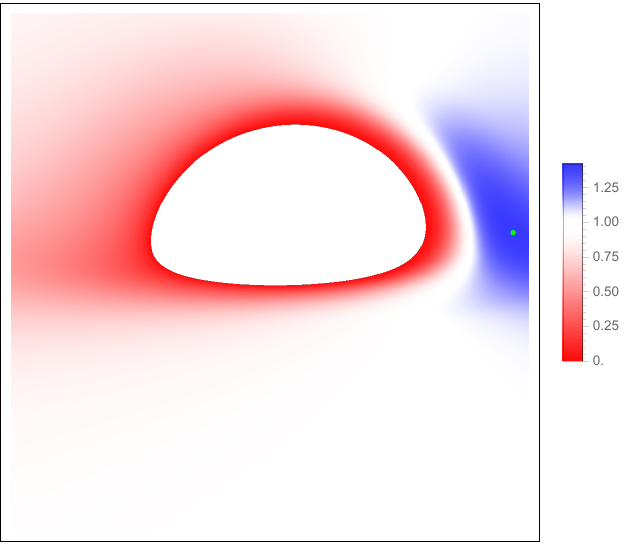}
        \caption{}
    \end{subfigure}
    \begin{subfigure}[b]{0.3\textwidth}
        \includegraphics[width=\textwidth]{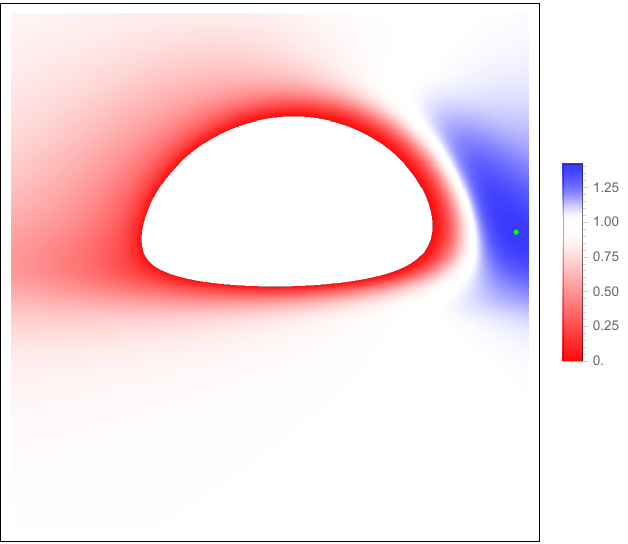}
        \caption{}
    \end{subfigure}
    \caption{The distributions of redshift and blueshift corresponding to the direct image of JP black hole images illuminated by a retrograde accretion flow, for a spin parameter $a = 0.9$ and an observer inclination angle $\theta_0 = 80^\circ$. The green markers indicate the locations of the maximum blueshift. Panels (a)-(e) show the distributions for $\epsilon_3 = 0\,, -2\,, -4\,, -6$, and $-8$, respectively.}
    \label{fig15}
\end{figure}
Fig. \ref{fig15} presents the redshift and blueshift distributions in the direct images of JP black holes with spin parameter $a = 0.9$, generated through the interaction between photons and the retrograde accretion flow in the accretion disk surrounding the black hole. Panels (a)-(e) of Fig. \ref{fig15} correspond to the redshift-blueshift maps of JP black holes with deviation parameters $\epsilon_3 = 0\,, -2\,, -4\,, -6$, and $-8$, respectively. When $\epsilon_3 = 0$, corresponding to a Kerr black hole illuminated by a retrograde thin accretion disk, regions of strong redshift are not confined to the immediate vicinity of the inner shadow boundary but instead extend over a relatively broad area surrounding the inner shadow. According to previous analyses, as the average radius of the inner shadow increases with the deviation parameter $|\epsilon_3|$, the inner shadow of a JP black hole correspondingly expands outward on the image plane. In the redshift-blueshift maps of the direct image produced by retrograde accretion flows in JP black hole images, as shown in Fig. \ref{fig15}, this outward expansion of the inner shadow causes the redshifted regions to shift outward in tandem with the growth of the inner shadow. However, as the magnitude of the deviation parameter $|\epsilon_3|$ increases, the overall redshift intensity associated with the retrograde accretion flow in the direct image gradually decreases during this outward expansion, and the regions exhibiting pronounced redshift progressively contract toward the inner shadow boundary, while the overall redshifted regions in the direct image become progressively weaker. This behavior indicates that the location of the maximum redshift in the direct images of JP black holes generated by retrograde accretion flows in the direct image gradually approaches the inner shadow boundary as $|\epsilon_3|$ increases, thereby leading to a reduction in the spatial extent of the strongly redshifted regions. The evolutionary behavior of the redshifted regions in JP black hole images illuminated by retrograde accretion flows with increasing deviation parameter $|\epsilon_3|$ is similar to that observed in the corresponding redshifted regions of JP black hole images illuminated by prograde accretion flows. The primary reason is that, as the deviation parameter $|\epsilon_3|$ increases, the inner shadow of the JP black hole gradually expands outward, causing the boundary of the inner shadow to move progressively farther away from the infinite-redshift surface near the event horizon. Consequently, the redshift intensity surrounding the inner shadow decreases, while the redshifted regions become increasingly concentrated near the boundary of the inner shadow. Meanwhile, the expansion of the inner shadow causes the blueshifted region to shift rightward in the redshift-blueshift image, and the position of the maximum blueshift moves accordingly with this displacement. However, throughout this process, the overall morphology of the blueshifted region associated with the retrograde accretion flow remains nearly unchanged as $|\epsilon_3|$ increases.

\begin{figure}[htbp]
    \centering
    \begin{subfigure}[b]{0.3\textwidth}
        \includegraphics[width=\textwidth]{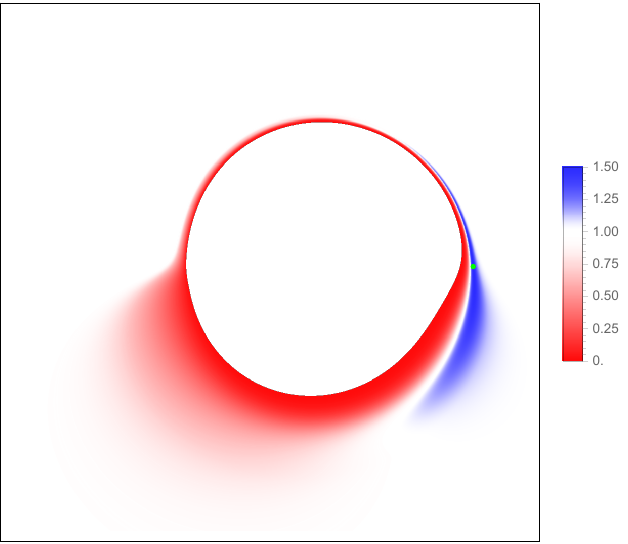}
        \caption{}
    \end{subfigure}
    \begin{subfigure}[b]{0.3\textwidth}
        \includegraphics[width=\textwidth]{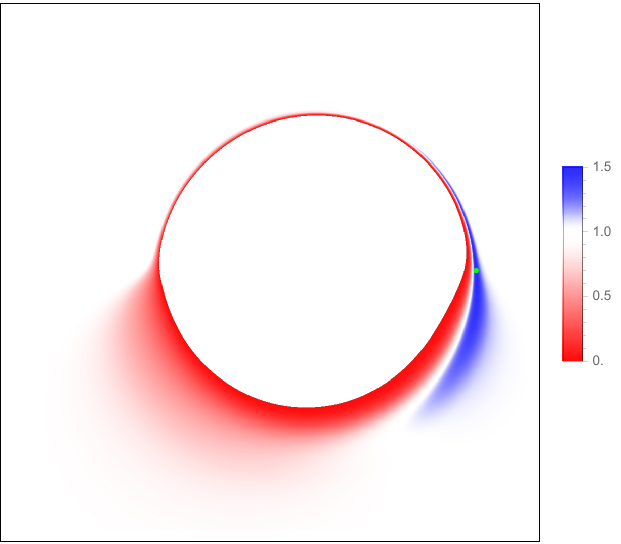}
        \caption{}
    \end{subfigure}
    \begin{subfigure}[b]{0.3\textwidth}
        \includegraphics[width=\textwidth]{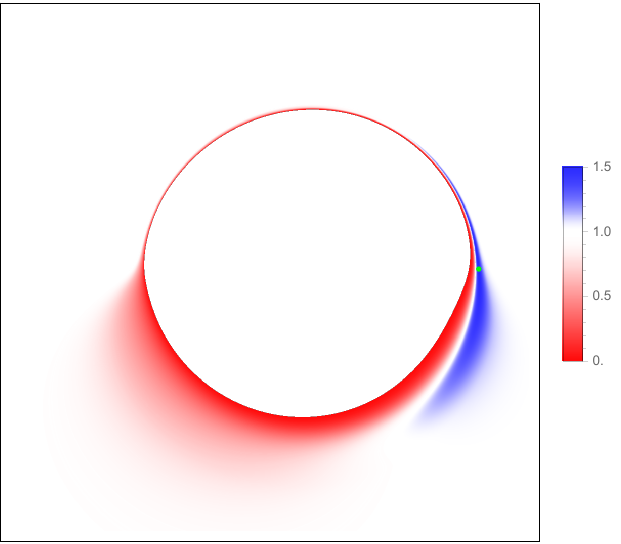}
        \caption{}
    \end{subfigure}
    \begin{subfigure}[b]{0.3\textwidth}
        \includegraphics[width=\textwidth]{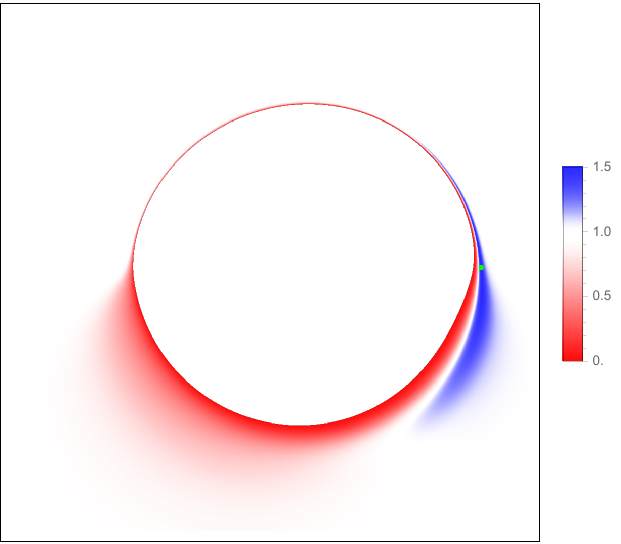}
        \caption{}
    \end{subfigure}
    \begin{subfigure}[b]{0.3\textwidth}
        \includegraphics[width=\textwidth]{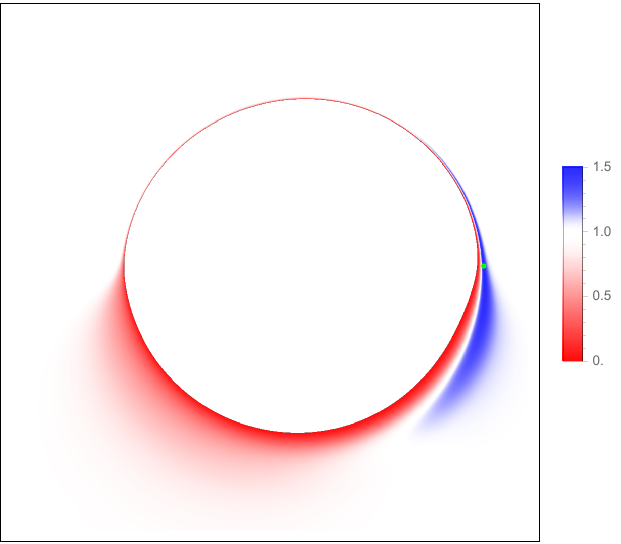}
        \caption{}
    \end{subfigure}
    \caption{The distributions of redshift and blueshift corresponding to the lensed image of JP black hole images illuminated by a retrograde accretion flow, for a spin parameter $a = 0.9$ and an observer inclination angle $\theta_0 = 80^\circ$. The green markers indicate the locations of the maximum blueshift. Panels (a)-(e) show the distributions for $\epsilon_3 = 0\,, -2\,, -4\,, -6$, and $-8$, respectively.}
    \label{fig16}
\end{figure}

In close analogy with the analysis of redshift and blueshift effects in JP black hole images produced by prograde accretion disks, after characterizing the redshifted and blueshifted regions in the direct images of JP black hole images generated by retrograde accretion disks, as well as the evolution of these regions and the locations of the maxima of the redshift and blueshift distributions with respect to the deviation parameter $\epsilon_3$, it becomes necessary to extend the analysis to the redshift and blueshift effects in lensed images produced by retrograde accretion disks in JP black hole images. Specifically, a systematic investigation of the spatial distributions and the locations of the maximum redshift and blueshift in the lensed images, together with the evolution of these regions and their corresponding extrema as functions of $\epsilon_3$, is required. Fig. \ref{fig16} presents the redshift and blueshift distributions, along with the corresponding locations of the extrema, in the lensed images of JP black holes produced by retrograde accretion flows in surrounding thin accretion disks.
Fig. \ref{fig16} presents the redshift and blueshift distributions produced in the lensed images of JP black hole illuminated by retrograde accretion flows, together with the locations corresponding to the extrema of the associated redshift and blueshift distributions. Panels (a)-(e) of Fig. \ref{fig16} present the redshift and blueshift distributions in the lensed images of JP black holes illuminated by retrograde accretion flows for deviation parameters $\epsilon_3 = 0\,, -2\,, -4\,, -6$, and $-8$, respectively. As illustrated in Fig. \ref{fig16}, the redshifted regions in the lensed images of JP black holes produced by retrograde accretion flows are predominantly concentrated along the boundary of the inner shadow. In the lower half of the inner shadow, the redshifted region is not strictly confined to the boundary of the inner shadow but extends into the surrounding area adjacent to the inner shadow. When the deviation parameter is set to $\epsilon_3 = 0$, corresponding to the redshift-blueshift distribution in the lensed image of a Kerr black holes illuminated by a retrograde thin accretion disk, the redshift along the boundary of the inner shadow reaches its maximum intensity, while the redshifted region in the lower half of the shadow exhibits the greatest spatial extension into the surrounding region. This behavior indicates that, for $\epsilon_3 = 0$, the redshift effect in the lensed images produced by retrograde accretion flows is characterized by both the strongest redshift and the largest redshifted area. As $|\epsilon_3|$ increases, although the redshifted regions remain concentrated near the boundary of the inner shadow and continue to extend into the surrounding region in the lower half of the inner shadow, the redshift intensity in the vicinity of the inner shadow gradually decreases, and the spatial extent of the redshifted region in the lower half correspondingly contracts. Consequently, with increasing $|\epsilon_3|$, both the magnitude and the spatial extent of the redshift in the lensed images produced by retrograde accretion flows are progressively reduced. Furthermore, Fig. \ref{fig16} shows that, in the lensed images of JP black holes produced by retrograde accretion flows, the blueshifted region and the location of the maximum blueshift exhibit characteristics that differ from those observed in the lensed images of JP black holes produced by prograde accretion flows in Fig. \ref{fig13}. For $\epsilon_3 = 0$, the blueshifted emission is confined to a narrow region adjacent to the inner shadow and extends clockwise along the boundary of the inner shadow. As the magnitude of the deviation parameter $|\epsilon_3|$ increases, the blueshifted region remains localized within a similarly narrow zone encircling the inner shadow and continues to exhibit a clockwise extension. Owing to the outward expansion of the inner shadow with increasing $\epsilon_3$, the corresponding blueshifted region in the lensed image gradually shifts rightward in concert with the expansion of the inner shadow. Notably, however, both the spatial extent and the intensity of the blueshifted region exhibit negligible variation as $\epsilon_3$ increases. This behavior implies that, for retrograde accretion flows, the distribution and magnitude of the blueshift in the lensed images of JP black holes are largely insensitive to the value of the deviation parameter $\epsilon_3$.

In JP black hole images illuminated by either prograde or retrograde accretion flows in surrounding accretion disks, the blueshifted regions in the redshift-blueshift maps corresponding to the direct and lensed images exhibit clearly identifiable locations of maximum blueshift. By employing the same methodology previously adopted to analyze the locations of maximum blueshift within the blueshifted regions of the redshift-blueshift maps corresponding to the direct and lensed images of JP black holes illuminated by prograde accretion flows, one can likewise systematically investigate the locations of maximum blueshift relative to the ISCO position in redshift-blueshift maps corresponding to both the direct and lensed images of JP black holes illuminated by retrograde accretion flows. Moreover, this approach enables a systematic investigation of the evolutionary behavior of the maximum blueshift locations relative to the ISCO position as functions of the deviation parameter $\epsilon_3$. Fig. \ref{fig15} presents the evolutionary behavior, with increasing deviation parameter $|\epsilon_3|$, of the relative distances between the locations of maximum blueshift and the ISCO position within the blueshifted regions of the redshift-blueshift maps corresponding to the direct and lensed images of JP black holes illuminated by retrograde accretion flows.
\begin{figure}[htbp]
    \centering
    \includegraphics[width=0.5\textwidth]{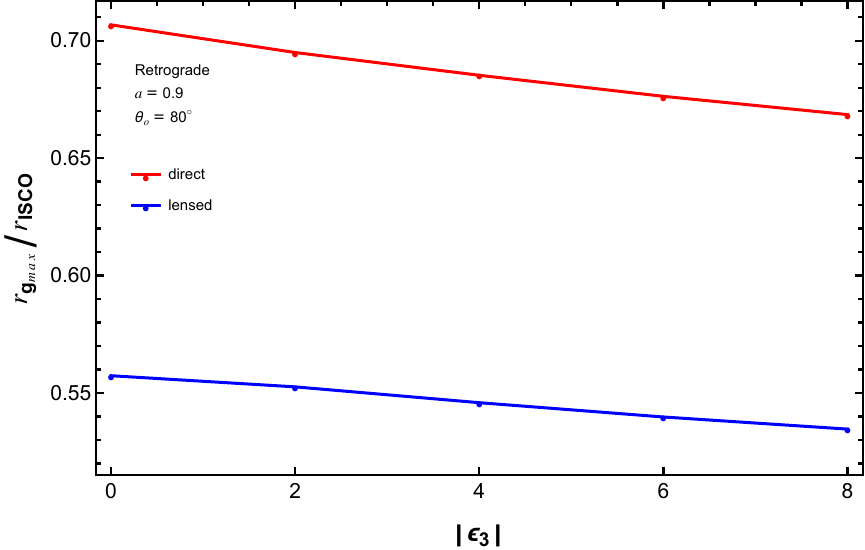}
    \caption{The evolution of the relative radial position of the maximum blueshift, $r_{g_{\text{max}}}/r_{\text{ISCO}}$, as a function of the deviation parameter $|\epsilon_3|$, for the direct and lensed images of JP black holes illuminated by a retrograde accretion disk.}
    \label{fig17}
\end{figure}
Fig. \ref{fig17} presents the dependence of the locations of maximum blueshift, measured relative to the ISCO radius, on the deviation parameter $\epsilon_3$ for both the direct and lensed images of a JP black hole with spin parameter $a = 0.9$, produced by retrograde accretion flows in the surrounding thin accretion disk and observed at an inclination angle of $\theta_0 = 80^\circ$. In Fig. \ref{fig17}, the red curve represents the evolution of the position of maximum blueshift in the direct image relative to the ISCO radius as the deviation parameter $|\epsilon_3|$ increases, while the blue curve shows the corresponding evolution for the lensed image with increasing $|\epsilon_3|$. When $\epsilon_3 = 0$, corresponding to the image of a Kerr black hole illuminated by retrograde accretion flows in the thin accretion disk, the locations of maximum blueshift generated through the interaction between photons and the retrograde accretion flow in the accretion disk in both the direct and lensed images lie inside the ISCO. As the deviation parameter $|\epsilon_3|$ increases, the ratios of the maximum blueshift locations to the ISCO radius in both the direct and lensed images decrease approximately linearly and at nearly identical rates, indicating that the maximum blueshift locations in both the direct and lensed images migrate inward relative to the ISCO with increasing $|\epsilon_3|$. However, the relatively small slopes of these curves indicate that, although the locations of maximum blueshift in both the direct and lensed images shift inward relative to the ISCO, the corresponding inward displacements remain comparatively limited. Consequently, the locations of maximum blueshift in both the direct and lensed images can be regarded as approximately insensitive to variations in the deviation parameter $\epsilon_3$. A comparative analysis of Figs. \ref{fig14} and \ref{fig17} further reveals that the locations of maximum blueshift in both the direct and lensed images of JP black holes produced by prograde accretion flows exhibit a strong dependence on the deviation parameter $\epsilon_3$, whereas the corresponding maximum blueshift locations associated with retrograde accretion flows are largely insensitive to variations in $\epsilon_3$.

\subsection{Imaging of JP black holes with non-closed event horizons and thin accretion disks}\label{nonclosedhorizonjpbh}

As discussed in Sec. \ref{jpmetricjpbh} regarding the spacetime constructed from the JP metric and the associated structure and properties of JP black holes, the event horizon of a JP black hole may exhibit two qualitatively distinct configurations, namely closed and non-closed geometries. Fig. \ref{fig1} (a) illustrates the distribution of these two distinct geometric configurations of the JP black hole event horizon within the parameter space spanned by the black hole spin parameter $a$ and the deviation parameter $\epsilon_3$ of the JP metric. In this figure, the white region corresponds to parameter values for which the JP black hole possesses a closed event horizon, while the cyan region represents the portion of parameter space associated with non-closed event horizon configurations. The black curve denotes the critical boundary separating these two regimes, indicating that when a point in the parameter space defined by $a$ and $\epsilon_3$ crosses this curve, the configuration of the JP black hole event horizon undergoes a transition between the closed and non-closed states. In Sec. \ref{jpblackholewithclosedeh}, the imaging structure, intensity distributions, as well as the redshift and blueshift effects in images of JP black holes with closed event horizons illuminated by both prograde and retrograde thin accretion disks were systematically analyzed. As shown in Fig. \ref{fig1} (a), within the admissible parameter space defined by the spin parameter $a$ and the deviation parameter $\epsilon_3$, JP black holes may possess either closed or non-closed event horizons. These distinct horizon configurations may give rise to qualitatively different observational signatures in JP black hole images. Accordingly, following the comprehensive investigation of the imaging properties of JP black holes with closed event horizons illuminated by surrounding thin accretion disks, it is essential to further conduct a systematic study of the imaging characteristics of JP black holes with non-closed event horizons illuminated by surrounding thin accretion disks in order to achieve a more complete and rigorous understanding of the imaging properties of JP black holes.

Similar to the construction of images of JP black holes with closed event horizons as observed from infinity, images of JP black holes with non-closed event horizons illuminated by thin accretion disks are also obtained using the numerical backward ray-tracing method. Before constructing images of JP black holes with non-closed event horizons surrounded by thin accretion disks using the backward ray-tracing method, it is necessary first to specify the spacetime location of the distant observer and the geometric configuration of the accretion disk surrounding the JP black hole, analogous to the preparatory procedure required for the numerical backward ray-tracing construction of images of JP black holes with closed event horizons. As in the case of JP black holes with closed event horizons illuminated by thin accretion disks, the radial coordinate of the distant observer is fixed at $r_0 = 100$ in the numerical backward ray-tracing calculations. However, the configuration of the thin accretion disks surrounding a JP black hole with a non-closed event horizons differs from that in the case of closed event horizon. For JP black holes with closed event horizons, the outer boundary of the accretion disk is set to $r_{\text{out}} = 20$, while the inner boundary is assumed to extend down to the event horizon of the JP black hole, i.e., $r_{\text{in}} = r_h$. In contrast, for JP black holes with non-closed event horizons, defining the radial extent of the accretion disk involves additional subtleties. Although the outer radius of the accretion disk can still be chosen as $r_{\text{out}} = 20$, identical to the case of closed event horizon, the inner boundary of the thin accretion disk cannot be extended to the event horizon, since a well-defined event horizon does not exist in the non-closed configuration. Because the JP metric is constructed as a deformation of the Kerr metric through the introduction of deviation parameters $\epsilon_3$, the resulting spacetime geometry departs from the Kerr solution in a controlled manner. Consequently, when specifying the radial extent of the accretion disk around JP black holes with non-closed event horizons, it is reasonable to assume that the inner radius of the accretion disk coincides with the event horizon radius of the corresponding Kerr black hole obtained in the limit where the deviation parameter vanishes, namely $r_{\text{in}} = r_h^{\text{Kerr}}$. 

Furthermore, when the event horizon of a JP black hole transitions from a closed to a non-closed configuration, the surrounding accretion disk exhibits a profound structural divergence between its prograde and retrograde components. For the prograde accretion flows, the accreting matter co-rotates with the spin direction of the JP black hole, and the corresponding accretion flows possess positive specific angular momentum ($\mathcal{L} > 0$). When the accretion flow is in the prograde state, the stable circular geodesics of the accreting matter surrounding a JP black hole with a non-closed event horizon disappear completely, thereby eliminating any physically stable bound states of the accreting matter. Consequently, prograde accretion flows become dynamically incapable of sustaining stable circular orbits around a JP spacetime with a non-closed event horizon. Conversely, for the retrograde accretion flows, the accreting matter counter-rotates with the spin direction of the JP black hole and is characterized by negative specific angular momentum ($\mathcal{L} < 0$). When the accretion flow is in the retrograde state, the stable circular orbits of the accreting matter near the ISCO continue to persist, thereby allowing the retrograde accretion flow to remain on gravitationally bound stable circular trajectories. Therefore, retrograde accretion flows can still stably orbit JP black holes with non-closed event horizons. The underlying physical mechanism driving this absolute prograde-retrograde asymmetry is rooted in the dramatic reshaping of the effective potential on the equatorial plane by the deformation parameter $\epsilon_3$. For prograde accretion flows, the orbital motion aligns with the frame-dragging effect along the $\phi$-direction of the BL coordinates. In the strong-field vicinity of JP black holes with non-closed event horizons, the amplified gravitational attraction overwhelms the intrinsically weakened centrifugal barrier of the co-rotating matter. This catastrophic imbalance destroys the potential well required for stable circular motion, causing the ISCO of the prograde branch to lose stability and migrate toward the central singularity. In contrast, the retrograde flow moves in direct opposition to the dragging of the rotating spacetime. The retrograde accretion flow dynamically interacts with the frame-dragging effect induced by the spin of the JP black hole, thereby giving rise to a strong dynamical repulsive mechanism that effectively enhances the local centrifugal barrier experienced by the counter-rotating matter. This enhanced repulsive effect enables the retrograde accretion flow to maintain a robust radial equilibrium against the gravitational attraction of the JP black hole within the region extending outward from its stable ISCO. Therefore, the complete suppression of the prograde accretion configuration is a direct consequence of the transition of the horizon topology to a non-closed state, which fundamentally destroys the stability of co-rotating circular motion. This dynamical bifurcation becomes particularly evident when constructing images of JP black holes with non-closed event horizons illuminated by thin accretion disks via the numerical backward ray-tracing method. To circumvent numerical divergences associated with the naked central singularity, an inner radial truncation boundary must be prescribed. According to the preceding comprehensive analysis, since the JP metric is constructed by introducing the deviation parameter $\epsilon_3$ into the Kerr metric, the radial position of the Kerr black hole event horizon may be adopted as the inner truncation boundary of the thin accretion disk. Owing to the introduction of the deviation parameter $\epsilon_3$, the JP metric becomes distorted relative to the Kerr geometry, causing the orbital radii associated with the prograde accretion flow to contract rapidly below this threshold toward the central singularity of the JP black hole. As a result, the prograde accretion branch is spatially filtered out from the exterior observable spacetime, whereas the retrograde branch remains robustly preserved due to its capacity to sustain localized orbital structures well outside the truncation boundary. Accordingly, in the subsequent analysis of the imaging properties of JP black holes illuminated by thin accretion disks with non-closed event horizons, only the retrograde accretion configuration will be considered.

\begin{figure}[htbp]
    \centering
    \begin{subfigure}[b]{0.3\textwidth}
        \includegraphics[width=\textwidth]{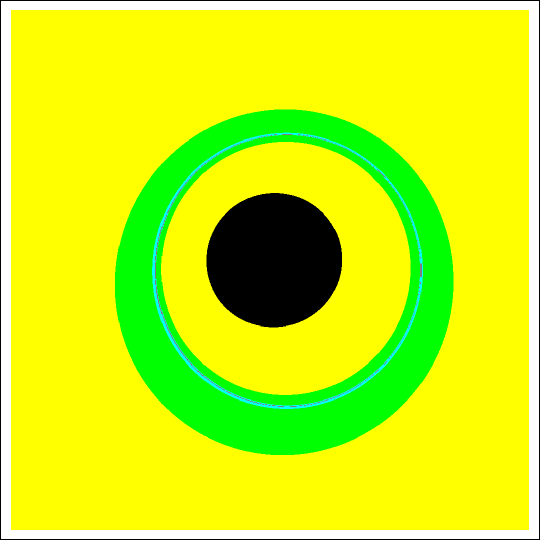}
        \caption{}
    \end{subfigure}
    \begin{subfigure}[b]{0.3\textwidth}
        \includegraphics[width=\textwidth]{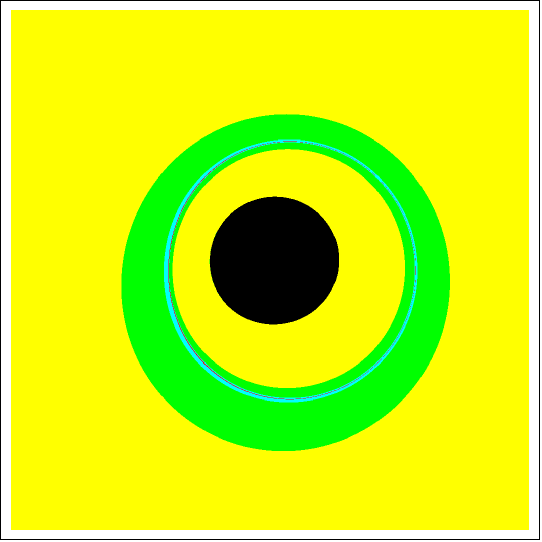}
        \caption{}
    \end{subfigure}
    \begin{subfigure}[b]{0.3\textwidth}
        \includegraphics[width=\textwidth]{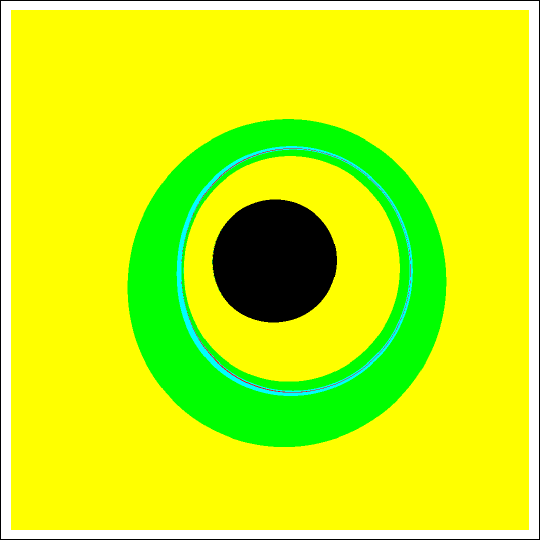}
        \caption{}
    \end{subfigure}
    \begin{subfigure}[b]{0.3\textwidth}
        \includegraphics[width=\textwidth]{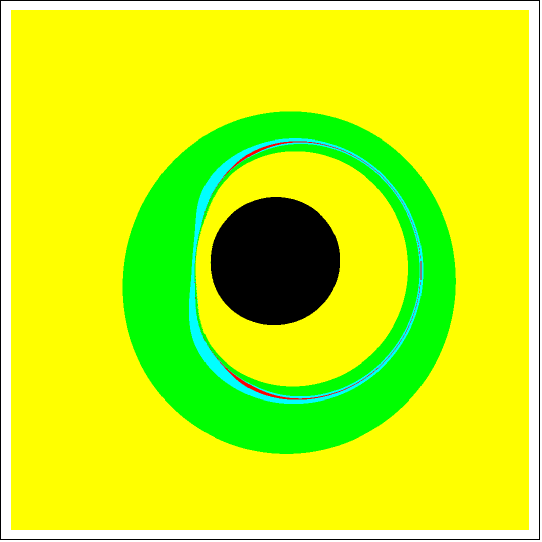}
        \caption{}
    \end{subfigure}
    \begin{subfigure}[b]{0.3\textwidth}
        \includegraphics[width=\textwidth]{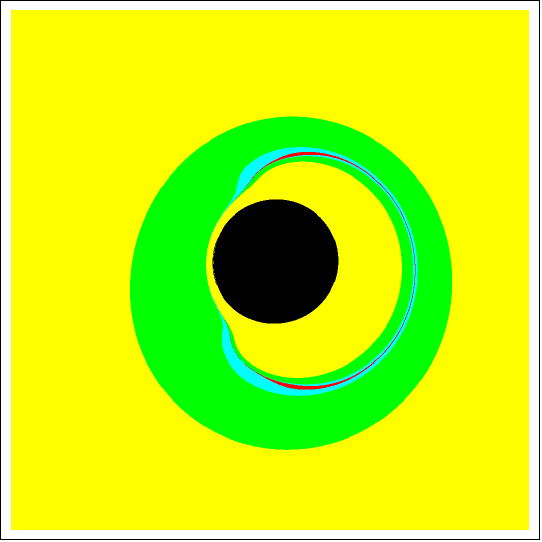}
        \caption{}
    \end{subfigure}
    \begin{subfigure}[b]{0.3\textwidth}
        \includegraphics[width=\textwidth]{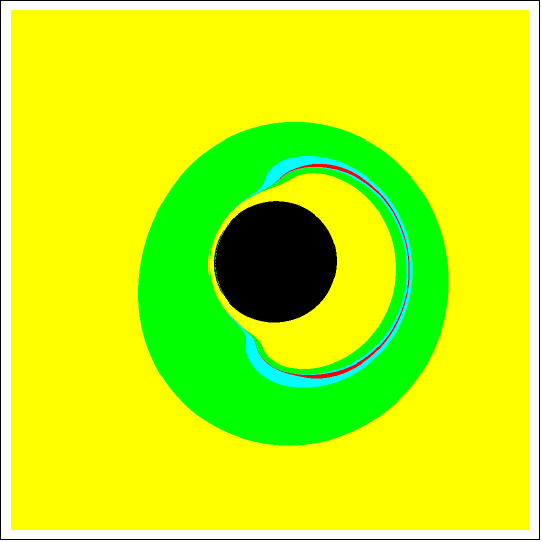}
        \caption{}
    \end{subfigure}
    \caption{Images of JP black holes with non-closed event horizons observed at an inclination angle of $\theta_0 = 17^\circ$. Different colors denote photon trajectories classified according to the number of intersections with the accretion disk. The black region represents the inner shadow, the yellow region the direct image, the green region the lensed image, and the blue and red regions the photon-ring structure. Panels (a)-(c) show the images for a spin parameter $a = 0.7$ with $\epsilon_3 = 4$, $6$, and $8$, respectively, while panels (d)-(f) show the corresponding images for $a = 0.9$.}
    \label{fig18}
\end{figure}

After specifying the location of the distant observer in the JP spacetime and the radial extent of the thin accretion disk surrounding the JP black hole, the numerical backward ray-tracing method can be employed to construct the images of JP black holes with non-closed event horizons illuminated by thin accretion disks containing retrograde accretion flows. The resulting JP black hole images are presented in Fig. \ref{fig18}.
Fig. \ref{fig18} presents images of JP black holes with non-closed event horizons illuminated by retrograde accretion flows in thin accretion disks, as observed by a distant observer at an inclination angle of $\theta_0 = 17^\circ$. The first row of Fig. \ref{fig18} shows images of JP black holes with a spin parameter $a = 0.7$, while the second row displays images of JP black holes corresponding to $a = 0.9$. From left to right, the columns of JP black hole images in Fig. \ref{fig18} correspond to deviation parameter values $\epsilon_3 = 4\,, 6$, and $8$, respectively. Similar to the methodology adopted in the analysis of the imaging properties of JP black holes with closed event horizons illuminated by thin accretion disks, the same approach is employed to further investigate the imaging properties of JP black holes with non-closed event horizons illuminated by thin accretion disks containing retrograde accretion flows. In particular, the distinct image components in the resulting JP black hole images are systematically classified and color-coded according to the number of interactions between photons and the thin accretion disk, thereby enabling a detailed and quantitative analysis of the imaging characteristics of JP black holes with non-closed event horizons. Adopting the same color-coding scheme as in Fig. \ref{fig3}, the black region at the center of each image in Fig. \ref{fig18} corresponds to the inner shadow of the JP black hole. The yellow regions denote the direct images formed by photons that cross the accretion disk once $(n = 1)$. The green regions correspond to the lensed images produced by photons that cross the accretion disk twice $(n = 2)$. The cyan and red regions indicate the photon ring, which is generated by photons that interact with the accretion disk three or more times $(n \ge 3)$.

As shown in the first row of Fig. \ref{fig18}, when the spin parameter of the JP black hole is $a = 0.7$, the morphology of the inner shadow remains essentially unchanged as the deviation parameter $\epsilon_3$ increases, and the area enclosed by the inner shadow decreases monotonically. Similarly, for the JP black hole with spin parameter $a = 0.9$ shown in the second row of Fig. \ref{fig18}, the configurations of the inner shadows remain essentially invariant, and the area of the inner shadow also decreases systematically with increasing $\epsilon_3$. A comparison with the results obtained in the previous section for JP black holes with closed event horizons indicates that the dependence of the area covered by the inner shadow of JP black holes with non-closed event horizons on the deviation parameter $\epsilon_3$ exhibits a trend opposite to that observed for the inner shadow area of JP black holes with closed event horizons. Specifically, for JP black holes with closed event horizons, the area occupied by the inner shadow in the JP black hole image increases as $\epsilon_3$ decreases, whereas for JP black holes with non-closed event horizons, the area of the inner shadow decreases monotonically with increasing $\epsilon_3$. On the other hand, for the photon ring surrounding the inner shadow in images of JP black holes with non-closed event horizons, the region enclosed by the photon ring decreases progressively as the deviation parameter $\epsilon_3$ increases in the first row of Fig. \ref{fig18}, corresponding to the case with spin parameter $a = 0.7$. Concurrently, the photon ring becomes increasingly stretched along the vertical axis on the image plane, evolving from an approximately circular shape into an elliptical configuration whose major axis is aligned with the vertical direction of the image plane. This behavior indicates that the evolution of the photon ring in JP black hole images with non-closed event horizons stands in clear contrast to that observed for black holes with closed event horizons. For JP black holes possessing closed event horizons, the overall shape of the photon ring remains essentially unchanged, and the area enclosed by the photon ring increases monotonically with decreasing $\epsilon_3$ as shown in the first row of Fig. \ref{fig3}, when the spin parameter of the JP black hole is fixed and the images of JP black holes are obtained at $\theta_0 = 17^\circ$. Furthermore, a comparison between the evolution of the photon ring in the JP black hole images with closed event horizons shown in the first row of Fig. \ref{fig3} for spin parameter $a = 0.9$ and observer inclination angle $\theta_0 = 17^\circ$, and the corresponding evolution of the photon ring in the JP black hole images with non-closed event horizons shown in the second row of Fig. \ref{fig18} under the same observational conditions, reveals that, as the spin parameter of the JP black hole increases from $a = 0.7$ to $a = 0.9$, the difference between the evolutionary behaviors of the photon-ring morphology with increasing deviation parameter $\epsilon_3$ in the non-closed and closed event horizon cases becomes progressively more significant. For JP black hole images with non-closed event horizons and spin parameter $a = 0.9$, the area enclosed by the photon ring decreases continuously as the deviation parameter $\epsilon_3$ increases, while the photon ring gradually deforms into an elliptical structure whose major axis remains aligned with the vertical direction of the image plane. Simultaneously, as the deviation parameter $\epsilon_3$ increases, the left side of the photon ring gradually shifts rightward, whereas the right side remains nearly unchanged, indicating that the ellipticity of the photon ring becomes increasingly pronounced with increasing $\epsilon_3$. The migration of the left portion of the photon ring toward the right side of the image plane causes the area enclosed by the photon ring in the JP black hole image to decrease progressively. Consequently, as the deviation parameter $\epsilon_3$ increases, the region of the direct image surrounding the inner shadow within the area enclosed by the photon ring gradually diminishes. When the deviation parameter satisfies $\epsilon_3 > 6$, the left portion of the photon ring intersects the inner shadow of the JP black hole, leading to a truncation of the photon ring by the inner shadow and rendering it discontinuous in the resulting image. This asymmetric deformation suggests that the non-closed horizon geometry induces a direction-dependent modification of photon trajectories near the shadow boundary.

\begin{figure}[htbp]
    \centering
    \begin{subfigure}[b]{0.3\textwidth}
        \includegraphics[width=\textwidth]{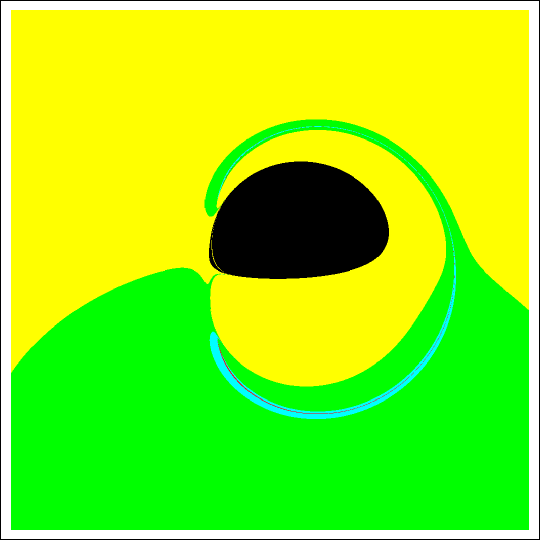}
        \caption{}
    \end{subfigure}
    \begin{subfigure}[b]{0.3\textwidth}
        \includegraphics[width=\textwidth]{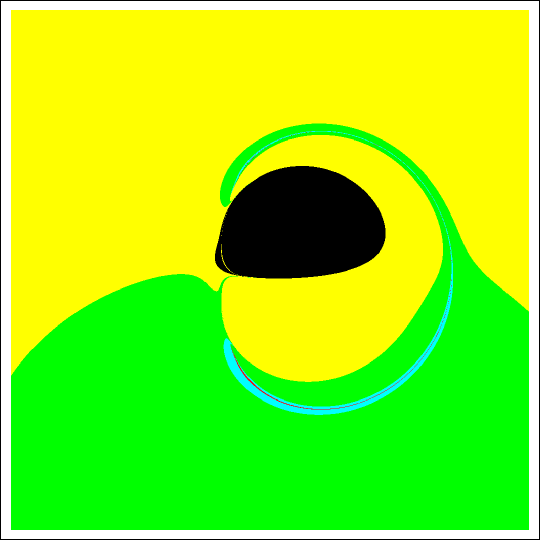}
        \caption{}
    \end{subfigure}
    \begin{subfigure}[b]{0.3\textwidth}
        \includegraphics[width=\textwidth]{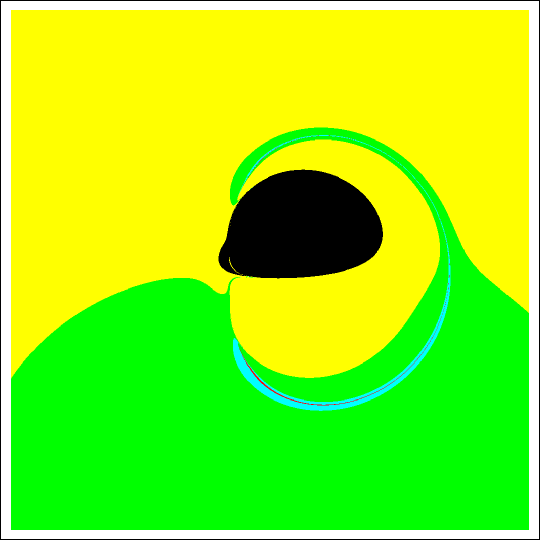}
        \caption{}
    \end{subfigure}
    \begin{subfigure}[b]{0.3\textwidth}
        \includegraphics[width=\textwidth]{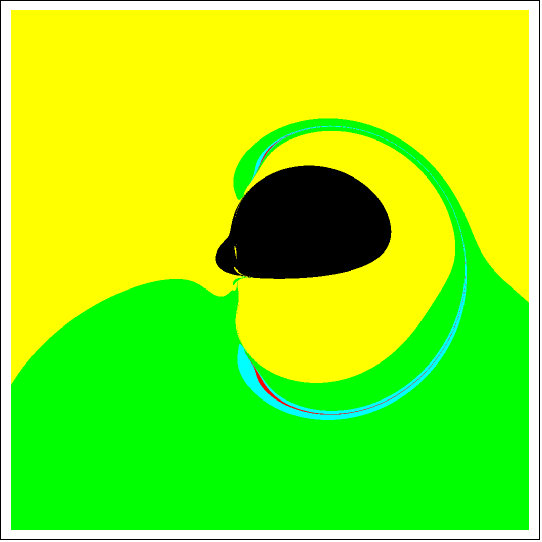}
        \caption{}
    \end{subfigure}
    \begin{subfigure}[b]{0.3\textwidth}
        \includegraphics[width=\textwidth]{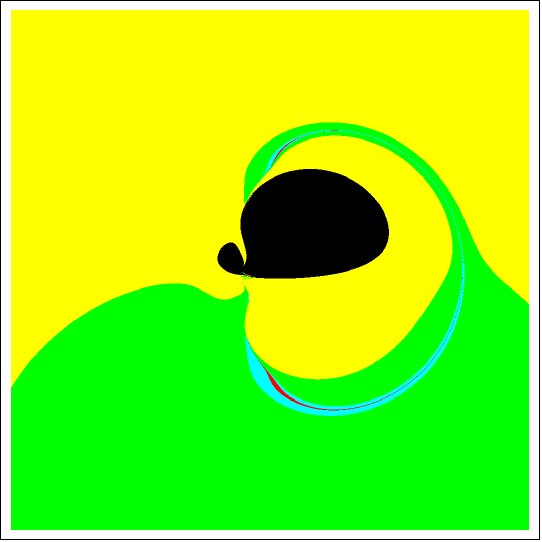}
        \caption{}
    \end{subfigure}
    \begin{subfigure}[b]{0.3\textwidth}
        \includegraphics[width=\textwidth]{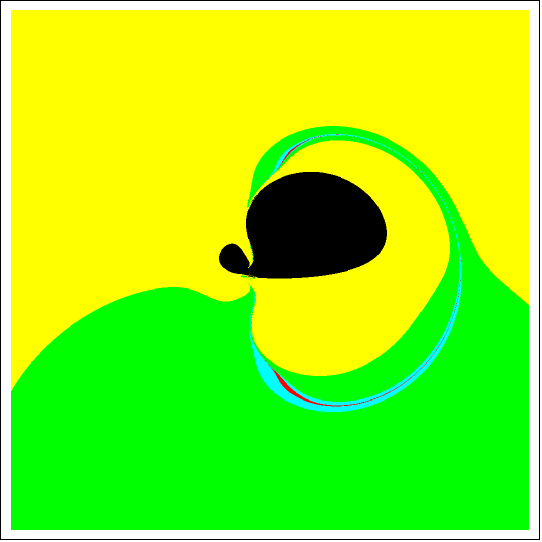}
        \caption{}
    \end{subfigure}
    \caption{Images of JP black holes with non-closed event horizons observed at an inclination angle of $\theta_0 = 80^\circ$. Panels (a)-(c) show the images for a spin parameter $a = 0.7$ with $\epsilon_3 = 4$, $6$, and $8$, respectively, while panels (d)-(f) show the corresponding images for $a = 0.9$.}
    \label{fig19}
\end{figure}

Furthermore, to achieve a comprehensive investigation of the image structures and imaging properties of JP black holes with non-closed event horizons, and to facilitate a direct comparison with the images of JP black holes possessing closed event horizons, the numerical backward ray-tracing method is likewise employed to construct images of JP black holes in the JP spacetime as observed by a distant observer at an inclination angle of $\theta_0 = 80^\circ$. The resulting images of JP black holes with non-closed event horizons at $\theta_0 = 80^\circ$ are presented in Fig. \ref{fig19}.
The first row of Fig. \ref{fig19} corresponds to images of JP black holes with non-closed event horizons and spin parameter $a = 0.7$, observed at an inclination angle of $\theta_0 = 80^\circ$, while the second row shows the corresponding images of JP black holes with spin parameter $a = 0.9$ at the same inclination angle. From left to right, the columns depict images of JP black holes with deviation parameter values $\epsilon_3 = 4, 6,$ and $8$, respectively. As illustrated in Fig. \ref{fig19}, for an observer inclination angle of $\theta_0 = 80^\circ$, the area of the inner shadow in the images of JP black holes with spin parameters $a = 0.7$ and $a = 0.9$ decreases monotonically as the deviation parameter $\epsilon_3$ increases. The position and extent of the lower boundary of the inner shadow remain nearly unchanged during this evolution, indicating that the reduction in the inner shadow area is predominantly attributable to the contraction of the remaining portions of the shadow boundary. Consequently, the shrinkage of the inner shadow primarily occurs in the upper and lateral regions of the shadow boundary, whereas the lower portion remains comparatively stable against variations in the deviation parameter $\epsilon_3$. This inward contraction of the region enclosed by the inner shadow in JP black hole images with increasing deviation parameter $\epsilon_3$ is consistent with the evolutionary trend of the inner shadow area observed in Fig. \ref{fig18} for an observer inclination angle of $\theta_0 = 17^\circ$.

Moreover, for the photon ring surrounding the inner shadow in images of JP black holes with non-closed event horizons observed at an inclination angle of $\theta_0 = 80^\circ$ in Fig. \ref{fig19}, the area enclosed by the photon ring decreases progressively as the deviation parameter $\epsilon_3$ increases for images of JP black holes with both spin parameters $a = 0.7$ and $a = 0.9$. This behavior is similar to the variation of the inner shadow area observed in JP black hole images at $\theta_0 = 17^\circ$, as shown in Fig. \ref{fig18}. However, in contrast to the JP black hole images presented in Fig. \ref{fig18}, at an inclination angle of $\theta_0 = 80^\circ$, the photon ring surrounding the inner shadow consistently retains an elliptical shape, with its major axis aligned along the vertical direction of the image plane. Moreover, the left segment of the photon ring is truncated not only by the inner shadow itself but also by the region of adjacent direct image. From left to right in Fig. \ref{fig19}, as the deviation parameter $\epsilon_3$ increases, the left boundary of the photon ring gradually shifts rightward in tandem with the contraction of the inner shadow, while the right boundary remains nearly stationary, resulting in an increasingly pronounced ellipticity of the photon ring. Figs. \ref{fig18} and \ref{fig19} illustrate the evolution of the photon ring structures of JP black holes with non-closed event horizons as functions of the deviation parameter $\epsilon_3$ for observer inclination angles of $\theta_0 = 17^\circ$ and $\theta_0 = 80^\circ$, respectively. A comparative analysis indicates that, for $\theta_0 = 17^\circ$, the photon ring surrounding the inner shadow of a JP black hole with spin parameter $a = 0.7$ remains closed as $\epsilon_3$ increases, whereas for $a = 0.9$ the photon ring progressively transitions from a closed to a non-closed configuration with increasing $\epsilon_3$. By contrast, when the inclination angle increases to $\theta_0 = 80^\circ$, the photon ring remains non-closed for both $a=0.7$ and $a=0.9$ over the entire range of $\epsilon_3$. Furthermore, a comparison with the JP black hole images possessing closed event horizons shown in the second row of Fig. \ref{fig3} demonstrates that, at an inclination angle of $\theta_0 = 80^\circ$, the photon ring remains closed for all values of $\epsilon_3$ when the event horizon is closed, whereas it becomes truncated for all values of $\epsilon_3$ when the event horizon is non-closed. This behavior indicates that the truncation of the photon ring in JP black hole images primarily originates from the absence of a well-defined event horizon.

Additionally, Fig. \ref{fig19} reveals a further salient imaging feature of JP black holes with non-closed event horizons as observed by a distant observer. In the first row of Fig. \ref{fig19}, corresponding to the images of JP black holes with spin parameter $a = 0.7$, a narrow direct image structure emerges immediately adjacent to the lower-left boundary of the inner shadow region. One end of this direct image structure intersects the lower boundary of the inner shadow, while the other gradually approaches the left boundary of the inner shadow and extends along it until eventually coinciding with the boundary of the black hole shadow. This narrow direct image segment effectively divides the inner shadow into two distinct regions. The right-hand region corresponds to the original inner shadow of the JP black hole, whereas the left-hand region can be regarded as an extension of the inner shadow associated with JP black holes possessing non-closed event horizons. Accordingly, this newly emerging region can be formally referred to as the extended inner shadow of a JP black hole with a non-closed event horizon. In the following discussion, the original shadow region enclosed by the primary shadow boundary is referred to as the original inner shadow, whereas the additional shadow-like region generated by the non-closed horizon geometry is denoted as the extended inner shadow. In the second row of Fig. \ref{fig19}, corresponding to the images of JP black holes with spin parameter $a = 0.9$, a similarly slender direct image feature is also present in the lower-left region of the inner shadow. As in the case with the spin parameter $a = 0.7$, this narrow direct image likewise separates the inner shadow in images of JP black holes into two distinct components, namely the original inner shadow and the extended inner shadow. As the deviation parameter $\epsilon_3$ increases, the narrow direct image feature within the inner shadow gradually broadens, and its spatial extent increases, thereby enhancing the separation between the original and extended inner shadows and causing the extended inner shadow to progressively shift toward the region of direct images located outside the photon ring. Moreover, as shown in Fig. \ref{fig19}, although the extended inner shadow gradually detaches from the original inner shadow and migrates toward the region of direct images exterior to the photon ring, the lower boundaries of the extended and original inner shadows remain continuously connected. Because the narrow direct image feature separating the two inner shadow components intersects the lower portion of the inner shadow and is connected to the region of lensed image in the lower half of the JP black hole image through a slender lensed image structure, the extended and original inner shadows remain connected through this intersection when the deviation parameter $\epsilon_3$ becomes sufficiently large.

\begin{figure}[htbp]
    \centering
    \begin{subfigure}[b]{0.4\textwidth}
        \includegraphics[width=\textwidth]{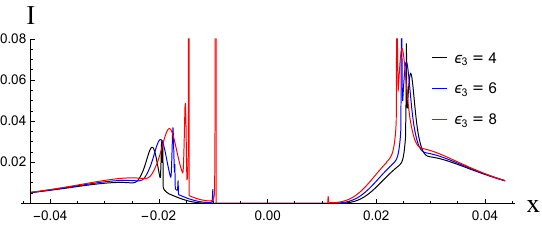}
        \caption{}
    \end{subfigure}
    \begin{subfigure}[b]{0.4\textwidth}
        \includegraphics[width=\textwidth]{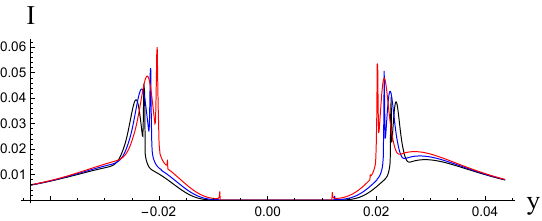}
        \caption{}
    \end{subfigure}
    \begin{subfigure}[b]{0.4\textwidth}
        \includegraphics[width=\textwidth]{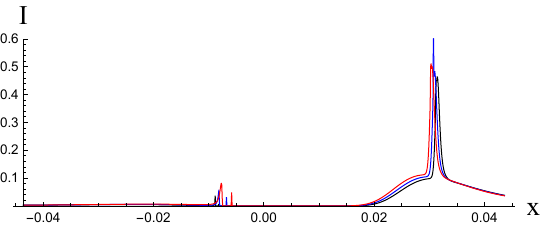}
        \caption{}
    \end{subfigure}
    \begin{subfigure}[b]{0.4\textwidth}
        \includegraphics[width=\textwidth]{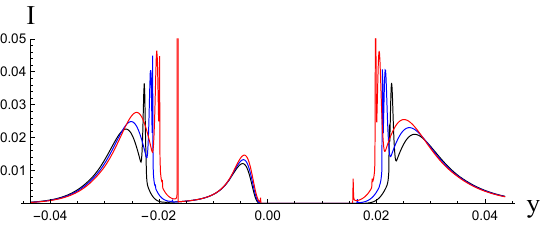}
        \caption{}
    \end{subfigure}
    \begin{subfigure}[b]{0.4\textwidth}
        \includegraphics[width=\textwidth]{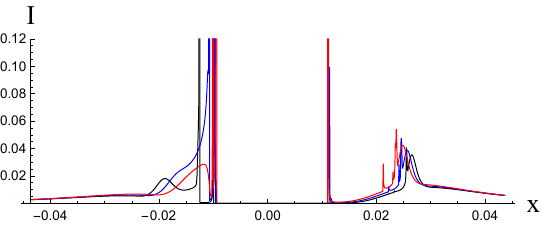}
        \caption{}
    \end{subfigure}
    \begin{subfigure}[b]{0.4\textwidth}
        \includegraphics[width=\textwidth]{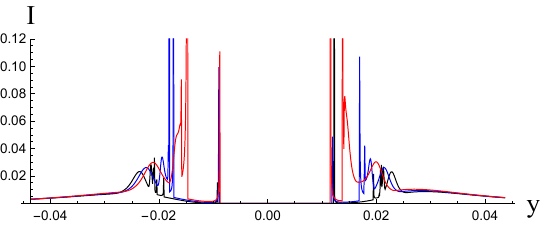}
        \caption{}
    \end{subfigure}
    \begin{subfigure}[b]{0.4\textwidth}
        \includegraphics[width=\textwidth]{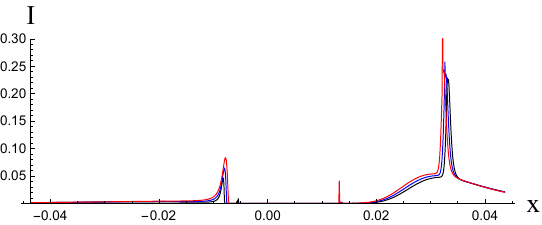}
        \caption{}
    \end{subfigure}
    \begin{subfigure}[b]{0.4\textwidth}
        \includegraphics[width=\textwidth]{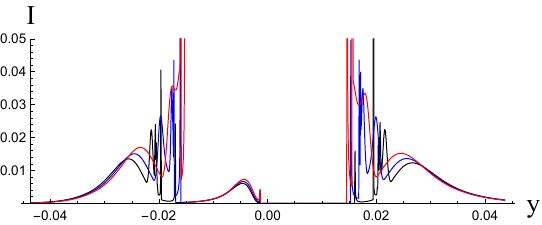}
        \caption{}
    \end{subfigure}
    \caption{The evolution of the intensity distributions along the $x$- and $y$-directions on the image plane for JP black holes with non-closed event horizons, for spin parameters $a = 0.7$ and $a = 0.9$ and inclination angles $\theta_0 = 17^\circ$ and $\theta_0 = 80^\circ$, as $\epsilon_3$ increases.}
    \label{fig20}
\end{figure}

After analyzing the dependence of the image structures and the corresponding morphological features of JP black holes with non-closed event horizons illuminated by thin accretion disks on the deviation parameter $\epsilon_3$, it should be emphasized that the images generated by JP black holes with non-closed event horizons encode only the structural information contained in the JP black hole images, with the image components classified according to the number of intersections between photons and the surrounding accretion disk. The corresponding intensity information, however, is not explicitly incorporated into these JP black hole images. Consequently, following the same methodological framework adopted for JP black holes with closed event horizons, a dedicated analysis of the dependence of the intensity distributions in the images of JP black holes with non-closed event horizons on the deviation parameter $\epsilon_3$ is required in order to provide a more complete and quantitatively accurate characterization of the imaging properties of JP black holes with non-closed event horizons. The dependence of the intensity distributions of JP black hole images with non-closed event horizons on the deviation parameter $\epsilon_3$, measured along the $x$- and $y$-axes of the Cartesian coordinate system on the image plane, is shown in Fig. \ref{fig20}.
Panels (a), (b), (e), and (f) of Fig. \ref{fig20} display the intensity distributions, as functions of the deviation parameter $\epsilon_3$, along the $x$- and $y$- axes of the Cartesian coordinate system on the image plane for JP black holes with non-closed event horizons and spin parameters $a = 0.7$ and $a = 0.9$ at an observer inclination angle of $\theta_0 = 17^\circ$. Panels (c), (d), (g), and (h) of Fig. \ref{fig20} present the corresponding results for an inclination angle of $\theta_0 = 80^\circ$. In the intensity distributions along the $x-$ and $y-$ axes of the Cartesian coordinate system on the image plane presented in Fig. \ref{fig20}, the black, blue, and red curves correspond to the intensity distribution profiles associated with deviation parameter values $\epsilon_3 = 4\,, 6$, and $8$, respectively. For the case of observer inclination angle $\theta_0 = 17^\circ$ and spin parameter $a = 0.7$, Fig. \ref{fig20} (a) shows that, as the deviation parameter $\epsilon_3$ increases, the two intensity peaks located on both sides of the image along the $x$-axis progressively shift toward $x = 0$. As discussed in the analysis of the imaging properties shown in Fig. \ref{fig18}, this behavior of the intensity peaks in Fig. \ref{fig20} (a) can be attributed to the simultaneous reduction in both the area of the inner shadow and the region enclosed by the photon ring as $\epsilon_3$ increases. Since the dominant intensity peaks in Fig. \ref{fig20} are associated with the radial locations of the photon ring, the positions of these peaks consequently migrate toward the image center with increasing $\epsilon_3$. Moreover, Fig. \ref{fig20} (a) indicates that, when the deviation parameter reaches $\epsilon_3 = 8$, an additional intensity peak appears near $x \simeq - 0.01$, located between the photon ring and the inner shadow of the JP black hole. This additional peak implies that, for sufficiently large values of $\epsilon_3$, corresponding to a sufficiently open event horizon configuration, photons originating from interactions with the surrounding accretion disk can escape to infinity from the region between the unstable circular photon orbit and the inner shadow and subsequently be detected by a distant observer. This behavior is in sharp contrast to the case of JP black holes with closed event horizons, for which photons originating inside the unstable circular photon orbit are typically captured by the black hole and therefore do not contribute to the observed image after interacting with the accretion disk. Accordingly, the intensity peak located near $x \simeq - 0.01$ in Fig. \ref{fig20} (a), which is generated by photons that interact with the accretion disk inside the unstable circular photon orbit and subsequently reach the distant observer, represents a characteristic signature specific to JP black holes with non-closed event horizons. On the other hand, the intensity distribution along the $y$-axis in Fig. \ref{fig20} (b) exhibits a comparatively simple and monotonic dependence on the deviation parameter $\epsilon_3$. As $\epsilon_3$ increases, the corresponding intensity peak undergoes only a systematic shift toward the image center, without the appearance of any additional characteristic peaks, in clear distinction from the behavior observed for the intensity distribution along the $x$-axis. Meanwhile, the peak intensity increases monotonically with increasing $\epsilon_3$.

The intensity distributions along the $x$- and $y$-axes for the JP black hole with spin parameter $a = 0.9$ at an inclination angle of $\theta_0 = 17^\circ$ are shown in Figs. \ref{fig20} (e) and (f). By comparing Figs. \ref{fig20} (a) and \ref{fig20} (e), as well as Figs. \ref{fig20} (b) and \ref{fig20} (f), it is found that, although the observer inclination angle is fixed at $\theta_0 = 17^{\circ}$, the intensity distributions for JP black holes with non-closed event horizons exhibit pronounced changes as the spin parameter increases from $a = 0.7$ to $a = 0.9$. In particular, for the intensity distribution along the $x-$ axis shown in Fig. \ref{fig20} (e), the evolution of the peak positions in the intensity profile with increasing deviation parameter $\epsilon_3$ indicates that the initially broadened intensity peak on the left side of the $x-$ axis gradually becomes concentrated near the coordinate $x = -0.01$. In close analogy with the additional characteristic intensity peaks observed between the photon ring and the inner shadow in Fig. \ref{fig20} (a), the intensity distribution along the $x-$ axis for the JP black hole image with spin parameter $a = 0.9$ shown in Fig. \ref{fig20} (e) likewise exhibits two distinct characteristic intensity peaks, located at $x = -0.01$ and $x = 0.01$ on the $x-$ axis of the image plane, respectively. Both intensity peaks originate from photons interacting with the accretion disk within the region between the photon ring and the inner shadow. For the $x < 0$ portion of the intensity distribution in Fig. \ref{fig20} (e), the intensity peak associated with the photon ring and the characteristic peak originating from the region between the photon ring and the inner shadow are clearly separated when $\epsilon_3 = 4$. As the deviation parameter increases to $\epsilon_3 = 8$, these two peaks merge near $x = -0.01$ and become effectively indistinguishable. In Fig. \ref{fig20} (e), for the case $\epsilon_3 = 6$, a prominent characteristic peak emerges in the $x > 0$ portion of the intensity distribution at $x = 0.01$. This intensity peak originates from photons interacting with the accretion disk within the region between the photon ring and the inner shadow. As $\epsilon_3$ is further increased to $8$, the position of this characteristic peak remains nearly fixed at $x = 0.01$, while its amplitude increases monotonically with increasing $\epsilon_3$. On the right-hand side of this characteristic peak, the intensity peak associated with the photon ring is observed. As shown in Fig. \ref{fig20} (e), the intensity peak associated with the photon ring gradually shifts toward $x = 0$, while its amplitude simultaneously increases as $\epsilon_3$ increases. An examination of the evolution of the left-hand side intensity peak along the $x$-axis with increasing $\epsilon_3$ further demonstrates that the photon ring structure persists in the images of JP black holes with non-closed event horizons throughout the entire range up to $\epsilon_3 = 8$. In the preceding analysis of the image morphology presented in Fig. \ref{fig18}, panels (d), (e), and (f) correspond to images of JP black holes with non-closed event horizons and the spin parameter $a = 0.9$ at the observer inclination angle $\theta_0=17^\circ$ for $\epsilon_3 = 4\,, 6$, and $8$, respectively. Although these images appear to suggest that the photon ring is truncated by the inner shadow and the surrounding direct image as $\epsilon_3$ increases, a more careful examination based on the intensity distributions demonstrates that the photon ring region in fact remains continuous and is not interrupted by either the inner shadow or the surrounding direct image. Instead, with increasing $\epsilon_3$, the region occupied by the photon ring progressively narrows and eventually becomes confined to the interface between the direct image surrounding the inner shadow and the adjacent lensed image, as shown in Fig. \ref{fig18}. In addition, the evolution of the intensity distributions along the $y$-axis with increasing deviation parameter $\epsilon_3$ is shown in Fig. \ref{fig20} (f). The intensity peaks on both sides of the $y$-axis gradually shift toward $y = 0$ as $\epsilon_3$ increases. This behavior is primarily attributable to the progressive reduction in the areas enclosed by the inner shadow and the photon ring of the JP black hole with a non-closed event horizons as the deviation parameter $\epsilon_3$ increases. Moreover, the amplitudes of the intensity peaks along the $y$-axis increase monotonically with increasing $\epsilon_3$. Additional characteristic intensity peaks are also visible in the vicinity of $y = -0.01$ and $y = 0.01$, located between the photon ring and the inner shadow of the JP black hole. Similar to the interpretation of the additional characteristic peaks identified in Fig. \ref{fig20} (a), these additional characteristic intensity peaks in Fig. \ref{fig20} (f) originate from the increasing openness of the JP black hole event horizon, which allows photons emitted from the region between the photon ring and the inner shadow to interact with the surrounding accretion disk and subsequently escape to a distant observer at infinity.

For an inclination angle of $\theta_0 = 80^\circ$, the intensity distributions along the $x-$ axis on the image plane for JP black holes with spin parameters $a = 0.7$ and $a = 0.9$ are shown in Figs.\ref{fig20} (c) and \ref{fig20} (g), respectively. As shown in these panels, the evolutionary behaviors of the peak structures in the intensity distributions along the $x-$ axis for JP black hole images with spin parameters $a = 0.7$ and $a = 0.9$ are essentially identical as the deviation parameter $\epsilon_3$ increases. The intensity peaks gradually become more pronounced with increasing $\epsilon_3$, while their positions along the $x-$ axis progressively shift toward the center of the coordinate axis as the deviation parameter $\epsilon_3$ increases. In the region $x < 0$ of Figs. \ref{fig20} (c) and \ref{fig20} (g), two intensity peaks are located near $x = -0.005$. According to the analysis of the image structures of JP black holes with non-closed event horizons presented in Fig. \ref{fig19} for both $a = 0.7$ and $a = 0.9$, the left peak is associated with the direct image located on the left side of the inner shadow, whereas the right peak originates from the narrow direct image inside the inner shadow that separates the inner shadow into the original inner shadow and the extended inner shadow. As the deviation parameter $\epsilon_3$ increases, the separation between these two peaks gradually increases, and the right peak systematically shifts toward the center of the original inner shadow. At the same time, the amplitudes of both peaks increase monotonically. These features indicate that, for both $a = 0.7$ and $a = 0.9$ at an inclination angle of $\theta_0 = 80^\circ$, a slender direct image strip is present on the lower-left side of the inner shadow, dividing the inner shadow into two components, namely the original inner shadow and the extended inner shadow. Furthermore, as the peak associated with the separating direct image moves inward toward the center of the original inner shadow with increasing $\epsilon_3$, the area of the extended inner shadow correspondingly increases. The simultaneous growth of the amplitudes of the two peaks also implies that the splitting effect induced by the direct image inside the inner shadow becomes progressively more pronounced as $\epsilon_3$ increases. In the region $x > 0$ of Figs. \ref{fig20} (c) and \ref{fig20} (g), distinct behaviors are observed for the intensity distributions of JP black holes with spin parameters $a = 0.7$ and $a = 0.9$. For the JP black hole with $a = 0.7$, only a single intensity peak associated with the photon ring is present. This intensity peak associated with the photon ring gradually shifts toward $x = 0$ and increases in amplitude as $\epsilon_3$ increases. In contrast, for the JP black hole with $a = 0.9$, the intensity distribution in the $x > 0$ region exhibits not only the intensity peak associated with the photon ring, but also an additional characteristic intensity peak located between the inner shadow and the photon ring. The emergence of this characteristic peak is primarily attributed to the fact that, when the event horizon of the JP black hole possesses a non-closed structure, photons located within the region between the inner shadow and the photon ring can still propagate to a distant observer after interacting with the accretion disk. As $\epsilon_3$ increases, the intensity peak associated with the photon ring likewise migrates toward $x = 0$, while its amplitude increases progressively. By contrast, the characteristic peak near the inner shadow remains nearly fixed at $x = 0.015$, whereas its amplitude increases monotonically with increasing $\epsilon_3$. For an inclination angle of $\theta_0 = 80^\circ$, the intensity distributions along the $y$-axis on the image plane for JP black holes with non-closed event horizons and spin parameters $a=0.7$ and $a=0.9$ are shown in Figs. \ref{fig20} (d) and \ref{fig20} (h), respectively. It is evident that, for both spin values, the intensity peaks associated with the photon ring on both sides of the $y$-axis progressively shift toward the image center as the deviation parameter $\epsilon_3$ increases, while their amplitudes increase simultaneously. In contrast, as the deviation parameter $\epsilon_3$ increases, the position of the intensity peak located near the image center remains essentially unchanged, whereas its amplitude increases monotonically.

\begin{figure}[htbp]
    \centering
    \begin{subfigure}[b]{0.3\textwidth}
        \includegraphics[width=\textwidth]{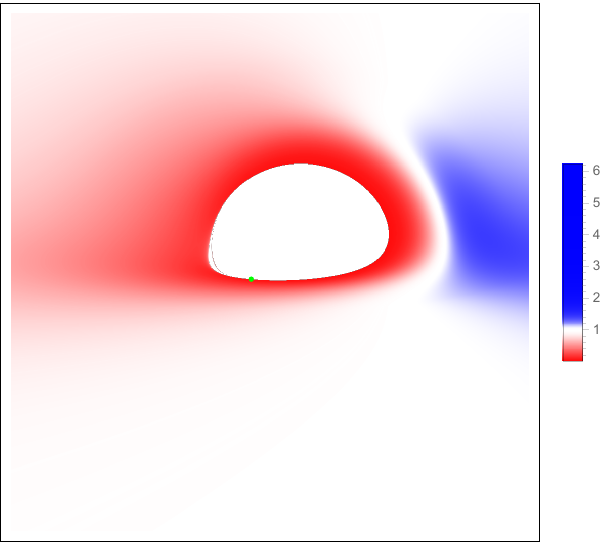}
        \caption{}
    \end{subfigure}
    \begin{subfigure}[b]{0.3\textwidth}
        \includegraphics[width=\textwidth]{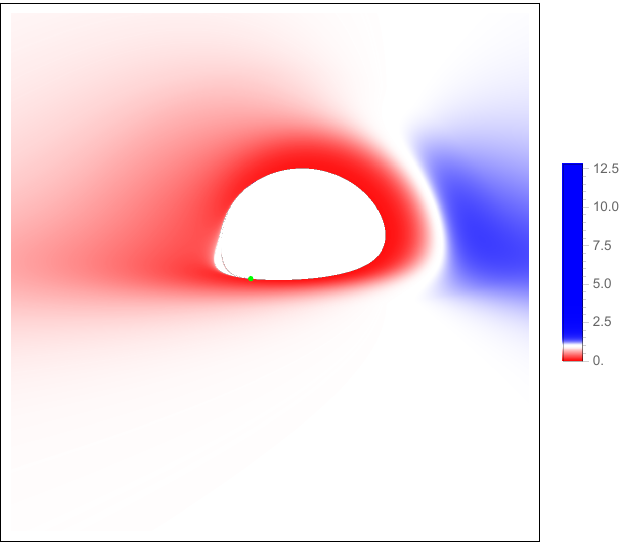}
        \caption{}
    \end{subfigure}
    \begin{subfigure}[b]{0.3\textwidth}
        \includegraphics[width=\textwidth]{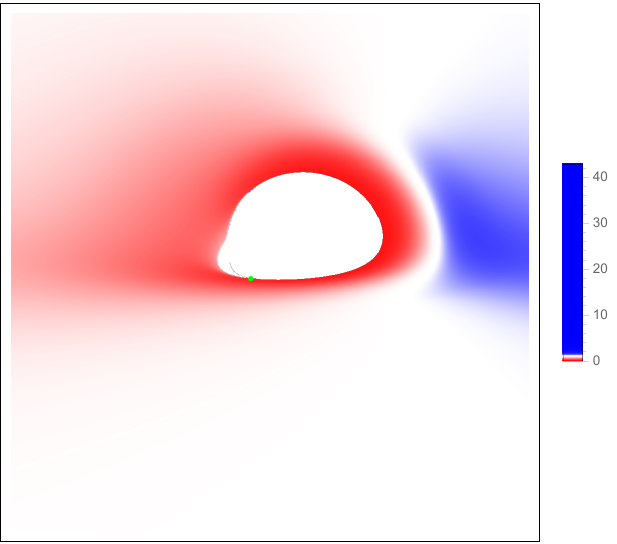}
        \caption{}
    \end{subfigure}
    \begin{subfigure}[b]{0.3\textwidth}
        \includegraphics[width=\textwidth]{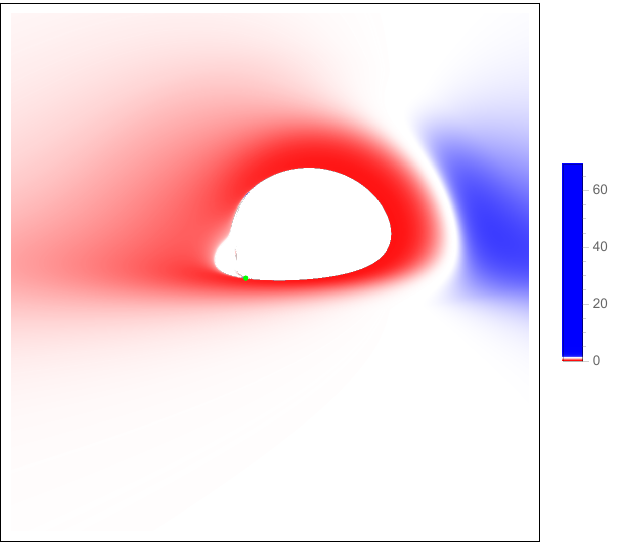}
        \caption{}
    \end{subfigure}
    \begin{subfigure}[b]{0.3\textwidth}
        \includegraphics[width=\textwidth]{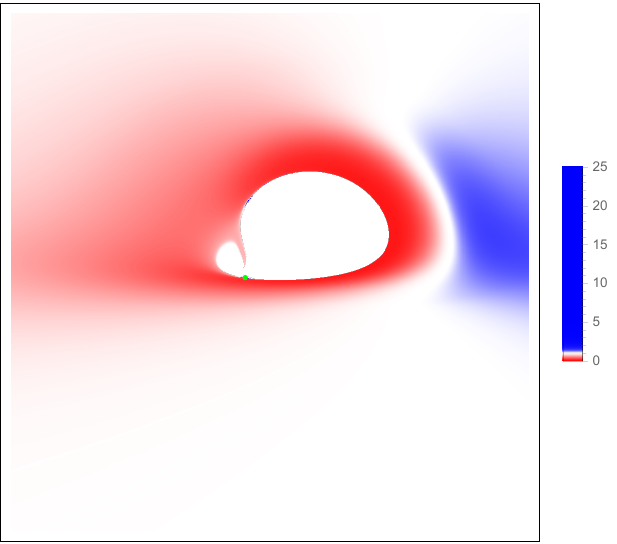}
        \caption{}
    \end{subfigure}
    \begin{subfigure}[b]{0.3\textwidth}
        \includegraphics[width=\textwidth]{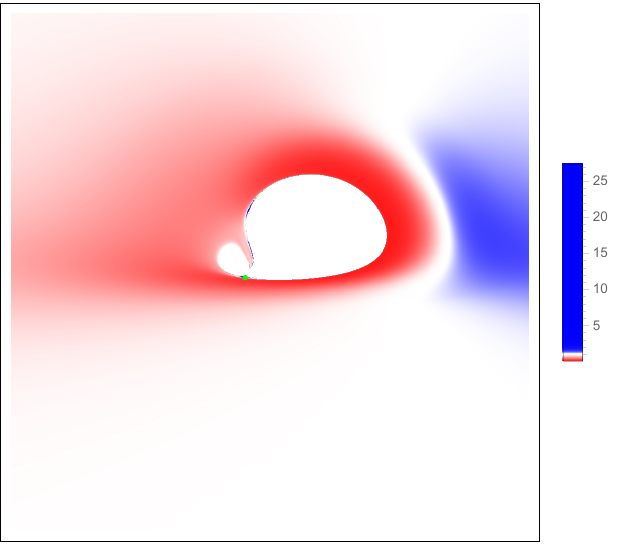}
        \caption{}
    \end{subfigure}
    \caption{The distributions of redshift and blueshift associated with the direct image of JP black holes with non-closed event horizons, for an observer inclination angle $\theta_0 = 80^\circ$. The green markers denote the locations of the maximum blueshift. Panels (a)-(c) correspond to $a = 0.7$ with $\epsilon_3 = 4$, $6$, and $8$, respectively, whereas panels (d)-(f) show the corresponding results for $a = 0.9$.}
    \label{fig21}
\end{figure}

Because the dynamical motion of the thin accretion disk surrounding a JP black hole influences the redshift and blueshift signatures imprinted on the black hole image through the interactions between photons and the accretion disk, it is necessary, following the same methodology adopted for investigating the imaging properties of JP black holes with closed event horizons, to further examine the redshift and blueshift properties in images of JP black holes with non-closed event horizons, as well as the evolutionary behaviors of the redshift and blueshift distributions as the deviation parameter $\epsilon_3$ varies, after completing the analysis of the image morphology and intensity distributions of JP black hole images with non-closed event horizons. Since only retrograde accretion flows can stably exist around JP black holes with non-closed event horizons, the analysis of the redshift and blueshift effects in JP black hole images can be restricted to the redshift and blueshift phenomena produced through the interaction between photons and the retrograde thin accretion disk. The resulting redshift-blueshift maps of JP black hole images with non-closed event horizons, observed at an observer inclination angle of $\theta_0 = 80^\circ$ and produced by interactions between photons and the retrograde thin accretion disk, are shown in Fig. \ref{fig21}.
The first row in Fig.\ref{fig21} displays the redshift and blueshift distributions produced by retrograde accretion flows around a JP black hole with the spin parameter $a = 0.7$, while the second row shows the corresponding distributions for a JP black hole with the spin parameter $a = 0.9$. From left to right, the columns in Fig.\ref{fig21} correspond to deviation parameter values $\epsilon_3 = 4\,, 6$, and $8$, respectively. For the redshift-blueshift images of the JP black hole in the first row in Fig. \ref{fig21}, the white region at the center of the redshift-blueshift map corresponds to the inner shadow of the JP black hole. As established by the structural analysis of the JP black hole images shown in Fig. \ref{fig19}, the area of the inner shadow decreases monotonically with increasing deviation parameter $\epsilon_3$. Consistently, the redshifted region surrounding the inner shadow in the redshift-blueshift maps contracts progressively as $\epsilon_3$ increases. Furthermore, as shown in the first row of Fig. \ref{fig19}, the red color becomes increasingly intense toward the boundary of the inner shadow and gradually weakens with increasing distance from the shadow boundary. This behavior indicates that photons interacting with the retrograde accretion disk experience stronger gravitational redshift when emitted closer to the inner shadow boundary, whereas the redshift effect diminishes at larger radii. In other words, the redshifted emission exhibits a pronounced radial gradient. Moreover, the deepest red color appears at the shadow boundary, implying that the location of the maximum redshift remains anchored to the edge of the inner shadow for JP black holes with non-closed event horizons, in close analogy with the case of JP black holes with closed event horizons shown in Figs. \ref{fig12} and \ref{fig13}. This result demonstrates that, although the event horizon of the JP black hole is non-closed, an infinite-redshift surface surrounding the black hole is still present. Furthermore, since the area of the inner shadow decreases with increasing $\epsilon_3$, the simultaneous contraction of the redshifted region indicates that the radial position of the infinite-redshift surface in JP spacetimes with non-closed event horizons shifts inward as $\epsilon_3$ increases, in tandem with the shrinking of the inner shadow. By contrast, the analysis of the imaging structures of JP black holes with non-closed event horizons at an inclination angle of $\theta_0 = 80^\circ$ in Fig. \ref{fig19} reveals the emergence of a narrow and elongated direct image in the lower-left region of the inner shadow. This elongated feature partitions the inner shadow into two distinct components, referred to as the original inner shadow and the extended inner shadow. With increasing deviation parameter $\epsilon_3$, this narrow direct image progressively migrates inward toward the original inner shadow, thereby leading to a systematic increase in the area of the extended inner shadow on the left side. As demonstrated in the first row of Fig. \ref{fig21}, the narrow direct image is also clearly identifiable in the corresponding redshift-blueshift maps. Moreover, these maps indicate that the elongated direct image responsible for separating the inner shadow is dominated by redshifted emission rather than blueshifted emission. In addition, as illustrated in the first row of Fig. \ref{fig21}, increasing the deviation parameter $\epsilon_3$ causes the narrow direct image inside the inner shadow to shift progressively toward the interior of the original inner shadow. This inward migration not only enlarges the area of the extended inner shadow but also leads to its gradual detachment from the region occupied by the original inner shadow. During this process, the red-colored region surrounding the extended inner shadow becomes progressively fainter, indicating a systematic decrease in the redshift intensity in the vicinity of the extended inner shadow. This behavior can be attributed to the fact that, although photons originating from the region of the extended inner shadow do not reach a distant observer, the region of the extended inner shadow moves progressively farther away from the infinite-redshift surface of the JP black hole as $\epsilon_3$ increases, thereby weakening the observed redshift effect around the extended inner shadow.

For the redshift-blueshift maps of the direct image shown in the second row of Fig. \ref{fig21} for the JP black hole with spin parameter $a = 0.9$, it is evident that the redshift intensity surrounding the extended inner shadow separated by the narrow direct image in Fig. \ref{fig21} (d) is weaker than that around the corresponding extended inner shadow in Fig. \ref{fig21} (a) for the case of $a = 0.7$ at $\epsilon_3 = 4$. This behavior indicates that increasing the spin parameter of the JP black hole from $a = 0.7$ to $a = 0.9$ leads to a reduction in the redshift intensity near the boundary of the extended inner shadow separated by the narrow direct image structure within the inner shadow. As can be inferred from the JP black hole images with non-closed event horizons shown in Fig. \ref{fig19} for an observer inclination angle $\theta_0 = 80^\circ$, the weakening of the redshift effect surrounding the extended inner shadow can be attributed to the increase in the spin parameter of the JP black hole. As the spin parameter increases, the direct image structure separating the original inner shadow from the extended inner shadow gradually shifts toward the interior of the JP black hole shadow, thereby causing the area occupied by the extended inner shadow to increase progressively while simultaneously increasing the separation between the extended inner shadow and the infinite-redshift surface of the JP black hole. Consequently, the redshift effect surrounding the extended inner shadow becomes progressively weaker. Moreover, the second row of Fig. \ref{fig21} shows that the red coloration around the extended inner shadow further fades as $\epsilon_3$ increases, implying a continued decrease in the redshift strength in this region. This trend can be attributed to the gradual broadening of the narrow direct image structure separating the extended inner shadow from the original inner shadow as $\epsilon_3$ increases, thereby enlarging the separation between the two components of the inner shadow and increasing the distance between the extended inner shadow and the infinite-redshift surface of the JP black hole. Consequently, in addition to the reduction in the redshift intensity surrounding the extended inner shadow caused by the increase in the JP black hole spin parameter, which drives the extended inner shadow farther away from the infinite-redshift surface, the increase in $\epsilon_3$ further suppresses the redshift intensity surrounding the extended inner shadow. This behavior is also supported by a comparison between panels (c) and (f) of Fig. \ref{fig21}, from which it can be observed that the red coloration surrounding the extended inner shadow in panel (f) is noticeably weaker than that in panel (c). This observation further confirms the preceding discussion regarding the dependence of the redshift intensity surrounding the extended inner shadow on both the spin parameter $a$ and the deviation parameter $\epsilon_3$.

Furthermore, a detailed examination of the blueshift signatures produced by photon interactions with the retrograde accretion flow in the accretion disk surrounding the black hole is required for JP black hole images with non-closed event horizons and spin parameters $a = 0.7$ and $a = 0.9$, as shown in Fig. \ref{fig21}. The redshift-blueshift maps in Fig. \ref{fig21} indicate that the blueshifted regions are predominantly located on the right-hand side of the inner shadow on the image plane. As the spin parameter increases from $a = 0.7$ to $a = 0.9$ and the deviation parameter $\epsilon_3$ varies from $4$ to $8$, the overall configuration of the blueshifted region outside the inner shadow remains nearly unchanged, implying that the blueshift configuration in the redshift-blueshift map is largely insensitive to both the spin parameter $a$ and the deviation parameter $\epsilon_3$. According to the analysis of the imaging structure of the JP black hole presented in Fig. \ref{fig19}, the area of the inner shadow decreases monotonically with increasing $\epsilon_3$. Consequently, in the redshift-blueshift maps of Fig. \ref{fig21}, the blueshifted region exterior to the inner shadow exhibits a slight global displacement toward the left as $\epsilon_3$ increases, following the contraction of the inner shadow. Furthermore, Fig. \ref{fig21} also presents the locations of maximum blueshift, which are denoted by the green markers in the figure. From the first row of Fig. \ref{fig21}, it is evident that, for the JP black hole with a non-closed event horizon and spin parameter $a = 0.7$, the position of the maximum blueshift is consistently located on the lower boundary of the inner shadow. As shown by the analysis of the imaging structure of the JP black hole with a non-closed event horizon in Fig. \ref{fig19}, the narrow direct image that separates the inner shadow into the original and extended inner shadow components intersects the lower boundary of the entire inner shadow. A comparative examination of the panels in Fig. \ref{fig21} further indicates that, as $\epsilon_3$ increases, the locations of maximum blueshift in images of JP black hole with both $a = 0.7$ and $a = 0.9$ consistently coincide with the intersection between the direct image structure separating the original inner shadow from the extended inner shadow of the JP black hole and the lower boundary of the entire inner shadow. This result indicates that, within the allowed parameter ranges of the spin parameter $a$ and the deviation parameter $\epsilon_3$, the intersection between the direct image structure separating the original inner shadow from the extended inner shadow and the lower boundary of the entire inner shadow uniquely determines the location of maximum blueshift in the direct images of JP black holes with non-closed event horizons.

Moreover, the panels in the second row of Fig. \ref{fig21} indicate that, for the JP black hole with spin parameter $a = 0.9$ and a non-closed event horizon, an additional blueshifted feature emerges near the boundary of the original inner shadow adjacent to the extended inner shadow as the deviation parameter $\epsilon_3$ increases. This feature develops concurrently with the progressive broadening of the narrow direct image that separates the original and extended inner shadows. This phenomenon can be attributed to the fact that, for JP black holes with non-closed event horizons, the broadening of the direct image structure separating the original inner shadow from the extended inner shadow with increasing $\epsilon_3$ introduces non-negligible blueshift contributions in addition to the dominant redshift component within the JP black hole images. Simultaneously, the emergence of the blueshift effect weakens the redshift contribution within the direct image structure separating the original inner shadow from the extended inner shadow in the JP black hole image. Consequently, the redshift effect near the interface between the separating direct image structure and the original inner shadow gradually diminishes, while the blueshift contribution becomes progressively dominant. This process ultimately leads to the formation of a localized blueshifted region near the boundary of the original inner shadow adjacent to the separating direct image structure in the JP black hole image. Therefore, according to the preceding analysis, the weakening of the redshift intensity observed near the boundary of the extended inner shadow in the redshift-blueshift maps shown in Fig. \ref{fig21} cannot be attributed solely to the increase in the distance between the extended inner shadow and the infinite-redshift surface of the JP black hole as the deviation parameter $\epsilon_3$ increases. An additional factor contributing to the reduction of the redshift intensity near the boundary of the extended inner shadow is that, during the process in which the extended inner shadow separates from the original inner shadow, additional blueshift contributions are generated within the direct image structure separating the original and extended inner shadows, thereby partially compensating for the redshift intensity near the boundary of the extended inner shadow. Moreover, since the blueshift effect is primarily introduced by the direct image structure separating the original and extended inner shadows, the weakening of the redshift effect becomes particularly pronounced in the region of the extended inner shadow adjacent to the separating direct image structure. This weakened redshift feature can be clearly observed in Fig. \ref{fig21}(f). Consequently, the attenuation of the redshift intensity around the extended inner shadow is jointly caused by its increasing distance from the infinite-redshift surface and the concomitant amplification of the blueshift contribution in the surrounding region.

\begin{figure}[htbp]
    \centering
    \begin{subfigure}[b]{0.3\textwidth}
        \includegraphics[width=\textwidth]{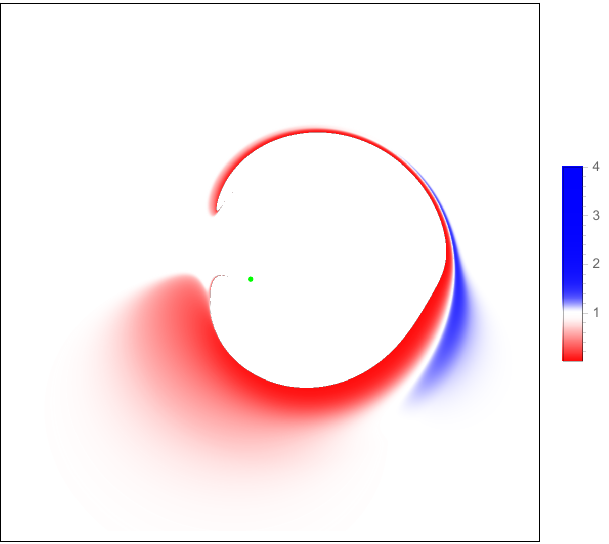}
        \caption{}
    \end{subfigure}
    \begin{subfigure}[b]{0.3\textwidth}
        \includegraphics[width=\textwidth]{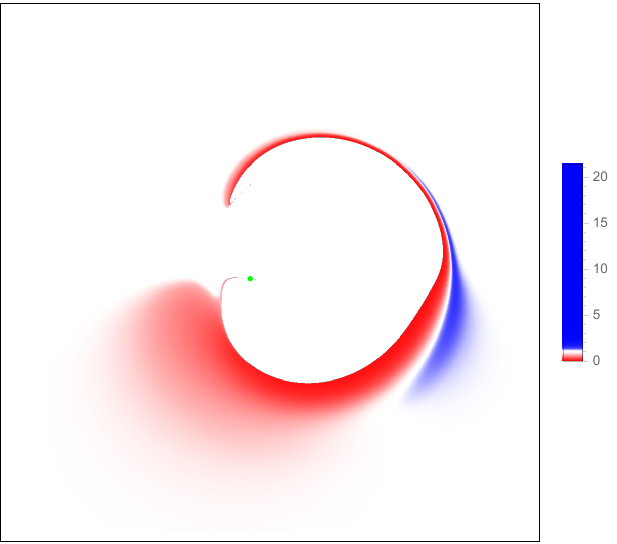}
        \caption{}
    \end{subfigure}
    \begin{subfigure}[b]{0.3\textwidth}
        \includegraphics[width=\textwidth]{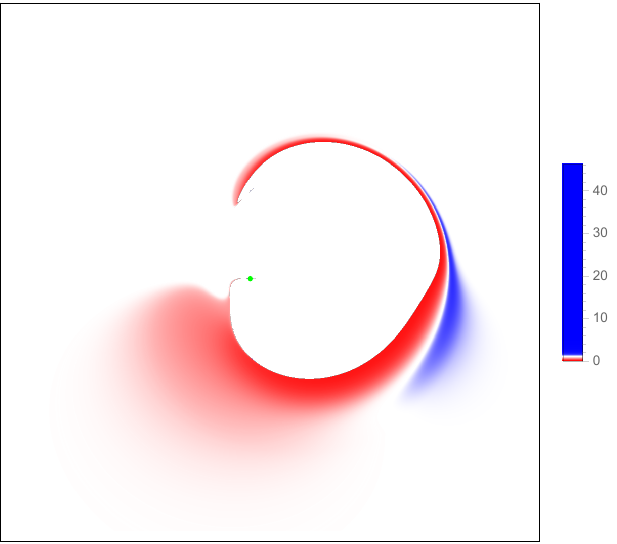}
        \caption{}
    \end{subfigure}
    \begin{subfigure}[b]{0.3\textwidth}
        \includegraphics[width=\textwidth]{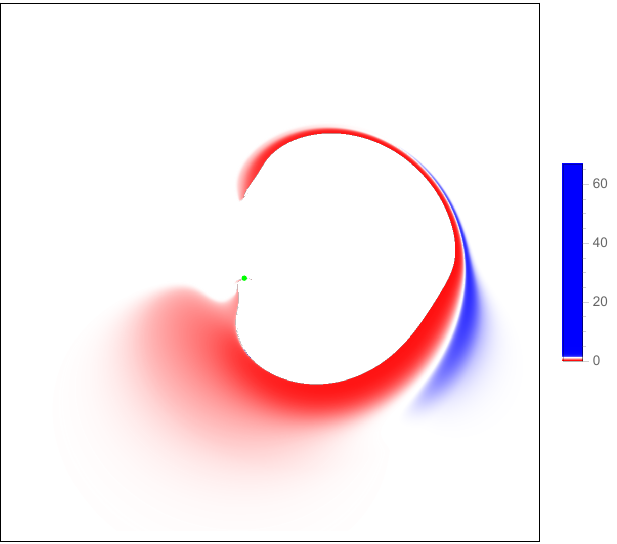}
        \caption{}
    \end{subfigure}
    \begin{subfigure}[b]{0.3\textwidth}
        \includegraphics[width=\textwidth]{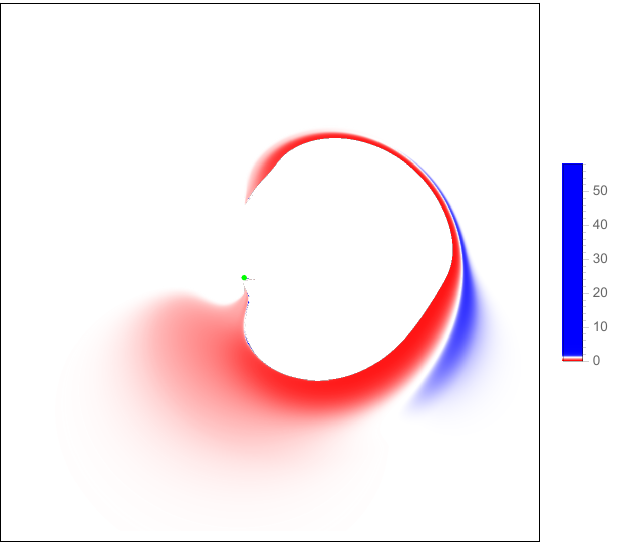}
        \caption{}
    \end{subfigure}
    \begin{subfigure}[b]{0.3\textwidth}
        \includegraphics[width=\textwidth]{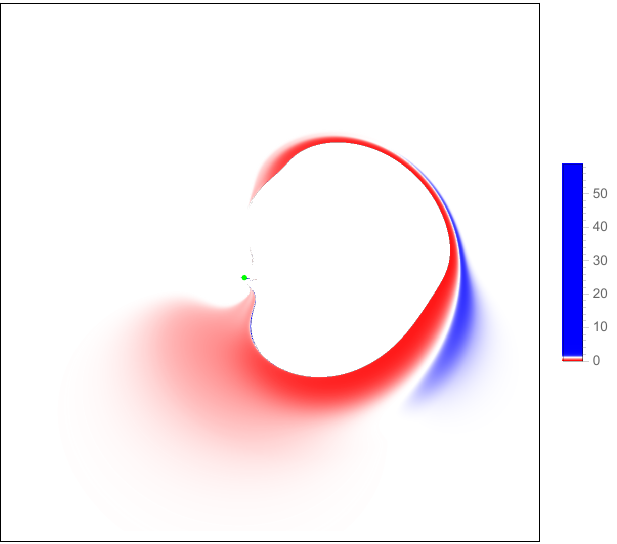}
        \caption{}
    \end{subfigure}
    \caption{The distributions of redshift and blueshift associated with the lensed image of JP black holes with non-closed event horizons, for an observer inclination angle $\theta_0 = 80^\circ$. The green markers denote the locations of the maximum blueshift. Panels (a)-(c) correspond to $a = 0.7$ with $\epsilon_3 = 4$, $6$, and $8$, respectively, whereas panels (d)-(f) show the corresponding results for $a = 0.9$.}
    \label{fig22}
\end{figure}

After examining the redshift and blueshift properties of the direct images produced through photon interactions with retrograde accretion flows in thin accretion disks around JP black holes with non-closed event horizons, and following the same methodology adopted for investigating the redshift and blueshift properties in the images of JP black holes with closed event horizons, one can further investigate the redshift and blueshift characteristics of the corresponding lensed images in order to provide a more complete characterization of the redshift and blueshift signatures in JP black hole images with non-closed event horizons. Fig. \ref{fig22} presents the distributions of redshifted and blueshifted regions in the lensed images of JP black holes with non-closed event horizons, generated through interactions between photons and retrograde accretion flows in the surrounding thin accretion disks.
In Fig. \ref{fig22}, the first row illustrates the redshift-blueshift maps in the lensed images for JP black holes with spin parameter $a = 0.7$, while the second row corresponds to the redshift-blueshift maps in the lensed images for JP black holes with spin parameter $a = 0.9$. From left to right, the columns in Fig. \ref{fig22} represent the results for deviation parameter values $\epsilon_3 = 4\,, 6$, and $8$, respectively.

As shown in the first row of Fig. \ref{fig22}, for the redshift-blueshift maps corresponding to the lensed images of JP black holes with non-closed event horizons and spin parameter $a = 0.7$, the area enclosed by the redshifted regions decreases slightly as the deviation parameter $\epsilon_3$ increases, while the redshift intensity represented by these regions also becomes progressively weaker. According to the discussion associated with Fig. \ref{fig19} regarding the evolution of the imaging structures of JP black holes with increasing deviation parameter $\epsilon_3$, the reduction in the redshifted regions and the weakening of the redshift intensity can primarily be attributed to the inward contraction of the infinite-redshift surface toward the interior of the inner shadow as $\epsilon_3$ gradually increases for JP black holes with non-closed event horizons. This evolution causes the relative distance between the redshifted regions in the JP black hole image and the infinite-redshift surface to increase progressively with increasing $\epsilon_3$, thereby leading to a gradual reduction in the extent of the redshifted regions together with a corresponding decrease in the redshift intensity within these regions. In addition, a narrow lensed image connected to the lower boundary of the entire inner shadow also exhibits redshift. With increasing $\epsilon_3$, the redshift intensity on the left side of this narrow lensed image progressively decreases. Since the region located to the left of the narrow lensed image structure connected to the lower boundary of the JP black hole inner shadow corresponds to the area surrounding the extended inner shadow of the JP black hole, the weakening of the redshift intensity in this region with increasing deviation parameter $\epsilon_3$ indicates that, consistent with the redshift intensity trend observed in the direct images of JP black holes with non-closed event horizons, the redshift intensity near the extended inner shadow in the corresponding lensed images also decreases progressively as the deviation parameter $\epsilon_3$ increases. Furthermore, when $\epsilon_3 = 8$, a prominent blueshift feature emerges at the endpoint of the narrow lensed image structure, corresponding to the intersection between the narrow lensed image and the JP black hole inner shadow. The appearance of this blueshift feature indicates that the reduction of the redshift intensity in the lensed image region surrounding the extended inner shadow of the JP black hole cannot be attributed solely to the gradual displacement of the extended inner shadow away from the infinite-redshift surface of the JP black hole. An additional contribution arises from the emergence of extra blueshift effects within the region connecting the lensed image structure to the lower boundary of the JP black hole inner shadow, which further suppresses the redshift intensity surrounding the extended inner shadow. A comparison between Fig. \ref{fig21} (c) and Fig. \ref{fig22} (c) reveals that, for the case $\epsilon_3 = 8$, although a blueshift effect likewise appears at the intersection between the narrow direct image structure separating the original and extended inner shadows and the lower boundary of the overall inner shadow in the direct image of the JP black hole with a non-closed event horizon and spin parameter $a = 0.7$, as shown in Fig. \ref{fig21} (c), this blueshift contribution exerts only a minor influence on the redshift effect surrounding the extended inner shadow. In contrast, Fig. \ref{fig22}(c) demonstrates that, when the deviation parameter increases to $\epsilon_3 = 8$, the weakening of the redshift effect near the narrow lensed image structure connecting the lensed image in the lower half of the image plane to the lower boundary of the JP black hole inner shadow becomes substantially more pronounced than the weakening of the redshift effect observed near the connection between the direct image structure separating the JP black hole inner shadow and the lower boundary of the inner shadow shown in Fig. \ref{fig21}(c). This result indicates that, as the deviation parameter $\epsilon_3$ increases, the emergence of a narrow direct image structure within the JP black hole inner shadow causes the inner shadow to separate into two distinct components, namely the original inner shadow and the extended inner shadow. During this separation process induced by the narrow direct image structure, additional blueshift effects are introduced both at the intersection between the narrow direct image structure and the lower boundary of the JP black hole inner shadow in the redshift-blueshift maps corresponding to the direct images, and at the intersection between the narrow lensed image structure connecting the lensed image in the lower half of the image plane to the lower boundary of the JP black hole inner shadow in the redshift-blueshift maps corresponding to the lensed images. Furthermore, the blueshift effect introduced by the narrow lensed image structure suppresses the surrounding redshift effect more significantly than the corresponding suppression produced by the narrow direct image structure.

As shown in the second row of panels in Fig. \ref{fig22}, for the redshift-blueshift maps corresponding to the lensed images of JP black holes with spin parameter $a = 0.9$, the evolutionary behavior is consistent with that observed in the corresponding lensed image redshift-blueshift maps for the case $a = 0.7$. Both the area occupied by the redshifted regions and the associated redshift intensity decrease slightly as the deviation parameter $\epsilon_3$ increases. This behavior can primarily be attributed to the gradual increase in the relative distance between the redshifted regions and the infinite-redshift surface of the JP black hole with a non-closed event horizon as the deviation parameter $\epsilon_3$ increases, thereby causing both the extent and intensity of the redshifted regions in the redshift-blueshift maps corresponding to the JP black hole lensed images to decrease progressively with increasing $\epsilon_3$. A comparison between the first and second rows of panels in Fig. \ref{fig22} further demonstrates that, as the spin parameter increases from $a = 0.7$ to $a = 0.9$, the redshifted region associated with the narrow lensed image structure connecting the lower boundary of the JP black hole inner shadow to the lensed image in the lower half of the JP black hole image gradually broadens. This behavior is primarily attributed to the progressive widening of the narrow lensed image structure connecting the JP black hole inner shadow to the lensed image in the lower half of the image plane as the spin parameter of the JP black hole increases, a feature that can be clearly identified in the JP black hole images shown in Fig. \ref{fig19}. Meanwhile, the region covered by the narrow lensed image structure continues to exhibit a dominant redshift effect in the corresponding redshift-blueshift maps. Furthermore, the redshift intensity of this narrow lensed structure decreases as the spin parameter of the JP black hole increases. In addition, from the panels in the second row of Fig. \ref{fig22}, the redshift intensity in the left-side region of the narrow lensed structure connected to the entire inner shadow also diminishes as $\epsilon_3$ increases. Since the region located to the left of the narrow lensed image structure connecting the lower boundary of the JP black hole inner shadow to the lensed image in the lower half of the image plane corresponds to the vicinity of the extended inner shadow, this behavior indicates that the redshift intensity of the lensed image region near the extended inner shadow likewise decreases progressively as $\epsilon_3$ increases. This phenomenon further demonstrates that, with increasing deviation parameter $\epsilon_3$, the extended inner shadow in the JP black hole images with spin parameter $a = 0.9$ progressively separates from the original inner shadow and consequently moves farther away from the infinite-redshift surface of the JP black hole, thereby leading to a gradual reduction in the redshift intensity surrounding the extended inner shadow. Moreover, for $\epsilon_3 = 8$, photons originating from the region immediately to the right of the narrow lensed structure and located at the interface between the direct and lensed images beneath the lower boundary of the inner shadow exhibit a discernible blueshift signature. This result indicates that, in the lensed images of a JP black hole with spin parameter $a = 0.9$, photons originating from the vicinity of the narrow lensed structure connected to the lower boundary of the entire inner shadow persistently exhibit progressively stronger blueshift signatures as the deviation parameter $\epsilon_3$ increases. This implies that the weakening of the redshift intensity within the narrow lensed image structure connecting the lower boundary of the JP black hole inner shadow to the lensed image in the lower half of the JP black hole image, as well as within the lensed image region located to the left of the narrow lensed image structure and adjacent to the extended inner shadow, cannot be attributed solely to the gradual increase in the distance between these regions and the infinite-redshift surface of the JP black hole as the deviation parameter $\epsilon_3$ increases. Another important contributing factor is that the blueshift contribution of photons within these regions becomes progressively enhanced with increasing deviation parameter, thereby partially suppressing the redshift effect in these regions. Furthermore, a comparison between panels (c) and (f) of Fig. \ref{fig22} further demonstrates that, as the JP black hole spin parameter increases from $a = 0.7$ to $a = 0.9$, the blueshift effect of the narrow lensed image structure connecting the lensed image structure to the lower boundary of the JP black hole inner shadow becomes increasingly enhanced. Consequently, the redshift effect at the interface between the lensed image structure and the lower boundary of the JP black hole shadow is further suppressed with increasing spin parameter. As the spin parameter increases from $a = 0.7$ to $a = 0.9$, the extended inner shadow progressively separates from the original inner shadow and recedes from the infinite-redshift surface, resulting in a systematic attenuation of the redshift intensity in the lensed images around the extended inner shadow. Furthermore, in the redshift-blueshift maps of the lensed images for both spin parameter $a = 0.7$ and $a = 0.9$, the blueshift contribution of photons within the narrow lensed structure adjacent to the extended inner shadow and connected to the lower boundary of the entire inner shadow systematically increases with $\epsilon_3$. This enhanced blueshift contribution partially offsets the redshift signal and consequently reduces the redshift intensity in the vicinity of the extended inner shadow in the lensed images. These results demonstrate that, in the redshift-blueshift maps of the lensed images of JP black holes, the redshifted regions that are most sensitive to variations in the spin parameter $a$ and the deviation parameter $\epsilon_3$ are predominantly localized in the vicinity of the extended inner shadow and within the narrow lensed structure connected to the lower boundary of the entire inner shadow.

As illustrated by the redshift-blueshift maps for the lensed images of JP black holes with non-closed event horizons in Fig. \ref{fig22}, the blueshifted emission produced by the interaction between photons and the retrograde accretion flow is predominantly confined to a narrow region on the right-hand side of the redshift-blueshift maps and immediately adjacent to the outer boundary of the redshifted region. A comparison among the individual panels of Fig. \ref{fig22} shows that the geometric configuration of the blueshifted region in the redshift-blueshift maps is nearly insensitive to variations in both the spin parameter $a$ of the JP black hole and the deviation parameter $\epsilon_3$. Moreover, the locations of maximum blueshift can still be identified in the redshift-blueshift maps corresponding to the lensed images of JP black hole images with non-closed event horizons. Accordingly, the locations of maximum blueshift in the redshift-blueshift maps associated with the lensed images of JP black holes for different values of the spin parameter $a$ and deviation parameter $\epsilon_3$ are denoted by green markers in each panel of Fig. \ref{fig22}. From panels (a)-(c) in Fig. \ref{fig22}, the maximum blueshift in the redshift-blueshift maps corresponding to the lensed images of a JP black hole with the spin parameter $a = 0.7$ is consistently located at the endpoint of the narrow lensed structure situated near the extended inner shadow within the redshifted region. Because this endpoint coincides with the intersection between the narrow lensed structure and the lower boundary of the entire inner shadow, the maximum blueshift in the lensed images is systematically anchored at this intersection. In addition, as $\epsilon_3$ increases, the location of the maximum blueshift remains fixed at this same intersection point. Furthermore, as shown in panels (d)-(f), corresponding to the lensed images of a JP black hole with spin parameter $a = 0.9$, although the narrow lensed structure becomes progressively wider, the maximum blueshift in the redshift-blueshift maps of the lensed images remains located at the endpoint of the broadened lensed structure. Since this endpoint continues to coincide with the intersection between the narrow lensed structure and the lower boundary of the entire inner shadow, the maximum blueshift is likewise fixed at this intersection and is essentially independent of the value of $\epsilon_3$. Therefore, although the position of the intersection between the narrow lensed structure and the lower boundary of the entire inner shadow varies with both the spin parameter $a$ and $\epsilon_3$, the maximum blueshift in the redshift-blueshift maps of the lensed images is always attained at this intersection point. Furthermore, according to the analysis of the imaging structure of JP black holes with non-closed event horizons in Fig. \ref{fig19}, the intersection point between the narrow direct image that separates the inner shadow and the lower boundary of the entire inner shadow is almost identical to the intersection point between the narrow lensed structure and the same lower boundary. Consistently, the redshift-blueshift maps of the direct image presented in Fig. \ref{fig21} indicate that the maximum blueshift in the redshift-blueshift maps of the direct images is also located at the intersection between the narrow direct image and the lower boundary of the entire inner shadow. These results indicate that, for JP black holes with non-closed event horizons, the locations of maximum blueshift in the direct and lensed images nearly coincide at the same position on the image plane when the spin parameter $a$ and the deviation parameter $\epsilon_3$ are fixed.

\subsection{Further disscussions for the extended inner shadow in images of JP black holes with non-closed event horizons}

In Sec. \ref{nonclosedhorizonjpbh}, considering spin parameters $a = 0.7$ and $a = 0.9$ and observer inclination angles $\theta_0 = 17^\circ$ and $\theta_0 = 80^\circ$, a comprehensive analysis of the imaging properties of JP black holes with non-closed event horizons was presented. The imaging structures, intensity distributions, and redshift-blueshift signatures in the resulting images were systematically examined. In addition, the dependence of the imaging structures, intensity distributions, and redshift-blueshift effects on the deviation parameter $\epsilon_3$ arising from the JP metric was also investigated. The results indicate that, in contrast to the imaging properties of JP black holes with closed event horizons, the image structures of JP black holes with non-closed event horizons exhibit significantly different morphological characteristics from those of the closed event-horizon cases for both spin parameters $a = 0.7$ and $a = 0.9$, as well as for the two observer inclination angles $\theta_0 = 17^\circ$ and $\theta_0 = 80^\circ$. As shown by the preceding analysis in Sec. \ref{nonclosedhorizonjpbh}, these differences in imaging features arise primarily from the transition of the event horizon of the JP black hole from a closed to a non-closed topology. In particular, an additional direct image structure appears within the entire inner shadow of the JP black hole due to the non-closed event horizon, causing the inner shadow to split into two distinct components, namely the original inner shadow and the extended inner shadow. The emergence of the extended inner shadow substantially alters the overall image morphology and exerts a pronounced influence on the surrounding intensity distribution as well as on the local redshift and blueshift patterns. However, the discussion in Sec. \ref{nonclosedhorizonjpbh} was restricted to image structures, intensity distributions, and redshift-blueshift features, and the physical origin of the extended inner shadow was not addressed. In this section, the formation mechanism of the extended inner shadow in the images of JP black holes with non-closed event horizons will be investigated in detail in order to clarify the underlying physical properties of these images.

As discussed in Sec. \ref{jpmetricjpbh}, the JP metric is obtained by introducing the deviation parameter $\epsilon_3$ into the Kerr metric. The presence of the deviation parameter $\epsilon_3$ therefore leads to a systematic departure of the JP spacetime from the Kerr geometry and changes its Petrov classification from type D to type I. As a consequence, the Carter constant is no longer well defined in the JP spacetime, and the Hamilton-Jacobi equation governing photon motion is, in general, no longer separable. The non-separability of the Hamilton-Jacobi equation precludes a fully rigorous analytical treatment of the construction of JP black hole images. Nevertheless, as discussed in Ref. \cite{Wang:2025ihg}, the Hamilton-Jacobi equation in the JP spacetime can be treated as approximately separable under suitable assumptions, which allows the construction of JP black hole images within an approximate analytical framework. Furthermore, following the same methodology adopted in the preceding sections for investigating the image structures and imaging properties of JP black holes with both closed and non-closed event horizons, the images of JP black holes can also be constructed through rigorous numerical backward ray-tracing calculations. Therefore, in the following discussion of the physical mechanism underlying the formation of the extended inner shadow in JP black hole images with non-closed event horizons, we will investigate the physical origin of the extended inner shadow from both the perspective of fully numerical backward ray-tracing calculations and that of the approximate analytical framework.

\subsubsection{Investigation on the formation mechanism of extended internal shadows through numerical methods}

We first investigate the physical origin of the extended inner shadow in JP black hole images with non-closed event horizons from the perspective of fully numerical backward ray-tracing simulations. By employing the backward ray-tracing numerical method, one can not only construct JP black hole images with high precision, but also accurately reconstruct the propagation trajectories of photons in the spacetime described by the JP metric during the process by which photons emitted from the accretion disk surrounding the JP black hole propagate toward a distant observer and contribute to specific pixels on the image plane. Therefore, by systematically analyzing the photon trajectories in the spacetime described by the JP metric corresponding to the pixels located within the regions occupied by different image structures in the JP black hole image, the physical mechanism underlying the emergence of the extended inner shadow in JP black holes with non-closed event horizons can be directly revealed through numerical backward ray-tracing calculations. In order to clearly characterize the photon trajectories corresponding to the various image structures in JP black hole images, particularly the propagation trajectories in spacetime associated with the photons corresponding to the region of the extended inner shadow, it is necessary to select a representative image in which the different image structures of the JP black hole image can be clearly distinguished. As can be observed from Fig. \ref{fig19} (f), for a JP black hole with spin parameter $a = 0.9$ and deviation parameter $\epsilon_3 = 8$, observed at an inclination angle $\theta_0 = 80^\circ$, the resulting JP black hole image exhibits an extended inner shadow that is nearly completely detached from the original inner shadow of the JP black hole. Since the various image structures associated with the non-closed event horizon in this JP black hole image exhibit clearly distinguishable morphological characteristics, this JP black hole image with a non-closed event horizon is particularly suitable for investigating the formation mechanism of the extended inner shadow in JP black hole images. To distinguish the propagation trajectories in the JP spacetime corresponding to the photons associated with pixels belonging to different image structures in the JP black hole image, the regions corresponding to the various image structures in the JP black hole image with a non-closed event horizon shown in Fig. \ref{fig19} (f) are labeled with numerical identifiers. For each region occupied by the labeled image structures in the JP black hole image, a representative pixel is selected from each numbered region on the image plane, and the trajectory of the corresponding photon in the spacetime described by the JP metric is reconstructed through numerical backward ray-tracing simulations. This method reconstructs the photon trajectories responsible for the formation of the different image structures in JP black hole images by backward tracing the propagation paths of photons from the image plane of the JP black hole image to their emission points on the accretion disk surrounding the JP black hole. Consequently, the formation mechanisms of the various image structures contained in the JP black hole image observed at infinity can be investigated within a systematic and physically transparent framework.

\begin{figure}[htbp]
    \centering
    \includegraphics[width=0.5\textwidth]{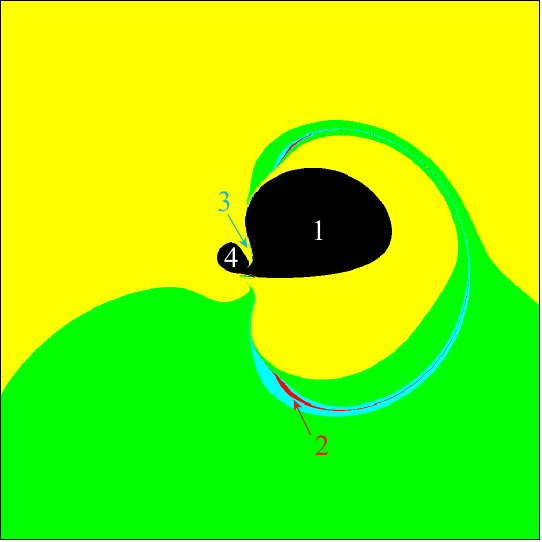}
    \caption{Identification of the distinct image structures in the image of a JP black hole with a non-closed event horizon, obtained for a spin parameter $a = 0.9$, deviation parameter $\epsilon_3 = 8$, and observer inclination angle $\theta_0 = 80^\circ$. Numerical labels are used to identify the different image components, with $1$, $2$, $3$, and $4$ denoting the original inner shadow, the photon ring, the direct image that separates the original and extended inner shadows, and the extended inner shadow, respectively.}
    \label{fig23}
\end{figure}
\begin{figure}[htbp]
    \centering
    \begin{subfigure}[b]{0.4\textwidth}
        \includegraphics[width=\textwidth]{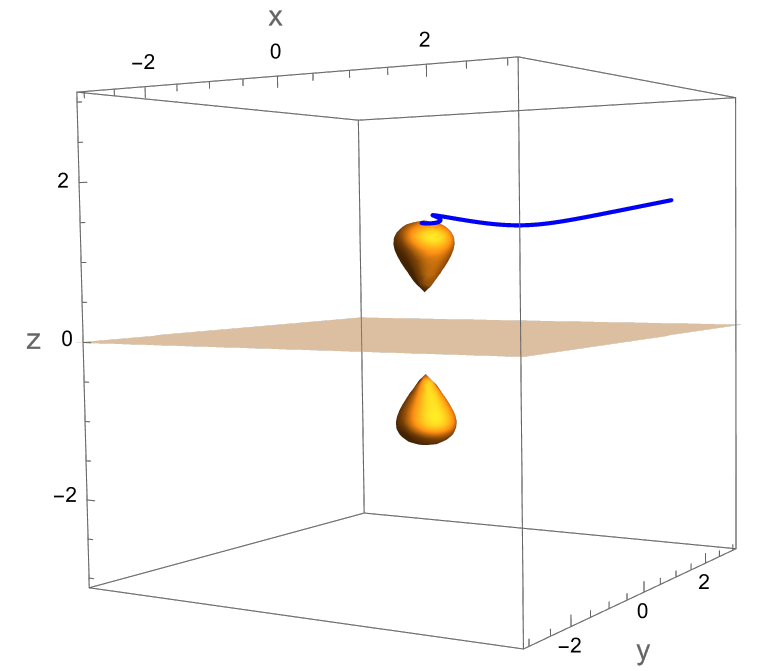}
        \caption{}
    \end{subfigure}
    \begin{subfigure}[b]{0.4\textwidth}
        \includegraphics[width=\textwidth]{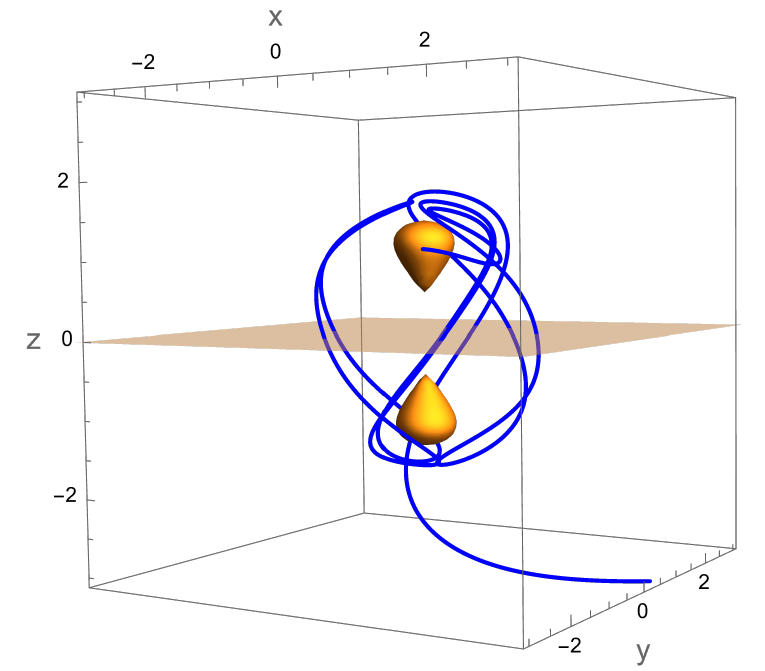}
        \caption{}
    \end{subfigure}
    \begin{subfigure}[b]{0.4\textwidth}
        \includegraphics[width=\textwidth]{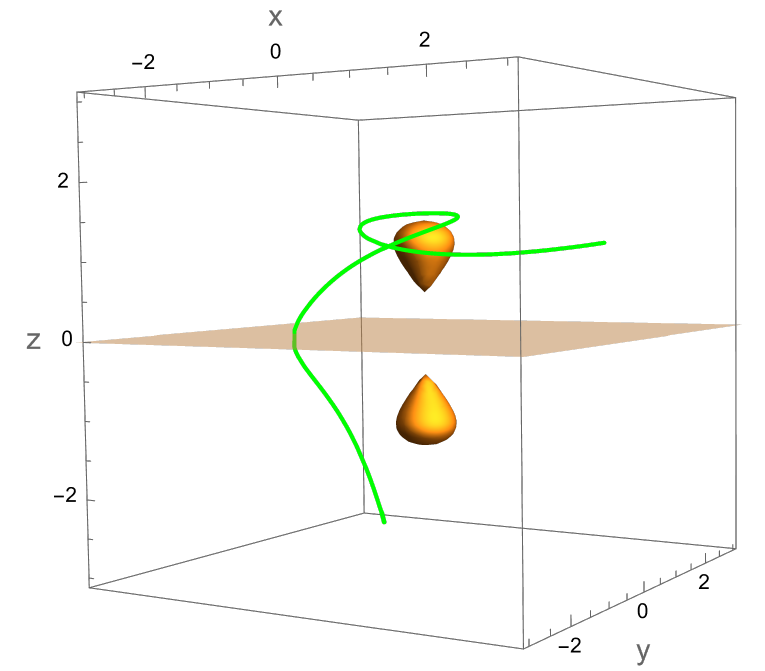}
        \caption{}
    \end{subfigure}
    \begin{subfigure}[b]{0.4\textwidth}
        \includegraphics[width=\textwidth]{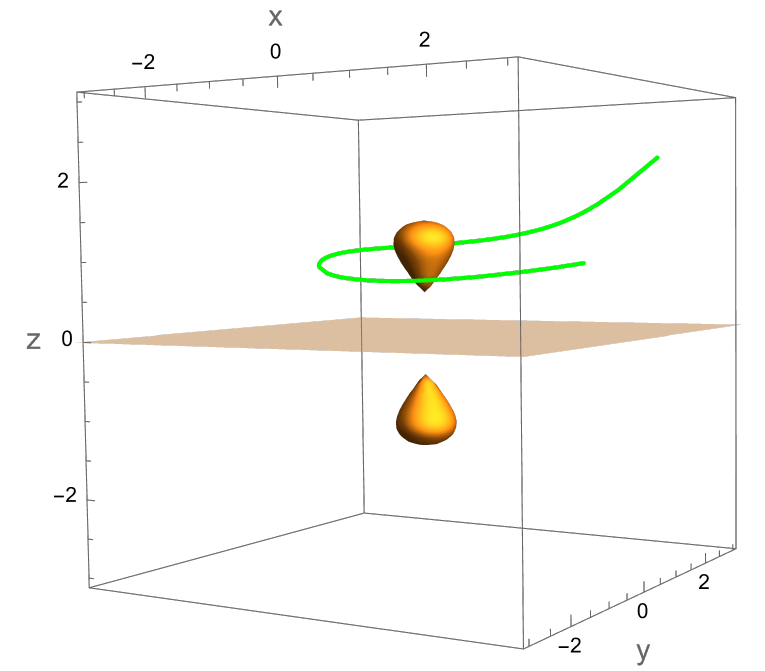}
        \caption{}
    \end{subfigure}
    \caption{Photon trajectories in the JP spacetime, displayed in the $xyz$-space, for representative pixels selected from regions $1$-$4$ of the JP black hole image. Panels (a)-(d) correspond to regions $1$-$4$, respectively. Region $1$ denotes the original inner shadow, region $2$ the photon ring, region $3$ the direct image separating the original and extended inner shadows, and region $4$ the extended inner shadow.}
    \label{fig24}
\end{figure}

Fig. \ref{fig23} shows the JP black hole image with a non-closed event horizon for spin parameter $a = 0.9$ and deviation parameter $\epsilon_3 = 8$ at an observer inclination angle $\theta_0 = 80^\circ$, corresponding to the JP black hole image shown in Fig. \ref{fig19} (f). The four representative imaging structures in Fig. \ref{fig23} are labeled numerically from $1$ to $4$. Region $1$ corresponds to the original inner shadow of the JP black hole. Region $2$ denotes the photon ring surrounding the direct image region adjacent to the inner shadow. Region $3$ represents the narrow direct image that partitions the entire inner shadow into the original and extended inner shadows. Region $4$ corresponds to the extended inner shadow in the JP black hole image. For each region occupied by the numerically labeled image structures in Fig. \ref{fig23}, a representative photon corresponding to a pixel within that region is selected, and its trajectory in the spacetime described by the JP metric is reconstructed using the numerical backward ray-tracing method. Figure \ref{fig24} illustrates the trajectories in the JP spacetime of four representative photons selected from the regions occupied by the four numerically labeled image structures identified in Fig. \ref{fig23}.
In each panel of Fig. \ref{fig24}, the yellow dumbbell-shaped structure depicts the configuration of the non-closed event horizon of the JP black hole in the three-dimensional space spanned by Cartesian coordinates $(x\,, y\,, z)$. The plane passing through the center of the dumbbell-shaped event horizon and oriented perpendicular to its symmetry axis corresponds to the equatorial plane on which the thin accretion disk surrounding the JP black hole with a non-closed event horizon resides. The blue and green curves shown in Fig. \ref{fig24} represent the trajectories of photons in the spacetime described by the JP metric, reconstructed through numerical backward ray-tracing calculations and visualized within a three-dimensional Cartesian coordinate system. Specifically, Fig. \ref{fig24} (a) illustrates the trajectory of a photon associated with Region $1$, which contributes to the formation of the original inner shadow. Fig. \ref{fig24} (b) shows the trajectory corresponding to Region $2$, associated with the photon ring. Fig. \ref{fig24} (c) presents the trajectory associated with Region $3$, corresponding to the narrow direct image that separates the original and extended inner shadows. Fig. \ref{fig24} (d) displays the trajectory associated with Region $4$, corresponding to the extended inner shadow.

As described above, the photon trajectory shown in Fig. \ref{fig24}(a) corresponds to a photon associated with a representative pixel selected from Region $1$ of the JP black hole image shown in Fig. \ref{fig23}, namely the region occupied by the original inner shadow of the JP black hole. This trajectory is reconstructed in the spacetime described by the JP metric through numerical backward ray-tracing calculations. Within the backward ray-tracing framework, this trajectory describes the propagation of a photon traced from Region $1$ on the image plane through the spacetime governed by the JP metric toward the JP black hole. As shown in the figure, when the photon propagates into the vicinity of the non-closed event horizon of the JP black hole, it undergoes a spiral motion around the upper segment of the dumbbell-shaped event horizon before ultimately being captured by the event horizon of the JP black hole. Throughout its entire propagation in the JP spacetime, the photon never crosses the equatorial plane on which the accretion disk resides, indicating that it does not interact with the accretion flow in the accretion disk at any stage of its trajectory. According to the theoretical foundation of the imaging formalism outlined in Sec. \ref{theoreticalframework}, only photons that intersect the accretion disk and interact with the accretion flow contribute to the observed intensity on the image plane of a distant observer, whereas photons that fail to cross the accretion disk do not produce any observable intensity. Consequently, the class of photons represented by Fig. \ref{fig24} (a) gives rise exclusively to dark regions in the JP black hole image. Furthermore, the photon trajectory reconstructed through the backward ray-tracing method and shown in Fig. \ref{fig24} (a) terminates at the event horizon of the JP black hole. Under time reversal, the corresponding future directed null geodesic would appear to originate from the vicinity of the non-closed event horizon of the JP black hole and propagate outward toward a distant observer. However, such a trajectory cannot correspond to a physically realizable photon signal observable at infinity, since the causal structure of the event horizon prohibits null geodesics from escaping either from the horizon itself or from its interior region. Consequently, no physical photon emerging from the event horizon or its interior can propagate along the trajectory represented in Fig. \ref{fig24} (a) and ultimately reach a distant observer. The absence of interaction between the photon and the accretion disk, together with the causal constraints imposed by the event horizon, guarantees that Region $1$ contributes no observable intensity. As a result, these two factors jointly account for the formation of the original inner shadow in the JP black hole image. It should be noted that Fig. \ref{fig24} (a) depicts only a representative photon trajectory reconstructed for Region $1$ through the backward ray-tracing procedure in the JP spacetime. As illustrated, the photon undergoes orbital motion around the upper segment of the dumbbell-shaped non-closed event horizon before ultimately being captured by the black hole. Owing to the reflection symmetry of the JP spacetime with respect to the equatorial plane, photons associated with Region $1$ may also follow symmetric trajectories that encircle the lower segment of the non-closed event horizon prior to being captured by the black hole.

The photon trajectory shown in Fig. \ref{fig24} (b) corresponds to the propagation trajectory in JP spacetime of a photon associated with a representative pixel selected from the photon ring structure identified as Region $2$ in the JP black hole image shown in Fig. \ref{fig23}. This photon trajectory is likewise reconstructed using the numerical backward ray-tracing method and serves to characterize the propagation properties of the photons responsible for the formation of the photon ring structure in the spacetime governed by the JP metric. As evidenced by the photon trajectory shown in Fig. \ref{fig24} (b), the formation of the photon ring structure in the image of a JP black hole with a non-closed event horizon can be attributed to the repeated oscillatory motion of photons within the region bounded by the upper and lower segments of the dumbbell-shaped non-closed event horizon. During this oscillatory process, the photons repeatedly traverse the accretion disk and undergo multiple interactions with the accreting matter, thereby generating the photon ring structure observed in the JP black hole image on the image plane of a distant observer. Furthermore, the reconstructed photon trajectory shown in Fig. \ref{fig24} (b) demonstrates that the motion of the photons responsible for the formation of the photon ring structure in the JP black hole image is simultaneously influenced by both the upper and lower segments of the dumbbell-shaped non-closed event horizon. In other words, the dynamical behavior of these photons in the vicinity of the JP black hole is governed by the combined gravitational influence of the two segments, rather than by either the upper or the lower segment independently. This result indicates that, during the multiple crossings of the accretion disk and the associated interactions with the accretion flow, the photon does not remain confined to the dynamical influence of only one segment of the event horizon. Instead, its trajectory repeatedly samples the gravitational fields associated with both segments of the dumbbell-shaped non-closed event horizon. For the representative trajectory shown in Fig. \ref{fig24} (b), the photon never remains confined to the vicinity of only one segment of the dumbbell-shaped non-closed event horizon after crossing the accretion disk. Rather, each subsequent passage through the accretion disk occurs after the photon has experienced the combined gravitational influence of both segments of the non-closed event horizon.

The photon trajectory shown in Fig. \ref{fig24} (c) corresponds to the propagation trajectory in JP spacetime of a photon associated with a representative pixel selected from Region $3$ in Fig. \ref{fig23}, namely the region occupied by the direct image structure that partitions the inner shadow of the JP black hole into the original inner shadow and the extended inner shadow. As illustrated by the photon trajectory shown in Fig. \ref{fig24} (c), the selected photon approaches the vicinity of the dumbbell-shaped non-closed event horizon of the JP black hole. Upon reaching this region, it undergoes orbital motion around the upper segment of the dumbbell-shaped non-closed event horizon before moving downward and intersecting the accretion disk located on the equatorial plane of the JP black hole, where it experiences a single interaction with the accretion flows. From the perspective of forward time evolution, the reconstructed trajectory indicates that the photon is emitted from the accretion disk located on the equatorial plane of the JP black hole, subsequently executes orbital motion around the upper segment of the dumbbell-shaped non-closed event horizon, and ultimately propagates to the image plane of a distant observer at infinity. Since the photon undergoes only a single interaction with the accreting matter, it contributes to the direct image component of the JP black hole image. Photons following this class of trajectories collectively form the direct image structure that partitions the inner shadow of the JP black hole on the image plane into two distinct components, namely the original inner shadow and the extended inner shadow. It should be noted that the trajectory presented in Fig. \ref{fig24} (c) represents only a representative photon selected from Region $3$ and reconstructed within the backward ray-tracing framework, in which the photon orbits the upper segment of the dumbbell-shaped non-closed event horizon before intersecting the accretion disk. Owing to the reflection symmetry of the dumbbell-shaped non-closed event horizon with respect to the equatorial plane, analogous trajectories also exist in which photons orbit the lower segment of the event horizon prior to intersecting the accretion disk.

Finally, a central objective of the present analysis is to elucidate the formation mechanism of the extended inner shadow in JP black hole images using backward ray-tracing numerical simulations. Fig. \ref{fig24} (d) presents the photon trajectory in the spacetime described by the JP metric, reconstructed through numerical backward ray-tracing calculations, of a photon associated with a representative pixel selected from the region occupied by the extended inner shadow, identified as Region $4$ in the JP black hole image shown in Fig. \ref{fig23}. As illustrated in Fig. \ref{fig24} (d), beginning from the representative pixel selected on the image plane of a distant observer, the corresponding photon is traced backward along the reverse direction of its null geodesic toward the upper segment of the dumbbell-shaped non-closed event horizon. After reaching the vicinity of the upper segment, the photon undergoes orbital motion around this region and subsequently escapes to spatial infinity along a trajectory in a direction different from the line of sight of the distant observer. Throughout its entire evolution in the JP spacetime, the selected photon neither intersects the accretion disk located on the equatorial plane nor is captured by the black hole. In the forward time physical interpretation, this trajectory corresponds to a photon arriving from spatial infinity, approaching the vicinity of the event horizon, orbiting under the influence of the local spacetime curvature near the upper segment of the dumbbell-shaped non-closed event horizon, and then escaping toward the distant observer. Owing to the reflection symmetry of the non-closed event horizon with respect to the equatorial plane, analogous trajectories also exist in which photons orbit the lower segment of the dumbbell-shaped event horizon before propagating to the distant observer. Because the photon does not undergo any interaction with the accretion disk throughout its propagation, it makes no contribution to the observable intensity of the JP black hole image. Consequently, the class of photon trajectories represented by Fig. \ref{fig24} (d), which neither intersects the accretion disk nor interacts with the accreting matter, does not contribute to the observable intensity distribution of the JP black hole image. Furthermore, under the corresponding forward-time interpretation, the photon trajectory represented in Fig. \ref{fig24} (d) originates from spatial infinity rather than from the accretion disk surrounding the JP black hole. Since the construction of JP black hole images is based on the assumption that only photons emitted from the accretion disk and subsequently propagating through the JP spacetime to the image plane of a distant observer contribute to the observed intensity distribution, photons originating from spatial infinity in the background spacetime of the JP black hole make no contribution to the observable intensity distribution of the resulting JP black hole image. Therefore, only photons emitted from the accretion disk surrounding the JP black hole and subsequently interacting with the accreting matter during their propagation can contribute to the observable intensity distribution of the JP black hole image detected by a distant observer. These two necessary conditions for image formation imply that the region on the image plane associated with photons following trajectories similar to that shown in Fig. \ref{fig24} (d) remains a dark region in the JP black hole image. Consequently, in addition to the original inner shadow of the JP black hole, the region on the image plane corresponding to this class of photon trajectories gives rise to an additional dark region in the image, which is identified as the extended inner shadow. The original inner shadow and the extended inner shadow originate from two fundamentally distinct classes of null geodesics. The former is generated by photon trajectories captured by the event horizon, whereas the latter is produced by photon trajectories that neither intersect the accretion disk nor terminate at the event horizon.

\subsubsection{Investigation on the formation mechanism of extended internal shadows through approximately analytical methods}

In the preceding section, the propagation trajectories in JP spacetime of the photons responsible for the formation of the various image structures in JP black hole images with non-closed event horizons were reconstructed through numerical backward ray-tracing simulations, thereby providing a direct explanation for the formation mechanism of the extended inner shadow in JP black hole images. By reconstructing the trajectories of photons associated with distinct image components in JP black hole images at an inclination angle of $\theta_0 = 80^\circ$, the photon dynamics in the spacetime governed by the JP metric were systematically analyzed. The results demonstrate that photons corresponding to the region of the extended inner shadow do not undergo any interaction with the accretion disk on the equatorial plane at any stage of the propagation process. Furthermore, the backward ray-tracing analysis shows that these photons are not associated with emission from the accretion disk surrounding the JP black hole. Instead, these photons are traced back to the asymptotic region of the JP spacetime. These two properties jointly imply that such photons do not contribute to the observable intensity distribution on the image plane of a distant observer. Consequently, the image-plane region associated with this class of photon trajectories appears as a dark region in the JP black hole image and is identified as the extended inner shadow. Therefore, unlike the original inner shadow, which is generated by photon trajectories captured by the event horizon, the extended inner shadow originates from a distinct class of photon trajectories that neither intersect the accretion disk nor terminate at the event horizon.

Moreover, previous investigations of the imaging characteristics of JP black holes have demonstrated that shadow images of JP black holes can be constructed not only through rigorous backward ray-tracing numerical simulations but also within an approximate analytical framework over an appropriate range of the deviation parameter. Although the analytical approach provides only an approximate description of the JP black hole shadow, it nevertheless reproduces the principal features of the resulting image morphology with reasonable accuracy. When employing an approximate analytical framework to investigate the shadow of JP black holes or the images produced by surrounding accretion disks, it is essential to formulate an analytical description of photon dynamics in the spacetime defined by the JP metric. Accordingly, the primary and most fundamental step in above an analytical treatment is the derivation of explicit expressions for the photon equations of motion, which are obtained from the Hamilton-Jacobi equation governing null geodesic dynamics in the spacetime. The JP metric is constructed by introducing a deviation parameter $\epsilon_3$ into the Kerr metric, with the magnitude of $\epsilon_3$ quantifying the departure from the Kerr geometry. The presence of this deviation parameter modifies the Petrov classification of the spacetime from type D to type I. Since the Kerr spacetime is of Petrov type D, the Carter constant is well defined in the Kerr geometry. In contrast, the introduction of the deviation parameter $\epsilon_3$ in the JP metric alters the algebraic structure of the spacetime, resulting in a Petrov type I classification. As a consequence, the hidden symmetry associated with the existence of the Carter constant is no longer preserved, and no exact Carter constant is known to exist in the JP spacetime.

Since the existence of a well-defined Carter constant in Kerr spacetime permits the exact separation of variables in the Hamilton-Jacobi equation governing null geodesics, the equations of motion for photons in the Kerr geometry can be derived in a fully analytical and self-consistent manner. However, the JP metric is constructed by introducing a deviation parameter $\epsilon_3$ into the Kerr solution, and the presence of this deviation parameter precludes the existence of an exactly defined Carter constant in the spacetime described by the JP metric. Nevertheless, the JP metric continuously reduces to the Kerr metric in the limit $\epsilon_3 \to 0$. In this limit, the algebraic structure of the JP spacetime likewise reduces to that of the Kerr geometry, and the exact Carter constant of Kerr spacetime is recovered. Consequently, when the deviation parameter satisfies the limiting condition $\epsilon_3 \to 0$, a separation constant that reduces to the Carter constant in the Kerr limit can be introduced within the JP spacetime. Furthermore, for sufficiently small values of the deviation parameter $\epsilon_3$, the Hamilton-Jacobi equation governing photon motion in the JP spacetime can be treated as approximately separable, and the photon equations of motion can be constructed within this approximate framework. In BL coordinates $(t\,, r\,, \theta\,, \phi)$ adapted to the JP metric, the photon equations of motion can be expressed as four first-order differential equations along the respective coordinate directions. The radial and polar components of these equations can be reformulated in terms of the radial potential $R (r)$ and the angular potential $\Theta (\theta)$, respectively. These two effective potentials play a crucial role in determining the structure of photon trajectories in the vicinity of the JP black hole. By exploiting the conserved quantities in JP spacetime, namely the energy $E$ and the axial angular momentum $L$, and introducing the Carter constant $K$ arising from the approximate separability of the Hamilton-Jacobi equation, one can define two dimensionless constants of motion, $\xi = L / E$ and $\eta = K / E^2$. These two dimensionless quantities are conventionally referred to as the photon impact parameters and characterize the null geodesics in the approximate analytical description of the JP spacetime. Specifically, photons possessing identical values of $\xi$ and $\eta$ follow the same null geodesic trajectory. In the Kerr limit, they reduce to the standard impact parameters associated with Kerr null geodesics. Accordingly, the radial potential $R(r)$ and the angular potential $\Theta(\theta)$ can be expressed in terms of these impact parameters as
\begin{equation}\label{epxressionsofpotentials}
    \begin{split}
        R (r) = & \left[ (r^2 + a^2) E - a L \right]^2 - \Delta \left[ K + (L - a E)^2 + m^2 r^2 \right] \\
        \Theta (\theta) = & K - \left(L^2 \csc^2 \theta - a^2 E^2 \right) \cos^2 \theta - \left(a^2 \cos^2 \theta + \frac{a^2 h \sin^2 \theta}{\Delta} \Sigma \right) m^2\,.
    \end{split}
\end{equation}
In addition, the explicit form of the photon equations of motion involves an additional function denoted by $\tilde{I} \left(r\,, \theta \right)$. Since this function depends simultaneously on both the radial coordinate $r$ and the polar coordinate $\theta$, it cannot be uniquely associated with either the radial or the angular sector of the photon dynamics. Consequently, the function $\tilde{I} \left(r\,, \theta \right)$ cannot be incorporated exclusively into either the radial potential $R(r)$ or the angular potential $\Theta(\theta)$. As a result, the photon equations of motion cannot be written in a form that is exactly separable into purely radial and purely angular components. Therefore, within the Hamilton-Jacobi formulation, the presence of $\tilde{I}(r,\theta)$ constitutes the direct manifestation of the non-separability of the photon dynamics in the JP spacetim. From a geometrical perspective, the appearance of this mixed coordinate-dependent function can be viewed as a manifestation of the departure of the JP spacetime from the Petrov type D structure characteristic of the Kerr geometry to a Petrov type I spacetime. Correspondingly, the hidden symmetry responsible for the existence of the Carter constant in Kerr spacetime is no longer preserved, thereby preventing an exact separation of the Hamilton-Jacobi equation.

Previous studies of the shadow properties of JP black holes have examined in detail the coordinate dependence of the function $\tilde{I} \left(r\,, \theta \right)$. The analysis demonstrates that, at radial distances sufficiently far from the event horizon of a JP black hole, the behavior of $\tilde{I} \left(r\,, \theta \right)$ is predominantly governed by the radial coordinate $r$, whereas its sensitivity to variations in the polar coordinate $\theta$ becomes negligible. Under this approximation, the function can be effectively incorporated into the radial potential $R (r)$, thereby enabling an approximate analytical construction of the shadow of JP black holes with closed event horizons. However, earlier investigations have shown that, when the event horizon of a JP black hole is non-closed, the shadow constructed using this approximate analytical approach exhibits significant deviations from the exact shadow obtained through backward ray-tracing simulations. This result indicates that, in the case of a non-closed event horizon, incorporating $\tilde{I} \left(r\,, \theta \right)$ solely into the radial potential $R (r)$ is insufficient to reproduce the correct shadow morphology of the JP black hole. In particular, the dependence of $\tilde{I} \left(r\,, \theta \right)$ on the polar coordinate $\theta$ becomes non-negligible once the event horizon departs from a closed configuration. Accordingly, in the analysis of images of JP black holes with non-closed event horizons surrounded by thin accretion disks, an approximate analytical treatment cannot assign $\tilde{I} \left(r\,, \theta \right)$ exclusively to either the radial potential $R (r)$ or the angular potential $\Theta (\theta)$. Instead, its simultaneous dependence on both coordinates must be retained in order to capture the essential features of photon dynamics in the vicinity of the non-closed event horizon. Moreover, since the accretion disk is assumed to extend down to the event horizon, the dependence of $\tilde{I} \left(r\,, \theta \right)$ on the polar coordinate $\theta$ should be carefully examined in the vicinity of the event horizon radius $r = r_h$. As illustrated in Fig. 2 of Ref. \cite{Wang:2025ihg}, as the radial coordinate approaches the radius of the event horizon $r_h$, the dependence of $\tilde{I} \left(r\,, \theta \right)$ on both $r$ and $\theta$ becomes increasingly significant. Therefore, in order to maintain consistency with the exact images of JP black hole with non-closed event horizons obtained from backward ray-tracing simulations, the approximate analytical treatment should retain the full coordinate dependence of $\tilde{I} \left(r\,, \theta \right)$.

To explicitly account for the fact that the function $\tilde{I} \left(r\,, \theta \right)$ depends simultaneously on both the radial coordinate $r$ and the polar coordinate $\theta$, and consequently influences both the radial potential $R (r)$ and the angular potential $\Theta (\theta)$, its contribution to the photon equations of motion can be incorporated in a symmetric manner into the formulations of the two effective potentials. In the absence of a rigorous criterion for uniquely assigning $\tilde{I} \left(r\,, \theta \right)$ to either sector, an equal partitioning provides a natural and symmetric approximation. In this approach, the original expressions for the radial and angular potentials given in Eq. (\ref{epxressionsofpotentials}) are reformulated by distributing $\tilde{I} \left(r\,, \theta \right)$ equally between the radial and angular potentials. Specifically, one half of $\tilde{I} \left(r\,, \theta \right)$ is incorporated into the radial potential $R (r)$, while the remaining half is included in the angular potential $\Theta (\theta)$. This equal partitioning reflects the assumption that the radial and angular dependences of $\tilde{I} \left(r\,, \theta \right)$ are of comparable importance and signifies that its contribution to the radial and angular dynamics is treated on an equal footing. Accordingly, the reformulated expressions for the radial and angular potentials can be written as
\begin{equation}\label{refexpofpotentials}
    \begin{split}
        R (r) = & \left[ (r^2 + a^2) E - a L \right]^2 - \Delta \left[ K + (L - a E)^2 + m^2 r^2 \right] + \frac{1}{2} \tilde{I} \left(r\,, \theta \right)\,,\\
        \Theta (\theta) = & K - \left(L^2 \csc^2 \theta - a^2 E^2 \right) \cos^2 \theta - \left(a^2 \cos^2 \theta + \frac{a^2 h \sin^2 \theta}{\Delta} \Sigma \right) m^2 + \frac{1}{2} \tilde{I} \left(r\,, \theta \right)\,.
    \end{split}
\end{equation}

\begin{figure}[htbp]
    \centering
    \includegraphics[width=0.8\textwidth]{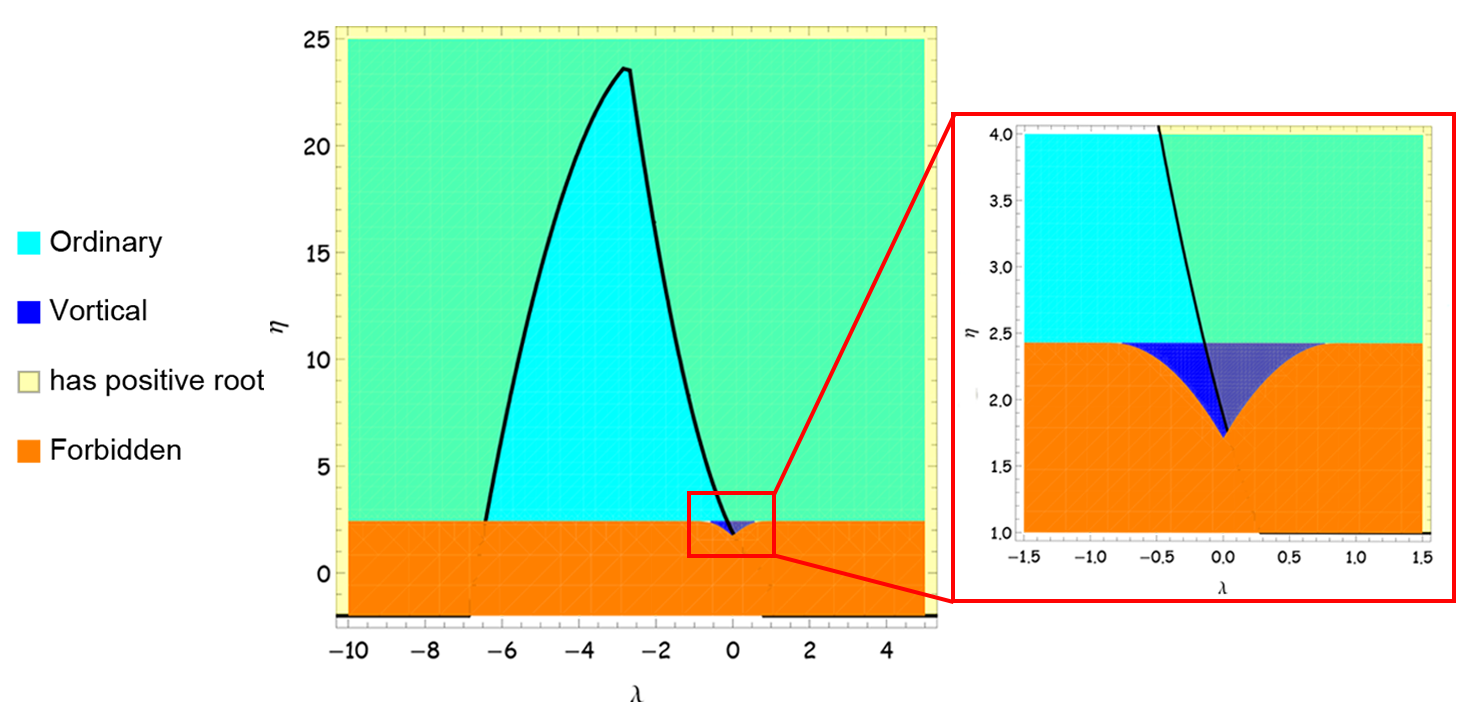}
    \caption{Parameter space of the photon impact parameters $(\xi\,, \eta)$. The different colored regions are separated by the boundary curves determined from the constraint relations between $\xi$ and $\eta$ obtained from the radial potential equations $R(r) = 0$ and $R'(r) = 0$, together with the angular potential equation $\Theta(\theta) = 0$.}
    \label{fig25}
\end{figure}

Furthermore, the reformulated expressions for the radial and angular potentials in Eq. (\ref{refexpofpotentials}) can be employed to analyze the structure of photon trajectories in the vicinity of JP black holes. From these reformulated potentials, two corresponding equations can be obtained, namely the radial potential equation $R (r) = 0$ and the angular potential equation $\Theta (\theta) = 0$. Solving these equations yields two sets of roots associated with the radial and angular potentials, respectively. From these sets of solutions, one can further deduce the constraint relations between the photon impact parameters $\xi$ and $\eta$ implied by the radial and angular potential equations. These constraint relations partition the parameter space spanned by $(\xi, \eta)$ into distinct regions. Since null geodesics in the JP spacetime are uniquely specified by the impact parameters $\xi$ and $\eta$, analyzing the geodesic structure corresponding to different regions in the $(\xi, \eta)$ parameter space provides an approximate analytical characterization of photon motion in the vicinity of the JP black hole. In particular, the constraint relation between the impact parameters $\xi$ and $\eta$ derived from the roots of the angular equation $\Theta (\theta) = 0$ is first established. Subsequently, the geodesic characteristics of photons associated with the distinct regions delineated in the $(\xi, \eta)$ parameter space are analyzed on the basis of this constraint relation. According to the constraint relation obtained from the angular potential equation $\Theta (\theta) = 0$, the parameter space spanned by $(\xi, \eta)$ can be partitioned into three distinct regions, as shown in Fig. \ref{fig25}.
In Fig. \ref{fig25}, the first region, depicted in orange, corresponds to the domain in the $(\xi, \eta)$ parameter space where the angular potential $\Theta (\theta)$ associated with photons of given impact parameters $\xi$ and $\eta$ satisfies the condition $\Theta (\theta) < 0$. This condition implies that no real angular motion is permitted, and therefore photons with impact parameters in this region cannot propagate along physically admissible trajectories. In other words, the photon geodesics corresponding to this region in the $(\xi, \eta)$ parameter space are not physically realizable. Moreover, in Fig. \ref{fig25}, the second region is represented by the green and light blue areas, while the third region is denoted by the dark blue area. In both the second and third regions, the angular potential satisfies the condition $\Theta(\theta) \geq 0$, indicating that the photon trajectories associated with the corresponding ranges of impact parameters are physically admissible. However, the photon geodesics corresponding to the second and third regions exhibit qualitatively distinct dynamical behaviors in the vicinity of the JP black hole. Photon trajectories associated with the ranges of impact parameters belonging to the second region in the $(\xi, \eta)$ parameter space are able to cross the equatorial plane of the JP black hole at $\theta = \pi/2$. This propagation behavior implies that these photons can intersect and interact with the thin accretion disk located on the equatorial plane surrounding the black hole. Consequently, these photons contribute to the intensity distribution observed on the image plane of a distant observer. The photon geodesics corresponding to the second region are commonly referred to as original geodesics. In contrast, photon trajectories associated with the ranges of impact parameters belonging to the third region are confined to a single hemisphere of the JP black hole spacetime. As a result, these photons remain restricted to one side of the spacetime and never cross the equatorial plane at $\theta = \pi/2$ during their propagation. Accordingly, the motion of these photons in the vicinity of the JP black hole is restricted to one of two disconnected angular domains, namely $\theta \in [0, \pi/2]$ or $\theta \in [\pi/2, \pi]$. Therefore, photons corresponding to impact parameters in the third region do not interact with the accretion disk and consequently make no contribution to the observable intensity on the image plane. The photon geodesics associated with the third region are commonly referred to as vortical geodesics.

The preceding analysis of the angular potential $\Theta (\theta)$ appearing in the photon equation of motion along the $\theta$ direction in BL coordinates demonstrates that the angular potential primarily governs whether a photon can cross the equatorial plane of the JP black hole during its propagation in the vicinity of the black hole. Consequently, it determines whether the photon can interact with the accreting matter in the accretion disk located on the equatorial plane. However, the geodesic behavior of photons near the black hole is not determined solely by the angular potential $\Theta (\theta)$. The radial potential $R (r)$, appearing in the radial component of the photon equations of motion, likewise imposes essential constraints on photon trajectories. To analyze the influence of the radial potential $R (r)$ on photon motion, one can derive the constraint relations between the photon impact parameters $\xi$ and $\eta$ from the roots of the radial equation $R (r) = 0$. These constraint relations provide insight into the restrictions imposed by the radial potential on photon trajectories in the vicinity of the JP black hole. The radial potential determines the existence of critical photon trajectories associated with unstable spherical photon orbits, which in turn define the critical curve observed on the image plane of a distant observer. This critical curve provides a natural classification of photon geodesics in the vicinity of the black hole. If the radial component of a photon trajectory crosses the critical boundary and evolves toward the capture region, the photon may become gravitationally bound and ultimately cross the event horizon, falling into the black hole. In contrast, if the corresponding photon trajectory remains in the escape region defined by the radial potential, the photon can escape to spatial infinity. The contrasting dynamical behavior of photons on either side of the critical curve determines whether a distant observer can receive photons and, consequently, whether these photons contribute to the observed intensity distribution in the black hole image. Since the dynamical behavior of photons in the vicinity of the critical curve is determined by the properties of the radial potential $R (r)$, a detailed investigation of the influence of $R (r)$ on the dynamical evolution of photon geodesics located on either side of the critical curve is crucial for understanding the formation of the distinct image structures present in the resulting JP black hole image. Since a photon geodesic is uniquely determined by the impact parameters $\xi$ and $\eta$, the constraints imposed by the radial potential can be analyzed in a manner analogous to that adopted for the angular potential $\Theta (\theta)$. In particular, the constraint relations among the photon impact parameters determined by the roots of the radial potential equation $R (r) = 0$ can be represented by corresponding curves in the $(\xi\,, \eta)$ parameter space, thereby providing a useful framework for investigating the constraints imposed by the radial potential on the dynamical evolution of photon trajectories. This representation enables a systematic and transparent identification of the regions in the $(\xi\,, \eta)$ parameter space corresponding to photons whose trajectories are associated with capture by the black hole and those corresponding to photons whose geodesics remain in the scattering region and ultimately escape to spatial infinity.

Since the dynamical behavior of photons differs on either side of the critical curve, any analysis of photon dynamics in the $(\xi\,, \eta)$ parameter space must first establish the constraint relation between the impact parameters $\xi$ and $\eta$ corresponding to the critical curve. This relation serves as the boundary separating regions associated with distinct photon dynamical behaviors in the $(\xi\,, \eta)$ parameter space. The radial location of the critical curve in the vicinity of the JP black hole is determined by the simultaneous conditions $R (r) = 0$ and $R'(r) = 0$, where the prime denotes differentiation with respect to the radial coordinate $r$. Solving these equations simultaneously yields the constraint relation between the photon impact parameters $\xi$ and $\eta$ corresponding to critical photon trajectories asymptotically approaching unstable spherical photon orbits. This constraint relation can be represented as a boundary curve in the parameter space spanned by $(\xi\,, \eta)$ and is illustrated in Fig. \ref{fig25} by the black solid line. As shown in Fig. \ref{fig25}, the photon geodesics associated with the ranges of impact parameters represented by the light-blue region enclosed by the black solid curve are characterized by radial motion that allows the corresponding trajectories to enter the capture region bounded by the critical curve. In contrast, the ranges of impact parameters represented by the green region outside the black solid curve are associated with photon trajectories whose radial motion remains entirely within the escape region exterior to the critical curve throughout their propagation. Moreover, photons associated with the light blue region belong to the class of trajectories that ultimately evolve toward capture by the black hole, whereas those corresponding to the green region belong to the class of trajectories that escape to spatial infinity. Within the backward ray-tracing framework, and in conjunction with the classification of regions in the $(\xi\,, \eta)$ parameter space determined by the constraint relation derived from the angular potential equation $\Theta (\theta) = 0$, photons associated with the light blue region are traced backward from the observer at infinity, intersect the accretion disk in the vicinity of the JP black hole, and are eventually captured by the black hole. Under the corresponding forward-time physical interpretation, the time-reversed counterparts of these trajectories would emerge from the vicinity of the event horizon and propagate outward toward the observer. However, these trajectories are excluded by the causal structure of the black hole spacetime. Consequently, photons associated with the light blue region do not contribute to the observable intensity distribution in the JP black hole image. Instead, these photons manifest as a dark region on the image plane, corresponding to the original inner shadow of the JP black hole. Therefore, the light blue region in the $(\xi\,, \eta)$ parameter space maps onto a dark domain on the image plane that constitutes the original inner shadow in the resulting image. In contrast, photons associated with the green region intersect the accretion disk and subsequently propagate to spatial infinity. Physically, these photons can be interpreted as being emitted from the accretion disk, undergoing geodesic motion in the vicinity of the JP black hole, and ultimately reaching the distant observer, thereby contributing to the observed intensity distribution on the image plane.

In addition, as shown in Fig. \ref{fig25}, the constraint curve associated with the critical curve extends into the dark blue region enclosed by the boundary derived from the constraint relation between the impact parameters $\xi$ and $\eta$ obtained from the roots of the angular potential equation $\Theta (\theta) = 0$. As illustrated in the enlarged panel on the right-hand side of Fig. \ref{fig25}, this constraint curve further subdivides the dark blue region in the $(\xi,\eta)$ parameter space into two distinct subregions located on either side of the boundary. As established in the preceding analysis of the angular potential, photons characterized by impact parameters within the dark blue region do not cross the equatorial plane and therefore do not interact with the accretion disk. For the left subregion of the dark blue domain, which lies inside the boundary defined by the constraint curve corresponding to the critical curve, the associated photon trajectories belong to the capture class and enter the capture region bounded by the critical curve. Consequently, photons corresponding to this range of impact parameters are ultimately captured by the JP black hole. Within the backward ray-tracing framework, these trajectories correspond to photons associated with the dark blue region that are traced backward from the image plane of a distant observer, which never intersect the accretion disk and eventually terminate at the event horizon of the black hole. Under the corresponding forward time interpretation, the time reversed counterparts of these photon trajectories would emerge from the vicinity of the event horizon and propagate toward the observer, which is excluded by the causal structure of the black hole spacetime. Consequently, photons associated with this subregion do not contribute to the observable intensity distribution and appear as a dark region on the image plane. Since the physical mechanism responsible for the capture of photon trajectories associated with the left dark blue subregion is identical to that governing the photon trajectories represented by the light blue region, photons in both regions are ultimately captured by the JP black hole because these trajectories enter the capture region bounded by the critical curve. Consequently, the photon trajectories associated with both the light blue region and the left dark blue subregion terminate at the black hole, and the regions on the image plane associated with both the light blue region and the left dark blue subregion jointly constitute the original inner shadow. By contrast, the right subregion of the dark blue domain lies outside the constraint curve corresponding to the critical curve. From the backward ray-tracing perspective, photon trajectories associated with this subregion originate from the distant observer, propagate through the vicinity of the JP black hole, and subsequently escape to spatial infinity along directions different from the line of sight of the distant observer, without ever intersecting the equatorial accretion disk. Within the adopted imaging framework, in which the observed image is assumed to be generated exclusively by photons emitted from and interacting with the thin accretion disk, photons corresponding to this class of trajectories do not contribute to the observable intensity distribution. Although the photons associated with the left and right dark blue subregions both produce dark features in the resulting JP black hole image, the physical mechanisms underlying the formation of these dark regions are fundamentally distinct. The dark region produced by the light blue region and the left dark blue subregion originates from photon trajectories that belong to the capture class and are ultimately absorbed by the black hole. In contrast, the photons associated with the right dark blue subregion neither intersect the accretion disk nor terminate at the event horizon. Instead, after propagating through the spacetime surrounding the JP black hole, these photons ultimately escape to spatial infinity. Consequently, the photon trajectories corresponding to this subregion also generate a dark region in the resulting JP black hole image. Since the shadow region generated by the impact parameters enclosed within the boundary of the critical curve corresponds to the original inner shadow of the JP black hole image, the dark region formed by photons associated with the impact parameters belonging to the right dark blue subregion is naturally identified as the extended inner shadow. Furthermore, the propagation properties of the photons associated with this region are fully consistent with those of the photon trajectory reconstructed through numerical backward ray-tracing simulations and shown in Fig. \ref{fig24} (d), which gives rise to the extended inner shadow. This agreement demonstrates that the approximate analytical treatment successfully captures the fundamental physical mechanism responsible for the formation of the extended inner shadow. Moreover, the close correspondence between the analytical prediction and the numerical results provides strong cross validation of the proposed interpretation of the origin of the extended inner shadow.

\section{Conclusions}\label{conlcusions}

The JP metric is constructed by introducing a set of deviation parameters into the Kerr metric. The motivation for formulating the JP metric is to establish a phenomenological spacetime geometry that extends the Kerr solution of classical GR through the inclusion of deviation parameters. Since these parameters are introduced directly into the Kerr geometry, the resulting JP metric does not, in general, represent an exact solution of the vacuum Einstein field equations. Nevertheless, the spacetime described by the JP metric preserves the fundamental stationarity and axisymmetry of the Kerr spacetime and is specifically designed to provide a framework for testing possible deviations from the Kerr geometry in the strong-field regime. The introduction of deviation parameters not only provides additional degrees of freedom in the spacetime geometry near the black hole, but also breaks the strict relation between the multipole moments and the mass and spin that characterizes Kerr black holes under the classical no-hair theorem. Consequently, the additional freedom introduced by these parameters enables tests of whether astrophysical black holes are fully described by the Kerr geometry. At the same time, the modification of the multipole structure provides a framework for examining the validity of the no-hair theorem in astrophysical contexts and for exploring whether astrophysical black holes may contain physical information beyond mass and spin. In the JP metric, the deviation parameters are denoted by $\epsilon_i$ with $i = 1\,, 2\,, 3\,, \cdots$. The first two parameters, $\epsilon_1$ and $\epsilon_2$, are strongly constrained by weak-field experimental tests and can be absorbed through a redefinition of the mass and spin of the JP black hole. For this reason, the influence of these two parameters on the spacetime structure is usually neglected. By contrast, the deviation parameter $\epsilon_3$ represents the leading physically relevant deviation from the Kerr geometry. Moreover, since $\epsilon_3$ enters the metric through terms proportional to $(M/r)^3$, its influence rapidly diminishes in the asymptotically distant region, ensuring consistency with existing weak-field observational constraints. However, in the strong-field region near the black hole, the effects of $\epsilon_3$ become substantially amplified. This property of the deviation parameter $\epsilon_3$ makes the JP metric a particularly useful framework for probing the spacetime geometry of astrophysical black holes and for testing the validity of the no-hair theorem. Therefore, the primary objective of this study is to investigate the extent to which the deviation parameter $\epsilon_3$ characterizes departures of the JP spacetime from the Kerr geometry and to examine how variations in this parameter modify the spacetime structure described by the JP metric. Variations in $\epsilon_3$ alter the spacetime geometry and consequently modify the gravitational field surrounding the black hole. As a result, the dynamical behavior of accretion flows in the surrounding accretion disk is expected to depend on the value of $\epsilon_3$. Since these dynamical properties are encoded in the observable appearance of JP black holes illuminated by surrounding accretion disks, the present study investigates how variations in the deviation parameter $\epsilon_3$ influence the observable images of JP black holes surrounded by thin accretion disks.

The introduction of the deviation parameter $\epsilon_3$ into the JP metric alters the Petrov classification of the spacetime from type D, characteristic of the Kerr spacetime, to type I. As a consequence of this change in the Petrov classification, a Carter constant can no longer be defined in the spacetime described by the JP metric. This implies that the Hamilton-Jacobi equation governing photon motion in the JP spacetime is generally non-separable, thereby preventing the construction of JP black hole images through a fully analytical approach. However, the backward ray-tracing method provides a rigorous numerical framework for accurately constructing images of JP black holes surrounded by thin accretion disks. Accordingly, in the present study, images of JP black holes illuminated by thin accretion disks are generated entirely through backward ray-tracing numerical simulations. This approach enables a systematic investigation of the image structures and observational characteristics of JP black hole images. Moreover, by analyzing the dependence of the resulting JP black hole images on the deviation parameter $\epsilon_3$, the evolution of the imaging features of JP black holes as functions of the deviation parameter $\epsilon_3$ can be examined. Furthermore, analyses of the spacetime geometry described by the JP metric show that the topology of the JP black hole event horizon can exhibit two qualitatively distinct configurations depending on the value of the deviation parameter $\epsilon_3$, namely closed and non-closed topologies. Accordingly, in the analysis of the image structures of JP black holes surrounded by thin accretion disks, the configurations corresponding to closed and non-closed event horizons are considered separately. For these two distinct scenarios, the imaging properties of JP black holes and the corresponding evolution of these imaging characteristics with varying deviation parameter $\epsilon_3$ are systematically investigated.

The image structures and associated characteristics of JP black holes with closed event horizons surrounded by thin accretion disks are first systematically investigated, together with the dependence of the imaging features on the deviation parameter $\epsilon_3$. Based on fully numerical backward ray-tracing simulations, images of JP black holes with closed event horizons illuminated by surrounding accretion disks are constructed. Analysis of the resulting images shows that, as the deviation parameter $|\epsilon_3|$ increases, both the region occupied by the inner shadow and the area enclosed by the photon ring gradually expand. Building upon this qualitative analysis of the image morphology, a quantitative investigation of the evolution of the inner shadow and photon ring with increasing $|\epsilon_3|$ is carried out by introducing the average radius on the image plane. Specifically, the dependence of the average radii of the inner shadow and the photon ring on the deviation parameter $\epsilon_3$ is evaluated for JP black holes with spin parameters $a = 0.7$ and $a = 0.9$, observed at inclination angles $\theta_0 = 17^\circ$ and $\theta_0 = 80^\circ$. These results are compared with the corresponding average radii of the inner shadow and photon ring in the image of a Kerr black hole with spin parameter $a = 0.9$ observed at the same inclination angles. The comparison shows that, for both $\theta_0 = 17^\circ$ and $\theta_0 = 80^\circ$, the average radii of the inner shadow and the photon ring in JP black hole images increase approximately linearly with the absolute value of the deviation parameter $|\epsilon_3|$. Furthermore, when $|\epsilon_3| > 4$, the relative deviation curves of the average radius of the inner shadow, measured with respect to the corresponding average radius in the image of a Kerr black hole with spin parameter $a = 0.9$, become nearly indistinguishable for JP black holes with spin parameters $a = 0.7$ and $a = 0.9$ at both observer inclination angles. This result indicates that, when the magnitude of the deviation parameter $|\epsilon_3|$ becomes sufficiently large, the evolution of the average radius of the inner shadow becomes increasingly insensitive to the spin parameter of the JP black hole. Nevertheless, a residual dependence on the observer inclination angle persists, as reflected by the remaining differences between the corresponding evolutionary curves obtained for $\theta_0=17^\circ$ and $\theta_0=80^\circ$. In contrast, the evolution of the average radius of the photon ring exhibits an even weaker dependence on both the spin parameter and the observer inclination angle. Specifically, for $|\epsilon_3| > 4$, the four evolutionary curves corresponding to the relative deviations of the average radius of the photon ring for JP black holes with spin parameters $a = 0.7$ and $a = 0.9$, observed at inclination angles $\theta_0 = 17^\circ$ and $\theta_0 = 80^\circ$, almost completely overlap. This behavior indicates that, for sufficiently large values of $|\epsilon_3|$, the evolution of the average radius of the photon ring is governed predominantly by the deviation parameter itself and becomes effectively independent of both the spin parameter and the observer inclination angle.

Building upon the analysis of the dependence of the inner shadow and the photon ring on the deviation parameter $|\epsilon_3|$ in JP black hole images with closed event horizons, the intensity distributions in these images and the evolutionary behavior of the intensity distributions with increasing $|\epsilon_3|$ are further investigated. Since the accretion disk surrounding a JP black hole may contain either prograde or retrograde accretion flows, and the dynamical properties of the accretion flow can significantly influence the observed intensity distribution, the effects of prograde and retrograde accretion flows are examined separately. For JP black holes surrounded by a prograde accretion disk, the intensity distributions on the image plane exhibit several characteristic features. For spin parameter $a = 0.7$, the intensity peaks generally decrease with increasing $|\epsilon_3|$ for both observer inclination angles $\theta_0 = 17^\circ$ and $\theta_0 = 80^\circ$. The only notable exception occurs in the intensity profile along the $x-$ axis for $\theta_0 = 80^\circ$, where the peak located in the region $x > 0$ exhibits a slight increase as $|\epsilon_3|$ increases. In addition, when the observer inclination angle is fixed at $\theta_0 = 80^\circ$, an additional intensity peak emerges near the center of the $y-$ axis. The magnitude of this peak decreases monotonically with increasing $|\epsilon_3|$, while its position gradually shifts toward the negative $y-$ direction. Meanwhile, the positions of the intensity peaks evolve systematically with increasing $|\epsilon_3|$, reflecting the corresponding changes in the image morphology induced by the deviation parameter. For JP black holes with spin parameter $a = 0.9$, the evolution of the intensity distributions exhibits a more complex dependence on $|\epsilon_3|$. Along the $x-$ axis, the intensity peaks generally increase with increasing $|\epsilon_3|$ when the observer inclination angle is $\theta_0 = 17^\circ$, although the left hand peak displays a non-monotonic behavior, increasing initially and subsequently decreasing as $|\epsilon_3|$ increases. For $\theta_0 = 80^\circ$, the evolution of the two intensity peaks along the $x-$ axis becomes markedly asymmetric. Specifically, the peak located in the region $x < 0$ decreases with increasing $|\epsilon_3|$, whereas the peak located in the region $x > 0$ exhibits the opposite behavior, increasing monotonically as $|\epsilon_3|$ increases. By contrast, the intensity peaks along the $y$-axis generally increase with increasing $|\epsilon_3|$ for both observer inclination angles. A particularly distinctive feature appears for $\theta_0 = 80^\circ$, where an additional intensity peak emerges near the center of the $y-$ axis. Unlike the corresponding central peak in the $a = 0.7$ case, the intensity of this additional peak first increases and subsequently decreases as $|\epsilon_3|$ increases, giving rise to an anomalous evolutionary behavior in the intensity distribution of rapidly rotating JP black holes viewed at high inclination angles. Furthermore, for a fixed observer inclination angle, the positions of the intensity peaks in the images of JP black holes with spin parameter $a = 0.7$ gradually converge toward those in the corresponding images with spin parameter $a = 0.9$ as $|\epsilon_3|$ increases. This behavior is consistent with the previously derived evolution of the average radii of the inner shadow and the photon ring. In particular, when $|\epsilon_3|$ becomes sufficiently large, the average radii of the inner shadow and the photon ring for JP black holes with $a = 0.7$ and $a = 0.9$ evolve in an almost identical manner for a given observer inclination angle. The convergence of the intensity peak positions therefore provides additional support for the conclusion that, at sufficiently large values of $|\epsilon_3|$, the growth rates of the average radii of the inner shadow and the photon ring become nearly independent of the spin parameter of the JP black hole. For the case of retrograde accretion flows, the overall intensity distributions exhibit qualitative features similar to those obtained in the prograde case. The principal difference is that, for both spin parameters $a = 0.7$ and $a = 0.9$, and for both observer inclination angles $\theta_0 = 17^\circ$ and $\theta_0 = 80^\circ$, all intensity peaks increase monotonically with increasing $|\epsilon_3|$. In particular, the anomalous behavior observed in the prograde case for the additional central peak in the $y-$ axis intensity distribution of JP black holes with $a = 0.9$ at $\theta_0 = 80^\circ$ is absent. Instead, this central peak also increases monotonically with increasing $|\epsilon_3|$, consistent with the general trend exhibited by the other intensity peaks in the image.

Since the boundaries of the inner shadow and the photon ring along the $x-$ and $y-$ directions can be clearly identified from the resulting intensity profiles of the JP black hole images, the quantities $d_x$ and $d_y$ are defined as the characteristic boundary scales of the inner shadow along the $x-$ and $y-$ axes, respectively, while $D_x$ and $D_y$ denote the corresponding boundary scales of the photon ring. In addition, $d_{\text{Kerr}}$ and $D_{\text{Kerr}}$ are introduced to represent the boundary scales of the inner shadow and the photon ring, respectively, as determined from the intensity distributions along the $x-$ and $y-$ axes of the image plane for a Kerr black hole with spin parameter $a = 0.9$. Based on these definitions, the relative deviations of the inner shadow and photon ring in JP black hole images with respect to their counterparts in Kerr black hole images are defined as $\delta_d=d/d_{\text{Kerr}}$ and $\delta_D=D/D_{\text{Kerr}}$, respectively. An examination of the evolution of these relative deviations with increasing deviation parameter $|\epsilon_3|$ reveals that, for JP black holes with spin parameters $a = 0.7$ and $a = 0.9$, illuminated by either prograde or retrograde accretion disks and observed at inclination angles $\theta_0=17^\circ$ and $\theta_0=80^\circ$, the relative deviations of the inner shadow and photon ring measured along the $y-$ direction exhibit evolutionary trends that closely follow those of the corresponding average radii as $|\epsilon_3|$ increases. In contrast, the relative deviations of the inner shadow and photon ring measured along the $x-$ direction exhibit evolutionary behaviors that differ markedly from those of the corresponding average radii as the deviation parameter $|\epsilon_3|$ increases. This result suggests that the evolution of the inner shadow and photon ring with increasing $|\epsilon_3|$ is determined primarily by their variation along the $y-$ direction of the image plane. Since the deviation parameter $\epsilon_3$ enters the JP metric solely through the function $h(r\,, \theta)$, and the departure of the JP metric from the Kerr metric is entirely determined by this function, the distinct evolutionary behaviors of the inner shadow and photon ring along the $x-$ and $y-$ directions can be attributed to the different properties of $h(r\,, \theta)$ in different regions of the JP spacetime. Within the BL coordinate system employed to describe the JP spacetime, the influence of the function $h(r\,, \theta)$ on the spacetime geometry is relatively weak in the vicinity of the rotational axis, whereas it becomes considerably stronger near the equatorial plane. Furthermore, owing to the nonzero angular momentum of the JP black hole, the frame-dragging effect generated by its rotation also reaches its maximum strength in the vicinity of the equatorial plane. Since both the influence of the function $h(r\,, \theta)$ and the frame-dragging effect are strongest in this region, the deformation induced by the deviation parameter and the rotational frame-dragging effect become strongly coupled near the equatorial plane. As a consequence, the motion of photons propagating in this region is jointly governed by these two effects. Accordingly, photons propagating in regions closer to the rotational axis are only weakly influenced by the deviation parameter, and geodesics of photons therefore remain close to the corresponding photon geodesics in the Kerr spacetime. By contrast, in regions closer to the equatorial plane, the combined effects of the deviation parameter and rotational frame dragging lead to substantially larger modifications of photon motion. Consequently, the photon geodesics in these regions depart significantly from those in the corresponding Kerr spacetime. On the image plane of the resulting JP black hole images, the structures measured along the $x-$ direction are more strongly influenced by photon trajectories probing the spacetime near the equatorial plane, whereas those measured along the $y-$ direction are more strongly influenced by photon trajectories probing regions closer to the rotational axis. Since the influence of frame dragging becomes significantly weaker toward the rotational axis, the spacetime geometry in this region is affected primarily by variations in the deviation parameter $\epsilon_3$. Consequently, the inner shadow and photon ring measured along the $y-$ direction undergo an approximately uniform rescaling with increasing $|\epsilon_3|$, while largely preserving the characteristic morphology of their Kerr counterparts. As a result, the evolution of the boundary scales of the inner shadow and photon ring along the $y-$ direction closely follows that of their corresponding average radii. In contrast, near the equatorial plane, the spacetime is simultaneously affected by the deformation introduced by the deviation parameter $\epsilon_3$ and the frame-dragging effect generated by the rotating JP black hole. The coupling between these two effects produces a distinct evolutionary behavior of the inner shadow and photon ring along the $x-$ direction, causing their dependence on $|\epsilon_3|$ to differ significantly from the corresponding evolution of average radii of the inner shadow and photon ring. Furthermore, the characteristic scales of the inner shadow and photon ring defined along the $x-$ and $y-$ directions of the image plane can be used to introduce the ellipticities of the inner shadow and photon ring, defined as $d_y/d_x$ and $D_y/D_x$, respectively. These quantities enable a quantitative analysis of the evolution of the morphologies of the inner shadow and photon ring with increasing deviation parameter $|\epsilon_3|$. The results indicate that, for an observer inclination angle of $\theta_0 = 17^\circ$, the inner shadows of JP black holes with spin parameters $a = 0.7$ and $a = 0.9$ remain nearly circular over the entire range of $|\epsilon_3|$ considered. In contrast, for $\theta_0 = 80^\circ$, the inner shadow assumes an elliptical shape with its major axis oriented along the $x-$ direction for both spin parameters. Moreover, the ellipticity of the inner shadow increases progressively with increasing $|\epsilon_3|$, indicating a gradual enhancement of its elongation along the $x-$ axis. The evolution of the inner shadow ellipticity further shows that the morphological evolution of the inner shadow with increasing $|\epsilon_3|$ is largely insensitive to the dynamical state of the accretion flow, namely whether the accretion disk is prograde or retrograde. By contrast, the ellipticity of the photon ring exhibits no clear systematic dependence on $|\epsilon_3|$. Moreover, the evolutionary trends of the photon ring ellipticity with increasing $|\epsilon_3|$ differ substantially between JP black hole images illuminated by prograde and retrograde accretion disks. This behavior indicates that the morphology of the photon ring is more sensitive to the dynamical state of the accretion flow than the morphology of the inner shadow.

Furthermore, since the photons forming JP black hole images interact with the accretion flow in the surrounding accretion disk before reaching a distant observer, these interactions produce observable redshift and blueshift effects in the resulting images. Consequently, a systematic examination of the redshift and blueshift characteristics of photons in JP black hole images is required to obtain a more complete characterization of the imaging properties of JP black holes. Moreover, the redshift and blueshift patterns formed on the image plane are directly determined by the dynamical state of the accretion flow in the surrounding accretion disk. Because the redshift and blueshift effects differ depending on whether photons interact with prograde or retrograde accretion flows, the redshift and blueshift properties of JP black hole images should be investigated separately for these two cases. In addition, the spatial configurations of the redshifted and blueshifted regions differ between the direct and lensed images. Accordingly, the spatial distributions of the redshifted and blueshifted regions in both the direct and lensed images are examined separately, and the dependence of these configurations on the deviation parameter $\epsilon_3$ is investigated. When the accretion flow in the accretion disk is in a prograde dynamical state, the redshifted regions in both the direct and lensed images expand outward in the corresponding redshift-blueshift maps as the deviation parameter $|\epsilon_3|$ increases. This behavior is closely associated with the progressive enlargement of the inner shadow with increasing $|\epsilon_3|$. During this process, the redshift intensity near the boundary of the inner shadow gradually decreases. This behavior occurs because, as the deviation parameter $|\epsilon_3|$ increases, both the inner shadow boundary and the infinite redshift surface of the JP black hole move outward, while the expansion rate of the inner shadow exceeds that of the infinite redshift surface. As a result, regions that previously exhibited strong redshift near the inner shadow boundary move farther away from the infinite redshift surface, leading to a reduction in the observed redshift intensity in the redshift-blueshift maps. In contrast, the blueshift region in the direct image changes only slightly with increasing $|\epsilon_3|$, undergoing primarily a global displacement toward the left side of the image as the inner shadow expands. The behavior of the blueshift region in the lensed image differs from that of the direct image. The blueshift region in the lensed image gradually extends along the boundary of the inner shadow in a counterclockwise direction as $|\epsilon_3|$ increases, eventually forming a narrow elongated structure located close to the boundary of the inner shadow. Since the location of the maximum blueshift can be identified in both the direct and lensed images, the evolution of this position with increasing $|\epsilon_3|$ can be further investigated. To characterize the location of the maximum blueshift in the redshift-blueshift maps, the radial position of the ISCO around the JP black hole is adopted as a reference. On this basis, the dependence of the maximum blueshift position relative to the ISCO on the deviation parameter $|\epsilon_3|$ can be systematically analyzed. The results show that, when $\epsilon_3 = 0$, the maximum blueshift in the direct image is located outside the ISCO radius, whereas the maximum blueshift in the lensed image lies inside the ISCO radius. As $|\epsilon_3|$ increases, the maximum blueshift positions in both images move inward toward the ISCO. In the range $|\epsilon_3| \in [0\,, 2]$, the maximum blueshift locations shift rapidly toward the interior of the ISCO. When $|\epsilon_3| > 2$, the maximum blueshift in the direct image moves inward approximately linearly with increasing $|\epsilon_3|$, whereas the maximum blueshift in the lensed image gradually approaches a nearly fixed position inside the ISCO. This behavior indicates that, with increasing $|\epsilon_3|$, the maximum blueshift position in the direct image progressively approaches that in the lensed image. When the accretion flow in the accretion disk is in a retrograde dynamical state, the evolution of the redshifted regions in the redshift-blueshift maps of both the direct and lensed images with increasing $|\epsilon_3|$ exhibits behavior similar to that observed in the prograde case. Specifically, the redshift regions expand outward as the inner shadow enlarges with increasing $|\epsilon_3|$, while the redshift intensity near the inner shadow boundary decreases during this expansion. However, the evolution of the blueshift regions with increasing $|\epsilon_3|$ differs from that observed in the prograde case. In the presence of retrograde accretion flow, the spatial configurations of the blueshift regions in both the direct and lensed images remain nearly unchanged as the deviation parameter $|\epsilon_3|$ increases, and the blueshift regions simply shift toward the right side of the image as the inner shadow expands. Furthermore, the evolution of the maximum blueshift positions relative to the ISCO radius in the redshift-blueshift maps of both the direct and lensed images is also examined for the retrograde case. The results show that the locations of the maximum blueshift in both the direct and lensed images differ from those obtained in the prograde scenario. When $\epsilon_3 = 0$, the maximum blueshift positions in both the direct and lensed images are located inside the ISCO. As the deviation parameter $|\epsilon_3|$ increases, the distances between the maximum blueshift positions and the ISCO increase approximately linearly with nearly identical slopes for both images. This behavior indicates that the maximum-blueshift positions in the direct and lensed images move progressively deeper inside the ISCO at nearly the same rate as the deviation parameter $|\epsilon_3|$ increases.

The value of the deviation parameter $\epsilon_3$ within different parameter ranges can lead to qualitatively distinct spacetime structures in the JP metric. As a consequence, the topology of the event horizon of the JP black hole may exhibit two distinct configurations, namely closed and non-closed topologies. Based on the analysis of the imaging structure and properties of JP black holes with closed event horizons, it is therefore necessary to further investigate the imaging structure and characteristics of JP black holes possessing non-closed event horizons. For JP black hole images with non-closed event horizons observed at an inclination angle $\theta_0 = 17^\circ$, the region covered by the inner shadow and the area enclosed by the photon ring for a JP black hole with spin parameter $a = 0.7$ decrease progressively as the deviation parameter $\epsilon_3$ increases. This behavior is opposite to that observed for JP black holes with closed event horizons, where both the inner shadow and the photon ring expand with increasing $|\epsilon_3|$. In addition, as the deviation parameter $\epsilon_3$ increases, the shape of the photon ring gradually transforms from a nearly circular structure into an elliptical configuration whose major axis is aligned with the $y-$ axis of the image plane. For JP black holes with spin parameter $a = 0.9$, the evolution of the inner shadow and photon ring follows a trend similar to that found for $a = 0.7$. Specifically, both the area of the inner shadow and the region enclosed by the photon ring decrease as $\epsilon_3$ increases, while the photon ring assumes an elliptical configuration with its major axis oriented along the $y$-axis. However, unlike the case with $a = 0.7$, the left portion of the photon ring in the image of a JP black hole with the spin parameter $a = 0.9$ gradually shifts toward the right as $\epsilon_3$ increases and eventually intersects the inner shadow, resulting in the apparent truncation of the photon ring by the inner shadow. When the observer inclination angle is fixed at $\theta_0 = 80^\circ$, the evolution of the inner shadow and the photon ring exhibits similar characteristics to those observed at $\theta_0 = 17^\circ$. Specifically, for both spin parameters $a = 0.7$ and $a = 0.9$, the region covered by the inner shadow and the area enclosed by the photon ring decrease as the deviation parameter $\epsilon_3$ increases. In addition, the photon ring consistently exhibits an elliptical morphology with its major axis aligned along the $y$-axis, while the left portion of the photon ring remains truncated by the inner shadow. Furthermore, in the JP black hole image obtained for spin parameter $a = 0.7$ at an inclination angle of $\theta_0 = 80^\circ$, an additional narrow direct image appears in the lower-left region within the inner shadow. This narrow direct image originates from the left boundary of the inner shadow and intersects the lower boundary of the inner shadow, thereby dividing the inner shadow into two distinct components, referred to as the original inner shadow and the extended inner shadow. As the spin parameter increases from $a = 0.7$ to $a = 0.9$, the direct image that separates the JP black hole inner shadow gradually shifts toward the interior of the inner shadow with increasing $\epsilon_3$, while its width simultaneously increases progressively. This evolution of the direct image with increasing deviation parameter $\epsilon_3$ results in an increasingly pronounced separation between the original inner shadow and the extended inner shadow of the JP black hole. When the deviation parameter reaches $\epsilon_3 = 8$, the original inner shadow and the extended inner shadow remain connected only through the intersection point between the separating direct image and the lower boundary of the inner shadow. In addition, the JP black hole image observed at $\theta_0 = 80^\circ$ shows that the lensed image located in the lower region of the image plane can also intersect the lower boundary of the inner shadow through a narrow elongated structure extending from the lensed image region. The intersection point between this narrow lensed structure and the inner shadow approximately coincides with the intersection point at which the separating direct image meets the lower boundary of the inner shadow.

Analogous to the analysis performed for JP black hole images with closed event horizons, the intensity distributions of JP black hole images with non-closed event horizons are further investigated on the basis of the previously discussed image structures. From the intensity distributions obtained for observer inclination angles $\theta_0 = 17^\circ$ and $\theta_0 = 80^\circ$, and for spin parameters $a = 0.7$ and $a = 0.9$, it can be observed that, as the deviation parameter $\epsilon_3$ increases, the peak intensities in the images generally exhibits an increasing trend, while the locations of the intensity peaks gradually shift toward the center of the $x-$ and $y-$ axes in the Cartesian coordinate system of the image plane. This behavior primarily arises because the region covered by the inner shadow and the area enclosed by the photon ring decrease as the deviation parameter $\epsilon_3$ increases. Moreover, in the intensity distributions of JP black hole images with non-closed event horizons, an additional intensity peak appears in the region between the photon ring and the inner shadow. This additional peak is absent from the intensity distributions of JP black hole images with closed event horizons. The presence of this peak indicates that photons associated with this region can still interact with the accretion flow in the surrounding accretion disk and subsequently propagate to a distant observer. This situation differs from that of JP black holes with closed event horizons, for which photons associated with the region between the photon ring and the event horizon are ultimately captured by the black hole and therefore cannot reach a distant observer. Consequently, the appearance of this additional intensity peak between the photon ring and the inner shadow constitutes a distinctive feature of JP black hole images with non-closed event horizons. Furthermore, the intensity distribution along the $x-$ axis for a JP black hole with spin parameter $a = 0.9$ observed at an inclination angle $\theta_0 = 17^\circ$ shows that, as the deviation parameter $\epsilon_3$ increases, the intensity peak located on the left side of the $x-$ axis gradually shifts toward $x = -0.01$. When the deviation parameter reaches $\epsilon_3 = 8$, the intensity peak becomes almost entirely concentrated near this position. This behavior indicates that, although the photon ring appears to be partially truncated by the inner shadow in the image of the JP black hole with the spin parameter $a = 0.9$ observed at the observer inclination angle of $\theta_0 = 17^\circ$, the photon ring surrounding the inner shadow nevertheless retains a continuous and topologically connected structure. In this case, the photon ring becomes concentrated near the boundary between the direct image and the lensed image on the left side of the inner shadow. In addition, the intensity distributions along the $x-$ axis for JP black hole images with spin parameters $a = 0.7$ and $a = 0.9$ observed at an inclination angle of $\theta_0 = 80^\circ$ reveal the presence of a relatively weak intensity peak on the left side of the $x-$ axis. This peak corresponds to the intensity contribution from the narrow direct image that separates the inner shadow of the JP black hole. Therefore, the intensity distributions provide further confirmation that, for an observer inclination angle $\theta_0 = 80^\circ$, an additional direct image indeed appears on the left side of the inner shadow. This direct image divides the inner shadow into two distinct regions, referred to as the original inner shadow and the extended inner shadow.

Furthermore, the redshift and blueshift effects appearing in JP black hole images with non-closed event horizons, produced when photons interact with the retrograde accretion flow in the surrounding accretion disk, should be investigated in greater detail. Since the resulting JP black hole images consist of both direct and lensed components that exhibit distinct redshift and blueshift signatures, the corresponding properties of the direct and lensed images should be examined separately. In the redshift-blueshift maps of the direct image, the maximum redshift remains concentrated near the boundary of the inner shadow, while the redshift intensity gradually decreases with increasing distance from this boundary. This behavior indicates that the spacetime still contains an infinite redshift surface even when the event horizon of the JP black hole is non-closed. The presence of this characteristic surface causes the strongest redshift to be concentrated near the inner shadow boundary. As the distance from this boundary increases, the separation from the infinite redshift surface also increases, resulting in a gradual reduction of the redshift intensity. Since the region covered by the inner shadow progressively decreases as the deviation parameter $\epsilon_3$ increases, the redshift region in the redshift-blueshift maps shifts toward the central region of the image plane with increasing $\epsilon_3$. In contrast, the blueshift region in the direct image is located on the right-hand side of the image, and its overall configuration remains nearly unchanged as $\epsilon_3$ increases. Because the area covered by the inner shadow decreases with increasing $\epsilon_3$, the blueshifted region undergoes a global displacement toward the right-hand side of the image. For JP black holes with spin parameter $a = 0.7$, the narrow direct image that separates the original inner shadow from the extended inner shadow gradually shifts inward as $\epsilon_3$ increases. As a result, the extended inner shadow occupies an increasingly larger region and progressively separates from the original inner shadow. During this separation process, the redshift intensity along the boundary of the extended inner shadow gradually decreases. This phenomenon arises because the extended inner shadow moves farther away from the infinite redshift surface as it separates from the original inner shadow, thereby weakening the redshift effect in the vicinity of the extended inner shadow. When the spin parameter increases from $a = 0.7$ to $a = 0.9$, the direct image that separates the original inner shadow and the extended inner shadow gradually broadens, causing the extended inner shadow to become increasingly distinct from the original inner shadow. As the deviation parameter increases, the redshift intensity around the extended inner shadow decreases further. When $\epsilon_3 = 8$, a localized blueshifted feature appears along the right boundary of the original inner shadow in the redshift-blueshift maps of the direct image. This result indicates that, for the case $a = 0.9$, the extended inner shadow moves farther away from the infinite redshift surface as $\epsilon_3$ increases, thereby weakening the redshift effect near the extended inner shadow. At the same time, the emergence of blueshift within the separating direct image partially compensates for the surrounding redshift contribution. Consequently, the combined influence of the increasing distance from the infinite redshift surface and the emergence of blueshift leads to a further reduction in the redshift intensity near the extended inner shadow. The redshift-blueshift maps of the direct image further show that, regardless of variations in the spin parameter or the value of the deviation parameter, the location of the maximum blueshift always occurs at the intersection between the direct image that separates the original inner shadow and the extended inner shadow and the lower boundary of the inner shadow. Based on this result, the redshift-blueshift properties of the lensed image for JP black holes with non-closed event horizons can also be examined. For JP black holes with the spin parameter $a = 0.7$, the redshift effect in the lensed image near the extended inner shadow similarly decreases as $\epsilon_3$ increases. This behavior is primarily attributed to the increasing distance between the extended inner shadow and the infinite redshift surface. As the spin parameter increases from $a = 0.7$ to $a = 0.9$, the redshift-blueshift maps of the lensed image show that the narrow lensed structure connecting the lensed image in the lower half of the image plane to the lower boundary of the inner shadow gradually broadens with increasing $\epsilon_3$. During this process, the redshift intensity near the extended inner shadow decreases further. When the deviation parameter reaches $\epsilon_3 = 8$, a localized blueshifted feature appears on the right-hand side of the narrow lensed structure adjacent to the lower boundary of the inner shadow. This result indicates that, for JP black holes with spin parameter $a = 0.9$, the decrease in the redshift intensity near the extended inner shadow in the lensed image results from the combined effects of the increasing distance from the infinite-redshift surface and the emergence of blueshift. In addition, the redshift-blueshift maps of the lensed image show that, for both spin parameters $a = 0.7$ and $a = 0.9$, the location of the maximum blueshift always occurs at the intersection between the narrow lensed structure and the lower boundary of the inner shadow as the deviation parameter $\epsilon_3$ increases. Since this intersection approximately coincides with the point at which the direct image that separates the original inner shadow from the extended inner shadow meets the lower boundary of the inner shadow, the maximum blueshift locations in the direct and lensed images correspond to essentially the same position in the JP black hole image.

Finally, the formation mechanism of the extended inner shadow in JP black hole images with non-closed event horizons is investigated using both the backward ray-tracing numerical method and an approximate analytical approach. First, from the perspective of numerical simulations, the backward ray-tracing method is employed to reconstruct photon trajectories in JP spacetime corresponding to the different image structures observed in the JP black hole images. The reconstructed photon trajectories associated with the extended inner shadow indicate that these photons originate from the extended inner shadow region on the image plane at infinity, propagate through JP spacetime toward the vicinity of the JP black hole, orbit around one segment of the dumbbell-shaped non-closed event horizon, and subsequently travel toward spatial infinity along trajectories directed away from the observer. During this propagation process, the photons do not intersect the accretion disk located on the equatorial plane and therefore do not interact with the accretion flow within the accretion disk. From the corresponding forward time physical interpretation, these photons originate from the asymptotic region of the JP spacetime, propagate toward the vicinity of the JP black hole, orbit around one segment of the dumbbell-shaped non-closed event horizon, and eventually escape to a distant observer without interacting with the accretion disk surrounding the JP black hole. In the imaging model adopted for JP black holes surrounded by thin accretion disks, it is assumed that all photons contributing to the observed JP black hole image are emitted from the accretion disk. Photons originating from the background spacetime therefore do not contribute to the construction of the JP black hole image. Moreover, only photons that intersect the accretion disk and interact with the accretion flow contribute to the intensity recorded on the image plane. However, the photon trajectories reconstructed through the backward ray-tracing method demonstrate that the photons corresponding to the extended inner shadow neither originate from the accretion disk nor intersect the accretion flow during their propagation. Consequently, these photons do not contribute to the observable intensity distribution in the JP black hole image and instead produce a dark region distinct from the original inner shadow. In contrast, photons corresponding to the original inner shadow follow trajectories that ultimately cross the event horizon of the JP black hole rather than escaping to spatial infinity. Therefore, this additional dark region, which is physically distinct from the original inner shadow, corresponds to the extended inner shadow in the JP black hole image. On this basis, the formation mechanism of the extended inner shadow is further investigated using an approximate analytical approach, and the resulting analytical predictions are used to verify and substantiate the physical interpretation obtained from the backward ray-tracing numerical simulations. Within the analytical framework, two parameters $\xi$ and $\eta$, referred to as the photon impact parameters, are constructed from the photon energy $E$, angular momentum $L$, and the Carter constant $K$. The geodesic motion of photons in the spacetime is uniquely characterized by these impact parameters. Accordingly, photon motion in JP spacetime can be analyzed by examining the radial potential $R (r)$ and the angular potential $\Theta (\theta)$ appearing in the photon geodesic equations, both of which are expressed in terms of the photon impact parameters. By analyzing the roots of the equations $R (r) = 0$ and $\Theta (\theta) = 0$, the constraint relations satisfied by the photon impact parameters can be obtained, thereby characterizing the allowed photon trajectories in JP spacetime. In addition to the radial potential $R (r)$ and the angular potential $\Theta (\theta)$, the photon equations of motion contain an additional function $\tilde{I} (r\,, \theta)$, which depends simultaneously on the radial coordinate $r$ and the polar coordinate $\theta$. The presence of this function constitutes the fundamental obstruction to the separability of the Hamilton-Jacobi equation governing photon motion in JP spacetime. To capture the physical behavior of photon motion in JP spacetime within an analytical framework and to obtain results that more closely reproduce the numerical results, the function $\tilde{I} (r\,, \theta)$ is treated as contributing to both the radial potential $R (r)$ and the angular potential $\Theta (\theta)$ rather than being assigned exclusively to either one. Accordingly, one half of the contribution of the function $\tilde{I} (r\,, \theta)$ is incorporated into the radial potential $R (r)$, while the remaining half is incorporated into the angular potential $\Theta (\theta)$. By analyzing the roots of the equations $\Theta (\theta) = 0$ and $R (r) = R^\prime (r) = 0$, the constraint relations between the photon impact parameters $\xi$ and $\eta$ can be derived. A parameter space spanned by the photon impact parameters $(\xi\,, \eta)$ can therefore be constructed. Within this parameter space, these constraint relations partition the space into several distinct regions, with each region corresponding to a different class of photon geodesics in JP spacetime. Within this partitioned parameter space, there exists a distinct region in which the photon geodesics determined by the corresponding impact parameters neither intersect the accretion disk on the equatorial plane nor originate from it, but instead are associated with photons traced back to the asymptotic region of the JP spacetime. Consequently, photons associated with this region of parameter space do not contribute to the observable intensity distribution of the JP black hole image. The null geodesics determined by the photon impact parameters corresponding to this region are classified as vortical geodesics. Photons propagating along vortical geodesics therefore generate a dark region in the JP black hole image that is physically distinct from the original inner shadow and corresponds to the extended inner shadow of the JP black hole. Therefore, the approximate analytical analysis yields the same physical interpretation for the formation mechanism of the extended inner shadow as that obtained from the backward ray-tracing numerical simulations. The consistency between the results derived from the numerical and analytical approaches provides mutual validation and demonstrates that the proposed mechanism responsible for the formation of the extended inner shadow in JP black hole images is physically well founded.

\section*{Acknowledgments}
This work is partially supported by the National Natural Science Foundation of China (NSFC) under Grant No. 12575048 and No. 12105015. X. -Y. Wang is also supported by the Guangdong Basic and Applied Basic Research Foundation under Grant No. 2023A1515012737 and the Talents Introduction Foundation of Beijing Normal University under Grant No. 111032109. M. Guo is also supported by the BNU Tang Scholar.

\bibliographystyle{utphys}

\bibliography{JP}
		
\end{document}